\documentclass[11pt, a4paper]{article}
\usepackage{amsmath, amssymb, amsthm}
\usepackage{graphicx}
\usepackage{booktabs}
\usepackage[margin=1in]{geometry}
\usepackage{hyperref}
\usepackage{bm}
\usepackage{empheq}
\usepackage{float}

\newcommand{\M}{\mathcal{M}}

\newcommand{\R}{\mathbb{R}}
\newcommand{\E}{\mathbb{E}}

\newcommand{\dd}{\mathrm{d}}
\newcommand{\affmark}[1]{\textsuperscript{#1}}

\title{\textbf{Frustrated Fields: Statistical Field Theory for \\
Frustrated Brownian Particles on 2D Manifolds}\thanks{All
calculations, numerical analysis, and manuscript
preparation were performed by Claude Code with Opus 4.6 working as an AI assistant under author's supervision. All remaining errors are my own. I would like to thank Charles Martin for valuable discussions. Python code for the simulations and figures used in this paper is available at \url{https://github.com/ighalp/frustrated-brownian-particles-manifolds}.}}

\author{Igor Halperin\affmark{*}}

\date{\today}

\begin{document}

\maketitle

\begin{center}
\affmark{*}Email: ighalp@gmail.com
\end{center}

\begin{abstract}
We develop a statistical field theory that describes the
large-$N$ limit of a system of Brownian particles with quenched
random pairwise interactions on a compact two-dimensional Riemannian manifold. The resulting Frustrated Fields (F2) model
is a non-linear field theory for a smooth self-interacting
density field $\rho$ on the manifold,
with local and non-local (in space and time)
self-interactions characteristic of spin-glass dynamics.
Particle simulations show \emph{adiabatic dimension reduction}:
on $S^2$, the density concentrates on a slowly precessing
great-circle ring whose orientation is a director
($\hat{\mathbf{n}} \sim -\hat{\mathbf{n}}$, even profile).
Conditioned on
this simulation-supported ring saddle and on a Born-Oppenheimer
separation between the slow orientation and the gapped density
fluctuations, symmetry fixes the low-energy dynamics to be the
nonlinear sigma model (NLSM) on the real projective plane
$S^2/\mathbb{Z}_2 = \mathbb{RP}^2$ (the $\mathbb{RP}^2$ NLSM
on the projective rotor space) in $(0+1)$ dimensions, governed
by a single low-energy constant, the rotational diffusion
coefficient $D_{\text{rot}}$. With $D_{\text{rot}}$ and the
static ring profile $f_0$ measured from particle simulations,
the resulting effective theory reproduces multiple independent
orientation- and density-sector diagnostics with no further
adjustable parameters.
\end{abstract}

\tableofcontents

%==============================================================================
\section{Introduction}
\label{sec:intro}
%==============================================================================

Many-body systems with quenched disorder and frustrated interactions give
rise to rich collective phenomena, from the ultrametric state space of
spin glasses to the aging dynamics of structural glasses. When such
systems are confined to curved surfaces, geometry adds a new ingredient:
the curvature of the manifold enters the interaction kernel through
geodesic distances and modifies the diffusion operator through the
Laplace-Beltrami structure. The question addressed in this paper is,
what statistical field theory describes the thermodynamic limit of such
a system on a two-dimensional manifold, and what universal low-energy
dynamics does it predict?

In a companion paper \cite{halperin2026frustrated} we studied a model
of \emph{frustrated Brownian particles} on two-dimensional Riemannian
manifolds: $N$ particles undergoing overdamped Langevin dynamics with
pairwise interactions linear in geodesic distance and with quenched
random couplings. The particle simulations show \emph{adiabatic
dimension reduction}: particles concentrate on lower-dimensional
submanifolds of the underlying geometry and adiabatically break the
rotational symmetries of the background. We use the word
\emph{adiabatically} (rather than \emph{spontaneously}) to signal
that at any instant the density has a definite dimension-reduced
configuration with a well-defined orientation, but that this
orientation itself performs a slow stochastic motion on the broken
symmetry manifold, so that the long-time-averaged state remains
rotationally invariant. On $S^2$ in particular, after a short stage
of non-equilibrium relaxation, the
density settles on a great-circle configuration whose orientation
$\hat{\mathbf{n}}(t) \in S^2/\mathbb{Z}_2$ executes a slow stochastic precession.

The present work addresses the thermodynamic limit $N \to \infty$
of this particle system. We first recast the $N$-body Langevin
dynamics at fixed disorder in its path-integral representation,
and then perform the disorder averaging over the quenched couplings
to arrive at a single-field statistical field theory for the density
$\rho(x,t)$, which we call the F2 (short for Frustrated Fields)
model. The construction is generic: it applies to any
compact two-dimensional Riemannian manifold (sphere, torus,
smoothly deformed variants of these, and bounded cylinders with
appropriate boundary conditions) and to a broad class of
pairwise interactions with long-range character on the manifold.
The formal derivation in Section~\ref{sec:large_N} is written
without boundary terms (the closed-manifold case); the bounded
cylinder requires the corresponding boundary terms to be retained
in the integration-by-parts steps.

The potentials tested in the present paper include the original
linear geodesic form $V(d) = d$ of
Ref.~\cite{halperin2026frustrated}, soft Coulomb logarithms
$\log(d + d_{\min})$, truncated geodesic-logs
$\max(U_{\min},\log d)$, the 2D chord Coulomb
$\log(2\sin(d/2))$ (the Green's function of the Laplace-Beltrami
operator on $S^2$), and the ordinary 3D Coulomb
$1/|\mathbf{x}_i - \mathbf{x}_j|$ for particles constrained to the
surface. The F2 construction itself is insensitive to this choice;
the universality study of Section~\ref{subsec:coulomb} identifies
the subclass of potentials whose force remains effective across the
full diameter of the manifold, which is the subclass that supports
ring formation.

The F2 model admits two equivalent presentations:
as an \emph{effective Dean-Kawasaki equation}, a nonlinear
functional Langevin equation for the smooth density $\rho(x,t)$
with a self-consistent two-time noise kernel inherited from the
disorder average; or as a single-field theory of $\phi^4$ type
in its lowest interaction structure but in fact non-polynomial\footnote{an Onsager-Machlup functional whose kinetic kernel
$\Omega^{-1}[\rho]$ is itself a functional of $\rho$ through the
self-consistency closure.}, whose self-interactions are local and
non-local in both space and time and share the formal structure
encountered in spherical spin glasses
\cite{mezard1987, castellani2005, cugliandolo1993, facoetti2019}.
Like Model B in the Hohenberg-Halperin classification
\cite{hohenberg1977} the dynamics is conserved, but unlike Model B
the noise is multiplicative and the
interaction is disorder-generated.

Section~\ref{sec:low_energy} specializes this generic framework to
the case where the two-dimensional manifold is the sphere $S^2$.
The particle simulations of
Ref.~\cite{halperin2026frustrated} show that the density
concentrates on a great-circle ring, adiabatically breaking the
rotational symmetry of the sphere. We take this
simulation-supported ring saddle, together with a Born-Oppenheimer
separation between the slow orientation and the gapped density
fluctuations, as inputs to the construction below; the saddle is
not derived from the F2 action by solving its Euler-Lagrange
equation.

With these inputs, the residual
$\mathrm{SO}(3) \to \mathrm{SO}(2)$ symmetry breaking and the
collective-coordinate construction for the slow ring orientation
$\hat{\mathbf{n}}(t)$ are enough on symmetry grounds to fix the
form of the low-energy dynamics of the F2 model: it is the
nonlinear sigma model (NLSM) on the real projective plane
$S^2/\mathbb{Z}_2 = \mathbb{RP}^2$ (the $\mathbb{RP}^2$ NLSM
on the projective rotor space) in $(0+1)$ dimensions; the
orientation is a director, since $\hat{\mathbf{n}}$ and
$-\hat{\mathbf{n}}$ describe the same density configuration of
the great-circle ring. The model is governed by a single
low-energy constant, the rotational diffusion coefficient
$D_{\text{rot}}$.

This symmetry-based route to the
effective theory is closely analogous to the construction of chiral
effective Lagrangians in quantum chromodynamics (QCD), the theory
of strong interactions. Strong couplings and non-perturbative
effects make extracting the low-energy dynamics of QCD directly
from the microscopic quark-gluon action extremely challenging;
instead, the low-energy theory of pions and
nucleons is built from the pattern of
spontaneous chiral-symmetry breaking
$\mathrm{SU}(2)_L \times \mathrm{SU}(2)_R \to \mathrm{SU}(2)_V$,
with a single low-energy constant, the pion decay constant $f_\pi$,
encoding the microscopic information that cannot be computed
analytically from the underlying theory. In our setting,
$D_{\text{rot}}$ plays the role of $f_\pi$, and its numerical value
is obtained from particle simulations. These simulations confirm
the $\mathbb{RP}^2$ NLSM prediction across a range of
independent diagnostics, and the universality study across the
several long-range potentials listed above supports the picture
that $D_{\text{rot}}$ is the only model-dependent quantity.

%The F2 model sits at the intersection of several lines of work.
%Its disorder-averaged action shares the non-local temporal kernel
%and time-reparametrization quasi-invariance of spherical $p$-spin
%and SYK-type dynamics \cite{cugliandolo1993, facoetti2019}. The
%collective coordinate $\hat{\mathbf{n}}(t)$ and its quantized
%angular-momentum levels provide a geometric analog of the skyrmion
%qubit of Psaroudaki and
%Panagopoulos~\cite{psaroudaki2021skyrmion}. The effective
%Dean-Kawasaki equation also admits a reading as a stochastic
%gradient flow in the Wasserstein space of probability measures
%\cite{villani2008, jordan1998, ambrosio2008}, extending optimal
%transport constructions to quenched disorder.

\paragraph{Outline.}
Section~\ref{sec:particle_model} reviews the frustrated Brownian
particle model on two-dimensional Riemannian manifolds.
Section~\ref{sec:large_N} develops a statistical field theory that describes the large-$N$ limit of this
particle system: we write the Langevin dynamics at fixed disorder
in its path-integral representation, average over the quenched
couplings, and arrive at the F2 model, presented equivalently as
the effective Dean-Kawasaki equation for the density field and as
the non-local $\phi^4$-like theory for $\rho(x,t)$ that underlies
the rest of the paper. Section~\ref{sec:low_energy} carries out the
low-energy reduction on $S^2$, deriving the $\mathbb{RP}^2$
NLSM effective theory for $\hat{\mathbf{n}}(t)$ through the
collective-coordinate decomposition, and discusses the analogy with
mean-field spin-glass theory.
Section~\ref{sec:numerical_implementation} describes the numerical
implementation of the $\mathbb{RP}^2$ NLSM dynamics, the comparison with
particle simulations, and the universality study with logarithmic
and Coulomb-type potentials on $S^2$. Section~\ref{sec:discussion}
discusses physical realizations, connections to other physics, and
open questions, and Section~\ref{sec:conclusion} summarizes the main
achievements and outlines future directions.
Appendix~\ref{app:covariant_langevin} reviews covariant Langevin
dynamics and path integrals on Riemannian manifolds;
Appendix~\ref{app:dean_kawasaki} derives the fixed-disorder
Dean-Kawasaki equation (the empirical-density SPDE for $N$
particles in a single realization of the quenched couplings);
Appendix~\ref{app:disorder_averaging_details} carries out the
disorder averaging and derives the effective (post-disorder-averaging)
Dean-Kawasaki equation of the F2 model;
Appendix~\ref{app:legendre_two_time} gives the derivation of the
two-time Legendre-polynomial correlator used in the density-sector
reduction of Section~\ref{subsec:corr_orientation}.

%==============================================================================
\section{Review: Frustrated Brownian Particles on Fixed Manifolds}
\label{sec:particle_model}
%==============================================================================

We begin by reviewing the microscopic particle model developed in \cite{halperin2026frustrated}, 
establishing notation and key results that will be generalized in subsequent sections.

\subsection{The Model}

Consider $N$ particles on a compact two-dimensional Riemannian manifold 
$(\M, g_{ij})$, with positions $\{q_n\}_{n=1}^N \subset \M$. The particles
interact through a potential energy:
\begin{equation}
    U[\{q_n\}] = \sum_{n < m} \phi_{nm} \, d_g(q_n,q_m)
    \label{eq:potential_particles}
\end{equation}
where $d_g(q_n,q_m)$ is the geodesic distance on $\M$ and $\phi_{nm} = \phi_{mn}$
are quenched random coupling constants drawn independently from a distribution
with zero mean and variance $J^2$:
\begin{equation}
    \E[\phi_{nm}] = 0, \quad \E[\phi_{nm}^2] = J^2
    \label{eq:coupling_stats}
\end{equation}
The geodesic distance $d_g(q_n,q_m)$ is non-smooth at the cut
locus of $\M$ (the antipodal point on $S^2$, the half-perimeter
loci on $T^2$) and on the diagonal $q_n = q_m$, where the
gradient $\nabla d_g$ is multi-valued or singular. The
singular sets of $V^i(x,y) = g^{ij}(x)\,\partial_j d_g(x,y)$
are of strictly positive codimension in
$\M\times\M$ (codimension one for the cut locus, codimension
two for the coincidence diagonal on the two-dimensional
manifold), so the singularities are integrable against any
continuous test field and against the smooth densities
$\rho$ that arise after coarse-graining or in the
disorder-averaged large-$N$ limit. We treat $V^i$ throughout
as a distribution that is well-defined when smeared against
such smooth densities; equivalently, the kernel can be
regularized at finite resolution (a small geodesic cutoff
near the diagonal and a small angular window around the cut
locus) and the regulator removed at the end. The cut-locus
contribution is therefore dropped on codimension grounds, not
on any vanishing assumption for the density.

Throughout this paper, Latin indices $i,j,k = 1,2$ denote curved
(coordinate) indices on the manifold, $a,b$ denote flat (frame)
indices, and $n,m$ label particles.

The manifold $\M$ is embedded in $\mathbb{R}^3$ via $X_\mu(q^1, q^2)$
($\mu = 1,2,3$), where $q_n^i$ ($i = 1,2$) are the intrinsic coordinates
of particle~$n$. The tangent vectors $e^\mu_i = \partial X_\mu/\partial q^i$
define the induced metric $g_{ij} = \sum_\mu e^\mu_i e^\mu_j$. We take
the thermal noise to originate in the embedding space $\mathbb{R}^3$ and
to be projected onto the tangent plane via the vielbein, yielding
manifold noise with covariance $\propto g^{ij}$. This is the standard
construction for overdamped Langevin dynamics on a Riemannian manifold
\cite{zinnjustin2002, ito1962, castrovillarreal2023}.

A short review of the theory of covariant overdamped Langevin dynamics
on Riemannian manifolds is presented in
Appendix~\ref{app:covariant_langevin}. As shown there, the covariant
Stratonovich form of the Langevin equation in intrinsic coordinates
reads
\begin{equation}
    \gamma\,\dd q_n^i = -g^{ij}(q_n)\frac{\partial U}{\partial q_n^j}\dd t + \sqrt{\Omega} \circ \dd W_n^i
    \label{eq:langevin_particles}
\end{equation}
where $\gamma$ is the friction coefficient, $T$ is the temperature, $g^{ij}$ is
the inverse metric tensor, and $\dd W_n^i$ is manifold Brownian motion with
covariance $\langle \dd W_n^i \dd W_m^j \rangle = g^{ij}(q_n) \delta_{nm} \dd t$.
The noise is generated via Cholesky decomposition
$\dd W_n^i = \sigma^{ia}(q_n)\,\dd\tilde{W}_n^a$ where
$\sigma^{ia}\sigma^{ja} = g^{ij}$ and the $\dd\tilde{W}^a$ are independent standard
Wiener processes. The Stratonovich convention ($\circ$) is the natural choice
for physical systems: it preserves the standard chain rule, so that
coordinate transformations follow ordinary calculus. The equivalent
It\^{o} form acquires a noise-induced drift
$m^i = (\Omega/2)\,e^j_a\,\partial_j\,e^i_a$, where
$e^i_a$ is the vielbein satisfying $e^i_a\,e^j_a = g^{ij}$ and
$\Omega = 2\gamma T$ is the noise strength parameter introduced
in Appendix~\ref{app:covariant_langevin}
(Eq.~\eqref{eq:omega_def}). This drift
arises when discretizing for numerical integration.

\paragraph{Dimensionless time and the Einstein relation.}
The Langevin equation \eqref{eq:langevin_particles} is written in
physical time. In the appendices and field-theoretic treatment, we
work in dimensionless time $\tau = t/\gamma$, which we continue to
denote $t$ for convenience. In dimensionless time the noise strength
is $\Omega = 2\gamma T$ (Eq.~\eqref{eq:omega_def}), which plays the
role of the physical-time noise strength $2D = 2T/\gamma$, where
$D = T/\gamma$ is the diffusion coefficient from the Einstein
relation ($k_B = 1$). The two are related by
$\Omega = 2\gamma^2 D$.

\subsection{Key Results from Particle Simulations}

Extensive numerical simulations reveal universal \emph{disorder-induced dimension 
reduction}: starting from generic initial conditions, particles spontaneously 
concentrate on lower-dimensional submanifolds. The specific structures depend 
on manifold topology. For the sphere $S^2$, particles collapse to a band near a great circle ($S^2 \to S^1$), breaking SO(3) $\to$ SO(2).
For the torus $T^2$, particles form two rings at opposite 
positions ($T^2 \to \coprod_2 S^1$), breaking SO(2)$\times$SO(2) $\to$ SO(2)$\times\mathbb{Z}_2$. For the bounded cylinder, particles localize into discrete clusters near boundaries, breaking SO(2) $\to \mathbb{Z}_2$.
%\begin{itemize}
%    \item \textbf{Sphere $S^2$}: Particles collapse to a band near a great circle 
%    ($S^2 \to S^1$), breaking SO(3) $\to$ SO(2).
%    
%    \item \textbf{Torus $T^2$}: Particles form two rings at opposite poloidal 
%    positions ($T^2 \to \coprod_2 S^1$), breaking SO(2)$\times$SO(2) $\to$ 
%    SO(2)$\times\mathbb{Z}_2$.
%    
%    \item \textbf{Bounded cylinder}: Particles localize into discrete clusters 
%    near boundaries, breaking SO(2) $\to \mathbb{Z}_2$.
%\end{itemize}
The symmetry-breaking direction slowly drifts via thermal noise, providing a
classical, dissipative analog of soft-mode dynamics from a continuously broken
symmetry, even though here the symmetry is only adiabatically rather than
spontaneously broken. These results motivate the field-theoretic treatment that
follows.

%==============================================================================
\section{Construction of the F2 Model}
\label{sec:large_N}
%==============================================================================

In this section we pass from the microscopic particle model to a
field-theoretic description. The presentation is \emph{general}:
neither the explicit form of the metric nor the explicit form of
the pair potential enters the derivation beyond general structural
properties. The construction applies to any two-dimensional closed
Riemannian manifold $(\M, g)$, with $g$ an arbitrary smooth
Riemannian metric, and to pairwise potentials of the form
\begin{equation}
U[\{q_n\}]
= \sum_{n<m}\phi_{nm}\,V\!\left(d_g(q_n, q_m)\right),
\label{eq:generic_pair_potential}
\end{equation}
where $V$ is any smooth function of the geodesic distance $d_g$
on $(\M, g)$ and $\phi_{nm}$ are the quenched random couplings.
The linear form $V(d) = d$ of
Ref.~\cite{halperin2026frustrated} is one instance in this class,
and further long-range choices (soft Coulomb, truncated
geodesic-log, 2D and 3D chord Coulomb on $S^2$) are examined in
the universality study of Section~\ref{subsec:coulomb}. The
derivation below carries through unchanged for any such choice;
only the explicit value of the interaction kernel in the effective
action depends on $V$ and on $g$.

The construction proceeds in four stages:
(i) we establish the large-$N$ scaling and introduce the empirical
density field;
(ii) we state the covariant Dean-Kawasaki equation for the density
at fixed quenched disorder, leaving the derivation to
Appendix~\ref{app:dean_kawasaki};
(iii) we express the dynamics as a path integral over the density
and an auxiliary noise field at fixed disorder;
(iv) we average over the quenched disorder and arrive at the F2
model, presented equivalently as an \emph{effective Dean-Kawasaki
equation} for the density or as a single-field nonlinear (more precisely, non-polynomial) theory
with self-interactions that are local and non-local in both space
and time.

\subsection{Empirical particle densities}
\label{subsec:scaling}

To obtain a well-defined large-$N$ limit, we adopt the standard
Sherrington-Kirkpatrick variance scaling: the couplings
$\phi_{nm}$ are taken Gaussian with variance $J^2/N$, where from
this point on $J$ denotes the $O(1)$ coupling after the $1/N$
rescaling. This ensures that the total interaction energy per
particle remains $O(1)$ as $N \to \infty$. The coupling
constants then satisfy
\begin{equation}
    \E[\phi_{nm}] = 0, \quad \E[\phi_{nm} \phi_{lp}] = \frac{J^2}{N}(\delta_{nl}\delta_{mp} + \delta_{np}\delta_{ml})
    \label{eq:coupling_statistics}
\end{equation}
The single-particle empirical density $ \rho_n $ for particle $n$ and 
the total empirical density $ \rho_N $ are defined as follows:
\begin{equation}
\label{eq:empirical_density}
\rho_n(x,t) = \frac{\delta^{(2)}(x - x_n(t))}{\sqrt{g(x)}}, \; \; \ 
 \rho_N(x, t) = \frac{1}{N} \sum_{n=1}^{N} \rho_n(x,t)    
\end{equation}
where $\sqrt{g(x)} = \sqrt{\det g_{ij}(x)}$ is the metric volume factor, ensuring
$\int_\M \rho_N \, \dd\mu_g = 1$ with $\dd\mu_g(x) = \sqrt{g(x)} \, \dd^2 x$.
%Define the empirical density as the normalized sum of delta functions on $\M$:
%\begin{equation}
%    \rho_N(x, t) = \frac{1}{N} \sum_{n=1}^{N} \frac{\delta(x - q_n(t))}{\sqrt{g(x)}}
%    \label{eq:empirical_density}
%\end{equation}
%where $\sqrt{g(x)} = \sqrt{\det g_{ij}(x)}$ is the metric volume factor, ensuring
%$\int_\M \rho_N \, \dd\mu_g = 1$ with $\dd\mu_g(x) = \sqrt{g(x)} \, \dd^2 x$.
The density $\rho_N$ is thus a probability distribution on $\M$, normalized to
unity. The single-particle and total empirical densities $\rho_n$
and $\rho_N$ are distributions, not pointwise functions: products
such as $\rho_n(x)\rho_n(y)$ at coincident $x = y$ and the
multiplicative noise $\sqrt{\rho}$ are distributionally
ill-defined at finite $N$ and acquire meaning only after some
regularization or in the large-$N$ limit. In the limit
$N \to \infty$, $\rho_N$ converges (in a suitable weak sense) to
a smooth density field $\rho(x,t)$, and from the disorder-averaged
generating functional onward we work with this smooth field; the
distributional nature of $\rho_n$ is needed only to derive the
single-particle SPDE \eqref{eq:single_particle_SPDE_main} and the
fixed-disorder MSRJD action and is not invoked in the
saddle-point analysis below. We write $\rho_N(x,t)$ using the
conventional field notation $x = (x^1, x^2)$, where the
components $x^1, x^2$ are intrinsic coordinates on $\M$. 

\subsection{Single-particle Dean-Kawasaki equation}
\label{subsec:dean_kawasaki}

For a fixed realization of the quenched couplings $\{\phi_{nm}\}$,
each particle obeys the covariant Stratonovich Langevin equation
(Appendix~\ref{app:dean_kawasaki},
Eq.~\eqref{eq:langevin_strat_dk}):
\begin{equation}
    \dd x_n^i = f_n^i\,\dd t + \sqrt{\Omega}\,e^i_a(x_n) \circ \dd\tilde{W}_n^a(t)
    \label{eq:dean_kawasaki_final}
\end{equation}
where $\Omega = 2\gamma T$ is the noise strength
(Eq.~\eqref{eq:omega_def}), $e^i_a(x)$ is the vielbein satisfying
$e^i_a\,e^j_a = g^{ij}$, $\tilde{W}_n^a$ are independent Wiener
processes, and the deterministic force is
$f_n^i = -(1/\gamma)\,g^{ij}\sum_m \phi_{nm}\nabla_j d_g(x,x_m)$.

Following the DK construction \cite{dean1996, illien2025} adapted
to Riemannian manifolds \cite{castrovillarreal2023}, 
%we define the
%single-particle empirical density
%(Appendix~\ref{app:dean_kawasaki},
%Eq.~\eqref{eq:single_particle_density}):
%\begin{equation}
%    \rho_n(x,t) = \frac{\delta^{(2)}(x - x_n(t))}{\sqrt{g(x)}}
%    \label{eq:dk_noise_final}
%\end{equation}
we derive the single-particle SPDE for the single-particle 
empirical density $ \rho_n(x,t) $
by applying the Stratonovich
chain rule to test functions
(Appendix~\ref{app:dean_kawasaki},
Eq.~\eqref{single_particle_SPDE}):
\begin{equation}
    \frac{\partial \rho_n}{\partial t} = -\nabla_i\!\left[\left(f_n^i + \sqrt{\Omega}\,e^i_a\eta^a\right)\rho_n\right]
    \label{eq:single_particle_SPDE_main}
\end{equation}
where $\eta^a(x,t) = dW^a/dt$ is the white noise field. We
emphasize that $\eta^a(x,t)$ attached to the
single-particle density $\rho_n$ is not an independent
Eulerian noise field with independent values at every $x$;
it is a distributional representation of the Brownian
increment of particle $n$, and acts on functions of $x$ only
through their value at the particle position $x_n(t)$. The
combination $\rho_n(x,t)\,\eta^a(x,t)$ should therefore be
read as $\delta^{(2)}(x - x_n(t))/\sqrt{g(x)} \cdot
dW_n^a/dt$, with all spatial covariance carried by the
$\delta$-function. The
divergence form ensures particle-number conservation:
$\partial_t\!\int_\M\!\rho_n\,d\mu_g = 0$. For the surface
Dean-Kawasaki equation on curved manifolds, see also
\cite{bell2026}; for field-theoretic simulations see
\cite{jin2025}; and for effects of geometry on active matter
dynamics see \cite{fily2016}. 
The SPDE (\ref{eq:single_particle_SPDE_main}) can also be written as 
\[ 
\partial_t\rho_n = -\mathcal{L}^{(\phi)}[\rho_n]
- \sqrt{\Omega}\,\nabla_i[\rho_n e^i_a\eta^a]
\]
where $ \mathcal{L}^{(\phi)}[\rho_n] $ stands for the 
deterministic drift operator at fixed disorder
(see Appendix~\ref{app:dean_kawasaki}, Eq.~\eqref{L_operator}):
\begin{equation}
    \mathcal{L}^{(\phi)}[\rho_n(x,t)] = \frac{1}{\gamma}\sum_m \phi_{nm}\int\!d\mu_g(y)\,\rho_m(y,t)\,\nabla_i\!\left[g^{ij}(x)\,\rho_n(x,t)\,\nabla_j d_g(x,y)\right]
    \label{eq:deterministic_operator}
\end{equation}
Unlike the original DK setting where identical pairwise potentials
allow direct summation over particles \cite{dean1996}, our quenched
random couplings $\phi_{nm}$ distinguish individual particles. This
prevents summation of the single-particle SPDEs into a single
equation for the mean density
$\rho_N = (1/N)\sum_n \rho_n$. Instead, as we describe next, the MSRJD path integral is
constructed for each particle, and disorder averaging is performed
at the level of the generating functional.

\subsection{MSRJD path integral at fixed disorder}
\label{sec:msrjd}
\label{sec:msrjd_fixed}

The single-particle SPDE \eqref{eq:single_particle_SPDE_main} is
a Langevin equation with multiplicative noise. The Martin-Siggia-Rose-Janssen-de~Dominicis (MSRJD) path integral
construction~\cite{martin1973, janssen1976, dedominicis1976, hunt1981}
proceeds by enforcing the equation of motion via a functional delta
function, introducing the response field $\hat{\rho}_n$, and
representing the Jacobian via Grassmann ghost fields
$\psi_n, \bar{\psi}_n$
(Appendix~\ref{app:dean_kawasaki},
Section~\ref{subsec:dk_as_langevin}).

The partition function
\begin{equation}
    Z = \int\prod_n\mathcal{D}[\eta_n]\,\mathcal{D}[\rho_n]\;\delta\!\left(\partial_t\rho_n + \mathcal{L}^{(\phi)}[\rho_n] + \sqrt{\Omega}\,\nabla_i[\rho_n e^i_a\eta^a]\right)\mathcal{J}[\rho_n]
    \label{eq:gen_func_fixed}
\end{equation}
satisfies $Z = 1$ by construction for every disorder realization,
provided the path-integral measure includes the Jacobian
$\mathcal{J}[\rho_n]$ from the change of variables $\eta\to\rho_n$
and the field configurations are constrained by a fixed initial
density $\rho_n(x,0) = \rho_n^{(0)}(x)$ specified independently of
the dynamics. We work throughout with fixed (non-averaged)
initial data, so $Z = 1$ is exact at each disorder realization
and no contribution from an initial-state measure appears in the
disorder average. This property eliminates the need for replica
methods when averaging over disorder
\cite{dedominicis1978, castellani2005}.

Representing the delta function via the response field $\hat{\rho}_n$,
computing the Jacobian via ghost fields, and integrating out the
Gaussian noise yields the MSRJD action for particle $n$
(Appendix~\ref{app:dean_kawasaki},
Eq.~\eqref{S_MSRJD_fixed_disorder}):
\begin{equation}
    S_n^{(\phi)} = \int\!\dd t\!\int\!\dd\mu_g(x)\left[\hat{\rho}_n\!\left(\partial_t\rho_n - \mathcal{L}^{(\phi)}[\rho_n]\right) + \bar{\psi}_n\!\left(\partial_t\psi_n - \mathcal{L}^{(\phi)}[\psi_n]\right) - \frac{\Omega}{2}\,g^{ij}G_i^{(n)}G_j^{(n)}\right]
    \label{eq:msrjd_action_fixed}
\end{equation}
The response field $\hat{\rho}_n$ is integrated along the
imaginary axis (equivalently, the convention
$\hat{\rho}_n \to i\hat{\rho}_n$ has been absorbed); after this
rotation the negative-quadratic noise term
$-\tfrac{\Omega}{2}g^{ij}G_i^{(n)}G_j^{(n)}$ produces a
positive-definite Gaussian weight in $G_i$, and the apparent sign
is conventional rather than indicative of an instability.
A central quantity is the composite boson-fermion current
(Appendix~\ref{app:dean_kawasaki}, Eq.~\eqref{G_current}):
\begin{equation}
    G_i^{(n)}(x,t) = \rho_n\nabla_i\hat{\rho}_n + (\nabla_i\bar{\psi}_n)\psi_n
    \label{eq:composite_current}
\end{equation}
which combines the bosonic current $\rho_n\nabla_i\hat{\rho}_n$ and
the ghost current $(\nabla_i\bar{\psi}_n)\psi_n$. Integration over
the Gaussian noise produces the term quadratic in $G_i^{(n)}$
(Appendix~\ref{app:dean_kawasaki}, Eq.~\eqref{S_noise}), encoding
the thermal noise correlations through the inverse metric $g^{ij}$.
The ghost sector $(\bar\psi_n,\psi_n)$ is retained to represent the
Jacobian determinant $\mathcal{J}[\rho_n]$ that enforces $Z = 1$;
the bosonic and ghost contributions to $G_i^{(n)}$ enter the
disorder-averaged action symmetrically through this combination,
and ghost determinants and $\rho$-functional determinants from
later integrations cancel pairwise as a consequence of the BRST
(supersymmetry) structure of $S_n^{(\phi)}$ that enforces $Z = 1$.
Ghosts therefore contribute through $G_i^{(n)}$ to physical
correlators in the disorder average; they are not dropped, only
repackaged.

%The action $S_n^{(\phi)}$ possesses a BRST symmetry generated by
%a constant Grassmann parameter $\epsilon$
%(Appendix~\ref{app:dean_kawasaki},
%Eqs.~\eqref{eq:susy_trans_1}--\eqref{eq:susy_trans_4}):
%$\delta\rho_n = \epsilon\psi_n$,
%$\delta\bar{\psi}_n = \epsilon\hat{\rho}_n$,
%$\delta\psi_n = \delta\hat{\rho}_n = 0$,
%which ensures $Z = 1$. The composite current $G_i^{(n)}$ is
%invariant under this transformation up to a total derivative on
%the closed manifold $\M$.

\subsection{Disorder averaging and mean-field decoupling}
\label{sec:disorder_average_Z}

Since $Z^{(\phi)} = 1$ for every disorder realization, disorder
averaging proceeds without replicas
\cite{dedominicis1978, castellani2005}. The standard approach in
spin-glass dynamics requires that averaging be performed on the
generating functional, not on the equation of motion.

\paragraph{Gaussian integration over disorder.}
The disorder enters the MSRJD action
\eqref{eq:msrjd_action_fixed} through the interaction drift
$\mathcal{L}^{(\phi)}$, which is linear in the couplings
$\phi_{nm}$. After integration by parts on $\M$
(Appendix~\ref{app:disorder_averaging_details}), the interaction
action takes the form
\begin{equation}
    S_{\text{int}}^{(\phi)} = \frac{1}{\gamma}\sum_{n,m}\phi_{nm}\int\!dz\;X_n(z)\,Y_m(z)
    \label{eq:S_int_phi}
\end{equation}
where $z = (x,y,t)$, $dz = dt\,d\mu_g(x)\,d\mu_g(y)$, and
$X_n(z) = G_i^{(n)}(x,t)\,V^i(x,y)$,
$Y_n(z) = \rho_n(y,t)$,
with the force kernel
$V^i(x,y) = g^{ij}(x)\nabla_j^{(x)}d_g(x,y)$.
The kernel $V^i(x,y)$ is non-smooth on the coincidence
diagonal $x = y$ and on the cut locus where
$\nabla d_g$ is multi-valued
(Sec.~\ref{subsec:scaling}); the disorder average below
involves products of such kernels, $V^i(x,y)V^j(x',y)$. These
products are integrable against the smooth densities and
dressed correlators of the large-$N$ theory by codimension:
the singular sets are codimension-1 (cut locus) or
codimension-2 (diagonal) submanifolds in $\M\times\M$, so the
distributional singularities can be smeared against any
smooth test field and the regulator removed at the end. A
more delicate treatment may be needed if finite-$N$
corrections involve coincident-point limits, which is beyond
the scope of the present mean-field analysis.
Since $S_{\text{int}}^{(\phi)}$ is linear in the zero-mean
Gaussian couplings $\phi_{nm}$, averaging
$\exp(-S_{\text{int}}^{(\phi)})$ produces
(Appendix~\ref{app:disorder_averaging_details},
Eq.~\eqref{eq:gaussian_avg_app})
\begin{equation}
    \E_\phi\!\left[e^{-S_{\text{int}}^{(\phi)}}\right] = \exp\!\left(\frac{1}{2}\E_\phi\!\left[\left(S_{\text{int}}^{(\phi)}\right)^2\right]\right)
    \label{eq:gaussian_disorder_avg}
\end{equation}
Evaluating the Gaussian contractions yields
(Appendix~\ref{app:disorder_averaging_details},
Eq.~\eqref{squared_terms})
\begin{equation}
    \frac{1}{2}\E_\phi\!\left[\left(S_{\text{int}}^{(\phi)}\right)^2\right] = \alpha^2 N\,\frac{1}{N^2}\sum_{n,m}\left(A_n^{(1)}B_m^{(1)} + A_n^{(2)}B_m^{(2)}\right)
    \label{eq:S_int_phi_ibp}
\end{equation}
where $\alpha^2 = J^2/(2\gamma^2)$ and the bilocal operators
$A_n^{(1,2)}$, $B_n^{(1,2)}$ are products of $X_n$ and $Y_n$
at two spacetime points $(z, z')$
(Appendix~\ref{app:disorder_averaging_details}). The original
potential $U = \sum_{n<m}\phi_{nm}\,d_g(q_n,q_m)$ contains no
diagonal couplings, so $\phi_{nn}$ never appears; the symmetric
covariance $\delta_{nl}\delta_{mp} + \delta_{np}\delta_{ml}$
reflects only the pair symmetry $\phi_{nm} = \phi_{mn}$ when the
sum in $S_{\text{int}}^{(\phi)}$ is extended to all
$(n,m)$ with $n \neq m$, and the factor of $2$ from this
symmetrization is absorbed into the prefactor of
\eqref{eq:S_int_phi_ibp}. \emph{Scaling audit at large $N$.}
After the SK rescaling $J^2 \to J^2/N$, the coupling
$\alpha^2 = J^2/(2\gamma^2) = O(1)$. Each $X_n$, $Y_n$
($A_n^{(a)}$, $B_n^{(a)}$ likewise) is built from
single-particle quantities and is $O(1)$ at fixed particle
label. The double sums $\sum_{n,m}A_n^{(a)}B_m^{(a)} = O(N^2)$
generically, so the combination $\alpha^2 N \cdot (1/N^2)
\sum_{n,m} = O(N) \cdot O(1) = O(N)$ at the level of the
disorder-averaged action density, in agreement with the
desired extensive scaling
$S_{\text{disorder}} = O(N)$. The $1/N^2$ prefactor inside the
sum is what makes the HS step well posed: the squared single
sums introduced below are $O(N^2)$ before division, so each HS
bilinear $\hat Q_a$ couples to an $O(1)$ density-like
combination after the $1/N^2$ normalization is absorbed.

\paragraph{Hubbard-Stratonovich linearization and large-$N$
decoupling.}
The double sums $(1/N^2)\sum_{nm}A_n B_m$ are rewritten as
squared single sums. Four bilocal Hubbard-Stratonovich (HS)
fields $\hat{Q}_a(z,z')$ ($a = 1, \ldots, 4$) linearize these
squared terms
(Appendix~\ref{app:disorder_averaging_details},
Eq.~\eqref{HS_transforms}). The HS branches with positive and
negative quadratic kernels require different integration
contours: the positive branches are integrated along the real
axis, while the negative branches are rotated by an explicit
factor of $i$, i.e.\ the HS field is treated as imaginary
(equivalently, integrated along the imaginary contour). With
this choice the saddle-point evaluation that follows is
well-defined. In the large-$N$ limit, the HS fields are then
replaced by their vacuum expectation values
(Appendix~\ref{app:disorder_averaging_details},
Eq.~\eqref{vacuum_averages}), which decouples the particle
labels. The result is a factorized generating functional
(Appendix~\ref{app:disorder_averaging_details},
Eq.~\eqref{Z_after_averaging}):
\begin{equation}
    \overline{Z}[J,\bar{J}] = \int\prod_{n=1}^N\mathcal{D}[\rho_n, \hat{\rho}_n, \psi_n, \bar{\psi}_n]\;e^{-\sum_n S_n^{\text{eff}} + \text{sources}}
    \label{eq:generating_functional}
\end{equation}
where the effective single-particle action is
(Appendix~\ref{app:disorder_averaging_details},
Eq.~\eqref{eq:S_n_eff}):
\begin{multline}
    S_n^{\text{eff}} = \int\!\dd t\,d\mu_g(x)\left[\hat{\rho}_n\partial_t\rho_n + \bar{\psi}_n\partial_t\psi_n - \frac{\Omega}{2}\,g^{ij}G_i^{(n)}G_j^{(n)}\right] \\
    - \alpha^2\iint\!dz\,dz'\left(\langle B^{(1)}\rangle A_n^{(1)} + \langle A^{(1)}\rangle B_n^{(1)} + \langle B^{(2)}\rangle A_n^{(2)} + \langle A^{(2)}\rangle B_n^{(2)}\right)
    \label{eq:S_disorder_vertex}
\end{multline}
The coefficients $\langle A^{(1,2)}\rangle$,
$\langle B^{(1,2)}\rangle$ are mesoscopic order parameters
defined as particle-averaged vacuum expectation values
(Appendix~\ref{app:disorder_averaging_details},
Eq.~\eqref{VEVs_for_AB}), determined self-consistently from the
generating functional \eqref{eq:generating_functional}. The
replacement of the HS fields by their VEVs is the leading-order
mean-field saddle of the integral over $\hat{Q}_a$: fluctuations
around this saddle are suppressed by powers of $1/N$ in the
bulk action, and contribute at subleading order to the effective
single-particle action. The present construction therefore
predicts the disorder-averaged $D_{\text{rot}}$; the large
sample-to-sample variation at $N = 400$
(Section~\ref{subsec:comparison}, coefficient of variation
$\approx 0.68$) is consistent with $1/N$ fluctuation corrections
still being non-negligible at the simulated sizes, and a
controlled $1/N$ expansion of $D_{\text{rot}}$ that would
quantify these effects is left to future work.
%This factorization is the phenomenon of \emph{propagation of
%chaos} \cite{mezard1987, castellani2005}: 
This demonstrates that 
in the large-$N$ limit,
the single-particle densities $\rho_1, \ldots, \rho_N$ become
independent and identically distributed random fields. The
quenched disorder is transformed into a self-interaction
and non-Markovian dynamics with memory. The empirical density
$\rho_N(x,t) = (1/N)\sum_n\rho_n(x,t)$ converges to the
expectation $\langle\rho(x,t)\rangle$ of a single representative
particle by the law of large numbers.

\subsection{Effective dynamics and consistency conditions}
\label{sec:msrjd_action}

The effective action \eqref{eq:S_disorder_vertex} can be rewritten
using dressed two-point functions
(Appendix~\ref{app:disorder_averaging_details},
Eqs.~\eqref{eq:two_point_functions}):
\begin{align}
    C_\rho(x,t,x',t') &= \frac{1}{N}\sum_n\langle\rho_n(x,t)\,\rho_n(x',t')\rangle \nonumber\\
    K_{ij}(x,t,x',t') &= \frac{1}{N}\sum_n\langle G_i^{(n)}(x,t)\,G_j^{(n)}(x',t')\rangle \label{eq:msrjd_action} \\
    R_i(x,t,x',t') &= \frac{1}{N}\sum_n\langle\rho_n(x,t)\,G_i^{(n)}(x',t')\rangle \nonumber
\end{align}
The expectation $\langle\,\cdot\,\rangle$ is the
disorder-averaged generating-functional expectation:
correlators are evaluated under the measure obtained after
the Gaussian average over $\phi_{nm}$ has been performed,
and simultaneously fold in the thermal noise of the
underlying Langevin dynamics. They are
\emph{disorder-averaged dynamical correlators of the
normalized MSRJD generating functional}, and the
corresponding objects in the simulations are
quenched-disorder-averaged dynamical correlators averaged
over realizations
(Section~\ref{subsec:comparison}). The annealed
$D_{\text{rot}}$, in which $\phi_{nm}$ is integrated over the
same Gaussian distribution as the dynamical degrees of
freedom and which differs from the disorder-averaged
$D_{\text{rot}}$ in the quenched/annealed gap, is not
computed in this work (Section~\ref{subsec:comparison}).
The dressed kernels $\hat{\mathcal{C}}^{ij}$,
$\hat{\mathcal{K}}$, $\hat{\mathcal{R}}^i$ are built from
$C_\rho$, $K_{ij}$, $R_i$ and the force kernel
$V^i(x,y) = g^{ij}(x)\nabla_j^{(x)}d_g(x,y)$
(Appendix~\ref{app:disorder_averaging_details},
Eqs.~\eqref{eq:two_point_functions}).
In terms of these, the effective action takes the form
(Appendix~\ref{app:disorder_averaging_details},
Eq.~\eqref{eq:S_n_eff_2}):
\begin{multline}
    S_n^{\text{eff}} = \underbrace{\int\!\dd t\,d\mu_g\;\hat{\rho}_n\partial_t\rho_n}_{S_{\text{det}}}
    + \int\!\dd t\,d\mu_g\;\bar{\psi}_n\partial_t\psi_n
    \underbrace{- \frac{\Omega}{2}\int\!\dd t\,d\mu_g\,g^{ij}G_i^{(n)}G_j^{(n)}}_{S_{\text{noise}}} \\
    \underbrace{- \alpha^2\iint\!dx\,dx'\,dt\,dt'\left(G_i^{(n)}\hat{\mathcal{C}}^{ij}G_j^{(n)} + \rho_n\hat{\mathcal{K}}\rho_n + 2G_i^{(n)}\hat{\mathcal{R}}^i\rho_n\right)}_{S_{\text{dis}}}
    \label{eq:S_eff_expanded}
\end{multline}
The labels $S_{\text{det}}$, $S_{\text{noise}}$, $S_{\text{dis}}$
indicate the roles of each contribution: the kinetic term
$S_{\text{det}}$ enforces the equation of motion for the density,
$S_{\text{noise}}$ encodes bare thermal noise correlations through
the inverse metric $g^{ij}$, and $S_{\text{dis}}$ contains the
disorder-induced non-local kernels coupling the particle to the
self-consistent mean field. The ghost kinetic term
$\bar{\psi}_n\partial_t\psi_n$ tracks the linearized density
dynamics. The minus sign on $S_{\text{noise}}$ and the sign of
the $G\hat{\mathcal{C}}G$ piece in $S_{\text{dis}}$ are
conventional after the imaginary rotation of $\hat\rho_n$ noted
above (Eq.~\eqref{eq:msrjd_action_fixed} discussion):
$G_i^{(n)} = \rho_n\nabla_i\hat\rho_n + \ldots$ contains a
factor of $i$ once $\hat\rho_n$ is rotated, so
$g^{ij}G_iG_j$ is purely imaginary and
$-\tfrac{\Omega}{2}g^{ij}G_iG_j$ furnishes a positive-definite
Gaussian weight on the rotated contour. The same applies to the
$G\hat{\mathcal{C}}G$ contribution in $S_{\text{dis}}$, since
$\hat{\mathcal{C}}$ is positive (a current-current correlator).
Stability of the Euclidean action against the response-field
direction is therefore not an issue once the contour is fixed.

\paragraph{Promotion of the composite field and effective PDE.}
The composite current $G_i^{(n)} = \rho_n\nabla_i\hat\rho_n
+ (\nabla_i\bar\psi_n)\psi_n$ is promoted to an independent
field via the exact functional identity
\[
1 = \int\!\mathcal{D}[G^{(n)}_i, \hat{G}^i_n]\;
\exp\!\Bigl(-i\!\int\!\hat{G}^i_n\bigl[G^{(n)}_i
- \rho_n\nabla_i\hat\rho_n
- (\nabla_i\bar\psi_n)\psi_n\bigr]\Bigr),
\]
which inserts a $\delta$-function constraint
$G^{(n)}_i = \rho_n\nabla_i\hat\rho_n
+ (\nabla_i\bar\psi_n)\psi_n$ together with its conjugate
response current $\hat{G}^i_n$
(Appendix~\ref{app:disorder_averaging_details},
Eq.~\eqref{identity_for_G}). The identity is exact: the
$\delta$-functional has unit functional Jacobian because the
defining relation is linear in $\hat\rho_n$ at fixed $\rho_n$
and bilinear in the ghosts, so no nontrivial determinant is
generated. After integrating out $\hat{\rho}_n$ and the ghost
fields, the response current $\hat{G}^i_n$ enforces a delta
function imposing the conservation
PDE~\eqref{eq:eom_rho}, and the ghost determinant from the
$\bar\psi\psi$ integration cancels exactly against the
$\rho$-functional determinant generated by the $\hat\rho_n$
integration; this cancellation is the BRST counterpart of the
$Z = 1$ normalization noted above
(Appendix~\ref{app:disorder_averaging_details},
Eq.~\eqref{integrate_psi_n}).
The density field then satisfies the PDE
(Appendix~\ref{app:disorder_averaging_details},
Eq.~\eqref{PDE_rho_n}):
\begin{equation}
    \frac{\partial\rho}{\partial t} + \nabla_i\!\left[\hat{G}^i[G,\rho]\,\rho\right] = 0
    \label{eq:eom_rho}
\end{equation}
where the effective drift $\hat{G}^i[G,\rho]$ is given by
(Appendix~\ref{app:disorder_averaging_details},
Eq.~\eqref{hat_G_from_G}):
\begin{equation}
    \hat{G}^i(x,t) = \int\!dx'\,dt'\left[\hat{\mathcal{A}}^{ij}(x,t,x',t')\,G_j(x',t') + 2\alpha^2\hat{\mathcal{R}}^i(x,t,x',t')\,\rho(x',t')\right]
    \label{eq:effective_drift}
\end{equation}
with the combined noise kernel
\begin{equation}
    \hat{\mathcal{A}}^{ij}(x,t,x',t') = \Omega\,g^{ij}(x)\,\delta(x-x')\,\delta(t-t') + 2\alpha^2\hat{\mathcal{C}}^{ij}(x,t,x',t')
    \label{eq:combined_kernel}
\end{equation}
This kernel combines bare thermal fluctuations (the $\Omega$
term) and disorder-induced current correlations (the
$\hat{\mathcal{C}}^{ij}$ term) into a single dressed kernel.
%Eq.~\eqref{eq:eom_rho} plays the role of the disorder-averaged
%Dean-Kawasaki equation: it has the divergence form of the
%single-particle SPDE \eqref{eq:single_particle_SPDE_main}
%(ensuring particle-number conservation), with $\hat{G}^i$
%replacing the stochastic force. 
The resulting generating
functional is
(Appendix~\ref{app:disorder_averaging_details},
Eq.~\eqref{eq:gen_functional_2}):
\begin{equation}
    Z[J,\bar{J}] = \prod_n\int\!\mathcal{D}[\rho_n]\,\mathcal{D}[G_i^{(n)}]\;\delta(\rho_n - \rho_n^G)\;e^{-\sum_n S_n(\rho_n, G_i^{(n)}) + \text{sources}}
    \label{eq:gen_func_rhoG}
\end{equation}
where the action depends on $\rho$ and $G_i$ through two
non-local kernels
(Appendix~\ref{app:disorder_averaging_details},
Eq.~\eqref{eq:action_2}):
\begin{equation}
    S(\rho, G_i) = \iint\!dx\,dx'\,dt\,dt'\left(\frac{1}{2}G_i\hat{\mathcal{A}}^{ij}G_j - \alpha^2\rho\hat{\mathcal{K}}\rho\right)
    \label{eq:action_rho_G}
\end{equation}
The first term shows that $G_i$ acts as a noise variable with
correlation determined by $\hat{\mathcal{A}}^{ij}$. The second
is a non-local density self-interaction driven by the quenched
disorder.

\paragraph{Self-consistency conditions.}
The two-point functions \eqref{eq:msrjd_action} obey
Schwinger-Dyson closure equations evaluated at the large-$N$
saddle of \eqref{eq:gen_func_rhoG}: the closure relations
below are not Gaussian identities valid for the full
interacting problem, but stationarity conditions in which the
dressed kernels $\hat{\mathcal{A}}$ and $\hat{\mathcal{K}}$
themselves depend on the same correlators through the
self-consistent disorder average. With this understanding,
the current-current correlator satisfies
(Appendix~\ref{app:disorder_averaging_details},
Eq.~\eqref{first_constraint}):
\begin{equation}
    \int\!dz''\;\hat{\mathcal{A}}^{ik}(z,z'')\,K_{kj}(z'',z') = \delta_j^i\,\delta(z-z')
    \label{eq:constraint_K}
\end{equation}
and the density-density correlator satisfies
(Appendix~\ref{app:disorder_averaging_details},
Eq.~\eqref{second_constraint}):
\begin{equation}
    2\alpha^2\int\!dz''\;C_\rho(z,z'')\,\hat{\mathcal{K}}(z'',z') = \delta(z-z')
    \label{eq:constraint_C}
\end{equation}
Eq.~\eqref{eq:constraint_C} should be read as the saddle-level
closure for $C_\rho$ at the same Schwinger-Dyson order; the PDE
constraint enforced by the Lagrange multiplier $\lambda$ is
solved at the saddle, and corrections from fluctuations around
the saddle are not retained. The dynamic evolution of these
correlators is governed by coupled integro-differential
equations
(Appendix~\ref{app:disorder_averaging_details},
Eqs.~\eqref{ODE_for_C_rho} and~\eqref{ODE_for_C_R_i}) that
encode the non-Markovian memory effects generated by the
quenched disorder.

\subsection{The effective Dean-Kawasaki equation}
\label{subsec:large_N_saddle}

The PDE constraint~\eqref{eq:eom_rho}, with the
drift~\eqref{eq:effective_drift} decomposed using the
kernel~\eqref{eq:combined_kernel}, is a stochastic PDE
for the density driven by the Gaussian field $G_i$
(Appendix~\ref{subsec:effective_rho_derivation},
Eq.~\eqref{eq:DK_stratonovich}):
\begin{equation}
\label{eq:effective_rho}
\partial_t\rho
= -\nabla_i\!\left[\left(\Omega\,g^{ij}G_j(x,t)
+ 2\alpha^2\!\int\!dz'\;
\hat{\mathcal{C}}^{ij}(z,z')G_j(z')
+ F^i[\rho]\right)\rho\right]
\end{equation}
where the self-consistent disorder drift is
\begin{equation}
\label{eq:self_consistent_drift}
F^i[\rho](x,t) = 2\alpha^2\int\!dx'\!\int_{-\infty}^{t}\!dt'\;
\hat{\mathcal{R}}^i(x,t,x',t')\,\rho(x',t')
\end{equation}
with the response kernel $\hat{\mathcal{R}}^i$ taken
retarded ($\hat{\mathcal{R}}^i(x,t,x',t') = 0$ for $t' > t$),
so that $F^i[\rho](x,t)$ depends only on the past history of
$\rho$ and the equation is causal. Here $G_j$ is a Gaussian
noise field with correlator $\langle G_i(z)G_j(z')\rangle
= \mathcal{W}_{ij}(z,z')$~\eqref{eq:constraint_K}.
This is the \emph{effective Dean-Kawasaki equation} for
the F2 model (Stratonovich convention, divergence form
$\partial_t\rho + \nabla_i J^i = 0$). It shares the
conservation structure of the standard Dean-Kawasaki
equation \cite{dean1996, illien2025} for $N$ identical
particles with a common pairwise potential $V(x,y)$. For the
empirical density $\rho_N(x,t) = N^{-1}\sum_n\rho_n(x,t)$
normalized to unit mass, the standard DK equation reads
\begin{equation}
\label{eq:standard_DK}
\partial_t\rho_N = -\nabla_i\!\left[
\rho_N\,b^i[\rho_N] \right]
+ \frac{1}{\sqrt{N}}\,
\nabla_i\!\left[\sqrt{2 D \rho_N}\;e^i_a\,\eta^a\right],
\end{equation}
with deterministic drift
$b^i[\rho_N] = -(1/\gamma)\,g^{ij}\nabla_j\!\int V(x,y)\rho_N(y)\,
d\mu_g(y)$, diffusion coefficient $D = T/\gamma$, vielbein
$e^i_a e^j_a = g^{ij}$, and an Eulerian space-time white
noise $\eta^a(x,t)$. The noise amplitude is $\sqrt{2D\rho_N}$
(this is the multiplicative $\sqrt{\rho}$ noise referred to
in the introduction), and the explicit $1/\sqrt{N}$ prefactor
reflects the unit-mass normalization of $\rho_N$. The
single-particle distributional SPDE
\eqref{eq:single_particle_SPDE_main} that we use in the MSRJD
construction has a different noise structure, as discussed
above: $\eta^a$ acts on $\rho_n$ only through the Brownian
increment of particle $n$.
The differences between \eqref{eq:effective_rho}
and \eqref{eq:standard_DK} are as follows.

\emph{Drift.}
In the standard DK equation, the drift
$-(1/\gamma)g^{ij}\nabla_j\int V\rho\,d\mu_g$ is local in
time: it depends on the density at the same instant $t$
through the instantaneous mean-field potential
$\int V(x,y)\rho(y,t)\,d\mu_g(y)$. In our
equation~\eqref{eq:effective_rho}, the drift
$F^i[\rho]$~\eqref{eq:self_consistent_drift} involves the
response kernel $\hat{\mathcal{R}}^i(x,t,x',t')$ integrated
over both $x'$ and $t'$. The integration over past times
$t' < t$ makes the drift \emph{non-Markovian}: the force on
the density at time $t$ depends on the density history
through a memory kernel. These memory effects, encoded in the
self-consistent two-point functions
$C_\rho$, $K_{ij}$, $R_i$
(Eqs.~\eqref{eq:constraint_K}--\eqref{eq:constraint_C}),
are the hallmark of glassy dynamics
\cite{cugliandolo1993, facoetti2019}.

\emph{Noise.}
The standard DK equation has a single noise term
$\sqrt{2D\rho_N}\,e^i_a\eta^a$ that is local in both
space and time (multiplicative $\sqrt{\rho}$ noise). Our
equation has two noise components: a \emph{local} term
$\Omega\,g^{ij}G_j(x,t)\,\rho$ (analogous to the standard
DK noise) and a \emph{non-local} term
$2\alpha^2\int\hat{\mathcal{C}}^{ij}(z,z')G_j(z')\,dz'
\cdot\rho$ that couples the density to the noise at other
spacetime points through the disorder-dressed current
correlator $\hat{\mathcal{C}}^{ij}$. The non-local noise
generates temporal correlations in the effective stochastic
force, absent in the standard DK equation.

\emph{Status.}
The standard DK equation is exact at any $N$: it is a
tautological rewriting of the $N$-particle Langevin dynamics
as a SPDE for the empirical density \cite{dean1996}. Our
equation~\eqref{eq:effective_rho} is a mean-field result,
valid at leading order in $1/N$ after disorder averaging of
the generating functional. At finite $N$, corrections from
the noise action $S_{\text{noise}} = O(1)$ produce
non-Gaussian density statistics \cite{velenich2008}.

%Fluctuations around the mean-field solution are controlled by
%the quadratic terms in the expanded action.
%Section~\ref{sec:low_energy} reduces the theory to
%orientation coordinates and derives the self-consistent
%equations for the ring profile and precession dynamics.

\subsection{Reduction to a single-field model}
\label{subsec:starting_point}

The generating functional~\eqref{eq:gen_func_rhoG} involves the
action $S(\rho, G_i)$~\eqref{eq:action_rho_G} subject to the
PDE constraint~\eqref{eq:eom_rho}. To enforce the constraint
for all $(x,t)$, we introduce a real Lagrange multiplier field
$\lambda(x,t)$ via the Fourier representation of the delta
functional and define the augmented action:
\begin{equation}
\label{eq:augmented_action_main}
S_\lambda[\rho, G_i, \lambda]
= S(\rho, G_i)
- i\!\int\!dx\,dt\;\lambda(x,t)\left[
\partial_t\rho + \nabla_i(\hat{G}^i[G,\rho]\,\rho)\right]
\end{equation}
%The starting point is the generating functional with the
%augmented
%action~\eqref{eq:augmented_action_main}
%(Section~\ref{subsec:large_N_saddle}):
%$Z = \int\mathcal{D}\rho\,\mathcal{D}G_i\,
%\mathcal{D}\lambda\;
%\exp(-S_\lambda[\rho,G,\lambda])$.
Since $S_\lambda$ is quadratic in $G_i$ (the noise action
$\frac{1}{2}G\hat{\mathcal{A}}G$ is Gaussian and the PDE
term $-i\lambda\nabla(\hat{G}\rho)$ is linear in $G$), we
integrate $G_i$ out exactly before introducing the
collective coordinate. The $G$-dependent part is
$\frac{1}{2}G^T\hat{\mathcal{A}}G + i\Sigma^T G$
(after integrating by parts in the PDE term), where
\begin{equation}
\label{eq:Sigma_def}
\Sigma^{j}(x,t)
= -\!\int\!dx'dt'\;
\hat{\mathcal{A}}^{ij}(x',t',x,t)\,\rho(x',t')\,
\nabla_i^{(x')}\lambda(x',t')
= \int\!dx'dt'\;\lambda(x',t')\,
\nabla_i^{(x')}\!\left[
\hat{\mathcal{A}}^{ij}(x',t',x,t)\,
\rho(x',t')\right],
\end{equation}
and the
second form follows by integration by parts (no boundary
terms on $S^2$). 
The Gaussian integral over $G_i$ 
%(shifting
%$G \to G - i\mathcal{W}\Sigma$) 
gives $(\det\hat{\mathcal{A}})^{-1/2}
\exp(-\frac{1}{2}\Sigma^T\mathcal{W}\Sigma)$
with $\mathcal{W}_{ij} = (\hat{\mathcal{A}}^{-1})_{ij}$.
The generating functional now takes the following form:
\begin{equation}
\label{eq:Z_rho_lambda}
Z = \text{const}\times\int\!\mathcal{D}\rho\,
\mathcal{D}\lambda\;
\exp\!\left(-S_{\text{eff}}[\rho, \lambda]\right),
\end{equation}
with the effective two-field action
\begin{equation}
\label{eq:S_eff_rho_lambda}
S_{\text{eff}}[\rho, \lambda]
= -\alpha^2\!\iint\rho\hat{\mathcal{K}}\rho
- i\!\int\!\lambda\left[\partial_t\rho
+ \nabla_i(F^i[\rho]\rho)\right]
+ \frac{1}{2}\!\int\!dx\,dt\,dx'dt'\;
\lambda(x,t)\,\Omega(x,t,x',t')\,\lambda(x',t'),
\end{equation}
where the noise-induced kernel is
\begin{align}
\label{eq:Omega_kernel}
\Omega(x,t,x',t') = \int\!&dx''\,dt''\,dx'''\,dt'''\;
\mathcal{W}_{ij}(x'',t'',x''',t''') \nonumber \\
&\times\nabla_k^{(x)}\!\left[
\hat{\mathcal{A}}^{ki}(x,t,x'',t'')\,\rho(x,t)\right]
\nabla_l^{(x')}\!\left[
\hat{\mathcal{A}}^{lj}(x',t',x''',t''')\,
\rho(x',t')\right].
\end{align}
\textit{Notation.} The same letter $\Omega$ denotes here, with
two spacetime arguments, the noise-induced two-point kernel
$\Omega(x,t,x',t')$, and earlier
(Eq.~\eqref{eq:omega_def}) with no arguments the bare scalar
noise strength $\Omega = 2\gamma T$. The two are distinct
objects: the kernel is a bilinear operator, the scalar a
constant; the number of arguments distinguishes them. From
this point through
Section~\ref{sec:numerical_implementation} the symbol $\Omega$
refers exclusively to the kernel; the scalar reappears only in
self-contained appendices. We keep the same letter
because the kernel reduces, in the limit of vanishing disorder
dressing, to a local form proportional to the bare $2\gamma T$.
%The three terms
%in~\eqref{eq:S_eff_rho_lambda} are: the density
%self-interaction (attractive, driving ring formation); the
%kinetic and drift term
%$-i\lambda[\partial_t\rho + \nabla(F\rho)]$ (linear in
%$\lambda$, with factor $i$ from the Fourier representation
%of the delta functional); and the noise-induced term
%$\frac{1}{2}\lambda\Omega\lambda$. 
Note that the kernel
$\Omega$~\eqref{eq:Omega_kernel} has a quadratic dependence on $\rho$  
through $\nabla_k[\hat{\mathcal{A}}^{ki}\rho]$.
The path integral~\eqref{eq:Z_rho_lambda} is now over
two fields $(\rho, \lambda)$ only; $G_i$ has been
integrated out exactly, and its effect is encoded in the
$\frac{1}{2}\lambda\Omega\lambda$ term.

%\paragraph{The role of $\lambda$ and the parallel with
%GJS.}
%The effective action~\eqref{eq:S_eff_rho_lambda} has two
%types of $\lambda$-dependence: the symplectic term
%$-i\lambda\partial_t\rho$ and the noise-induced term
%$\frac{1}{2}\lambda\Omega\lambda$. The symplectic term
%identifies $\lambda$ as a conjugate momentum for $\rho$,
%analogous to $\pi$ for $\phi$ in
%GJS~\cite{gervais1975collective}. In the GJS kink
%theory, the canonical pair $(\phi, \pi)$ has the
%symplectic term $\int\pi\dot\phi\,dx$ and the
%Hamiltonian~(GJS Eq.~(2.15)) is quadratic in $\pi$:
%$H = (P + \int\pi\chi')^2/(2M_0) + \ldots$. Our action
%has the same structure, with
%$-i\lambda\partial_t\rho$ as the symplectic term and
%$\frac{1}{2}\lambda\Omega\lambda$ as the quadratic
%kinetic energy of the momentum.
%
%In the GJS Hamiltonian formulation, two FP subsidiary
%conditions ($F_1$ for $\phi$, $F_2$ for $\pi$, GJS
%Eqs.~(2.5) and~(2.8)) are needed because $\phi$ and $\pi$
%are independent canonical variables. The same logic would
%apply to our $(\rho, \lambda)$ system. However, since
%$S_{\text{eff}}$ is quadratic in $\lambda$, there is a
%simpler alternative: integrate $\lambda$ out exactly,
%reducing to a single-field theory for $\rho$ alone, with
%only one FP condition per collective coordinate.

\paragraph{Integrating out $\lambda$: reduction to a
single-field theory.}
Since $S_{\text{eff}}[\rho, \lambda]$ is quadratic in
$\lambda$ (linear + quadratic terms), we can integrate
$\lambda$ out exactly by completing the square, just as we
integrated out $G_i$. We introduce the total time
derivative operator
\begin{equation}
\label{eq:cov_deriv}
\mathcal{D}_t\rho(x,t)
\equiv \partial_t\rho(x,t)
+ \nabla_i\!\left(F^i[\rho]\,\rho\right),
\end{equation}
which combines the bare time derivative with the
self-consistent drift. The $\lambda$-dependent part
of~\eqref{eq:S_eff_rho_lambda} is then
$-i\lambda\,\mathcal{D}_t\rho
+ \frac{1}{2}\lambda\Omega\lambda$.
%Completing the square:
%\begin{equation}
%-i\lambda\,\mathcal{D}_t\rho
%+ \tfrac{1}{2}\lambda\Omega\lambda
%= \tfrac{1}{2}(\lambda - i\Omega^{-1}\mathcal{D}_t\rho)
%\,\Omega\,
%(\lambda - i\Omega^{-1}\mathcal{D}_t\rho)
%+ \tfrac{1}{2}\mathcal{D}_t\rho\,\Omega^{-1}\,
%\mathcal{D}_t\rho.
%\end{equation}
%Shifting $\lambda \to \lambda
%- i\Omega^{-1}\mathcal{D}_t\rho$, 
The
Gaussian integral over $\lambda$ gives
$(\det\Omega)^{-1/2}
\exp(-\frac{1}{2}\mathcal{D}_t\rho\,\Omega^{-1}\,
\mathcal{D}_t\rho)$, producing the single-field effective
action
\begin{equation}
\label{eq:S_single_field}
S_{\text{eff}}[\rho]
= \frac{1}{2}\!\iint\mathcal{D}_t\rho\,\Omega^{-1}\,
\mathcal{D}_t\rho
- \alpha^2\!\iint\rho\hat{\mathcal{K}}\rho,
\end{equation}
where $\Omega^{-1}$ is the inverse of the noise-induced
kernel~\eqref{eq:Omega_kernel}. In writing
\eqref{eq:S_single_field} we have omitted the functional
determinant $\frac{1}{2}\log\det'\Omega[\rho]$ produced by the
Gaussian $\lambda$-integration. Because $\Omega$ is
state-dependent through factors $\nabla[\hat{\mathcal{A}}\rho]$,
this term is not generally a constant; it contributes at
one-loop order to the saddle and fluctuation Hessian. We
neglect it in the leading-saddle / Freidlin-Wentzell
approximation used throughout this section, and absorb its
low-energy effect into the phenomenological constants used
below; in particular, since $D_{\text{rot}}$ is fitted from
simulation in Section~\ref{subsec:comparison}, any one-loop
measure correction to it is implicit in the fitted value.
The kernel $\Omega[\rho]$ also inherits a zero mode from the
conservation of total density: the constant mode
$\delta\rho(x,t) = \text{const}$ has $\mathcal{D}_t\rho = 0$,
so $\lambda$ couples only to mass-conserving fluctuations.
The inverse $\Omega^{-1}$ should accordingly be understood on
the subspace orthogonal to this conserved zero mode, with the
constant mode either gauge-fixed at the level of the
Lagrange-multiplier integral or trivially decoupled in the
saddle expansion below; the rest of the construction is
unaffected. This is a nonlinear
field theory for $\rho$ with a ``kinetic'' term
$\frac{1}{2}(\mathcal{D}_t\rho)\Omega^{-1}
(\mathcal{D}_t\rho)$ built from the total time
derivative~\eqref{eq:cov_deriv}. The effective action (\ref{eq:S_single_field}) can also be written in an 
expanded form:
\begin{equation}
\label{eq:S_single_expanded}
S_{\text{eff}}[\rho] = 
\iint \left[ \frac{1}{2} (\partial_t\rho)\,\Omega^{-1}\,
(\partial_t\rho) +
 (\partial_t\rho)\,\Omega^{-1}\,
\nabla_i(F^i\rho) +
\frac{1}{2} \nabla_i(F^i\rho)\,
\Omega^{-1}\,\nabla_j(F^j\rho) -
\alpha^2  \rho\hat{\mathcal{K}}\rho \right]
\end{equation}
%\begin{align}
%\label{eq:S_single_expanded}
%S_{\text{eff}}[\rho]
%&= \frac{1}{2}\iint(\partial_t\rho)\,\Omega^{-1}\,
%(\partial_t\rho)
%&& \text{(kinetic term, state-dependent
%``mass'' $\Omega^{-1}[\rho]$)} \nonumber \\
%&\quad + \iint(\partial_t\rho)\,\Omega^{-1}\,
%\nabla_i(F^i\rho)
%&& \text{(drift-kinetic cross term)} \nonumber \\
%&\quad + \frac{1}{2}\iint\nabla_i(F^i\rho)\,
%\Omega^{-1}\,\nabla_j(F^j\rho)
%&& \text{(non-local self-interaction)} \nonumber \\
%&\quad - \alpha^2\iint\rho\hat{\mathcal{K}}\rho.
%&& \text{(quadratic density interaction)}
%\end{align}
The first term is the kinetic energy with a
state-dependent mass $\Omega^{-1}[\rho]$ (the kernel
$\Omega$~\eqref{eq:Omega_kernel} depends on $\rho$ through
the factors $\nabla_k[\hat{\mathcal{A}}^{ki}\rho]$, which
makes $\Omega$ bilinear in $\rho$). The second term is a
drift-kinetic cross coupling between the time evolution
and the self-consistent drift. The third term is a
\emph{non-local nonlinear self-interaction} of the
density: since the drift
$F^i[\rho]$~\eqref{eq:self_consistent_drift} is linear in
$\rho$, the combination $F^i\rho$ is quadratic in $\rho$,
producing a quartic in $\rho$ contribution to the term 
$\nabla(F\rho)\,\Omega^{-1}\,\nabla(F\rho)$ for a fixed $\Omega^{-1}$. However, the inverse kernel $\Omega^{-1}$ also depends on 
$ \rho $ as the initial kernel $\Omega$ is bilinear in $ \rho $. 
%at leading order, i.e.\ when $\Omega^{-1}$ is
%treated as a reference kernel evaluated at some
%configuration. The residual $\rho$-dependence of
%$\Omega^{-1}$ (through $\Omega$'s bilinear $\rho$ factors)
This generates an infinite tower of higher-order corrections
$O(\rho^6), O(\rho^8), \ldots$, via the formal expansion
$\Omega^{-1}[\rho] = \Omega_0^{-1}
- \Omega_0^{-1}\,\Omega_1\,\Omega_0^{-1}
+ \ldots$ with $\Omega_1$ linear and $\Omega_2$ quadratic
in $\delta\rho = \rho - \rho_0$. The fourth term is the
quadratic self-interaction of the density generated by the
disorder-averaged pair potential. Together, the third and
fourth terms realize the F2 model as a nonlinear (non-polynomial)
 theory for $\rho$ that can be expanded into an infinite series 
of terms that are non-local in both space and time 
through their respective kernels
$\hat{\mathcal{K}}$ and $\Omega^{-1}$.
The resulting path integral 
\begin{equation}
\label{eq:Z_single_field}
Z = \text{const}\times\int\!\mathcal{D}\rho\;
\exp\!\left(-S_{\text{eff}}[\rho]\right),
\end{equation}
describes a nonlinear non-local field theory for the single field $\rho(x,t)$.
%The EL equation for $\rho$ and the collective-coordinate
%decomposition are presented in
%Sections~\ref{subsec:faddeev_popov}--\ref{subsec:zero_modes}
%below.

%==============================================================================
\section{Low-Energy Effective F2 Theory on the sphere $S^2$}
\label{sec:ring_precession}
\label{sec:low_energy}
%==============================================================================

Section~\ref{sec:large_N} constructed the F2 model for a generic
compact two-dimensional Riemannian manifold (closed, or with
boundary subject to appropriate boundary conditions) and for any pair
potential depending on the geodesic distance. In this section we
specialize to the case in which the manifold is the sphere $S^2$.
On the sphere, the high degree of symmetry (the full
$\mathrm{SO}(3)$ rotation group acts isometrically on the
manifold and hence on all ingredients of the F2 action) provides
a powerful organizing principle for the low-energy reduction: as
we show below, symmetry under $\mathrm{SO}(3)$ constrains the
form of the reduced theory strongly enough that the low-energy
action is fixed up to a single low-energy constant, the
rotational diffusion coefficient $D_{\text{rot}}$.

As described in Ref.~\cite{halperin2026frustrated}, after a short
stage of non-equilibrium relaxation the particle density on $S^2$
settles on a great-circle ring whose orientation
$\hat{\mathbf{n}}(t) \in S^2$ adiabatically breaks
$\mathrm{SO}(3) \to \mathrm{SO}(2)$ and executes a slow stochastic
precession. In the long-time (low-energy) limit, the spatial
degrees of freedom are slaved to this orientation: the
infinite-dimensional F2 field theory reduces to a finite-dimensional
stochastic mechanics on the orientation manifold
$\mathbb{RP}^2 = S^2/\mathbb{Z}_2$ (with $S^2$ used as the
gauge-fixed double cover; see Section~\ref{subsec:low_energy_ansatz}).
This reduction is analogous to the mean-field limit of spin glass
theory, where the spatial dependence is dropped and the field
theory reduces to a single-site (quantum-mechanical) problem with
self-consistent memory kernels
\cite{cugliandolo1993, facoetti2019}. In our setting, the proper framework is a Born-Oppenheimer
(adiabatic) approximation: the fast degrees of freedom (density
excitations along and across the ring) equilibrate on the
timescale $1/m^2_{\text{gap}}$, while the slow degree of freedom
$\hat{\mathbf{n}}(t)$ obeys an effective $\mathbb{RP}^2$-NLSM
dynamics whose parameters are determined by the fast
equilibrium. We carry out the reduction systematically using
the method of collective coordinates in the path integral,
starting from the generating functional~\eqref{eq:gen_func_rhoG}
and arriving at the nonlinear sigma model (NLSM) on the
real projective plane $S^2/\mathbb{Z}_2 = \mathbb{RP}^2$
(the $\mathbb{RP}^2$ NLSM on the projective rotor space)
in $(0+1)$ dimensions (with the signed inertia-tensor
eigenvector providing a continuous $S^2$-valued orientation
$\hat{\mathbf{n}}(t)$, defined and discussed in
Section~\ref{subsec:low_energy_ansatz}).

\paragraph{The collective coordinate method and
its adaptation to the F2 model.}
The method we use is the collective-coordinate
expansion around extended-particle states developed by
Gervais, Jevicki, and Sakita
\cite{gervais1975collective} (GJS) for the quantization
of solitons in nonlinear field theories; we follow the
textbook treatments of
Refs.~\cite{rajaraman1982, coleman_aspects} throughout. In the GJS
treatment, the kink solution $\phi_0(x - X)$ of the
$\phi^4$ theory is parametrized by a collective
coordinate $X(t)$ (the kink position), and the field is
decomposed as $\phi(t,x) = \phi_0(x - X(t)) + \eta(t,x)$
with $\eta$ the fluctuation. A Faddeev-Popov identity
extracts $X$ from the path integral and fixes the gauge
by requiring $\eta$ to have no projection onto the zero
mode $\phi_0'(x - X)$ (the derivative of the kink profile
with respect to the invariant combination $x - X$). The
zero mode arises because the equation of motion is
translationally invariant, so $\phi_0(x - X)$ is a
solution for every $X$, and differentiating with respect
to $X$ gives a null eigenvector of the linearized
operator.

In our F2 model, the role of the kink is played by the
ring density $\rho_{\hat{\mathbf{n}}}(x)
= f_0(\hat{\mathbf{n}}\cdot\mathbf{x})$, and
the collective coordinate is the orientation
$\hat{\mathbf{n}}(t) \in S^2$. The structural
correspondence is:
\begin{center}
\begin{tabular}{l|l}
GJS (kink, translation) & F2 (ring, rotation) \\
\hline
Invariant: $x - X$ &
Invariant: $\hat{\mathbf{n}}\cdot\mathbf{x}$ \\
Solution: $\phi_0(x - X)$ &
Solution: $f_0(\hat{\mathbf{n}}
\cdot\mathbf{x})$ \\
Collective coord: $X \in \mathbb{R}$ &
Collective coord: $\hat{\mathbf{n}} \in S^2$ \\
Broken generators: 1 (translation) &
Broken generators: 2 (SO(3)/SO(2)) \\
Zero mode: $\phi_0' = -\partial\phi_0/\partial X$ &
Zero modes: $\psi_a
= \partial\rho_{\hat{\mathbf{n}}}/\partial n^a$ \\
FP: $\int\phi_0'\eta\,dx = 0$ (1 cond.) &
FP: $\int\psi_a\delta\rho\,d\mu = 0$ (2 cond.) \\
FP det: $M_0 = \int(\phi_0')^2$ &
FP det: $G_0 = \pi\!\int\!(f_0')^2\sin\theta\,d\theta$ \\
\end{tabular}
\end{center}
The key difference is that the GJS kink lives on
$\mathbb{R}$ and breaks a one-parameter translational
symmetry, producing one zero mode, while our ring lives on
$S^2$ and breaks SO(3) to SO(2), producing two zero modes
(the two independent tilts of the ring). Correspondingly,
the invariant combination changes from the linear
$x - X$ (translation) to the bilinear
$\hat{\mathbf{n}}\cdot\mathbf{x}$ (rotation). The
Faddeev-Popov procedure, the eigenfunction expansion with
primed sums excluding zero modes, and the integration over
the collective coordinate all carry over from GJS with
this substitution.

%% ============================================================
\subsection{Euler-Lagrange equations and ring solutions}
\label{subsec:low_energy_ansatz}

%\paragraph{The ring and the collective coordinate}

As described in \cite{halperin2026frustrated}, after a
short stage of non-equilibrium relaxation, which proceeds via
instanton-like non-perturbative mechanisms, the particle density
concentrates on a ring (great circle) on $S^2$. The ring
orientation adiabatically breaks the full SO(3) symmetry of the model
down to SO(2) (azimuthal rotations about the ring normal).
The manifold of degenerate ring configurations is
$\text{SO}(3)/\text{SO}(2) \simeq S^2$, parametrized by
the unit normal $\hat{\mathbf{n}}(t)$, and slowly explored by its adiabatic diffusion. 

\paragraph{The Euler-Lagrange equation for the soliton
profile.}
A saddle-point (``classical'') density $\rho_0$ of the
single-field action would satisfy the Euler-Lagrange equation
\begin{equation}
\label{EL_equation_rho_0}
\frac{d}{dt} \frac{\delta S_{\text{eff}}}{ \delta(\partial_t\rho)}
- \frac{\delta S_{\text{eff}}}{ \delta(\rho)} = 0
\end{equation}
We do not solve \eqref{EL_equation_rho_0}. It is an
integro-differential equation in which the kernels
$\Omega^{-1}[\rho]$, $\hat{\mathcal{K}}$, and $F^i[\rho]$ are
themselves functionals of $\rho$ through the disorder-averaged
self-consistency, and its analysis under any explicit ansatz
is left to future work. We use \eqref{EL_equation_rho_0}
below only as the saddle equation that the
simulation-supported ring profile $f_0$ would have to satisfy.
For the action~\eqref{eq:S_single_field},
$\delta S_{\text{eff}}/\delta(\partial_t\rho)
= \Omega^{-1}\mathcal{D}_t\rho$ (from the
$\frac{1}{2}\mathcal{D}_t\rho\,\Omega^{-1}\,
\mathcal{D}_t\rho$ term), and
$\delta S_{\text{eff}}/\delta\rho$ receives
contributions from the density self-interaction
$-\alpha^2\hat{\mathcal{K}}\rho$, the drift-dependent
terms in $\mathcal{D}_t\rho$, and the $\rho$-dependence
of $\Omega^{-1}$ itself (through
$\nabla_k[\hat{\mathcal{A}}^{ki}\rho]$ in the
kernel~\eqref{eq:Omega_kernel}). The EL equation (\ref{EL_equation_rho_0}) takes the following form:
\begin{equation}
\label{eq:EL_explicit}
\frac{d}{dt}\!\left[\Omega^{-1}\mathcal{D}_t\rho\right]
+ 2\alpha^2\!\int\!\hat{\mathcal{K}}\rho
- \frac{\delta}{\delta\rho}\left[
\frac{1}{2}\mathcal{D}_t\rho\,\Omega^{-1}\,
\mathcal{D}_t\rho\right]_{\text{explicit}} = 0,
\end{equation}
where the last term denotes the variation of the kinetic
term with respect to the explicit $\rho$-dependence
(through $F^i[\rho]$ in $\mathcal{D}_t\rho$ and through
$\Omega^{-1}[\rho]$), at fixed $\partial_t\rho$.

% \paragraph{The ring in the adapted frame}
%
% The analysis of ring dynamics is simplified by rotating
% to the adapted frame where $\hat{\mathbf{n}}(t)$ is the
% north pole. This rotation can be performed using
% Cayley-Klein parameters
% \cite{pennestri2016,cottingham2001} or Euler angles
% $(\alpha, \beta, \gamma)$.
% If $(\Theta, \Phi)$ are the lab-frame spherical
% coordinates and $(\theta, \phi)$ are the adapted-frame
% coordinates, the passive transformation
% $\mathbf{x} = R^T\mathbf{x}'$ gives:
% \begin{align}
% \label{eq:coord_transform_sec4}
% \cos\theta
%   &= \sin\alpha\,\sin\Theta\,\cos(\Phi - \beta)
%    + \cos\alpha\,\cos\Theta \nonumber \\
% \sin\theta\,\cos\phi
%   &= \cos\alpha\,\sin\Theta\,\cos(\Phi - \beta)
%    - \sin\alpha\,\cos\Theta \\
% \sin\theta\,\sin\phi
%   &= \sin\Theta\,\sin(\Phi - \beta) \nonumber
% \end{align}
% where $(\alpha, \beta)$ are the polar and azimuthal angles
% of $\hat{\mathbf{n}}$ and $\gamma = 0$ (Faddeev-Popov gauge
% fixing of Appendix~D). The ring band is
% $\pi/2 - \Delta\theta \leq \theta \leq \pi/2
% + \Delta\theta$, $0 \leq \phi < 2\pi$.

\paragraph{The ring density as a function of
$\hat{\mathbf{n}}\cdot\mathbf{x}$.}
A note on notation. We write $\mathbf{x}\in\mathbb{R}^3$
(bold) for the lab-frame Euclidean position vector of a
point on the physical sphere, with $|\mathbf{x}|^2 = 1$;
this is the 3D embedding-space coordinate. Where we
need it as the argument of fields such as
$\rho(x, t)$, the same point is alternatively
parametrized by intrinsic 2D coordinates $x = (x^1, x^2)$
on $S^2$, in the convention of
Appendix~\ref{app:covariant_langevin}. Likewise
$\hat{\mathbf{n}}\in\mathbb{R}^3$ (bold) is the 3D unit
vector specifying the ring orientation (equivalently, a
point on the orientation sphere
$S^2_{\hat{\mathbf{n}}}$); we use the same symbol for
both. The dot product $\hat{\mathbf{n}}\cdot\mathbf{x}$
is the standard $\mathbb{R}^3$ inner product of the two
3-vectors.

The ring density $\rho_{0}(x)$ depends on $\mathbf{x}$
and $\hat{\mathbf{n}}$ only through the scalar product
$\hat{\mathbf{n}}\cdot\mathbf{x}$. This follows from the
residual SO(2) symmetry: the ring is invariant under
azimuthal rotations about $\hat{\mathbf{n}}$, and the
only scalar that can be formed from $\mathbf{x}$ and
$\hat{\mathbf{n}}$ that is invariant under these
rotations is $\hat{\mathbf{n}}\cdot\mathbf{x}$ (because
the SO(2) rotations about $\hat{\mathbf{n}}$ preserve
$\hat{\mathbf{n}}\cdot\mathbf{x}$ while mixing the
transverse components of $\mathbf{x}$). We
\emph{posit}, motivated by the particle simulations of
Ref.~\cite{halperin2026frustrated}, the ring-saddle ansatz
\begin{equation}
\label{eq:rho_n}
\rho_{0}(x,t)
= f_0(\hat{\mathbf{n}}\cdot\mathbf{x}),
\end{equation}
where $f_0$ is a function of a single scalar variable
(the profile of the ring). Substitution
of~\eqref{eq:rho_n} into the EL
equation~\eqref{EL_equation_rho_0} does not yield a closed
scalar equation for $f_0$ alone: it yields a projected
self-consistency problem coupling $f_0$ to the
disorder-averaged two-time correlators $C_\rho$, $K_{ij}$,
and $R_i$ that obey their own Schwinger-Dyson closures.
Existence and stability of an SO(2)-symmetric saddle of the
full coupled system are likewise not established here; we
treat \eqref{eq:rho_n} as a simulation-supported ansatz and
use its empirical validity (a Gaussian-like profile of width
$\sigma \approx 4.8^\circ$, Section~\ref{subsec:comparison})
as a fixed input to the construction below.

\paragraph{Target space: $\mathbb{RP}^2$ vs.\ $S^2$.}
Strictly, the simulation-supported profile~\eqref{eq:rho_n}
is even, $f_0(u) = f_0(-u)$, since a great-circle band has
the same density on the two hemispheres. The configurations
labelled by $\hat{\mathbf{n}}$ and $-\hat{\mathbf{n}}$
therefore describe the same physical density, and the
physical orientation space is $\mathbb{RP}^2 = S^2/\mathbb{Z}_2$
rather than $S^2$. The orientation field $\hat{\mathbf{n}}$
is therefore a \emph{director} in the standard
$\mathbb{RP}^{N-1}$ NLSM sense: an $N = 3$ component unit
vector with the $\mathbb{Z}_2$ identification
$\hat{\mathbf{n}} \sim -\hat{\mathbf{n}}$, equivalent to
either the lattice $\mathbb{Z}_2$-gauged $O(N)$ formulation
of Hasenfratz--Niedermayer~\cite{lammert1995} or the
Landau--de~Gennes $Q$-tensor description
$Q_{ij} = n_i n_j - \delta_{ij}/3$
familiar from nematic liquid
crystals~\cite{degennesprost} (see also~\cite{nakahara2003geometry}
for the underlying topology of $\mathbb{RP}^2$). Components $n_i$ are
well-defined only in a local representative (a lift of
$\mathbb{RP}^2$ to $S^2$); the strictly physical content lies
in $\mathbb{Z}_2$-invariant combinations such as
$Q_{ij}$, $|q(\tau)|$, and even-$\ell$ Legendre modes.

Throughout the rest of this section we
work with the \emph{signed inertia-tensor eigenvector}, by
which we mean the following: the smallest-eigenvalue
eigenvector of the empirical inertia tensor
$I_{ij}(t) = (1/N)\sum_n x_n^i(t)x_n^j(t)$ is a unit line
in $\mathbb{R}^3$ (defined only up to an overall sign,
since both $\pm\hat{\mathbf{n}}$ are eigenvectors with the
same eigenvalue) and represents a point in $\mathbb{RP}^2$;
to track this line as a continuous-in-time function, we pick
a sign at the first recorded snapshot and at each subsequent
snapshot choose the sign that minimises the angular jump from
the previous step (continuity convention). The result is a
continuous map
$\hat{\mathbf{n}}: [0, T] \to S^2$ whose equivalence class
$[\hat{\mathbf{n}}](t) \in \mathbb{RP}^2$ is the strictly
physical orientation. In topological language,
$\hat{\mathbf{n}}(t)$ is a \emph{lift} of the physical
$\mathbb{RP}^2$-valued trajectory $[\hat{\mathbf{n}}](t)$
through the double-cover map
$\pi: S^2 \to \mathbb{RP}^2 = S^2/\mathbb{Z}_2$ — a
continuous $S^2$-valued curve such that
$\pi(\hat{\mathbf{n}}(t)) = [\hat{\mathbf{n}}](t)$ at every
time~\cite{mermin1979,nakahara2003geometry}. Two such lifts exist for any given
trajectory on $\mathbb{RP}^2$ (related by the global sign flip
$\hat{\mathbf{n}} \to -\hat{\mathbf{n}}$ at the initial
time), and the continuity convention is the standard
$\mathbb{Z}_2$ gauge fixing that picks one of them. We use
the term ``lift'' in this sense throughout the paper, in
the convention of Mermin's classification of defects in
ordered media~\cite{mermin1979}. The signed inertia-tensor eigenvector
is the object used to define the signed orientation
autocorrelation $q(\tau) = \hat{\mathbf{n}}(t)\cdot
\hat{\mathbf{n}}(t+\tau)$, and more generally any $S^2$-formula
appearing in the body of this paper.
Strictly $\mathbb{RP}^2$-invariant diagnostics use $|q(\tau)|$
or the tensor correlator
$\langle n_i(t)n_j(t+\tau)\rangle$; in the diffusive regime
of the present simulations $q(\tau)$ stays positive and the
signed and unsigned diagnostics agree. Density-sector observables built
from $\rho_0$ are $\mathbb{Z}_2$-invariant and depend only on
the even Legendre modes (consistent with the absence of
odd-$\ell$ contributions in the closed form for $C_\rho$
derived in Section~\ref{subsec:corr_orientation}). The local
effective action and the leading kinetic term are unchanged
by the global identification, so the SO(3)-based reduction of
this section goes through unchanged; the global target-space
topology is $\mathbb{RP}^2$.\footnote{The faithfully acting
symmetry of the $\mathbb{RP}^2$ NLSM is $\mathrm{SO}(3)$
rather than $\mathrm{O}(3)$. Although the kinetic action
$|\dot{\hat{\mathbf{n}}}|^2$ is invariant under all of
$\mathrm{O}(3)$, the parity element $-I \in \mathrm{O}(3)$
is precisely the gauge identification taking $S^2$ to
$\mathbb{RP}^2 = S^2/\mathbb{Z}_2$ and so acts as the
identity on the target. The faithful symmetry on the
physical orientation manifold is therefore
$\mathrm{O}(3)/\{I,-I\} = \mathrm{SO}(3)$. The standard
``$\mathrm{O}(3)$ NLSM'' of the particle-physics literature
has target $S^2$, where parity acts non-trivially; the
director-target version with target $\mathbb{RP}^2$ and
faithful symmetry $\mathrm{SO}(3)$ matches the conventional
naming of nematic-type sigma models.}

The $\mathbb{Z}_2$-invariance is not a kinematic accident of
the inertia-tensor pipeline but a consequence of the
microscopic dynamics. The Langevin equation
\eqref{eq:langevin_particles} with symmetric quenched
couplings $\phi_{nm} = \phi_{mn}$ obeys detailed balance with
respect to the Boltzmann distribution
$\propto e^{-U/T}$, and the deterministic pairwise force
exerts zero net torque on a great-circle ring (a property of
the linear-in-distance potential proved in
Ref.~\cite{halperin2026frustrated}). The slow orientation
$\hat{\mathbf{n}}(t)$ therefore evolves by purely thermal,
parity-symmetric Wiener noise, with no persistent
circulation along the ring and no net angular velocity. This
is what allows us to integrate out the fast in-ring motion
into a slow theory on $\mathbb{RP}^2$, and is the dynamical
content of the no-Berry-phase argument in
Section~\ref{subsec:corr_orientation}: an extension of the
F2 model with explicit time-reversal breaking (active
particles, chirally driven probes, antisymmetric quenched
couplings) would generate a non-zero average circulation, an
antisymmetric component of the kinetic kernel, and lift the
target back to $S^2$ with a Berry-phase term in the
effective action.

\paragraph{Topological aspects of the $\mathbb{RP}^2$ target.}
\label{para:rp2_topology}
With $\mathbb{RP}^2 = S^2/\mathbb{Z}_2$ the
\emph{real projective plane} as the physical target manifold,
the resulting effective theory is the standard
\emph{$\mathbb{RP}^2$ nonlinear sigma model} (henceforth
$\mathbb{RP}^2$ NLSM) familiar from the soft-matter and
lattice-QFT literature on director / nematic order
parameters~\cite{lammert1995, degennesprost}: the same local
$SO(3)$-invariant kinetic action as the $O(3)$ NLSM on $S^2$
but with the $\mathbb{Z}_2$ identification
$\hat{\mathbf{n}} \sim -\hat{\mathbf{n}}$ in the global target.
We adopt this name in what follows. The non-trivial topology of
the orientation manifold, $\pi_1(\mathbb{RP}^2) = \mathbb{Z}_2$,
has three consequences for the $\mathbb{RP}^2$ NLSM that are
otherwise scattered across the rest of the paper.

(i) \emph{Two homotopy classes of closed orientation loops.}
A trajectory $\hat{\mathbf{n}}: [0,T] \to \mathbb{RP}^2$ with
$[\hat{\mathbf{n}}(0)] = [\hat{\mathbf{n}}(T)]$ falls into one
of two homotopy classes labelled by $w \in \mathbb{Z}_2$:
$w = 0$ (the lifted $\hat{\mathbf{n}}(t) \in S^2$ returns to
$+\hat{\mathbf{n}}(0)$) or $w = 1$ (the lifted trajectory
returns to $-\hat{\mathbf{n}}(0)$, a non-contractible
``$\pi$-rotation'' loop).
The relevant topological diagnostic is the $\mathbb{Z}_2$-valued
Wilson line $W_C = \exp(i\pi w[C])$ for closed orientation
loops $C$. In the diffusive regime of the present
simulations ($q(\tau) > 0$ throughout the run, see
Section~\ref{subsec:comparison}), the lifted dynamics stays
in the trivial class $w = 0$ and the topology has no
quantitative consequence for the orientation diagnostics
reported below. On the diffusive side, the same structure is
captured by the method-of-images representation of the heat
kernel,
$P_{\mathbb{RP}^2}(\mathbf{x},\mathbf{y};\tau)
= P_{S^2}(\mathbf{x},\mathbf{y};\tau)
+ P_{S^2}(\mathbf{x},-\mathbf{y};\tau)$,
which makes the spectral sum even-$\ell$ only by
construction. The lowest strictly $\mathbb{Z}_2$-invariant
relaxation rate is therefore $\lambda_2 = 6\,D_{\text{rot}}$,
three times the gauge-fixed $\ell = 1$ rate
$\lambda_1 = 2\,D_{\text{rot}}$ that controls the signed
autocorrelation $q(\tau)$ on the $S^2$-lift.

(ii) \emph{Two quantization sectors.}
At the path-integral level the partition function decomposes
as $Z = \sum_{w \in \mathbb{Z}_2} e^{i\theta w}\,Z_w$, with
$\theta \in \{0, \pi\}$ a quantization choice that distinguishes
two inequivalent quantum theories on $\mathbb{RP}^2$:
$\theta = 0$ (trivial, single-valued wavefunctions, Hilbert
space spanned by even-$\ell$ spherical harmonics) and
$\theta = \pi$ (spinor-like, double-valued wavefunctions,
Hilbert space spanned by odd-$\ell$ spherical harmonics).
This is the same algebraic structure that distinguishes
integer-spin from half-integer-spin representations of
$\mathrm{SU}(2)$, here arising as a quantization ambiguity on
the orientation manifold itself rather than from an internal
spin. The F2 model lives in the $\theta = 0$ sector: its
density profile is parity-even, all physical observables are
$\mathbb{Z}_2$-invariant, and no microscopic spin-1/2 is
present. Quantum implementations of the orientation
(Section~\ref{subsec:connections}) therefore inherit the
even-$\ell$ Hilbert space; the odd-$\ell$ sector would
require a separate physical realization that supplies a
spin-1/2-like degree of freedom on the orientation.

(iii) \emph{Selection rule on angular momentum harmonics.}
The two facts above are unified by the same selection rule:
single-valued functions on $\mathbb{RP}^2$ are spanned by
spherical harmonics of even $\ell$ only. Concretely, the
density profile $f_0$ has Legendre coefficients $c_\ell = 0$
for odd $\ell$, the density-density correlator
$C_\rho(\gamma,\tau)$ contains only even-$\ell$ Legendre
modes (Eq.~\eqref{eq:rp2_selection_rule} below; the relaxation
spectrum $\lambda_\ell = \ell(\ell+1)D_{\text{rot}}$ for
$\ell = 0, 2, 4, \ldots$ is the eigenvalue spectrum of the
Laplace-Beltrami operator on $\mathbb{RP}^2$), and any
quantum implementation has accessible level structure
$E_\ell = \ell(\ell+1)\hbar^2/(2\mathcal{I})$ for even $\ell$.
The signed orientation autocorrelation
$q(\tau) = \hat{\mathbf{n}}(t)\cdot\hat{\mathbf{n}}(t+\tau)$
appears to violate this rule (it is the $\ell = 1$ projection
of the dot product of the lifted vectors); but $q(\tau)$ is a
property of the gauge-fixed $S^2$-lift, not of the physical
$\mathbb{RP}^2$-trajectory, and the strictly physical
even-$\ell$ analog is the tensor correlator
$\langle n_i(t)n_j(t+\tau)\rangle$, which lives in the
$\ell = 2$ sector.

For our (0+1)-dimensional model, $\pi_2(\mathbb{RP}^2) =
\mathbb{Z}$ has no role. It would matter for a spatial
extension of the F2 sigma model, where the orientation field
is parameterised by points on a 2D substrate; there
$\pi_1(\mathbb{RP}^2) = \mathbb{Z}_2$ supports topologically
stable \emph{half-integer disclinations} of the director field
(the standard $\pi$-rotation defects of nematic liquid
crystals; two such defects annihilate to a smooth configuration
as $1 + 1 = 0 \mod 2$), and $\pi_2$ supports hedgehog and
skyrmion-like textures with quantized charge in $\mathbb{Z}$,
which becomes $\mathbb{Z}/\mathbb{Z}_2 = \mathbb{N}$ once
equivalence under the non-trivial element of $\pi_1$ is taken
into account. These are open targets for future work beyond the
(0+1)-dimensional reduction of the present paper.

\paragraph{Gauge-theory reformulation: $\mathbb{RP}^2$ NLSM
as $S^2$ NLSM coupled to a $\mathbb{Z}_2$ gauge field.}
The $\mathbb{Z}_2$ identification
$\hat{\mathbf{n}} \sim -\hat{\mathbf{n}}$ on the orientation
target can be recast as a discrete $\mathbb{Z}_2$ gauge
symmetry of a lifted $S^2$ description. In the standard
lattice/QFT formulation~\cite{lammert1995}, the model is
written as an $O(N)$ unit-vector $\hat{\mathbf{n}}_i \in S^{N-1}$
on each (discrete) time slice $i$, together with
$\mathbb{Z}_2$ link variables $U_{ij} = \pm 1$ on each
time-step bond, with the action
$S = -\beta \sum_{\langle i j\rangle} U_{ij}\,
\hat{\mathbf{n}}_i\!\cdot\!\hat{\mathbf{n}}_j$ and the local
$\mathbb{Z}_2$ gauge transformation
$\hat{\mathbf{n}}_i \to \epsilon_i \hat{\mathbf{n}}_i$,
$U_{ij} \to \epsilon_i\epsilon_j U_{ij}$ ($\epsilon_i = \pm 1$).
Summing over the $\mathbb{Z}_2$ gauge field $U_{ij}$
projects onto $\mathbb{Z}_2$-gauge-invariant configurations
and reproduces the $\mathbb{RP}^{N-1}$ NLSM partition
function. In the continuum (0+1)-dimensional limit relevant
here ($N = 3$), the $\mathbb{Z}_2$ gauge field is a
flat connection on the time line and its only physical
degree of freedom is its $\mathbb{Z}_2$ holonomy along
closed orientation loops, exactly the winding class
$w \in \pi_1(\mathbb{RP}^2) = \mathbb{Z}_2$ of
Eq.~\eqref{eq:Z_RP2_sectors}.

The signed-inertia-tensor-eigenvector construction we use
in this paper is the operational realization of the
\emph{temporal} gauge fixing of this $\mathbb{Z}_2$ gauge: the
continuity convention picks $U_{ij} = +1$ on every bond
along the trajectory, leaving a single $S^2$-valued field
$\hat{\mathbf{n}}(t)$ as the gauge-fixed representative.
The dictionary between the two equivalent geometric
formulations of the same $\mathbb{Z}_2$ structure, expressed
in the continuum (0+1)-dimensional language used throughout
our paper, is
\begin{center}
{\small
\begin{tabular}{@{}p{0.18\textwidth}|p{0.40\textwidth}|p{0.34\textwidth}@{}}
\hline
& Lifted $S^2$ with $\mathbb{Z}_2$ gauge
& Projective $\mathbb{RP}^2$ target \\
\hline
Field
& $\hat{\mathbf{n}}(t) \in S^2$
& $[\hat{\mathbf{n}}(t)] \in \mathbb{RP}^2$ \\
Local gauge invariance
& $\hat{\mathbf{n}}(t) \to \epsilon(t)\,\hat{\mathbf{n}}(t)$
with $\epsilon(t) = \pm 1$
& none (target identifies $\pm\hat{\mathbf{n}}$) \\
Loop holonomy
& $\mathbb{Z}_2$-valued holonomy of the lifted curve along
closed orientation loops
& winding class $w \in \pi_1(\mathbb{RP}^2)$ \\
$\theta$-angle
& $\theta$-term weighing the two homotopy classes of the lift
& quantization sector $\theta \in \{0, \pi\}$ \\
Gauge fixing
& continuity convention on the lifted trajectory
& used in our construction of the signed orientation
vector $\hat{\mathbf{n}}$ \\
\hline
\end{tabular}
}
\end{center}
The $\mathbb{Z}_2$ link variables $U_{ij}$ of the lattice
ancestor in Ref.~\cite{lammert1995} are not carried in the
continuum: in (0+1) dimensions a flat $\mathbb{Z}_2$
connection has no local degrees of freedom, and its only
physical content is the $\mathbb{Z}_2$ holonomy along closed
orientation loops, recorded in the "Loop holonomy" row.
Our concrete construction of $\hat{\mathbf{n}}(t)$ from the
inertia tensor with the continuity sign convention is then
exactly the gauge-fixed special case in which the lifted
trajectory is single-valued throughout, equivalently the
particular continuous lift of $[\hat{\mathbf{n}}(t)]$
selected by the same continuity rule on the right column.

Every topological feature of the $\mathbb{RP}^2$ structure
(the $\mathbb{Z}_2$ identification, the
$\pi_1 = \mathbb{Z}_2$ winding class, the two quantization
sectors $\theta \in \{0, \pi\}$, the lift as gauge fixing) is
the same algebraic structure expressed in three equivalent
languages: target quotient ($\mathbb{RP}^2$ NLSM), lifted
gauge field ($S^2$ NLSM with $\mathbb{Z}_2$ gauge), or
projective director ($Q$-tensor). The third language
trades $\hat{\mathbf{n}}$ for the symmetric traceless
tensor $Q_{ij} = n_i n_j - \delta_{ij}/3$, which has the
same algebraic form regardless of which of the first two
formulations one starts from: it is gauge-invariant in the
lifted formulation and a well-defined function on
$\mathbb{RP}^2$ in the projective one (since
$Q_{ij}(\hat{\mathbf{n}}) = Q_{ij}(-\hat{\mathbf{n}})$),
which is precisely why $Q_{ij}$ is the standard order
parameter of nematic liquid-crystal
theory~\cite{degennesprost,lammert1995}. The physical
content is the same in all three formulations, and the
$\mathbb{Z}_2$-gauge symmetry is what guarantees that the
gauge-fixed calculation we use reproduces the
gauge-invariant physical observables.

In summary, the $\mathbb{RP}^2$ NLSM differs from the standard
$\mathrm{O}(3)$ NLSM on $S^2$ only at the global level: the
two share the same local $SO(3)$-invariant kinetic action, but
the $\mathbb{Z}_2$ identification $\hat{\mathbf{n}} \sim
-\hat{\mathbf{n}}$ replaces $O(3)$ by $SO(3)$ as the
faithfully-acting symmetry, restricts the single-valued
Hilbert space to even-$\ell$ harmonics (so the lowest
physical relaxation rate becomes $\lambda_2 = 6\,D_{\text{rot}}$
rather than $\lambda_1 = 2\,D_{\text{rot}}$), introduces a
non-trivial $\pi_1(\mathbb{RP}^2) = \mathbb{Z}_2$ winding class
with two quantization sectors $\theta \in \{0, \pi\}$ (only the
$\theta = 0$ sector is realized in the F2 model), and produces
a different topological-defect content in spatial extensions
(half-integer disclinations from $\pi_1$, hedgehog charges in
$\mathbb{N}$ rather than $\mathbb{Z}$ from $\pi_2$ modulo the
$\pi_1$-action).

\paragraph{Connection to the random-matrix view of the same
dynamics.}
A complementary description of the same FBP dynamics is
developed in the companion random-matrix manuscript
\cite{halperin2026rmt}, which studies the spectral structure
of the $N\times N$ geodesic-distance matrix $M_{ij}(t) =
\arccos(\mathbf{x}_i(t)\cdot\mathbf{x}_j(t))$ between
particle positions. The Funk-Hecke decomposition of the
$\arccos$ kernel contains \emph{only odd-$\ell$} Legendre
polynomials, because $\arccos(t) - \pi/2$ is odd in
$t = \cos d$. The resulting BBS quasi-multiplets organise
the spectrum of $M(t)$ into rank-$(2\ell+1)$ blocks at
$\ell = 1, 3, 5, \ldots$, with a kernel-amplitude scaling
$\Lambda_\ell \propto 1/(2\ell+1)$. The dominant geometric
content sits in the $\ell = 1$ block, which is, up to a
constant, the empirical inertia tensor: its smallest-eigenvalue
eigenvector is exactly the signed inertia-tensor eigenvector
$\hat{\mathbf{n}}(t)$ used here.

The two parity restrictions, $f_0$ being a function of
even-$\ell$ on $\mathbb{RP}^2$ and the $\arccos$ kernel
having only odd-$\ell$ Legendre content, are not in tension
because they constrain different objects. The even-$\ell$
selection rule of \eqref{eq:rp2_selection_rule} applies to
single-point functions of the orientation
$\hat{\mathbf{n}} \in \mathbb{RP}^2$ and is a topological
consequence of the $\mathbb{Z}_2$ identification
$\hat{\mathbf{n}} \sim -\hat{\mathbf{n}}$. The odd-$\ell$
expansion of the $\arccos$ kernel is a parity property of
the specific function $\arccos(t) - \pi/2$ on $[-1, 1]$,
applied to two-point functions of unidentified particle
positions $\mathbf{x}_n \in S^2$; particles are not
$\mathbb{Z}_2$-identified, only the ring orientation is. The
two views are connected through the $\ell = 1$ block of the
distance matrix: this odd-$\ell$ block of $M(t)$ extracts the
inertia-tensor eigenvector, whose sign ambiguity makes it a
director, whose dynamics is the
$\mathbb{RP}^2$ NLSM studied in the present paper. The higher odd-$\ell$ multiplets of $M(t)$
($\ell = 3, 5, \ldots$) are spectral signatures of the gapped
density fluctuations that the BO reduction of
Section~\ref{subsec:low_energy_ansatz} integrates out.
%The
%adapted-frame colatitude is related by
%$\cos\theta = \hat{\mathbf{n}}\cdot\mathbf{x}$, so
%$f_0(\hat{\mathbf{n}}\cdot\mathbf{x})
%= f_0(\cos\theta)$; writing the argument as $\cos\theta$
%or as $\hat{\mathbf{n}}\cdot\mathbf{x}$ is a matter of
%convention, and we use the latter to emphasize the
%dependence on the collective coordinate.
This is the rotational analog of the GJS kink
$\phi_0(x - X)$: the kink depends on $x$ and $X$ through
the translationally invariant combination $x - X$; the
ring depends on $\mathbf{x}$ and $\hat{\mathbf{n}}$
through the rotationally invariant combination
$\hat{\mathbf{n}}\cdot\mathbf{x}$.

%\paragraph{Substitution of the ring ansatz and the
%nonlinear ODE.}
%The classical field depends on time only through
%$\hat{\mathbf{n}}(t)$:
%$\rho_0(x,t) = f_0(\hat{\mathbf{n}}(t)\cdot\mathbf{x})$,
%with
%$\partial_t\rho_0 = \dot{n}^\alpha\psi_\alpha$ and
%$\mathcal{D}_t\rho_0
%= \dot{n}^\alpha\psi_\alpha
%+ \nabla_i(F^i[f_0]\,f_0)$.
%Substituting into~\eqref{eq:EL_explicit} and using the
%azimuthal isometry ($f_0$ and all kernels depend on
%$\mathbf{x}$ only through
%$u = \hat{\mathbf{n}}\cdot\mathbf{x}$), the EL equation
%reduces to a nonlinear integro-differential equation for
%$f_0(u)$:
%\begin{equation}
%\label{eq:f0_ODE}
%\frac{d}{dt}\!\left[\Omega^{-1}\left(
%\dot{n}^\alpha\psi_\alpha
%+ \nabla(F\,f_0)\right)\right]
%+ 2\alpha^2\!\int\!\hat{\mathcal{K}}f_0
%- \left[\frac{\delta}{\delta\rho}
%\frac{1}{2}\mathcal{D}_t\rho\,\Omega^{-1}\,
%\mathcal{D}_t\rho\right]_{\rho = f_0} = 0.
%\end{equation}
%The non-locality in time enters through $\Omega^{-1}$
%(which involves the propagator $\mathcal{W}_{ij}$ at
%different times via~\eqref{eq:Omega_kernel}), making
%this an ODE with a memory kernel. At leading adiabatic
%order ($\dot{n} \to 0$), the $\partial_t\rho_0$ terms
%are subleading and the equation reduces to the
%quasi-stationary balance
%$2\alpha^2\int\hat{\mathcal{K}}f_0
%= [\delta(\frac{1}{2}\nabla(Ff_0)\Omega^{-1}
%\nabla(Ff_0))/\delta\rho]$
%between the density self-interaction and the
%noise-induced diffusion.

\paragraph{The Born-Oppenheimer logic.}
The ansatz
$\rho_0(x,t) = f_0(\hat{\mathbf{n}}(t)\cdot\mathbf{x})$
uses the Born-Oppenheimer (BO) approximation in two
complementary ways. First, the saddle profile $f_0$ is
determined by the Euler-Lagrange
equation~\eqref{eq:EL_explicit} obtained at
\emph{static} $\hat{\mathbf{n}}$: the equation balances
the static drift $\nabla_i(F^i[\rho_0]\rho_0)$ and the
density self-interaction kernel and ignores any time
variation of $\hat{\mathbf{n}}$. The resulting $f_0$ is
the instantaneous ring profile that an arbitrarily slow
$\hat{\mathbf{n}}(t)$ would track. Second, in the
subsequent derivation of the effective dynamics for the
slow orientation $\hat{\mathbf{n}}(t)$
(Sections~\ref{subsec:faddeev_popov}--\ref{subsec:markov_limit}),
the gapped density fluctuations around $f_0$ are
integrated out at each time slice, with the kernels
evaluated at the instantaneous orientation. The
dimensionless small parameter controlling this two-stage
procedure is
\begin{equation}
\label{eq:BO_epsilon}
\epsilon = \frac{D_{\text{rot}}}{m^2_{\text{gap}}}\ll 1,
\end{equation}
the ratio of the rotational-diffusion rate
$D_{\text{rot}}$ to the spectral gap $m^2_{\text{gap}}$
of the second-variation operator that controls the
relaxation of the gapped density fluctuations
(Section~\ref{subsec:zero_modes}); equivalently
$\tau_{\text{fast}}/\tau_{\text{slow}}$ with
$\tau_{\text{fast}} \sim 1/m^2_{\text{gap}}$ and
$\tau_{\text{slow}} \sim 1/D_{\text{rot}}$. Sub-leading
BO corrections arise from retaining the time variation
of $\hat{\mathbf{n}}$ during the fast integration:
expanding $\hat{\mathbf{n}}(t')$ for
$t' \in [t-\tau_{\text{fast}},t]$ around
$\hat{\mathbf{n}}(t)$ generates terms involving
$\dot{\hat{\mathbf{n}}}, \ddot{\hat{\mathbf{n}}},\ldots$
in the effective action, suppressed by powers of
$\epsilon$.

\paragraph{What SO(3)-invariants can appear in the BO
corrections.}
The disorder-averaged single-field action of
Section~\ref{subsec:starting_point} is SO(3)-invariant
on $S^2$, and the effective orientation action
$S[\hat{\mathbf{n}}]$ inherits this symmetry (the
explicit derivation is given in
Section~\ref{subsec:faddeev_popov}). Under
$\hat{\mathbf{n}}\to R\hat{\mathbf{n}}$ with $R\in
\mathrm{SO}(3)$, time derivatives transform as
$\dot{\hat{\mathbf{n}}}\to R\dot{\hat{\mathbf{n}}}$,
$\ddot{\hat{\mathbf{n}}}\to R\ddot{\hat{\mathbf{n}}}$,
\ldots, and the squared norm
$|\dot{\hat{\mathbf{n}}}|^2 = \dot{\hat{\mathbf{n}}}
\cdot\dot{\hat{\mathbf{n}}}$ is SO(3)-invariant since
$(R\dot{\hat{\mathbf{n}}})\cdot(R\dot{\hat{\mathbf{n}}})
= \dot{\hat{\mathbf{n}}}\cdot R^T R\,\dot{\hat{\mathbf{n}}}
= |\dot{\hat{\mathbf{n}}}|^2$ using $R^T R = I$. The
same applies to $|\ddot{\hat{\mathbf{n}}}|^2$,
$|\dot{\hat{\mathbf{n}}}|^4$,
$\dot{\hat{\mathbf{n}}}\cdot\ddot{\hat{\mathbf{n}}}$, and
all higher-derivative invariants.

The unit-vector constraint $|\hat{\mathbf{n}}|^2 = 1$
removes $\hat{\mathbf{n}}$-only invariants: any
SO(3)-invariant function of $\hat{\mathbf{n}}$ alone
must be a function of $|\hat{\mathbf{n}}|^2 = 1$, i.e.\ a
constant. Differentiating the constraint gives the
identities $\hat{\mathbf{n}}\cdot\dot{\hat{\mathbf{n}}}
= 0$ and $\hat{\mathbf{n}}\cdot\ddot{\hat{\mathbf{n}}}
= -|\dot{\hat{\mathbf{n}}}|^2$, so even ``mixed''
contractions like $(\hat{\mathbf{n}}\cdot\dot{\hat{\mathbf{n}}})^2$
or $(\hat{\mathbf{n}}\cdot\ddot{\hat{\mathbf{n}}})^2$
reduce to functions of pure-derivative invariants. Every
admissible SO(3)-invariant scalar therefore contains at
least one time derivative.

\emph{Can corrections take the form $f(\hat{\mathbf{n}})\,
g(\dot{\hat{\mathbf{n}}})$ that vanishes at
$\dot{\hat{\mathbf{n}}} = 0$?} Only with
$f(\hat{\mathbf{n}}) = $ const. A non-constant
$f(\hat{\mathbf{n}})$ such as $(\hat{\mathbf{n}}\cdot
\hat{\mathbf{e}}_3)$ or $(\hat{\mathbf{n}}\cdot
\hat{\mathbf{n}}_*)$ singles out a preferred lab-frame
direction, breaking SO(3); the symmetry argument forbids
it at any loop order. Once $f$ is forced to be constant,
$g$ must itself be an SO(3)-invariant of
$\dot{\hat{\mathbf{n}}}$ alone, hence a function of
$|\dot{\hat{\mathbf{n}}}|^2$. The same reasoning at
higher derivative order gives constants times
$|\ddot{\hat{\mathbf{n}}}|^2$,
$|\dot{\hat{\mathbf{n}}}|^4$, etc. All admissible
corrections vanish identically when $\hat{\mathbf{n}}(t)$
is static (so $\dot{\hat{\mathbf{n}}} = \ddot{\hat{\mathbf{n}}}
= \cdots = 0$); they cannot anchor $\hat{\mathbf{n}}$ to
a preferred orientation. The leading correction to the
Markovian NLSM~\eqref{eq:NLSM_action} is therefore an
$|\dot{\hat{\mathbf{n}}}|^2$-renormalization of $\kappa$,
followed by higher-derivative SO(3)-invariants
$|\ddot{\hat{\mathbf{n}}}|^2,
|\dot{\hat{\mathbf{n}}}|^4,\ldots$ suppressed by powers
of $\epsilon$, all of which we drop at the
leading-derivative order.

\paragraph{Order-of-magnitude estimate of $\epsilon$.}
A spectral calculation of $m^2_{\text{gap}}$ as the lowest
nonzero eigenvalue of $\mathbf{M}[\rho_0]$ would require
solving the saddle equation for $f_0$ and diagonalizing the
resulting Hessian, neither of which is carried out in this
work. We use instead a phenomenological identification: the
relaxation rate of the gapped density fluctuations is read
off the band-formation timescale measured in the particle
simulations of Ref.~\cite{halperin2026frustrated}, which
report $\tau_{\text{fast}} \approx 5$ for our parameters
($T = 0.4$, $N = 400$). Identifying
$m^2_{\text{gap}} \simeq 1/\tau_{\text{fast}}\simeq 0.2$
and using the collective rate
$D_{\text{rot}} \approx 0.003$ measured in the same
simulations (Table~\ref{tab:Drot_disorder},
Section~\ref{subsec:comparison}),
\begin{equation}
\label{eq:eps_estimate}
\epsilon \;=\; \frac{D_{\text{rot}}}{m^2_{\text{gap}}}
\;\simeq\; \frac{0.003}{0.2}
\;\sim\; 1.5\times 10^{-2},
\end{equation}
well below unity, supporting the BO truncation. We treat
this estimate as an empirical consistency check rather than
a controlled spectral calculation: a numerical
diagonalization of $\mathbf{M}[\rho_0]$ around the measured
ring profile would turn the inferred timescale into a
genuine eigenvalue and is left to future work.

%% ============================================================
\subsection{Field decomposition and collective variables}
\label{subsec:faddeev_popov}

\paragraph{Decomposition of $\rho$.}
Following GJS~\cite{gervais1975collective} (Eq.~(2.4)),
we decompose the density field in the single-field path
integral~\eqref{eq:Z_single_field} into a saddle-point
part and a fluctuation,
\begin{equation}
\label{eq:field_decomp}
\rho(\mathbf{x}, t)
= f_0\!\left(\hat{\mathbf{n}}(t)\cdot\mathbf{x}\right)
+ \delta\rho(\mathbf{x}, t),
\end{equation}
%where $\mathbf{x}\in\mathbb{R}^3$  with $|\mathbf{x}|^2 = 1$ is the 3D  Euclidean position vector for a point on the sphere $ S^2 $.
The saddle-point density depends on $\mathbf{x}$ only through
the scalar invariant
$u \equiv \hat{\mathbf{n}}(t)\cdot\mathbf{x}$ and is
therefore axially symmetric about $\hat{\mathbf{n}}(t)$, while
the fluctuation $\delta\rho(\mathbf{x}, t)$ is an
arbitrary scalar function on $S^2$ at each $t$. 

Substituting $\rho = \rho_0 + \delta\rho$ into
$S_{\text{eff}}$ and Taylor-expanding around the saddle
gives
\begin{equation}
\label{eq:S_action_expansion}
S_{\text{eff}}[\rho]
= S_{\text{eff}}^{(0)}[\hat{\mathbf{n}}]
+ \underbrace{S_{\text{eff}}^{(1)}[\delta\rho]}_{=\,0
\text{ (saddle)}}
+ S_{\text{eff}}^{(2)}[\delta\rho]
+ O(\delta\rho^3),
\end{equation}
where
$S_{\text{eff}}^{(0)}[\hat{\mathbf{n}}]
= S_{\text{eff}}[\rho_0]$ is the action evaluated on the
saddle (a functional of $\hat{\mathbf{n}}(t)$ through
$\rho_0 = f_0(\hat{\mathbf{n}}(t)\cdot\mathbf{x})$), the
linear term vanishes by the saddle condition, and
$S_{\text{eff}}^{(2)}[\delta\rho]$ is the
quadratic-in-fluctuation piece. The latter will be
expanded in eigenfunctions of the second variational
derivative of $S_{\text{eff}}$ at the saddle. To set up
notation, we represent the Euler-Lagrange equation 
(\ref{EL_equation_rho_0}) as a point-wise constraint on the functional derivative of the effective action 
\begin{equation}
\label{eq:E_def}
\mathcal{E}[\rho](\mathbf{x},t)
\;\equiv\; \frac{\delta S_{\text{eff}}}{\delta\rho(\mathbf{x},t)},
\qquad \mathcal{E}[\rho_0](\mathbf{x},t) = 0\ \text{for all }(\mathbf{x},t),
\end{equation}
and its
functional derivative that produces the second-variation kernel:
\begin{equation}
\label{eq:M_def_42}
\mathbf{M}(\mathbf{x},t;\mathbf{y},t')
\;\equiv\;
\frac{\delta\mathcal{E}[\rho](\mathbf{x},t)}
{\delta\rho(\mathbf{y},t')}\bigg|_{\rho_0}
\;=\;
\frac{\delta^2 S_{\text{eff}}}
{\delta\rho(\mathbf{x},t)\,\delta\rho(\mathbf{y},t')}
\bigg|_{\rho_0},
\end{equation}
in terms of which
\begin{equation}
\label{eq:S2_kernel_42}
S_{\text{eff}}^{(2)}[\delta\rho]
= \tfrac{1}{2}\!\iint d\mu(\mathbf{x})\,dt\,
d\mu(\mathbf{y})\,dt'\;
\delta\rho(\mathbf{x},t)\,
\mathbf{M}(\mathbf{x},t;\mathbf{y},t')\,
\delta\rho(\mathbf{y},t').
\end{equation}
For the purposes of this paper we use only the formal
definition of $\mathbf{M}[\rho_0]$ and its residual
$\mathrm{SO}(2)_{\hat{\mathbf{n}}(t)}$ symmetry, which
organizes its eigenbasis and produces the spectral
structure used in
Section~\ref{subsec:zero_modes}; an explicit
semi-detailed form of $\mathbf{M}[\rho_0]$ in $\delta\rho$
can be obtained from~\eqref{eq:S_single_field} but is not
needed for what follows.

\paragraph{Zero modes from the orientation degeneracy.}
Zero modes arise from the collective-coordinate
construction in the standard way (GJS~\cite{gervais1975collective};
Rajaraman~\cite{rajaraman1982}, Ch.~8;
Coleman~\cite{coleman_aspects}, Ch.~7): take the
variation of the soliton with respect to its collective
coordinate and re-express the result through spatial
derivatives of the soliton profile. For the kink
$\phi_0(x - X_0)$ this gives
$-\partial\phi_0/\partial X_0 = \phi_0'(x - X_0)$, the
spatial derivative along the broken direction.

For the ring, the collective coordinate is the lab-frame
unit 3-vector $\hat{\mathbf{n}}\in\mathbb{R}^3$,
$|\hat{\mathbf{n}}|^2 = 1$, equivalently a point on the
orientation sphere $S^2_{\hat{\mathbf{n}}}$ (we reserve
$S^2$ without subscript for the physical manifold on
which the particles live). An admissible variation is a
tangent vector $\delta\hat{\mathbf{n}} \in
T_{\hat{\mathbf{n}}}S^2_{\hat{\mathbf{n}}} =
\{\mathbf{v}\in\mathbb{R}^3 :
\mathbf{v}\cdot\hat{\mathbf{n}} = 0\}$, naturally a
3-vector in the lab-frame plane orthogonal to
$\hat{\mathbf{n}}$. Since
$f_0(\hat{\mathbf{n}}\cdot\mathbf{x})$ solves
\eqref{eq:EL_explicit} for every
$\hat{\mathbf{n}}\in S^2_{\hat{\mathbf{n}}}$, the
corresponding field variation at fixed lab-frame
$\mathbf{x}\in S^2$ is, to linear order,
\begin{equation}
\label{eq:psi_alpha}
\psi[\delta\hat{\mathbf{n}}](\mathbf{x})
\;\equiv\;
f_0\!\bigl((\hat{\mathbf{n}}+\delta\hat{\mathbf{n}})\cdot
\mathbf{x}\bigr)
- f_0(\hat{\mathbf{n}}\cdot\mathbf{x})
\;=\;
f_0'(\hat{\mathbf{n}}\cdot\mathbf{x})\,
(\delta\hat{\mathbf{n}}\cdot\mathbf{x})
\;+\;O(|\delta\hat{\mathbf{n}}|^2),
\end{equation}
the rotational analog of $\phi_0'(x - X_0)$. Every dot
product is the standard $\mathbb{R}^3$ inner product;
the $f_0'$ prefactor is the spatial derivative of the
ring profile and $\delta\hat{\mathbf{n}}\cdot\mathbf{x}$
is the linear function that selects the broken direction.

Eq.\eqref{eq:psi_alpha} is the basis-independent
parametrization of the zero modes by tangent vectors
$\delta\hat{\mathbf{n}} \in T_{\hat{\mathbf{n}}}
S^2_{\hat{\mathbf{n}}}$, a two-dimensional space, so two
independent zero modes are present. To write them as a
discrete pair we pick any orthonormal basis
$\{\hat{\mathbf{e}}_a(\hat{\mathbf{n}})\}_{a = 1,2}$ of
$T_{\hat{\mathbf{n}}}S^2_{\hat{\mathbf{n}}}$
and 
label\footnote{The pair $\{\hat{\mathbf{e}}_1, \hat{\mathbf{e}}_2\}
\subset \mathbb{R}^3$ together with $\hat{\mathbf{n}}$
form an orthonormal basis of
$\mathbb{R}^3$. Following Appendix~\ref{app:covariant_langevin}, Latin
indices $a, b$ run $1, 2$ on the tangent plane (flat
frame), curved physical-manifold indices are $i, j$, and
embedding-space (lab) indices are
$\alpha, \beta \in \{1, 2, 3\}$.} 

\begin{equation}
\label{eq:psi_a_basis}
\psi_a(\mathbf{x};\hat{\mathbf{n}})
\equiv \psi[\hat{\mathbf{e}}_a(\hat{\mathbf{n}})]
(\mathbf{x})
= f_0'(\hat{\mathbf{n}}\cdot\mathbf{x})\,
(\hat{\mathbf{e}}_a\cdot\mathbf{x}),
\qquad a = 1, 2.
\end{equation}
The basis $\{\hat{\mathbf{e}}_a(\hat{\mathbf{n}})\}$ depends on
$\hat{\mathbf{n}}$, but every observable that enters the
Faddeev-Popov procedure (the gauge condition, the Gram
matrix, the kinetic-term coefficient) is a scalar
contraction in $a$ and is insensitive to that choice. The
zero modes are compatible with the density normalization
$\int_\M \rho_0\,d\mu_g = 1$: rotations of $\hat{\mathbf{n}}$
are isometries of $S^2$ and preserve the volume form, so
$\partial_{n^a}\!\int\rho_0\,d\mu_g = \int\psi_a\,d\mu_g = 0$
identically, and the FP gauge surface
$\int\psi_a\,\delta\rho\,d\mu_g = 0$ is therefore
automatically compatible with mass conservation
$\int\delta\rho\,d\mu_g = 0$.
%For the same reason no Cartan-frame or vielbein
%machinery on $S^2_{\hat{\mathbf{n}}}$ is needed at this
%stage; the lab-frame 3-vector treatment of
%$\hat{\mathbf{n}}$ and $\delta\hat{\mathbf{n}}$ is
%sufficient. Non-equal-time correlators built from
%$\hat{\mathbf{n}}(t)$ are evaluated using the lab-frame
%3-vector $\hat{\mathbf{n}}(t)$ itself, never via
%frame components.

The $\psi_a$ are the zero modes (zero-frequency normal
modes) of the second-variation operator
$\mathbf{M}[\rho_0]$, generated by the SO(3)/SO(2)
rotational degeneracy of the ring saddle. Following
Rajaraman~\cite{rajaraman1982} (Sec.~5.5), we
distinguish these zero modes from Nambu-Goldstone
bosons: the $\psi_a$ are isolated, discrete-spectrum
eigenvectors of $\mathbf{M}[\rho_0]$ at zero
eigenvalue, whereas Nambu-Goldstone bosons appear as
the lower limit of a continuum of excitations, with
their zero frequency the bottom of a gapless branch
rather than an isolated point. Zero modes of this kind
are a feature of quantizing a non-trivial
spatio-temporally structured saddle (soliton or
instanton) within a theory that is itself invariant
under a continuous symmetry; here the relevant symmetry
is the SO(3) rotational invariance of the effective F2
action, broken by the choice of ring orientation
$\hat{\mathbf{n}}$ at the saddle~\cite{rajaraman1982}. The
orientation $\hat{\mathbf{n}}(t)$ promotes this
rotational degeneracy to a slow collective coordinate
whose dynamics, treated in
Section~\ref{subsec:markov_limit}, realizes the
rotational diffusion of the broken direction itself.

Under $\mathrm{SO}(2)_{\hat{\mathbf{n}}}$ the prefactor
$f_0'(\hat{\mathbf{n}}\cdot\mathbf{x})$ is invariant
while $\delta\hat{\mathbf{n}}$ rotates in the
two-dimensional vector representation, so the doublet
$\{\psi_a\}$ carries the same representation. The
complex combinations $\psi_\pm \equiv \psi_1 \pm i\psi_2$
are eigenvectors of the
$\mathrm{SO}(2)_{\hat{\mathbf{n}}}$ generator with
eigenvalues $m = \pm 1$. They span the kernel of
$\mathbf{M}$ and are excluded from the gapped sector by
the spectral expansion~\eqref{eq:spectral_expansion} of
Section~\ref{subsec:zero_modes}.

\paragraph{Zero-eigenvalue property.}
The zero modes $\psi_a$ are eigenfunctions of the
second-variation kernel $\mathbf{M}$~\eqref{eq:M_def_42}
with eigenvalue zero, so they sit in the kernel of the
quadratic action: $S_{\text{eff}}^{(2)}[\psi_a] = 0$.
The argument is as follows.

Conditional on the ring-saddle ansatz~\eqref{eq:rho_n} being
an actual family of exact saddles of the F2 action (something
we do not derive from
\eqref{EL_equation_rho_0}), the saddle field
$\mathcal{E}[\rho_0]$~\eqref{eq:E_def} vanishes pointwise for
\emph{every} $\hat{\mathbf{n}}\in S^2_{\hat{\mathbf{n}}}$, so
differentiating this identity along an admissible direction
$\delta\hat{\mathbf{n}}\in T_{\hat{\mathbf{n}}}
S^2_{\hat{\mathbf{n}}}$ gives, by the functional chain
rule applied to a functional ($\mathcal{E}[\rho]$) whose
argument $\rho_0(\hat{\mathbf{n}})$ depends on the
parameter $\hat{\mathbf{n}}$,
\begin{equation}
\label{eq:M_psi_zero}
0 = \delta_{\delta\hat{\mathbf{n}}}
\mathcal{E}[\rho_0](\mathbf{x},t)
= \int\! d\mu(\mathbf{y})\,dt'\;
\frac{\delta\mathcal{E}[\rho](\mathbf{x},t)}
{\delta\rho(\mathbf{y},t')}\bigg|_{\rho_0}\,
\delta_{\delta\hat{\mathbf{n}}}
\rho_0(\mathbf{y},t')
= \int\! d\mu(\mathbf{y})\,dt'\;
\mathbf{M}(\mathbf{x},t;\mathbf{y},t')\,
\psi[\delta\hat{\mathbf{n}}](\mathbf{y},t')
\end{equation}
where we used Eq.(\ref{eq:psi_alpha}) in he last equation.
As this relation holds for any $ (x,t) $, it implies the relation
$\mathbf{M}\,\psi[\delta\hat{\mathbf{n}}] = 0$.
Multiplying this relation by $\psi[\delta\hat{\mathbf{n}}](\mathbf{x},t)$
and integrating against $d\mu(\mathbf{x})\,dt$
yields $S_{\text{eff}}^{(2)}[\psi[\delta\hat{\mathbf{n}}]]
= 0$. This demonstrates that the zero modes are eigenvectors of $\mathbf{M}$ with eigenvalue zero.

Physically, since $S_{\text{eff}}$ takes the same value
on every orientation (SO(3) equivariance), displacing
the saddle along $S^2_{\hat{\mathbf{n}}}$ costs no
action; the directions of zero cost are tangent to the
orbit of $f_0$ under SO(3) and are spanned by the zero
modes~\eqref{eq:psi_alpha}. This is the rotational
analog of the GJS kink zero mode $\phi_0'(x - X_0)$.

\paragraph{The Faddeev-Popov identity and the change of
variables.}
The decomposition~\eqref{eq:field_decomp} is not yet
unique: shifting $\hat{\mathbf{n}}$ by an infinitesimal
$\delta\hat{\mathbf{n}}\in T_{\hat{\mathbf{n}}}
S^2_{\hat{\mathbf{n}}}$ while adjusting
$\delta\rho \to \delta\rho - \psi[\delta\hat{\mathbf{n}}]$
leaves $\rho$ unchanged. This would introduce redundant
multiple integration over physically equivalent
configurations upon a naive change of variables in the
path integral from $\rho$ to the combination
$(\hat{\mathbf{n}}, \delta\rho)$. This redundancy is
resolved by the Faddeev-Popov procedure
\cite{gervais1975collective, rajaraman1982,
coleman_aspects}, where one enforces the constraint that
fluctuations be orthogonal to both zero modes:
\begin{equation}
\label{eq:FP_gauge}
F_a[\rho;\hat{\mathbf{n}}]
= \int_{S^2}\psi_a(x;\hat{\mathbf{n}})\,
\bigl[\rho(x,t) - \rho_0(x;\hat{\mathbf{n}})\bigr]\,
d\mu_g(x),
\qquad a = 1,2,
\end{equation}
%We next introduce the pair $\{\hat{\mathbf{e}}_1, \hat{\mathbf{e}}_2\}
%\subset \mathbb{R}^3$ entering $\psi_a$ is an orthonormal
%basis of the tangent plane $T_{\hat{\mathbf{n}}}S^2$,
%i.e.\ of the two-dimensional subspace of $\mathbb{R}^3$
%orthogonal to $\hat{\mathbf{n}}$ ($\hat{\mathbf{e}}_a
%\cdot\hat{\mathbf{n}} = 0$, $\hat{\mathbf{e}}_a\cdot
%\hat{\mathbf{e}}_b = \delta_{ab}$); together with
%$\hat{\mathbf{n}}$ they form an orthonormal basis of
%$\mathbb{R}^3$.
and the FP identity reads
\begin{equation}
\label{eq:FP_identity}
1 = \int\!\mathcal{D}\hat{\mathbf{n}}(t)\;
\delta\!\left(F_a[\rho;\hat{\mathbf{n}}]\right)\,
\det\!\left(\frac{\partial F_a}{\partial n^b}\bigg|_\rho
\right).
\end{equation}
Setting $F_a = 0$ pins $\delta\rho$ to be transverse to
the zero modes. Here $n^a$ ($a = 1, 2$) are local
coordinates on $S^2_{\hat{\mathbf{n}}}$ adapted to the
tangent-plane basis $\{\hat{\mathbf{e}}_a\}$, with
$\partial\hat{\mathbf{n}}/\partial n^a = \hat{\mathbf{e}}_a$
and $\partial\rho_0/\partial n^a = \psi_a$
(by~\eqref{eq:psi_alpha}). Differentiating $F_a$ at
\emph{fixed $\rho$}, both $\psi_a(x;\hat{\mathbf{n}})$
and $\rho_0(x;\hat{\mathbf{n}})$ vary with
$\hat{\mathbf{n}}$:
\begin{equation}
\label{eq:FP_jacobian}
\frac{\partial F_a}{\partial n^b}\bigg|_\rho
=
\int\!\frac{\partial\psi_a}{\partial n^b}\,
\delta\rho\,d\mu_g \;-\;\int\!\psi_a\,\psi_b\,d\mu_g.
\end{equation}
At the saddle $\delta\rho = 0$, the FP Jacobian reduces
to the constant Gram matrix
$-G_{ab}(\hat{\mathbf{n}}) = -\int\psi_a\,\psi_b\,d\mu_g$.
A direct computation in coordinates adapted to
$\hat{\mathbf{n}}$ (Appendix~\ref{app:FP_expansion})
gives
\begin{equation}
\label{eq:Gram_constant}
G_{ab} = G_0\,\delta_{ab},
\qquad
G_0 = \pi\!\int_{-1}^{1}\! f_0'(u)^2\,(1-u^2)\,du,
\end{equation}
manifestly an $\hat{\mathbf{n}}$-independent number; the
constant $\det G_{ab} = G_0^2$ absorbs into the overall
normalization of $Z$.

The $\delta\rho$-dependent piece of the FP
Jacobian~\eqref{eq:FP_jacobian} is handled by writing
$\partial F_a/\partial n^b\big|_\rho
= -G_{ab} + L_{ab}[\delta\rho]$ with
$L_{ab}[\delta\rho]
= \int(\partial\psi_a/\partial n^b)\,\delta\rho\,d\mu_g$
and exponentiating via $\ln\det = \mathrm{Tr}\ln$. The
result is an additive contribution to the action of
linear and quadratic terms in $\delta\rho$
(Appendix~\ref{app:FP_expansion}, Eq.~\eqref{eq:FPapp_logdet}),
\begin{equation}
\label{eq:logdet_FP}
\ln\det\!\left(\frac{\partial F_a}{\partial n^b}\right)
= \ln G_0^2
- \!\int\! J\,\delta\rho\,d\mu_g\,dt
- \tfrac{1}{2}\!\iint\!\delta\rho\,K\,\delta\rho
+ O(L^3),
\end{equation}
where the source $J(\mathbf{x}) = G_0^{-1}\,\nabla^2_{S^2}
f_0(\hat{\mathbf{n}}\cdot\mathbf{x})$ is the spherical
Laplacian of the saddle profile (an
$\mathrm{SO}(2)_{\hat{\mathbf{n}}}$-singlet) and the
kernel $K(\mathbf{x};\mathbf{y})$ is an SO(3)-invariant
two-point object built from $f_0', f_0'', u_x, u_y,
\mathbf{x}\cdot\mathbf{y}$; explicit forms are derived
in Appendix~\ref{app:FP_expansion}.

Changing variables from $\rho$ to $(\hat{\mathbf{n}},
\delta\rho)$ via~\eqref{eq:field_decomp} (the Jacobian is
unity since $\rho$ depends linearly on $\delta\rho$ at
fixed $\hat{\mathbf{n}}$), substituting the action
expansion~\eqref{eq:S_action_expansion} (linear term
vanishing by the saddle condition) and the FP-Jacobian
expansion~\eqref{eq:logdet_FP} (the constant $\ln G_0^2$
absorbed into the overall normalization), and combining
the two quadratic-in-$\delta\rho$ pieces into the single
renormalized kernel $\tilde{\mathbf{M}} = \mathbf{M} + K$,
the path integral~\eqref{eq:Z_single_field} factorizes
as
\begin{equation}
\label{eq:Z_FP_factorized}
Z \;\propto\; \int\!d\hat{\mathbf{n}}(t)\,
e^{-S_{\text{eff}}^{(0)}[\hat{\mathbf{n}}]}
\int\!\mathcal{D}'\delta\rho\;
\exp\!\Bigl(
-\tfrac{1}{2}\!\iint\!\delta\rho\,\tilde{\mathbf{M}}
[\hat{\mathbf{n}}]\,\delta\rho
-\!\int\!J[\hat{\mathbf{n}}]\,\delta\rho
+ O(\delta\rho^3)
\Bigr),
\end{equation}
\begin{equation}
\label{eq:measure_factorization}
\mathcal{D}\rho
= d\hat{\mathbf{n}}\,\mathcal{D}'\delta\rho,
\end{equation}
where the prime on $\mathcal{D}'\delta\rho$ denotes the
projection imposed by $\delta(F_a)$, restricting the
fluctuation measure to directions transverse to $\psi_1,
\psi_2$.%\footnote{The two FP constraints
%$F_a[\rho;\hat{\mathbf{n}}] = 0$ are
%\emph{integrated} (one pair of overlaps with the zero
%modes), not pointwise in spacetime; the
%delta-functions in~\eqref{eq:FP_identity} are therefore
%realized by restricting the measure to the codimension-2
%subspace transverse to $\psi_1, \psi_2$, with no
%conjugate response field needed (in contrast to the
%MSRJD construction in
%Appendix~\ref{subsec:dk_as_langevin}, where the
%stochastic equation of motion has to be enforced at every
%spacetime point and a response field is required).
%For the same reason, the FP Jacobian is a finite
%$2 \times 2$ determinant rather than an
%infinite-dimensional functional determinant, and is
%exponentiated directly via $\ln\det = \mathrm{Tr}\ln$
%without introducing Grassmann ghost fields.}
The orientation
measure
$d\hat{\mathbf{n}} \equiv d\mu_{S^2}(\hat{\mathbf{n}})$
is the SO(3)-invariant (round) measure on the lifted $S^2$
of unit normals (the double cover of the physical
$\mathbb{RP}^2$ orientation manifold,
Section~\ref{subsec:low_energy_ansatz}), with total mass
$4\pi$ on $S^2$; $d\mu_g$ continues to denote the induced
volume measure on the physical manifold. The factorized
form~\eqref{eq:Z_FP_factorized} is the analogue, for the
orientation manifold, of the GJS path integral over the
kink centroid $X(t)$ with the fluctuation $\delta\phi$
orthogonal to $\phi_0'(x - X)$.

%% ============================================================
\subsection{Reduction to
the quantum-mechanical path integral}
\label{subsec:zero_modes}

\paragraph{The orientation path integral.}
Performing the Gaussian integration over $\delta\rho$ on
the gauge surface $\mathcal{D}'\delta\rho$
in~\eqref{eq:Z_FP_factorized}, by completing the square
in the linear source $J$, reduces the path integral to
one over the orientation alone:
\begin{equation}
\label{eq:Z_oneloop}
Z \;\propto\; \int\!d\hat{\mathbf{n}}(t)\;
\exp\!\bigl(-S[\hat{\mathbf{n}}]\bigr),
\end{equation}
\begin{equation}
\label{eq:S_n_full}
S[\hat{\mathbf{n}}]
= S_{\text{eff}}^{(0)}[\hat{\mathbf{n}}]
+ \tfrac{1}{2}\log\det'\tilde{\mathbf{M}}[\hat{\mathbf{n}}]
- \tfrac{1}{2}\!\int\!J\,\tilde{\mathbf{M}}^{-1}\,J
+ (\text{higher-loop bulk corrections}),
\end{equation}
where $\tilde{\mathbf{M}} = \mathbf{M} + K$ is the
FP-renormalized second-variation kernel, and the prime
restricts to the gapped sector orthogonal to the zero
modes $\psi_a$. Below we compute the
saddle action $S_{\text{eff}}^{(0)}$ explicitly, use the
spectral decomposition of $\mathbf{M}[\hat{\mathbf{n}}]$
to give meaning to the one-loop determinant, and then
use SO(3) symmetry to constrain the surviving terms.

\paragraph{The leading term
$S_{\text{eff}}^{(0)}[\hat{\mathbf{n}}]$.}
The saddle action $S_{\text{eff}}^{(0)}[\hat{\mathbf{n}}]
= S_{\text{eff}}[\rho_0]$ depends on $\hat{\mathbf{n}}(t)$
through $\rho_0 = f_0(\hat{\mathbf{n}}(t)\cdot\mathbf{x})$.
Because $\mathcal{D}_t\rho_0 = \dot{n}^a\psi_a
+ \nabla_i(F^i[\rho_0]f_0)$ is linear in $\dot{n}^a$ (the
drift piece is $\dot{n}$-independent), the kinetic
density $\tfrac{1}{2}\mathcal{D}_t\rho_0\,
\Omega^{-1}[\rho_0]\,\mathcal{D}_t\rho_0$ is exactly
quadratic in $\dot{n}^a$:
\begin{align}
\label{eq:adiabatic_expansion}
S_{\text{eff}}^{(0)}[\hat{\mathbf{n}}]
= V_0(\hat{\mathbf{n}})
+ \frac{1}{2}\!\int_0^T\!dt\int_0^T\!dt'\;
\dot{n}^a(t)\,M_{ab}(t,t')\,\dot{n}^b(t'),
\end{align}
where the static piece is
\begin{align}
\label{eq:V0_def}
V_0(\hat{\mathbf{n}})
&= \tfrac{1}{2}\!\iint
\nabla_i(F^i f_0)(x)\,
\Omega^{-1}[\rho_0](x,x')\,
\nabla_j(F^j f_0)(x')\,d\mu d\mu'
\nonumber\\
&\quad{}-\;\alpha^2\!\iint
f_0(\hat{\mathbf{n}}\!\cdot\!\mathbf{x})\,
\hat{\mathcal{K}}(x,x')\,
f_0(\hat{\mathbf{n}}\!\cdot\!\mathbf{x}')\,d\mu d\mu',
\end{align}
and the rotational moment-of-inertia kernel is
$M_{ab}(t,t') = \iint\psi_a(x)\,\Omega^{-1}[\rho_0]
(x,t;x',t')\,\psi_b(x')\,d\mu(x)d\mu(x')$, generally
non-local in time. Residual
$\mathrm{SO}(2)_{\hat{\mathbf{n}}}$ isotropy reduces it
to $M_{ab}(t,t') = \kappa(t-t')\,\delta_{ab}$, with a
scalar memory kernel $\kappa(\tau)$. The expansion
terminates exactly: no higher-order terms in $\dot{n}$
appear because $\mathcal{D}_t\rho_0$ is linear in
$\dot{n}$. The cross term $\dot{n}^a\iint\psi_a\,
\Omega^{-1}\,\nabla(F f_0)$ (a would-be Berry phase)
vanishes by representation theory of the residual
$\mathrm{SO}(2)_{\hat{\mathbf{n}}}$: the zero modes
$\psi_a$ carry the 2D vector representation, the drift
divergence $\nabla(F f_0)$ is an
$\mathrm{SO}(2)_{\hat{\mathbf{n}}}$-singlet (it depends
on $\mathbf{x}$ only through
$\hat{\mathbf{n}}\cdot\mathbf{x}$), and
$\Omega^{-1}[\rho_0]$ commutes with the
$\mathrm{SO}(2)_{\hat{\mathbf{n}}}$ generator; the
overlap of a vector with a singlet through a
generator-commuting kernel vanishes by representation
orthogonality. The argument relies on $\Omega^{-1}$ being
symmetric under the $\mathrm{SO}(2)_{\hat{\mathbf{n}}}$
that fixes $\hat{\mathbf{n}}$, which itself follows from
the disorder average and the parity-symmetric construction
of the noise kernel (no chiral or time-reversal-odd input
in the F2 action). In a stochastic problem with explicit
parity- or time-reversal-breaking sources, an antisymmetric
component of the kinetic kernel could in principle survive
and would generate a Berry-phase term; the present model
admits no such source, and the absence of a Berry phase is
a consequence of this microscopic time-reversal property
together with the SO(2) representation argument above.

\paragraph{One-loop determinant from the spectral
decomposition of $\mathbf{M}[\hat{\mathbf{n}}]$.}
The second-variation operator
$\mathbf{M}[\rho_0]$~\eqref{eq:M_def_42} is a
self-adjoint integro-differential operator on
$L^2(S^2)$, built from the saddle profile $\rho_0$ via
the SO(3)-equivariant primitives of the F2 model
($\Omega^{-1}[\rho_0]$, $F^i[\rho_0]$, $\hat{\mathcal{K}}$,
$\hat{\mathcal{A}}^{ij}$, $\mathcal{W}_{ij}$, $\nabla$,
$d\mu_g$). We expand the fluctuation in its orthonormal
eigenbasis,
\begin{equation}
\label{eq:spectral_expansion}
\delta\rho(\mathbf{x}, t)
= \sum_{k} c_k(t)\,\eta_k(\mathbf{x};\hat{\mathbf{n}}(t)),
\qquad
\mathbf{M}[\hat{\mathbf{n}}]\,\eta_k = \omega_k^2\,\eta_k,
\end{equation}
which diagonalizes the quadratic action
$S_{\text{eff}}^{(2)} = \tfrac{1}{2}\sum_k\omega_k^2\,c_k^2$.
Because $\rho_0 = f_0(\hat{\mathbf{n}}\cdot\mathbf{x})$
is $\mathrm{SO}(2)_{\hat{\mathbf{n}}}$-invariant,
$\mathbf{M}$ commutes with the
$\mathrm{SO}(2)_{\hat{\mathbf{n}}}$ generator and the
eigenbasis $\{\eta_k\}$ block-diagonalizes by the
$\mathrm{SO}(2)_{\hat{\mathbf{n}}}$ representation that
each $\eta_k$ carries: singlets, the 2D vector
representation, and higher representations. The kernel
of $\mathbf{M}$ contains exactly the two zero modes
$\psi_a$ in the vector representation; everything else
is gapped, with spectral gap $m^2_{\text{gap}} > 0$ that
controls the relaxation time of the fast density
fluctuations and underwrites the BO reduction of
Section~\ref{subsec:low_energy_ansatz}. The Gaussian
integral over the gapped modes therefore produces
\begin{equation}
\tfrac{1}{2}\log\det{}'\tilde{\mathbf{M}}[\hat{\mathbf{n}}]
\;=\; \tfrac{1}{2}\!\sum_{k:\,\omega_k^2>0}
\log\tilde{\omega}_k^2[\hat{\mathbf{n}}],
\end{equation}
each (renormalized) eigenvalue
$\tilde{\omega}_k^2$ of $\tilde{\mathbf{M}} =
\mathbf{M} + K$ being a functional of $\hat{\mathbf{n}}(t)$
through $\rho_0$.

\paragraph{Symmetry constraints on $V_0$ and the
fluctuation determinant.}
Each contribution to $S[\hat{\mathbf{n}}]$
in~\eqref{eq:S_n_full} is built from SO(3)-equivariant
primitives: the disorder-averaged
action~\eqref{eq:S_single_field} is SO(3)-invariant; the
FP gauge condition transforms covariantly under
simultaneous rotation $(\rho,\hat{\mathbf{n}})\to
(R\rho,R\hat{\mathbf{n}})$; the constrained measure
$\mathcal{D}'\delta\rho$, the FP Jacobian, and the
orientation measure $d\hat{\mathbf{n}}$ are all
SO(3)-invariant. Hence the integrated-out result
satisfies
\begin{equation}
\label{eq:S_SO3_invariance}
S[R\hat{\mathbf{n}}] = S[\hat{\mathbf{n}}]
\quad\text{for all }R\in\mathrm{SO}(3),
\end{equation}
with the corresponding transformation of time
derivatives. Combined with the unit-vector constraint
$|\hat{\mathbf{n}}|^2 = 1$, which forbids any non-constant
SO(3)-invariant function of $\hat{\mathbf{n}}$ alone, the
individual contributions to $S[\hat{\mathbf{n}}]$ are
constrained as follows. The
$\dot{\hat{\mathbf{n}}}$-independent piece
$V_0(\hat{\mathbf{n}})$ of $S_{\text{eff}}^{(0)}$ is an
SO(3)-invariant function of $\hat{\mathbf{n}}$ alone and
therefore reduces to a constant
$V_0(\hat{\mathbf{n}})\equiv V_0$ (the rotational analog
of the kink-mass constancy of soliton physics:
GJS~\cite{gervais1975collective};
Rajaraman~\cite{rajaraman1982}, Sec.~8.3;
Coleman~\cite{coleman_aspects}, Ch.~7). The fluctuation
determinant
$\tfrac{1}{2}\log\det'\tilde{\mathbf{M}}[\hat{\mathbf{n}}]$
and the source-shift
$\tfrac{1}{2}\int J\,\tilde{\mathbf{M}}^{-1}J$ depend on
$(\hat{\mathbf{n}},\dot{\hat{\mathbf{n}}})$ only through
SO(3)-invariants and contribute at most to kinetic-term
coefficients. Higher-loop corrections from $V_3,
V_4, \ldots$ are SO(3)-invariant for the same reason.
None of these terms can generate an
$\hat{\mathbf{n}}$-dependent potential. The conclusion is
exact and all-orders in the loop expansion. Beyond
the leading BO order, corrections take the form of
SO(3)-invariant higher-derivative kinetic terms
$|\ddot{\hat{\mathbf{n}}}|^2,
|\dot{\hat{\mathbf{n}}}|^4,\ldots$, suppressed by powers
of the BO parameter
$\epsilon = D_{\text{rot}}/m^2_{\text{gap}}\ll 1$
(Section~\ref{subsec:low_energy_ansatz}); none of them
generate an orientational potential.

\paragraph{The non-Markovian effective action.}
Putting the pieces together, $V_0$ is constant and the
linear-in-$\dot{n}^a$ cross term vanishes (no Berry
phase), so only the kinetic kernel
$M_{ab}(t,t') = \kappa(t-t')\,\delta_{ab}$ survives in
$S_{\text{eff}}^{(0)}$, with the scalar memory kernel
\begin{equation}
\label{eq:M_alpha_beta}
\kappa(\tau) = \iint\psi_1(x)\,
\Omega^{-1}[\rho_0](x,t;x',t+\tau)\,
\psi_1(x')\,d\mu(x)\,d\mu(x').
\end{equation}
The Faddeev-Popov correction $K$
in~\eqref{eq:logdet_FP} enters $V_0$ and the one-loop
determinant $\frac{1}{2}\log\det'\tilde{\mathbf{M}}$ but
does not appear in the kinetic kernel $\kappa(\tau)$ at
this order: $\kappa$ is built from $\Omega^{-1}[\rho_0]$
on the zero-mode subspace, on which $K$ acts trivially by
construction of the FP gauge surface. Higher-loop
corrections of $\kappa$ from coupling to the gapped
fluctuations are absorbed into the phenomenological value
of $D_{\text{rot}}$ that we fit from simulation
(Tier 2, Section~\ref{subsec:comparison}). Folding in
the SO(3)-invariance of
$\det'\tilde{\mathbf{M}}[\hat{\mathbf{n}}]$, the full
one-loop effective action is the non-Markovian $\mathbb{RP}^2$ NLSM
model:
\begin{equation}
\label{eq:S_eff_nonmarkov}
S[\hat{\mathbf{n}}]
= \frac{1}{2}\int_0^T\!dt\int_0^T\!dt'\;
\kappa(t-t')\,
\dot{\hat{\mathbf{n}}}(t)\cdot
\dot{\hat{\mathbf{n}}}(t'),
\qquad |\hat{\mathbf{n}}|^2 = 1.
\end{equation}
No potential, no Berry phase; the only dynamical content
is the non-Markovian kinetic term, with $\kappa(\tau)$
capturing the finite relaxation time $\tau_{\text{fast}}
\sim 1/m^2_{\text{gap}}$ of the gapped density
fluctuations.

%% ============================================================
\subsection{The Markovian limit: $\mathbb{RP}^2$ NLSM
in 0+1 dimensions}
\label{subsec:markov_limit}

\paragraph{From non-Markovian to Markovian.}
When the memory kernel decays fast
($\tau_{\text{fast}} \ll \tau_c$, the orientation
decorrelation time), the non-Markovian
action~\eqref{eq:S_eff_nonmarkov} reduces to a
Markovian form by replacing
$\kappa(\tau) \to \kappa\,\delta(\tau)$, where
$\kappa = \int_0^\infty\kappa(\tau)\,d\tau$ is the
time-integrated stiffness. The Markovian replacement is an
\emph{additional} dynamical approximation, not a consequence
of SO(3) symmetry: symmetry alone fixes the tensor
structure $M_{ab}(t,t') \propto \delta_{ab}$ and the absence
of a Berry phase, but does not force
$\kappa(\tau)$ to be $\delta$-correlated. The empirical
exponential autocorrelation observed in the particle
simulations of Section~\ref{subsec:comparison} is the
empirical signature that the inequality
$\tau_{\text{fast}} \ll \tau_c$ holds in the regime tested. This gives the effective
action:
\begin{equation}
\label{eq:NLSM_action}
S[\hat{\mathbf{n}}]
= \frac{\kappa}{2}\int_0^T\!dt\;
|\dot{\hat{\mathbf{n}}}|^2,
\qquad |\hat{\mathbf{n}}|^2 = 1,
\end{equation}
with $|\dot{\hat{\mathbf{n}}}|^2
= \delta_{ij}\dot{n}^i\dot{n}^j$ and
$D_{\text{rot}} = 1/(2\kappa)$.
\begin{equation}
\label{eq:D_ab_explicit}
\kappa = \int_0^\infty\!\kappa(\tau)\,d\tau,
\qquad
D_{\text{rot}} = \frac{1}{2\kappa}.
\end{equation}
The relation $D_{\text{rot}} = 1/(2\kappa)$ follows from
matching $\frac{1}{2}\kappa|\dot{\hat{\mathbf{n}}}|^2$
to the standard MSRJD form
$\frac{1}{4D}\int\dot{x}^2\,dt$. The matching is carried out
in flat-space form using the embedding-space norm
$|\dot{\hat{\mathbf{n}}}|^2$. Passing to a fully covariant
path integral on $S^2$ produces additional curvature
corrections to the path-integral measure (Faddeev-Popov-type
$\sqrt{g}$ insertions and a $\hbar$-order Riemann-curvature
piece familiar from quantum mechanics on Riemannian
manifolds, Appendix~\ref{app:covariant_langevin}); these are
subleading in the slow-orientation limit and do not affect
the leading diffusion coefficient $D_{\text{rot}}$, but
should be retained if higher-derivative corrections to the
sigma-model action are pursued.

Equation~\eqref{eq:NLSM_action} is the
$\mathbb{RP}^2$ NLSM in 0+1 dimensions (the
classical-stochastic relaxation analog of free-particle
dynamics on $\mathbb{RP}^2$, identical at the level of the
local action to free Brownian motion on $S^2$ when expressed
via the signed inertia-tensor eigenvector of
Section~\ref{subsec:low_energy_ansatz}). It is the Markovian approximation
of the non-Markovian action~\eqref{eq:S_eff_nonmarkov}. The Markovian $\mathbb{RP}^2$ NLSM
predicts exponential autocorrelation
$C(\tau) = e^{-2D_{\text{rot}}\tau}$ and a power
spectrum $S(\omega) \propto 1/\omega^2$. The particle
simulations confirm both predictions when the
orientation is extracted via the inertia tensor method
(Section~\ref{subsec:comparison}).\footnote{An earlier
analysis in~\cite{halperin2026frustrated} reported
$S(\omega) \propto \omega^{-1.6}$ using the
instantaneous angular momentum
$\mathbf{L}(t)/|\mathbf{L}(t)|$ as a proxy for the
orientation. This shallower exponent is an artifact of
the fast thermal noise in $\mathbf{L}$; the inertia
tensor method filters out this noise and recovers the
Markovian $\omega^{-2}$.}

The scalar correlator
$q(\tau) = \langle\hat{\mathbf{n}}(t)\cdot
\hat{\mathbf{n}}(t')\rangle$ contains all two-point
information about the orientation. The full tensor
correlator
$\langle n_i(t)\,n_j(t')\rangle
= \frac{1}{3}\delta_{ij}\,q(\tau)$
is proportional to $\delta_{ij}$ by the SO(3) isotropy
of the diffusion: all off-diagonal components vanish,
and all diagonal components equal $q(\tau)/3$.

The exponential decay of $q(\tau)$ and the saturation of
the MSD at $2$ are consequences of diffusion on the
\emph{compact} manifold $S^2$, not of a drift term. On
a flat, infinite plane, pure diffusion
($\dot{x} = \sqrt{2D}\xi$) gives
$\langle x^2\rangle = 2Dt$ (linear growth, no decay of
correlations). Exponential relaxation in flat space
requires a restoring drift (the Ornstein-Uhlenbeck
process). On $S^2$, the compactness plays the role of
the restoring force: a diffusing orientation eventually
explores the full sphere and becomes uniformly
distributed, causing $q(\tau) \to 0$. The decay rate is
$2D_{\text{rot}} = D_{\text{rot}}\,\ell(\ell+1)$ evaluated
at $\ell = 1$: for diffusion on $S^2$ the $\ell$-th
spherical harmonic relaxes at rate $D_{\text{rot}}\,
\ell(\ell+1)$, equal to $D_{\text{rot}}$ times the
eigenvalue $\ell(\ell+1)$ of the negative Laplace-Beltrami
operator, and the components of $\hat{\mathbf{n}}$ itself
are the $\ell = 1$ spherical harmonics.
The Stratonovich SDE
(Section~\ref{subsec:orientation_sde}) has no drift at
all; the $-2D_{\text{rot}}n_i$ term in the Ito form
\eqref{eq:nhat_ito} is the geometric
Stratonovich-to-Ito correction needed to maintain
$|\hat{\mathbf{n}}|^2 = 1$, not a physical force.

\paragraph{The resulting path integral.}
\begin{equation}
\label{eq:Z_reduced}
Z = \text{const}\times
\int\!\mathcal{D}\hat{\mathbf{n}}\;
\exp\!\left(-\frac{\kappa}{2}\int_0^T\!dt\;
|\dot{\hat{\mathbf{n}}}|^2\right).
\end{equation}
Equation~\eqref{eq:Z_reduced} is the path integral over the
\emph{lifted} signed orientation
$\hat{\mathbf{n}}(t) \in S^2$ defined by the continuity
convention of Section~\ref{subsec:low_energy_ansatz}. The
physical $\mathbb{RP}^2$-NLSM partition function is obtained
by including the topological constraints from
$\pi_1(\mathbb{RP}^2) = \mathbb{Z}_2$: closed orientation
trajectories on $\mathbb{RP}^2$ split into the two homotopy
classes $w \in \mathbb{Z}_2$ of
Section~\ref{subsec:low_energy_ansatz}, and the partition
function decomposes accordingly,
\begin{equation}
\label{eq:Z_RP2_sectors}
Z_{\mathbb{RP}^2}^{(\theta)}
\;=\; \sum_{w\,\in\,\mathbb{Z}_2}
e^{i\theta w}\, Z_w,
\qquad
Z_w \;=\; \int_{[w]}\!\mathcal{D}\hat{\mathbf{n}}\;
\exp\!\left(-\frac{\kappa}{2}\!\int_0^T\!dt\;
|\dot{\hat{\mathbf{n}}}|^2\right),
\end{equation}
where $\int_{[w]}$ denotes the integral over closed
trajectories on the lifted $S^2$ in homotopy class $w$
(trajectories that return to $+\hat{\mathbf{n}}_0$ for
$w = 0$ or to $-\hat{\mathbf{n}}_0$ for $w = 1$), and
$\theta \in \{0, \pi\}$ labels the two inequivalent
quantizations of the model on $\mathbb{RP}^2$. The F2 model
is the $\theta = 0$ sector,
$Z_{\mathbb{RP}^2}^{(\theta = 0)} = Z_{w=0} + Z_{w=1}$, the
positive-density branch corresponding (at the heat-kernel
level) to the method-of-images symmetrization
$P_{\mathbb{RP}^2}(\mathbf{x},\mathbf{y};T)
= P_{S^2}(\mathbf{x},\mathbf{y};T)
+ P_{S^2}(\mathbf{x},-\mathbf{y};T)$
on the lifted cover. The $\theta = \pi$ counterpart is the
signed combination $Z_{w=0} - Z_{w=1}$, a spinor-like sector
that does not correspond to any positive probability density
and is therefore not realized in our classical-stochastic
problem. At the level of correlators, the F2 sector is
characterized by the even-$\ell$ selection rule
\eqref{eq:rp2_selection_rule}, already built into the
closed form of $C_\rho$ derived in
Section~\ref{subsec:corr_orientation}. In the
$\mathbb{Z}_2$-gauge-theory language of
Section~\ref{subsec:low_energy_ansatz}
(``Gauge-theory reformulation''), the relationship between
Eq.~\eqref{eq:Z_reduced} and Eq.~\eqref{eq:Z_RP2_sectors}
becomes transparent. Eq.~\eqref{eq:Z_reduced} is the lifted
path integral in the gauge fixed by the continuity
convention. With closed-loop (partition-function) boundary
conditions on $\mathbb{RP}^2$, the lifted trajectory returns
to $\pm\hat{\mathbf{n}}_0$ on $S^2$, and the path integral
sums over both possibilities, yielding $Z_{w = 0} + Z_{w = 1}$.
This is precisely $Z_{\mathbb{RP}^2}^{(\theta = 0)}$ from
Eq.~\eqref{eq:Z_RP2_sectors}, the unweighted (method-of-images)
sum over the two homotopy classes of the lift. The full
Eq.~\eqref{eq:Z_RP2_sectors} extends Eq.~\eqref{eq:Z_reduced}
by attaching the topological phase $e^{i\theta w}$ to each
class: $\theta = 0$ is the F2 sector (positive density),
$\theta = \pi$ gives the signed combination
$Z_{w = 0} - Z_{w = 1}$, the spinor-like sector that does
not correspond to a positive probability density. The body
uses Eq.~\eqref{eq:Z_reduced} as a notational shorthand for
the $\theta = 0$ projective partition function and projects
to $\mathbb{RP}^2$ at the level of observables; for the
$\mathbb{Z}_2$-gauge-invariant quantities we actually compute
(density correlator, $|q(\tau)|$, the tensor correlator
$\langle n_i n_j\rangle$), Eq.~\eqref{eq:Z_reduced} and
Eq.~\eqref{eq:Z_RP2_sectors} at $\theta = 0$ agree.

\paragraph{Hierarchy of approximations.}
The hierarchy is: (i) the leading large-$N$
disorder-averaged F2 field theory; (ii) the non-Markovian $\mathbb{RP}^2$-NLSM action~\eqref{eq:S_eff_nonmarkov} (the leading
BO/one-loop effective action, conditional on the ring
saddle); (iii) the Markovian $\mathbb{RP}^2$ NLSM
action~\eqref{eq:NLSM_action} (additional approximation,
$\kappa(\tau) \to \kappa\delta(\tau)$, simpler but less
accurate at high frequencies).

%% ============================================================
%% TIER 3: CORRELATION REDUCTION AND SELF-CONSISTENCY
%% ============================================================

\subsection{Reduction of correlation functions}
\label{subsec:corr_orientation}

In the BO long-time limit, the density correlator
$C_\rho(x,t;x',t')
= \langle\rho(x,t)\rho(x',t')\rangle$ (the primary
observable of the single-field theory) reduces to an
explicit function of the single parameter
$D_{\text{rot}}$ (Markovian) or $\kappa(\tau)$
(non-Markovian). The reduction uses the Legendre expansion
of the band profile $f_0$ and the spectral decomposition
of Brownian motion on $S^2$.

\paragraph{Legendre expansion of $f_0$ and the
$\mathbb{RP}^2$ selection rule.}
Since $f_0(\hat{\mathbf{n}}\cdot\mathbf{x})$ depends on
$\mathbf{x}$ only through
$\hat{\mathbf{n}}\cdot\mathbf{x}$, it admits an expansion
in Legendre polynomials $P_\ell(u)$ (the standard
orthogonal polynomials on $[-1,1]$ with normalization
$\int_{-1}^{1}P_\ell P_{\ell'}\,du
= 2\delta_{\ell\ell'}/(2\ell+1)$):
\begin{equation}
\label{eq:f0_legendre}
f_0(\hat{\mathbf{n}}\cdot\mathbf{x})
= \sum_{\ell=0}^{\infty}c_\ell\,
P_\ell(\hat{\mathbf{n}}\cdot\mathbf{x}),
\qquad
c_\ell = \frac{2\ell+1}{2}\int_{-1}^{1}
f_0(u)\,P_\ell(u)\,du.
\end{equation}
The $\mathbb{RP}^2$ structure of the orientation manifold
(Section~\ref{subsec:low_energy_ansatz}, ``Topological
aspects of the $\mathbb{RP}^2$ target'') imposes a
selection rule on this expansion: $f_0$ is invariant under
$\hat{\mathbf{n}}\to-\hat{\mathbf{n}}$, equivalently
$f_0(u) = f_0(-u)$, so
$P_\ell(-u) = (-1)^\ell P_\ell(u)$ implies
\begin{equation}
\label{eq:rp2_selection_rule}
c_\ell = 0 \quad\text{for odd } \ell;
\qquad
f_0(\hat{\mathbf{n}}\cdot\mathbf{x})
= \sum_{\ell\,\text{even}}c_\ell\,
P_\ell(\hat{\mathbf{n}}\cdot\mathbf{x}).
\end{equation}
The same selection rule holds for any $\mathbb{Z}_2$-invariant
function on the orientation, since single-valued functions on
$\mathbb{RP}^2$ are spanned by even-$\ell$ spherical
harmonics. As a concrete example, the Gaussian band profile
$f_0(u) = \mathcal{N}\exp(-u^2/(2\sigma^2))$, with $\sigma$
the ring width and $\mathcal{N}$ the normalization, gives
\begin{equation}
\label{eq:c_ell_gaussian}
c_\ell = \frac{2\ell+1}{2}\,\mathcal{N}
\int_{-1}^{1}\exp\!\left(-\frac{u^2}{2\sigma^2}\right)
P_\ell(u)\,du,
\end{equation}
with $c_{2k+1} = 0$ by~\eqref{eq:rp2_selection_rule}. For a
narrow ring ($\sigma \ll 1$), many even Legendre modes
contribute and the sum must be truncated at
$\ell_{\max} \sim 1/\sigma$ for accurate evaluation; for a
broad ring ($\sigma \sim 1$), the expansion is dominated by
the first few even terms.

\paragraph{Two-time correlator of Legendre polynomials
(Markovian).}
For \emph{Markovian} isotropic rotational diffusion on
$S^2$ with diffusion coefficient $D_{\text{rot}}$, the
stationary two-time correlator of Legendre polynomials
evaluated at fixed points
$\mathbf{x}, \mathbf{x}' \in S^2$ is
\begin{equation}
\label{eq:Legendre_two_time}
\left\langle P_\ell(\hat{\mathbf{n}}(t)\cdot\mathbf{x})
\,P_{\ell'}(\hat{\mathbf{n}}(t')\cdot\mathbf{x}')
\right\rangle
= \frac{\delta_{\ell\ell'}}{2\ell+1}\,
e^{-\ell(\ell+1)D_{\text{rot}}|\tau|}\,
P_\ell(\mathbf{x}\cdot\mathbf{x}')
= \frac{\delta_{\ell\ell'}}{2\ell+1}\,
[q(\tau)]^{\ell(\ell+1)/2}\,
P_\ell(\mathbf{x}\cdot\mathbf{x}'),
\end{equation}
where $q(\tau) = \langle\hat{\mathbf{n}}(t)\cdot
\hat{\mathbf{n}}(t')\rangle = e^{-2D_{\text{rot}}|\tau|}$
is the $\ell=1$ orientation correlator. The compact
``$[q(\tau)]^{\ell(\ell+1)/2}$'' form is simply a
rewriting of the Markovian exponential
$e^{-\ell(\ell+1)D_{\text{rot}}|\tau|}$; it does
\emph{not} generalize to arbitrary $q(\tau)$. 
In the
non-Markovian regime (memory kernel
$\kappa(\tau) \ne \kappa\,\delta(\tau)$), the higher-$\ell$
spherical harmonics relax with their own memory
structure rather than as simple powers of the $\ell=1$
correlator, and the right-hand side
of~\eqref{eq:Legendre_two_time} must be replaced by a
distinct two-time function $q_\ell(\tau)$ for each
$\ell$. A compact derivation in the Markovian case,
using the addition theorem for spherical harmonics and
the heat-kernel expansion on $S^2$, is given in
Appendix~\ref{app:legendre_two_time}.

\paragraph{Density-density correlator (Markovian).}
Combining~\eqref{eq:f0_legendre}
and~\eqref{eq:Legendre_two_time}, in the Markovian limit
of the orientation dynamics:
\begin{equation}
\label{eq:Crho_reduced}
C_\rho(\mathbf{x},t;\mathbf{x}',t')
= \left\langle
f_0(\hat{\mathbf{n}}(t)\cdot\mathbf{x})\,
f_0(\hat{\mathbf{n}}(t')\cdot\mathbf{x}')
\right\rangle
= \sum_{\ell\,\text{even}}\frac{c_\ell^2}{2\ell+1}\,
e^{-\ell(\ell+1)D_{\text{rot}}|\tau|}\,
P_\ell(\mathbf{x}\cdot\mathbf{x}').
\end{equation}
The sum runs over even $\ell$ only: $c_\ell^2 = 0$ for odd
$\ell$ by the $\mathbb{RP}^2$ selection
rule~\eqref{eq:rp2_selection_rule}. Equivalently,
$C_\rho$ is built from products of two parity-even
factors $f_0$ and is therefore intrinsically
$\mathbb{Z}_2$-invariant in $\hat{\mathbf{n}}$, so it
descends to a function on $\mathbb{RP}^2$ and can only
contain even-$\ell$ harmonics; this is the density-sector
analog of the heat kernel on $\mathbb{RP}^2$ being the
even-$\ell$ projection of the heat kernel on $S^2$. Each
even Legendre mode $\ell$ of the density correlator decays
exponentially at its own rate $\ell(\ell+1)D_{\text{rot}}$,
an eigenvalue of the Laplace--Beltrami operator on
$\mathbb{RP}^2$ (which is the even-$\ell$ subset of the
spectrum on $S^2$). The full space-time structure of
$C_\rho$ in the Markovian regime is fixed by the single
dynamical parameter $D_{\text{rot}}$ and the even-$\ell$
band-profile coefficients
$c_\ell$~\eqref{eq:f0_legendre}. The closed
form~\eqref{eq:Crho_reduced} presupposes Markovian
isotropic diffusion of $\hat{\mathbf{n}}(t)$ on $S^2$, so
agreement with simulation does not by itself \emph{infer}
Markovianity; once $D_{\text{rot}}$ has been fixed from a
single observable, agreement of the full
$\ell$-resolved Legendre spectrum with
\eqref{eq:Crho_reduced} \emph{tests} Markovianity in the
density sector, the test that is carried out in
Section~\ref{subsec:comparison}. 

\subsection{Langevin equations for the orientation vector}
\label{subsec:orientation_sde}

The equations of motion for $\hat{\mathbf{n}}(t)$ follow
from the effective action by the standard MSRJD
correspondence. The general (non-Markovian) case gives
the GLE below; the Markovian limit gives a simple SDE.

\paragraph{Markovian SDE.}
In the Markovian limit~\eqref{eq:NLSM_action},
$\hat{\mathbf{n}}$ undergoes pure diffusion on
$S^2_{\hat{\mathbf{n}}}$ with no deterministic drift (in
tangent-plane coordinates $n^a$, $a = 1,2$):
\begin{equation}
\label{eq:langevin_tangent}
\dot{n}^a = \sqrt{2D_{\text{rot}}}\,\xi^a(t),
\end{equation}
where $D_{\text{rot}} = 1/(2\kappa)$ and $\xi^a$ is unit
white noise with
$\langle\xi^a(t)\xi^b(t')\rangle
= \delta^{ab}\delta(t-t')$.

\paragraph{Embedding-space form.}
In embedding-space coordinates, the
Stratonovich SDE on $S^2$ is:
\begin{equation}
\label{eq:markov_sde}
d\hat{\mathbf{n}} = \sqrt{2D_{\text{rot}}}\,
(\mathbf{I} - \hat{\mathbf{n}}\hat{\mathbf{n}}^T)
\circ d\mathbf{W}(t),
\end{equation}
where $P_{ij} = \delta_{ij} - n_in_j$ is the
tangent-plane projector ensuring
$|\hat{\mathbf{n}}|^2 = 1$. This is isotropic Brownian
motion on $S^2$ with no preferred direction. The It\^o
form:
\begin{equation}
\label{eq:nhat_ito}
dn_i = -2D_{\text{rot}}\,n_i\,dt
+ \sqrt{2D_{\text{rot}}}\,(\delta_{ij} - n_in_j)\,dW_j.
\end{equation}
The term $-2D_{\text{rot}}n_i$ is the geometric
Stratonovich-to-It\^o correction (proportional to the
mean curvature vector of $S^2 \subset \mathbb{R}^3$),
not a physical force. The embed-and-project simulations
of Ref.~\cite{halperin2026frustrated} implement
\eqref{eq:nhat_ito} numerically without writing this
drift explicitly: the post-step renormalization
$\hat{\mathbf{n}}\to\hat{\mathbf{n}}/|\hat{\mathbf{n}}|$
that enforces the unit-norm constraint reproduces the
$-2D_{\text{rot}}\hat{\mathbf{n}}$ drift to leading order
in $dt$.
%
%\paragraph{External fields.}
%If the SO(3) symmetry is broken by an external
%perturbation (e.g., a gravitational or electric field in
%the polymer-bubble realization of
%Section~\ref{sec:discussion}), a potential
%$V(\hat{\mathbf{n}})$ is generated, adding a drift
%$-D_{\text{rot}}\nabla_{S^2}V$ to the SDE:
%\begin{equation}
%\label{eq:sde_general}
%d\hat{\mathbf{n}} = -D_{\text{rot}}
%P(\hat{\mathbf{n}})\nabla_{S^2}V\,dt
%+ \sqrt{2D_{\text{rot}}}\,P(\hat{\mathbf{n}})
%\circ d\mathbf{W}.
%\end{equation}
%Without an external field, $V = 0$ and this reduces to
%the pure diffusion~\eqref{eq:markov_sde}.

\paragraph{GLE (non-Markovian).}
In the non-Markovian regime, the memory kernel
$\kappa(\tau)$~\eqref{eq:M_alpha_beta} replaces the
local stiffness $\kappa\delta(\tau)$. The generalized
Langevin equation is most cleanly written in
tangent-plane coordinates $n^a$ ($a = 1, 2$) at the
current orientation, in which it follows directly from
the quadratic action~\eqref{eq:adiabatic_expansion}:
\begin{equation}
\label{eq:GLE_tangent}
\int_0^t d\tau'\;\kappa(t - \tau')\,
\dot{n}^a(\tau')
\;=\; \zeta^a(t),
\qquad
\left\langle\zeta^a(t)\,\zeta^b(t')\right\rangle
\;=\; \kappa(|t - t'|)\,\delta^{ab},
\end{equation}
where $\zeta^\alpha$ is the tangent-plane noise,
satisfying the standard fluctuation-dissipation relation
for a linear GLE~\cite{kubo1966}. The $\delta^{\alpha\beta}$ in the
noise correlator is the residual-$\mathrm{SO}(2)$
isotropy among the two tangent directions at
$\hat{\mathbf{n}}$; no function of
$\hat{\mathbf{n}}(t)$ enters, because tangent-plane
coordinates are adapted to the current orientation.

In the Markovian limit
$\kappa(\tau) \to \kappa\,\delta(\tau)$,
Eq.~\eqref{eq:GLE_tangent} collapses to
$\kappa\,\dot{n}^\alpha(t) = \zeta^\alpha(t)$ with
$\langle\zeta^\alpha(t)\,\zeta^\beta(t')\rangle =
\kappa\,\delta^{\alpha\beta}\,\delta(t - t')$. Rescaling
to a unit-amplitude white noise
$\xi^\alpha = \zeta^\alpha/\sqrt{\kappa}$ (so that
$\langle\xi^\alpha\xi^\beta\rangle =
\delta^{\alpha\beta}\delta(t - t')$) and using
$D_{\text{rot}} = 1/(2\kappa)$, Eq.~\eqref{eq:GLE_tangent}
reduces to
$\dot{n}^\alpha = \sqrt{2 D_{\text{rot}}}\,\xi^\alpha$,
which is exactly the Markovian
SDE~\eqref{eq:langevin_tangent}.

The embedding-space version of~\eqref{eq:GLE_tangent}
can be obtained by contracting with the tangent-frame
vectors $e^\alpha_i(\hat{\mathbf{n}})$; the memory
integral then introduces a two-time tangent
projector $\sum_\alpha e^\alpha_i(\hat{\mathbf{n}}(t))\,
e^\alpha_j(\hat{\mathbf{n}}(\tau'))$ between the current
and past tangent spaces, which reduces to the ordinary
$\delta_{ij} - n_i n_j$ only at coincident times. The
tangent-plane form~\eqref{eq:GLE_tangent} is therefore
the simplest invariant statement of the non-Markovian
orientation dynamics.

%% ============================================================
\subsection{The effective theory: what is fixed by
symmetry and what requires computation}
\label{subsec:three_tiers}
\label{subsec:self_consistent_eqs}

The derivation of
Sections~\ref{subsec:starting_point}--\ref{subsec:orientation_sde}
fixes the leading low-energy form of the effective theory for
the F2 model on the sphere $S^2$ under the stated assumptions
(simulation-supported ring saddle, BO separation, leading
saddle / one-loop measure). The result is the
$\mathbb{RP}^2$ NLSM in 0+1 dimensions
(Eq.~\eqref{eq:NLSM_action} or its non-Markovian
generalization~\eqref{eq:S_eff_nonmarkov}),
with a single scalar stiffness
$\kappa = 1/(2D_{\text{rot}})$ (Markovian) or a single scalar
memory kernel $\kappa(\tau)$ (non-Markovian). Five symmetry
constraints fix this form uniquely:
(i) adiabatic SO(3)$\to$SO(2)$_{\hat{\mathbf{n}}}$ symmetry
breaking $\Rightarrow$ a unit-vector collective coordinate,
and round metric / single scalar $\kappa$ by SO(2) isotropy;
(ii) SO(3) invariance of the underlying disorder-averaged
theory $\Rightarrow$ no $\hat{\mathbf{n}}$-dependent
potential at any loop order
(Section~\ref{subsec:faddeev_popov});
(iii) the $\mathbb{Z}_2$ identification
$\hat{\mathbf{n}} \sim -\hat{\mathbf{n}}$ inherited from the
even ring profile $f_0(u) = f_0(-u)$ (the director
property) $\Rightarrow$ the collective coordinate lives on
the projective sphere
$\hat{\mathbf{n}}(t) \in \mathbb{RP}^2 = S^2/\mathbb{Z}_2$,
single-valued physical observables are spanned by even-$\ell$
spherical harmonics
(Eq.~\eqref{eq:rp2_selection_rule}), and the model is the
$\mathbb{RP}^2$ NLSM rather than the standard
O(3) NLSM on $S^2$;
(iv) representation orthogonality
($\mathrm{SO}(2)_{\hat{\mathbf{n}}}$ vector
$\times$ singlet through a generator-commuting kernel
vanishes) $\Rightarrow$ no Berry phase, for any ring
profile;
(v) MSRJD structure $\Rightarrow$ Onsager-Machlup form.
No other parameters can appear at leading order in the BO
parameter $\epsilon$~\eqref{eq:BO_epsilon}.

This situation parallels the construction of low-energy
effective theories elsewhere in physics, where the form is
dictated by the symmetry-breaking pattern while the
numerical values of the low-energy constants require
additional input. The relevant comparison is:
\begin{center}
\begin{tabular}{l|l|l}
& QCD & F2 model \\
\hline
Fundamental theory &
quarks + gluons &
$\rho$ on $S^2$ (after $G$, $\lambda$ integrated out) \\
Symmetry breaking &
$\text{SU}(2)_L\!\times\!\text{SU}(2)_R
\to \text{SU}(2)_V$ &
$\text{SO}(3) \to \text{SO}(2)$, with $\mathbb{Z}_2$ director identification \\
Target manifold &
$\mathrm{SU}(2)$ (pion field) &
$\mathbb{RP}^2 = S^2/\mathbb{Z}_2$ (projective rotor space) \\
Soft modes &
Nambu-Goldstone pions &
zero-mode/orientation diffusion (director) \\
Effective theory &
chiral Lagrangian &
$\mathbb{RP}^2$ NLSM \\
Low-energy constant &
$f_\pi$ &
$D_{\text{rot}}$ (or $\kappa(\tau)$) \\
\end{tabular}
\end{center}
The numerical values of $D_{\text{rot}}$ and the band
profile $f_0(\theta)$ in the F2 model can be obtained at
three levels.

\paragraph{Tier 1: structural (from symmetry).}
The form of the effective theory is fully specified by the
constraints above and yields parameter-free results: the
effective action, the SDE, and the GLE. Given any value of
$D_{\text{rot}}$, the theory then predicts all
orientation observables, including the autocorrelation
$q(\tau) = e^{-2D_{\text{rot}}|\tau|}$ (Markovian), the
MSD, and the FDR $X(\tau)$.

\paragraph{Tier 2: phenomenological (inputs from
simulation).}
$D_{\text{rot}}$ and $f_0(\theta)$ are treated as
empirical inputs measured from particle simulations of the
full F2 model, the analog of extracting $f_\pi$ from pion
data. With these two inputs, the effective theory predicts
all other orientation observables, and the density
correlator follows from~\eqref{eq:Crho_reduced}. For the
simulations of Section~\ref{subsec:comparison}, $f_0$ is a
Gaussian with $\sigma = 4.8^\circ$
(Fig.~\ref{fig:ring_frame}) and
$D_{\text{rot}} = 0.003 \pm 0.002$
(Table~\ref{tab:Drot_disorder}); the resulting predictions
match the particle simulation on all seven tested
diagnostics.

\paragraph{Tier 3: fully microscopic (self-consistency
loop).}
A Tier-3 computation would yield $f_0$ and $D_{\text{rot}}$
ab-initio from the microscopic F2 parameters
$(\Omega, \alpha^2, N, \Delta\theta)$ by solving a coupled
nonlinear system: (i)~the ring profile is fixed by the EL
equation~\eqref{eq:EL_explicit} at the saddle, which
involves the drift $F^u$ and the noise kernel $\Omega$;
(ii)~$F^u$ and $\Omega$ are functionals of the density
correlator $C_\rho$, which in the BO limit is expressed
through $f_0$ and $q(\tau)$~\eqref{eq:Crho_reduced};
(iii)~the stiffness $\kappa$~\eqref{eq:M_alpha_beta} is
the overlap integral of $\Omega^{-1}$ against the
zero-mode direction $\psi_1$ at the saddle. Together,
these conditions form a closed nonlinear loop in the
scalar functions $\bar{C}(\tau)$, $\bar{R}(\tau)$,
$\bar{K}(\tau)$, and $f_0(\theta)$, the projection of the
F2 Schwinger-Dyson hierarchy
(Section~\ref{subsec:large_N_saddle}) onto orientation
space, and constitute the lattice-QCD analog for the F2
problem. The system has not been solved here because the
dressed kernels are non-local on $S^2$ and their
projection couples all Legendre modes $c_\ell$; moreover,
the quenched disorder induces large sample-to-sample
fluctuations (coefficient of variation $\sim 0.68$ for
$D_{\text{rot}}$, Table~\ref{tab:Drot_disorder}), so the
disorder-averaged self-consistency need not represent the
typical quenched behavior.

At the structural level, this construction is formally
parallel to that of the spherical $p$-spin
glass~\cite{sompolinsky1981, bouchaud1990classical,
cugliandolo1993, cugliandolo1997}. The normalized spin
vector $\boldsymbol{\sigma}(t)/\sqrt{N}$ plays the role of
$\hat{\mathbf{n}}(t)$, the $p$-spin two-point function
$C(\tau) = (1/N)\langle\boldsymbol{\sigma}(t)
\cdot\boldsymbol{\sigma}(t')\rangle$ plays the role of
$q(\tau)$, and the Cugliandolo-Kurchan memory kernel
$\propto C^{p-1}$~\cite{cugliandolo1993, cugliandolo1997}
plays the role of the band-weighted force kernel generated
by the disorder average. The essential difference is that
the $p$-spin model is infinite-range and spatially
featureless, whereas the F2 model lives on a physical
two-dimensional surface and its spatial structure is
encoded in the ring profile
$f_0(\hat{\mathbf{n}}\cdot\mathbf{x})$, which enters the
effective theory only through the overlap integrals that
define $\kappa$.
Further physical analogs of this construction (nuclear
rotational bands, molecular rotation, and the Haldane
$\mathrm{O}(3)$ NLSM) are discussed in
Section~\ref{subsec:connections}.

%==============================================================================
\section{Numerical Implementation of the Effective $\mathbb{RP}^2$-NLSM Dynamics}
\label{sec:numerical_implementation}
%==============================================================================

The effective $\mathbb{RP}^2$ NLSM derived in
Section~\ref{sec:low_energy} for the F2 model on $S^2$
reduces the orientation dynamics to the Ito
SDE~\eqref{eq:nhat_ito} on $S^2$ with a single parameter
$D_{\text{rot}}$. Following the
Tier-2 approach
(Section~\ref{subsec:three_tiers}), $D_{\text{rot}}$ is
measured from particle simulations rather than computed
from the self-consistency equations. This section
describes the numerical implementation.\footnote{Python
code for the simulations and figures presented in this
section is available at
\url{https://github.com/ighalp/frustrated-brownian-particles-manifolds},
in the \texttt{F2\_model/} folder, with documentation
explaining how each figure and its underlying data are
reproduced.}

\begin{figure}[ht]
\centering
\includegraphics[width=\textwidth]{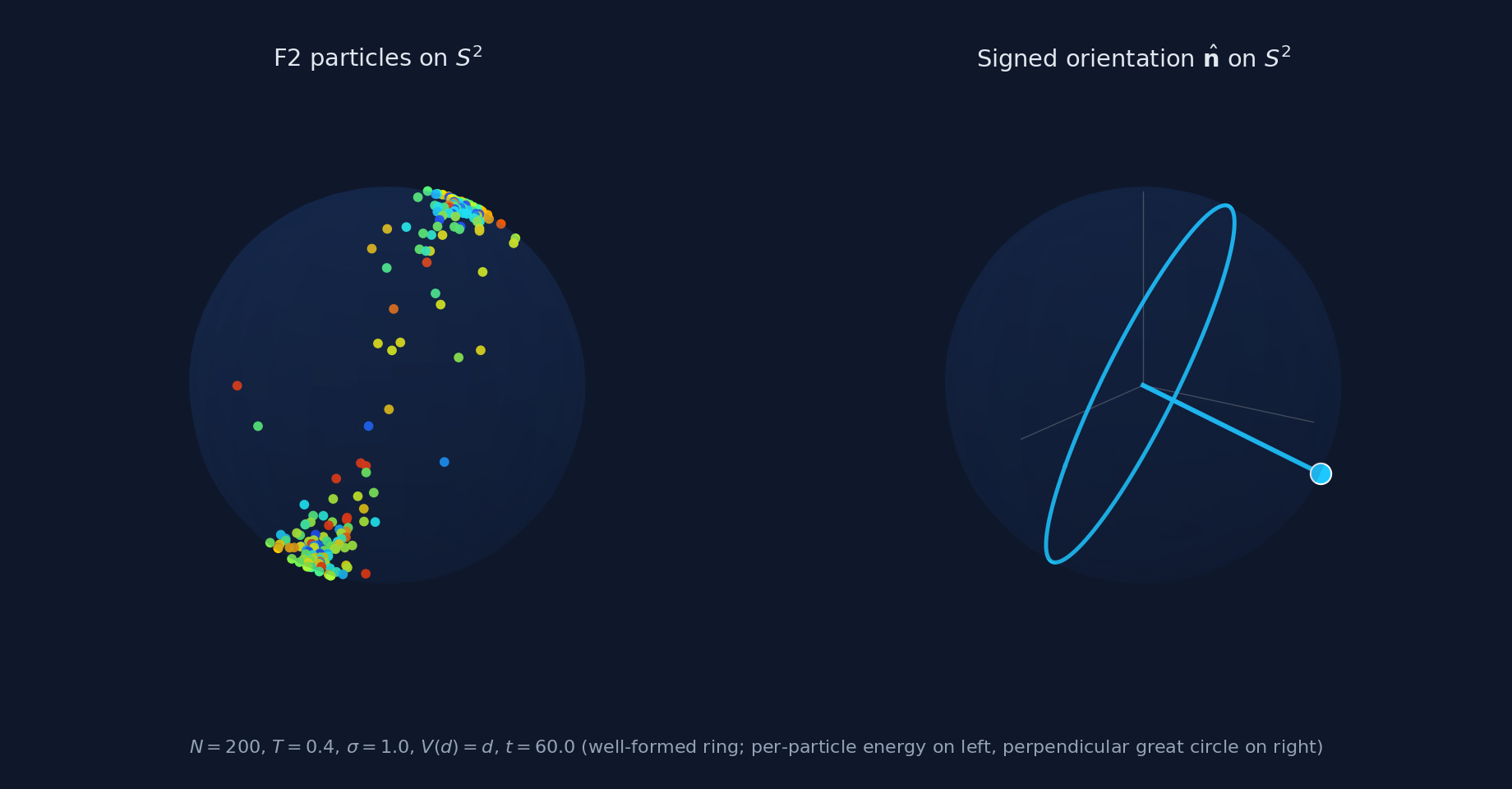}
\caption{Snapshot of an F2 simulation at $t = 60$ from a Big-Bang
initial condition ($N = 200$, $T = 0.4$, $\sigma = 1$, $V(d) = d$,
Gaussian quenched couplings). Left panel: particle positions on
$S^2$ once the ring has formed, colored by per-particle potential
energy $E_i = \sum_{j} \phi_{ij}\, d(x_i, x_j)$ on the same
blue-to-red rainbow used by the WebGL simulator. Right panel: the
signed orientation $\hat{\mathbf{n}}(t)$, extracted as the
smallest-eigenvalue eigenvector of the inertia tensor with the
continuity sign convention, drawn together with the director line
from the origin and the great circle perpendicular to
$\hat{\mathbf{n}}$, which is the locus along which the ring is
concentrated. The full animated version, showing the Big-Bang
spread and ring formation in log-time and the long-time precession
of the ring across two independent disorder realizations, is
\texttt{gifs/f2\_signed\_orientation.gif} in the project
repository
(\url{https://github.com/ighalp/frustrated-brownian-particles-manifolds}).}
\label{fig:ring_snapshot}
\end{figure}

\subsection{Simulating the $\mathbb{RP}^2$ NLSM}
\label{subsec:markov_numerics}

The $\mathbb{RP}^2$-NLSM SDE in embedding-space Ito
form~\eqref{eq:nhat_ito} is:
$dn_i = -2D_{\text{rot}}n_i\,dt
+ \sqrt{2D_{\text{rot}}}(\delta_{ij} - n_in_j)\,dW_j$.
Given $D_{\text{rot}}$, the simulation proceeds by
Euler-Maruyama: at each time step,
(i) generate a noise increment
$\Delta W_j \sim \mathcal{N}(0, \Delta t)$ in
$\mathbb{R}^3$;
(ii) apply the Ito drift $-2D_{\text{rot}}n_i\Delta t$
and the projected noise
$\sqrt{2D_{\text{rot}}}(\delta_{ij}-n_in_j)\Delta W_j$;
(iii) renormalize
$\hat{\mathbf{n}} \to \hat{\mathbf{n}}/|\hat{\mathbf{n}}|$.
From the simulated trajectory, the observables
$q(\tau) = \langle\hat{\mathbf{n}}(t)\cdot
\hat{\mathbf{n}}(t+\tau)\rangle$, MSD, and the power
spectrum are computed and compared with the particle
simulation (Section~\ref{subsec:comparison}).

The value $D_{\text{rot}} = 0.0031$ used in the
$\mathbb{RP}^2$-NLSM simulation is measured from a long
particle simulation ($N = 400$, $T = 0.4$, $t = 200$) by
fitting the autocorrelation
$q(\tau) = e^{-2D_{\text{rot}}\tau}$ to the
inertia-tensor-based orientation data. No self-consistency
equations, ring profile computation, or noise kernel
evaluation is needed: the single measured parameter
$D_{\text{rot}}$ fully specifies the effective theory.

\paragraph{Consistency of the simulation pipeline with the
$\mathbb{RP}^2$ topology.}
Two design choices in the simulation pipeline make it
automatically consistent with the projective-target
identification of
Section~\ref{subsec:low_energy_ansatz}, without requiring
any explicit topological-sector accounting. First,
Eq.~\eqref{eq:nhat_ito} is integrated as Brownian motion
on the lifted $S^2$ rather than directly on $\mathbb{RP}^2$;
this is exactly equivalent to the method-of-images
construction of Brownian motion on $\mathbb{RP}^2$ (the
projection $\hat{\mathbf{n}}\to[\hat{\mathbf{n}}]$ takes
the $S^2$-Brownian motion to the $\mathbb{RP}^2$-Brownian
motion), and any $\mathbb{Z}_2$-invariant observable
computed from the lifted trajectory agrees with its
$\mathbb{RP}^2$ counterpart. Second, the inertia-tensor
extraction in the particle simulation returns an unsigned
line in $\mathbb{R}^3$ — a point in $\mathbb{RP}^2$ — and
the continuity-based sign convention used to track
$\hat{\mathbf{n}}(t)$ from one snapshot to the next (pick
the sign that minimises the angular jump from the previous
step) is the operational realization of the gauge-fixed
lift. In the diffusive regime resolved by the present
simulations, $q(\tau) > 0$ throughout the run, the lifted
trajectory does not enter the $w = 1$ winding class, and
the classical $\theta = 0$ sector (positive density, no
internal spinor structure) is the only sector that
contributes. The simulation therefore reproduces the
$\mathbb{RP}^2$-NLSM dynamics correctly without an
explicit sum over topological sectors; the
$\mathbb{Z}_2$ structure shows up only as the
even-$\ell$-only selection rule of
Eq.~\eqref{eq:rp2_selection_rule} on the physical
density-sector observables (verified at the
$100$-realization level in
Section~\ref{subsec:comparison}, density correlator).

\subsection{Comparison with particle simulations}
\label{subsec:comparison}

\paragraph{Ring profile: theory vs.\ simulation.}
The ring profile $f_0(\theta)$ and the stiffness
$\kappa$ are determined by the self-consistency problem
described in Section~\ref{subsec:self_consistent_eqs},
which couples the ring profile, the noise kernel
$\Omega$, and the drift $F^u$ into a system that must
be solved simultaneously. This Tier-3 computation has
not yet been carried out.

\begin{figure}[H]
\centering
\includegraphics[width=\textwidth]{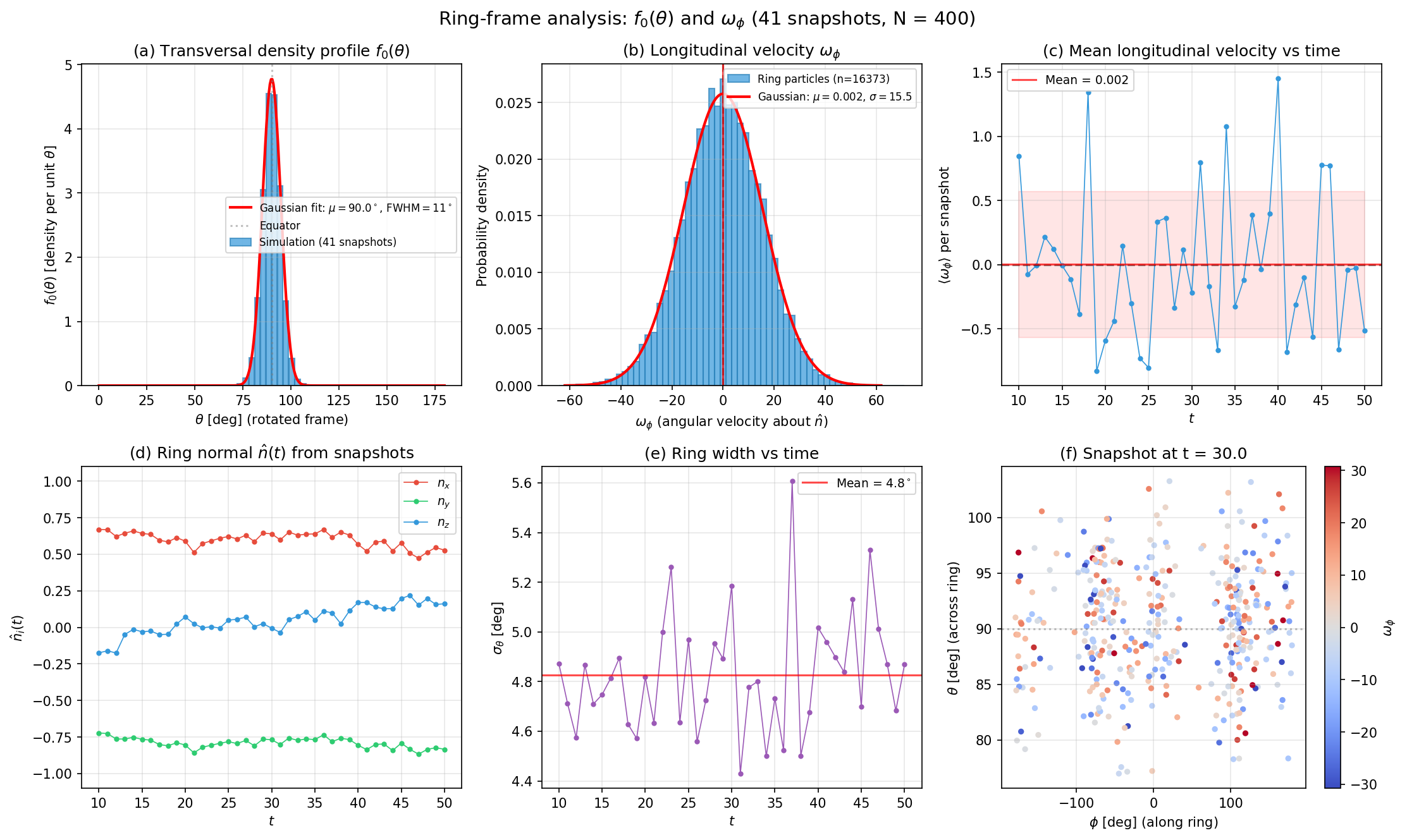}
\caption{Ring-frame analysis from $41$ snapshots
($N = 400$, $T = 0.4$).
(a) Transversal density profile $f_0(\theta)$ in the
rotated frame, aggregated over all snapshots. The
Gaussian fit (red) gives $\mu = 90.0^\circ$
(equatorial), $\sigma = 4.8^\circ$, FWHM $= 11^\circ$.
Nearly all particles ($399/400$) lie within the ring.
(b) Distribution of longitudinal angular velocity
$\omega_\phi$ (velocity along the ring about
$\hat{\mathbf{n}}$): symmetric Gaussian with mean
$= 0.002 \pm 0.121$ and $\sigma = 15.5$, consistent
with zero drift
($|\text{mean}/\text{std}| = 10^{-4}$).
(c) Mean $\omega_\phi$ per snapshot: fluctuates about
zero with no trend.
(d) Ring normal $\hat{\mathbf{n}}(t)$ from the inertia
tensor at each snapshot: slow diffusive evolution.
(e) Ring width $\sigma_\theta$ is stable at
$\approx 4.8^\circ$ throughout the NESS.
(f) Single-snapshot scatter plot
$(\phi, \theta)$ colored by $\omega_\phi$: no systematic
azimuthal pattern.}
\label{fig:ring_frame}
\end{figure}

In the absence of the full self-consistent solution,
the ring profile is measured directly from the particle
simulation via the multi-snapshot ring-frame analysis
(Fig.~\ref{fig:ring_frame}). The measured profile is
Gaussian with $\sigma = 4.8^\circ$
(full width at half maximum, FWHM, $= 11^\circ$),
centered at the equator in the
rotated frame (Section~\ref{subsec:comparison},
``Ring density profile''). This measured $f_0(\theta)$
provides the microscopic input needed for computing
the stiffness $\kappa$~\eqref{eq:M_alpha_beta} and
the Legendre coefficients
$c_\ell$~\eqref{eq:c_ell_gaussian}, without requiring
the self-consistent solution of the full kernel system.

\paragraph{Orientation dynamics and comparison with
particle simulations.}
We solve the SDE~\eqref{eq:nhat_ito} and compare with
particle simulations. The orientation
$\hat{\mathbf{n}}(t)$ is extracted from the particle positions
via the inertia tensor (eigenvector with the smallest eigenvalue
of $(1/N)\sum_n\mathbf{x}_n\mathbf{x}_n^T$).

\paragraph{Estimator for $q(\tau)$ from a single long
trajectory.}
The $\mathbb{RP}^2$ NLSM predicts the orientation autocorrelation
as a disorder- and ensemble-averaged two-point function
$q(\tau) = \langle\hat{\mathbf{n}}(0)\cdot
\hat{\mathbf{n}}(\tau)\rangle$. Given a single trajectory
$\hat{\mathbf{n}}(t)$ of duration $T_{\rm run}$ at fixed disorder,
we estimate it by the time-translation average
\begin{equation}
\label{eq:q_estimator}
\widehat{q}(\tau)
= \frac{1}{T_{\rm run} - \tau}\int_0^{T_{\rm run} - \tau}
\hat{\mathbf{n}}(t) \cdot \hat{\mathbf{n}}(t + \tau)\,dt,
\end{equation}
and fit the decay rate $2 D_{\text{rot}}$ to the
short-lag exponential $\widehat{q}(\tau) \simeq
e^{-2 D_{\text{rot}}\tau}$.
Isotropic diffusion on $S^2$ at rate $D_{\text{rot}}$
is ergodic and mixing on the compact target space, so as
$T_{\rm run} \to \infty$ the time
average~\eqref{eq:q_estimator} converges to the
stationary ensemble average uniform on $S^2$, with a
statistical error that scales as $1/\sqrt{T_{\rm run}/\tau_c}$ at
each lag, where $\tau_c = 1/(2 D_{\text{rot}})$. An
interval of duration $T_{\rm run}$ thus contains $\sim T_{\rm run}/\tau_c$
effectively independent orientation samples.

Our long single-trajectory run at $N = 400$ ($T = 0.4$) has
$T_{\rm run} = 200$ and $\tau_c \approx 164$, so
$T_{\rm run}/\tau_c \approx 1.2$: the trajectory contains roughly
one orientation-correlation time beyond the transient.
This is enough to determine $D_{\text{rot}}$ from the
short-lag slope of $\widehat{q}(\tau)$ and to resolve
the MSD, the power spectrum, the Markovian-velocity
diagnostic, and the memory kernel at the lags where each
dominates, all of which depend on the well-sampled
$\tau \ll \tau_c$ regime. Point-by-point agreement of
$\widehat{q}(\tau)$ near $\tau \sim \tau_c$, where the
curve approaches zero and the sampling noise dominates,
is \emph{not} tested at this level of statistics. A
stronger test uses the disorder-averaged estimator
$\langle q(\tau)\rangle_{\text{disorder}}$ obtained by
pooling many independent realizations, which is both the
object directly predicted by the F2 theory and
statistically far better-sampled; the partial
10-realization ensemble of
Table~\ref{tab:Drot_disorder} already constrains the
mean $D_{\text{rot}}$, and a larger ensemble dedicated to
the full curves $\langle q(\tau)\rangle$ and
$\langle C_\rho(\gamma,\tau)\rangle$ is deferred to a
follow-up analysis.

\begin{figure}[H]
\centering
\includegraphics[width=\textwidth]{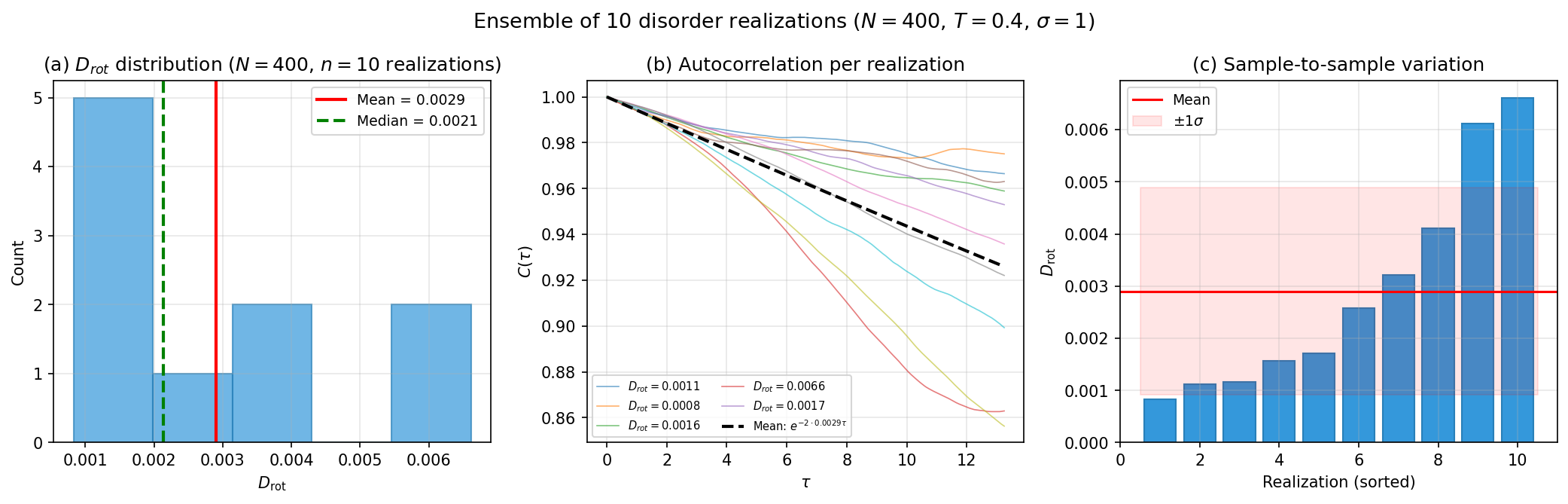}
\caption{Ensemble of $10$ disorder realizations at
$N = 400$, $T = 0.4$.
(a) Distribution of $D_{\text{rot}}$ across
realizations.
(b) Orientation autocorrelation $C(\tau)$ for each
realization (colored) and the ensemble mean (black
dashed).
(c) Sample-to-sample variation of $D_{\text{rot}}$
(sorted).}
\label{fig:ensemble_N400}
\end{figure}

To estimate $D_{\text{rot}}$ and quantify its
sample-to-sample fluctuations across disorder, we ran $10$
independent particle simulations at $N = 400$, $T = 0.4$,
$\sigma = 1$, $\gamma = 1$ ($dt = 0.005$, total simulation
time $t = 50$), each with a different realization of the
quenched couplings $\phi_{nm}$
(Table~\ref{tab:Drot_disorder},
Fig.~\ref{fig:ensemble_N400}). The orientation
$\hat{\mathbf{n}}(t)$ was extracted at each recorded step via
the inertia tensor and $D_{\text{rot}}$ fitted from the
short-lag slope of $\widehat{q}(\tau)$. The ensemble gives
$\langle D_{\text{rot}}\rangle = 0.0029 \pm 0.0020$, with a
coefficient of variation $\approx 0.68$, a hallmark of
quenched disorder. The long single-trajectory value
$D_{\text{rot}} = 0.0031$ ($\tau_c = 164$) extracted from the
$t = 200$ run lies close to the ensemble mean.
\begin{table}[H]
\centering
\begin{tabular}{c|c}
Quantity & Value \\
\hline
$D_{\text{rot}}$ range & $0.0008$--$0.0066$ (8-fold variation) \\
$\langle D_{\text{rot}}\rangle$ (10 samples) &
$0.0029 \pm 0.0020$ \\
median $D_{\text{rot}}$ & $0.0021$ \\
$\tau_c = 1/(2\langle D_{\text{rot}}\rangle)$ & $172$ \\
Coeff.\ of variation & $0.68$ \\
\end{tabular}
\caption{Rotational diffusion coefficient from $10$
independent particle simulations at $N = 400$, $T = 0.4$,
$\sigma = 1$, $\gamma = 1$, $dt = 0.005$. Each sample
uses a different realization of the quenched couplings.
$\hat{\mathbf{n}}(t)$ extracted via the inertia tensor.}
\label{tab:Drot_disorder}
\end{table}

\paragraph{Quenched vs.\ disorder-averaged dynamics.}
Three different objects appear in the comparison and should be
distinguished. The \emph{annealed} $D_{\text{rot}}$ would be
extracted from a theory in which $\phi_{nm}$ is integrated over
the same Gaussian distribution as the dynamical degrees of
freedom; this is not what we compute. The \emph{disorder-averaged}
$D_{\text{rot}}$ predicted by the F2 field theory
(Section~\ref{subsec:large_N_saddle}) is built from the
disorder-averaged generating functional, in which $\phi_{nm}$ is
fixed for the dynamics and then integrated at the level of
expectation values, and is the right object to compare with the
ensemble mean over realizations. Finally, each \emph{quenched}
particle simulation in Table~\ref{tab:Drot_disorder} runs at a
fixed $\phi_{nm}$ and yields a sample-specific
$D_{\text{rot}}$. The coefficient of variation $\approx 0.68$
across the $10$ samples is consistent with non-self-averaging
of $D_{\text{rot}}$ at $N = 400$: typical and disorder-averaged
values can differ, and the F2 prediction is for the latter. A
controlled separation of annealed, quenched-typical, and
disorder-averaged dynamics, including any $1/N$ corrections to
self-averaging, is left to future work.

\begin{figure}[H]
\centering
\includegraphics[width=\textwidth]{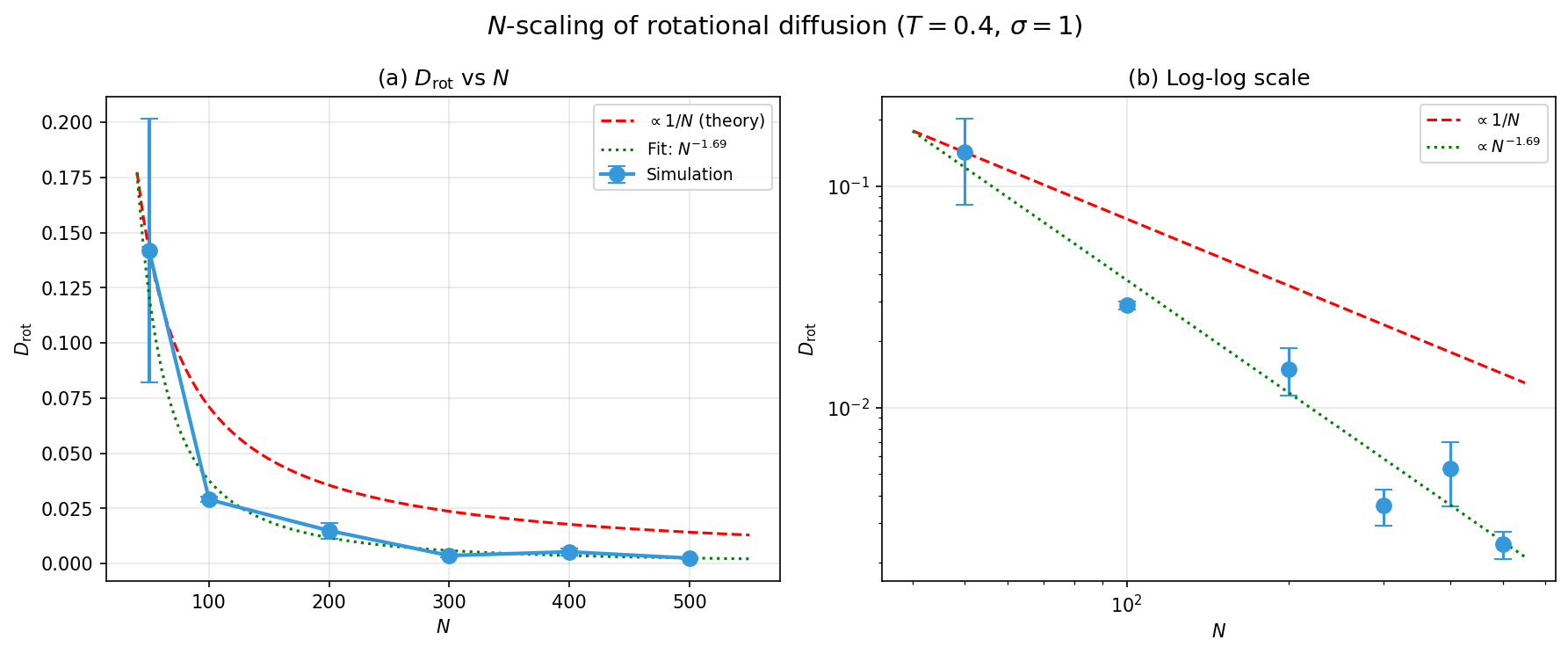}
\caption{$N$-scaling of $D_{\text{rot}}$ from particle
simulations ($T = 0.4$, $\sigma = 1$, $3$ disorder
realizations per $N$). (a) Linear scale. (b) Log-log
scale. The theoretical $1/N$ scaling (red dashed) and
the power-law fit $N^{-1.69}$ (green dotted) are shown.
$D_{\text{rot}}$ decreases with $N$, but faster than the
theoretical $1/N$.}
\label{fig:n_scaling}
\end{figure}

\paragraph{$N$-scaling of $D_{\text{rot}}$.}
The collective averaging mechanism
(Section~\ref{subsec:comparison}, ``Why the orientation
evolves slowly'') suggests that $D_{\text{rot}}$ should
decrease with $N$, since rotating the ring requires
coherently displacing more particles. To test this, we
ran particle simulations at $N = 50$, $100$, $200$,
$300$, $400$, $500$ ($T = 0.4$, $\sigma = 1$,
$\gamma = 1$), with $3$ independent disorder
realizations at each $N$
(Table~\ref{tab:Drot_N},
Figure~\ref{fig:n_scaling}).
\begin{table}[H]
\centering
\begin{tabular}{c|c|c|c}
$N$ & $\langle D_{\text{rot}}\rangle$ & $\pm$ sem
& $\tau_c$ \\
\hline
50  & 0.142 & 0.060 & 3.5 \\
100 & 0.029 & 0.001 & 17 \\
200 & 0.015 & 0.004 & 34 \\
300 & 0.0036 & 0.0007 & 139 \\
400 & 0.0053 & 0.0017 & 95 \\
500 & 0.0024 & 0.0003 & 207 \\
\end{tabular}
\caption{$D_{\text{rot}}$ measured from particle
simulations at varying $N$ ($T = 0.4$, $\sigma = 1$,
$\gamma = 1$), averaged over $3$ disorder realizations.}
\label{tab:Drot_N}
\end{table}

The available data show a clear decrease of $D_{\text{rot}}$
with $N$, but do not yet determine the asymptotic exponent.
A finite-range power-law fit to the simulated points gives
$D_{\text{rot}} \sim N^{-1.69 \pm 0.2}$ over the range
$N = 50$--$500$; the $N^{-1.69}$ curve should be read only as
a finite-sample fit over the simulated range, not as an
asymptotic exponent. Three disorder realizations per $N$
produce a noisy per-$N$ mean (the $N = 400$ point lies above
the $N = 300$ point, for instance), and the $N = 400$ value
in this small scan also differs from the $10$-realization
ensemble mean in Table~\ref{tab:Drot_disorder}. The reliable
qualitative conclusion is that $D_{\text{rot}}$ decreases
significantly with $N$; the asymptotic exponent is open.

Several mechanisms could plausibly produce the observed
faster-than-$1/N$ decrease at the simulated sizes:
correlated collective barriers that slow the orientation
diffusion in a way that grows with $N$; quenched pinning of
the ring profile $f_0$ by specific disorder realizations,
producing an effective inertia larger than the
annealed-collective-averaging $\kappa \propto N$; or a
finite-size crossover toward a different asymptotic exponent
at $N \gg 500$. The coefficient of variation $\approx 0.68$
at $N = 400$ (Section~\ref{subsec:comparison},
Table~\ref{tab:Drot_disorder}) shows that $D_{\text{rot}}$
is not yet self-averaging at the simulated sizes. Settling
the exponent and identifying its mechanism requires a
quenched calculation of $D_{\text{rot}}$ and a larger scan
in $N$ with more realizations per $N$, both left to future
work.

\paragraph{Diffusive vs.\ subdiffusive precession.}
The effective theory predicts purely diffusive precession
(MSD exponent $\alpha = 1$,
$\langle|\hat{\mathbf{n}}(t)
- \hat{\mathbf{n}}(0)|^2\rangle = 4D_{\text{rot}}t$)
for $t \ll \tau_c$.
Ref.~\cite{halperin2026frustrated} reported
$\alpha = 0.9 \pm 0.3$ from five independent
realizations at $N = 400$, with individual values
ranging from $0.55$ to $1.13$. The subdiffusive
character ($\alpha < 1$) in some realizations reflects
trapping in specific disorder landscapes: each quenched
set of couplings $\phi_{nm}$ creates a rugged free
energy landscape for the orientation, with barriers that
slow the diffusion below the purely diffusive rate. The
disorder-averaged theory smooths out these barriers,
producing the clean exponential autocorrelation and
linear MSD. The diffusive character
($\alpha \approx 1$) is a property of the
\emph{ensemble average}, not of individual quenched
realizations.

\paragraph{Ring density profile and longitudinal
velocity.}
To characterize the ring structure underlying the SO(3)
reduction, we recorded $41$ snapshots of particle
positions and velocities in the ring frame (the frame
where $\hat{\mathbf{n}} = \hat{z}$) at intervals of
$\Delta t = 1$ over $t = 10$ to $50$ (after ring
formation). Figure~\ref{fig:ring_frame} presents the
results.

The density profile $f_0(\theta)$ is well approximated
by a Gaussian centered at the equator ($\theta_0 =
90^\circ$) with $\sigma = 4.8^\circ$ (FWHM $= 11^\circ$).
This is the microscopic ring profile that enters the
stiffness $\kappa$~\eqref{eq:M_alpha_beta} through the
zero-mode overlap integral. The narrow width
($\sigma / \pi \approx 0.027$) is consistent with the
adiabatic separation assumed in the derivation: a
tightly localized ring forces transverse fluctuation
modes to vary on the angular scale $\sigma$, so their
intrinsic eigenvalues scale as $1/\sigma^2$, much larger
than $D_{\text{rot}}$. This is a geometric (mode-curvature)
indicator of stiffness in the transverse sector and is
distinct from the dynamical relaxation rate
$m^2_{\text{gap}} \simeq 1/\tau_{\text{fast}}$ entering
the BO parameter
$\epsilon$~\eqref{eq:BO_epsilon}.

The longitudinal velocity $\omega_\phi$ has zero mean
drift ($|\text{mean}/\text{std}| = 10^{-4}$), providing
direct confirmation at the particle level that no
Berry-phase precession occurs. The large standard
deviation $\sigma_\omega \approx 15.5$ reflects the fast
thermal motion of individual particles along the ring,
which averages out in the collective orientation
$\hat{\mathbf{n}}(t)$.

The ring-frame analysis confirms three properties
assumed in the $\mathbb{RP}^2$-NLSM reduction of Section~\ref{sec:low_energy}:
(i) the density is stationary in the co-rotating frame
(panel~e: constant ring width across all $41$ snapshots);
(ii) the density is azimuthally symmetric
(panel~f: uniform $\phi$-distribution, no azimuthal
structure in $\omega_\phi$);
(iii) the profile $f_0(\theta)$ is well approximated by
a Gaussian in $\theta$, or equivalently in the variable
$u = \cos\theta \approx \pi/2 - \theta$ near the
equator, with $\sigma_u \approx 0.084$ rad. This is the
form assumed in eq.~\eqref{eq:c_ell_gaussian}.

\paragraph{Why the orientation evolves slowly.}
Two effects produce the timescale separation
quantified by the BO parameter $\epsilon$ of
Section~\ref{subsec:low_energy_ansatz}: (i) the
\emph{zero-torque property} of pairwise interactions
($\sum_{ij}\phi_{ij}\mathbf{x}_i\times\hat{\mathbf{t}}_{ij}
= 0$, proven in
Ref.~\cite{halperin2026frustrated}), so that no
deterministic force rotates the ring and only thermal
noise drives the orientation; and (ii) \emph{collective
averaging}, since rotating the ring requires coherently
displacing all $N$ particles, with stiffness
$\kappa = 1/(2D_{\text{rot}}) \approx 167$ set by the
zero-mode overlap integral
$\iint\psi_1\,\Omega^{-1}\,\psi_1$~\eqref{eq:M_alpha_beta}.

\begin{figure}[H]
\centering
\includegraphics[width=\textwidth]{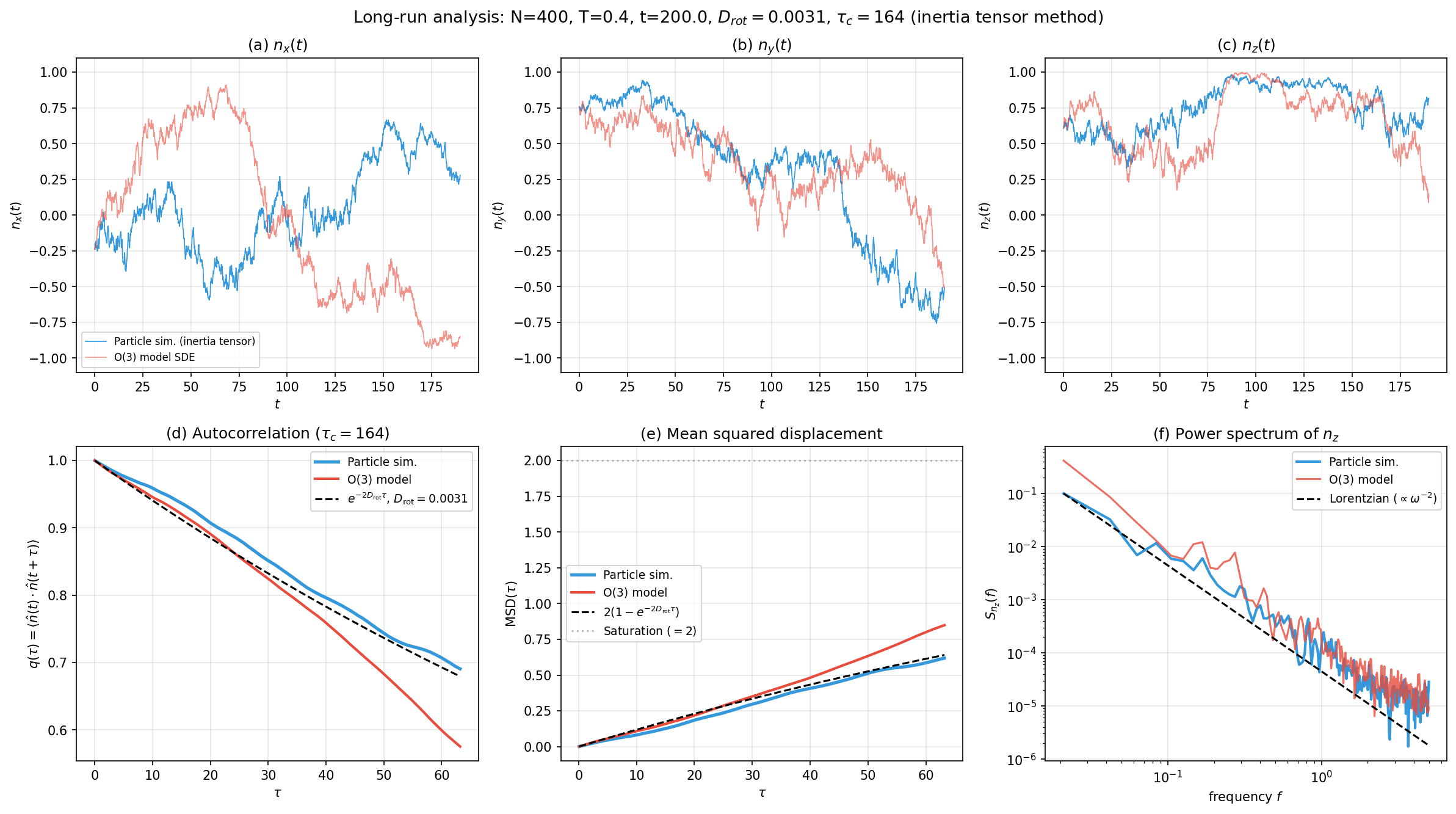}
\caption{Comparison of the effective SO(3)
model SDE (red) with the particle simulation (blue),
$N = 400$, $T = 0.4$, $t = 200$. Both start from the
same initial $\hat{\mathbf{n}}$; the $\mathbb{RP}^2$-NLSM SDE uses
$D_{\text{rot}} = 0.0031$ measured from the particle
simulation's autocorrelation, with the same fine
time step $dt = 0.005$ and recording cadence.
Top row: components (a) $n_x(t)$, (b) $n_y(t)$,
(c) $n_z(t)$ from both simulations, showing matching
volatility and similar diffusive character.
Bottom row: (d) autocorrelation
$q(\tau)$ matching $e^{-2D_{\text{rot}}\tau}$ with
$\tau_c = 164$;
(e) MSD saturating at $2$;
(f) power spectrum of $n_z$, both matching the
Markovian $\omega^{-2}$.}
\label{fig:o3_model_comparison}
\end{figure}

\paragraph{Direct simulation of the $\mathbb{RP}^2$ NLSM.}
To provide a quantitative comparison, we simulate the
effective $\mathbb{RP}^2$-NLSM SDE~\eqref{eq:nhat_ito} directly and
compare with the particle simulation, using the
inertia tensor method to extract $\hat{\mathbf{n}}(t)$ in
both cases. A long run ($t = 200$, $N = 400$, $T = 0.4$)
gives $D_{\text{rot}} = 0.0031$ ($\tau_c = 164$) from
the autocorrelation fit.
Figure~\ref{fig:o3_model_comparison} shows the results.

\begin{figure}[H]
\centering
\includegraphics[width=\textwidth]{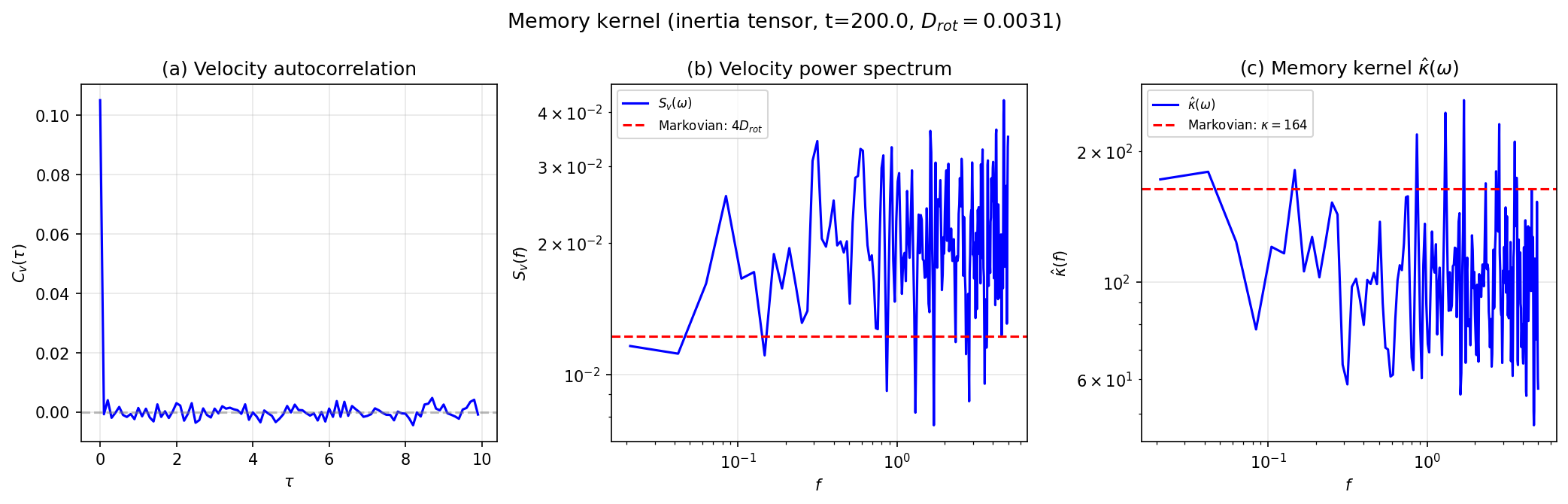}
\caption{Memory kernel extraction from particle
simulation ($N = 400$, $T = 0.4$, $t = 200$),
using the inertia tensor method.
(a) Velocity autocorrelation $C_v(\tau)$: decays to zero
within one time step
($C_v(\Delta t)/C_v(0) = -0.007$), consistent with
white noise.
(b) Velocity power spectrum $S_v(\omega)$:
approximately flat, matching the Markovian prediction
$4D_{\text{rot}}$ (red dashed).
(c) Memory kernel $\hat{\kappa}(\omega)$: approximately
constant at $\kappa \approx 164$, confirming Markovian
dynamics.}
\label{fig:memory_kernel}
\end{figure}

The $\mathbb{RP}^2$ NLSM reproduces the particle simulation at the
level of all tested observables: autocorrelation decay
rate, MSD, absence of drift, isotropic coverage of
$S^2$, and power spectrum. When the orientation is
extracted correctly via the inertia tensor, the power
spectrum follows the Markovian $\omega^{-2}$
(Fig.~\ref{fig:o3_model_comparison}, panel~f); the
velocity autocorrelation decays to zero within one time
step (Fig.~\ref{fig:memory_kernel}, panel~a,
$C_v(\Delta t)/C_v(0) = -0.007$); and the memory kernel
$\hat{\kappa}(\omega)$ is approximately constant
(Fig.~\ref{fig:memory_kernel}, panel~c).
A direct comparison of the increment volatility confirms
the quantitative match: the RMS of the orientation
increment $|\Delta\hat{\mathbf{n}}|$ per time step is
$0.032$ for the particle simulation and $0.035$ for the $\mathbb{RP}^2$-NLSM SDE, a difference of $8\%$. This small discrepancy
may arise from finite-sample effects (single
realization, finite trajectory length), from the mild
smoothing inherent in the inertia tensor method
(which averages over $N = 400$ particle positions),
or from the difference between quenched disorder
(particle simulation) and disorder-averaged dynamics
 $\mathbb{RP}^2$ NLSM). The Markovian $\mathbb{RP}^2$ NLSM
model~\eqref{eq:NLSM_action} is therefore the correct
effective theory at all timescales resolved by the
inertia tensor.

\paragraph{Markovianity test via the memory kernel.}
The non-Markovian effective
action~\eqref{eq:S_eff_nonmarkov} predicts that the
velocity power spectrum
$S_v(\omega) = \int\langle\dot{\hat{\mathbf{n}}}(t)\cdot
\dot{\hat{\mathbf{n}}}(t+\tau)\rangle\,
e^{-i\omega\tau}\,d\tau$
is related to the memory kernel by
$\hat{\kappa}(\omega) = 2/S_v(\omega)$. For Markovian
dynamics, $\hat{\kappa}(\omega) = \kappa$ is constant and
$S_v$ is white noise; any $\omega$-dependence signals
non-Markovian memory.

Figure~\ref{fig:memory_kernel} shows the memory kernel
extracted from the long particle simulation ($t = 200$),
using the inertia tensor method for
$\hat{\mathbf{n}}(t)$.

The result is clear: the orientation dynamics is
\emph{Markovian} when measured correctly with the inertia
tensor. The velocity autocorrelation $C_v(\tau)$ decays
to zero within one time step
($C_v(\Delta t)/C_v(0) = -0.007 \approx 0$), the
velocity power spectrum is approximately flat (white
noise), and $\hat{\kappa}(\omega)$ is constant.
The inertia tensor method uses all $N$ particle
positions to determine the ring plane, filtering out
fast thermal noise and producing a clean measurement of
the slow collective coordinate.

In the hierarchy of
Section~\ref{subsec:markov_limit}, the non-Markovian
action~\eqref{eq:S_eff_nonmarkov} is the fundamental
result of the BO reduction, and the Markovian
action~\eqref{eq:NLSM_action} is its
$\kappa(\tau) \to \kappa\delta(\tau)$ limit. The data are
consistent with the Markovian limit over the time resolution
and frequency window accessible to the inertia-tensor
estimator: no statistically significant memory-kernel
structure is resolved in the present simulations. We do not
infer global Markovianity from this; we infer that the
Markovian $\mathbb{RP}^2$ NLSM is consistent with the orientation
data within the bandwidth and statistics of the
inertia-tensor measurement.

\paragraph{Density-density correlator from a
100-realization ensemble.}
All diagnostics above use only the scalar orientation
$\hat{\mathbf{n}}(t)$ extracted by the inertia tensor.
A more stringent test of the $\mathbb{RP}^2$ NLSM is the full
spatial density-density correlator
\[
C_\rho(\gamma, \tau)
= \left\langle\rho(x, t)\,\rho(x', t + \tau)\right\rangle,
\]
with $\gamma = \arccos(x\cdot x')$ the great-circle
angle between the two spatial arguments on $S^2$. Its $\mathbb{RP}^2$-NLSM prediction~\eqref{eq:Crho_reduced} depends on only
two ingredients: the ring profile $f_0$ (through its
Legendre coefficients $c_\ell$) and the orientation
autocorrelation $q(\tau) = e^{-2 D_{\text{rot}}\tau}$.
Both ingredients are measured directly from the same
particle data, so the comparison is parameter-free
\emph{once these two low-energy inputs are fixed from
simulation}: the test verifies that the $\mathbb{RP}^2$-NLSM reduction
propagates correctly into the density sector (a closure
test), not that the F2 theory predicts $C_\rho$ from the
microscopic action alone, which would require computing
$f_0$ and $D_{\text{rot}}$ from first principles. In
what follows we make the construction explicit.

\emph{Ensemble protocol.}
We ran $100$ independent particle simulations at
$N = 400$, $T = 0.4$, $\sigma = 1$, each with an
independent realization of the quenched couplings
$\phi_{nm}$ and of the initial configuration, using
$dt = 0.0025$, an equilibration time
$t_{\text{eq}} = 10$, and a tracking time
$t_{\text{track}} = 40$ per realization, with snapshots
recorded at $\Delta t_{\text{rec}} = 0.25$ ($160$
snapshots per realization). All $100$ runs formed rings.

\begin{figure}[H]
\centering
\includegraphics[width=\textwidth]{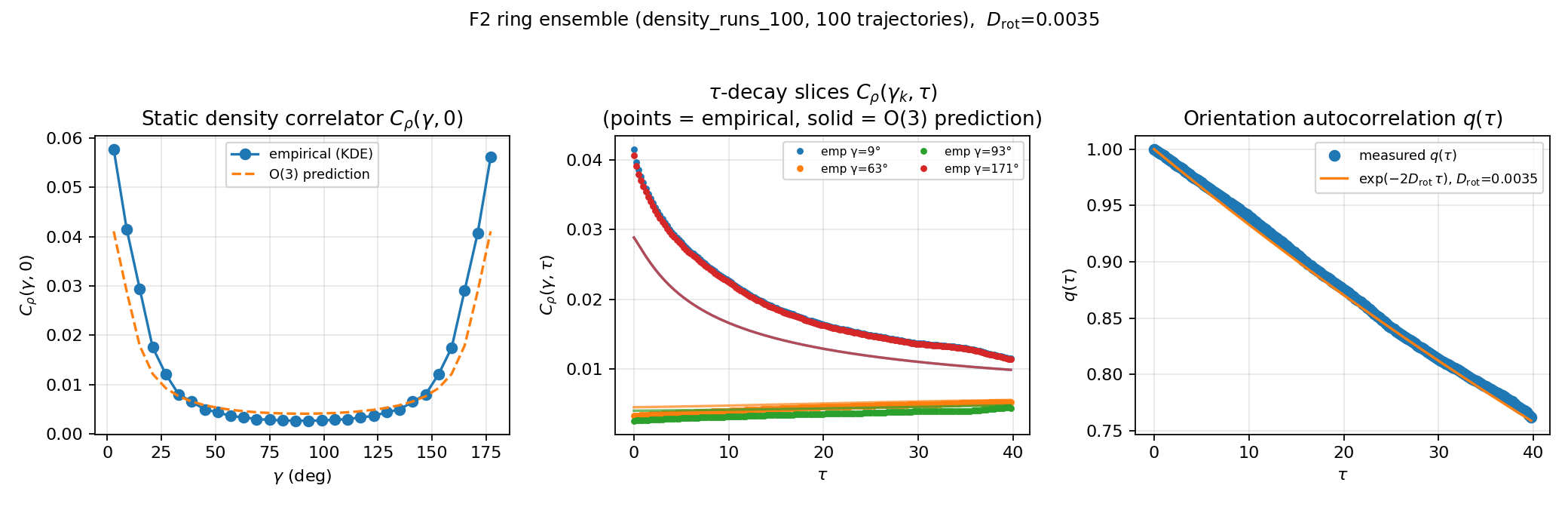}
\caption{Disorder-averaged density-density correlator
on $S^2$ from $100$ independent particle simulations
($N = 400$, $T = 0.4$, $dt = 0.0025$; per realization
$t_{\text{eq}} = 10$ and $t_{\text{track}} = 40$ with
$160$ snapshots; $\sigma_{\text{KDE}} = 0.1$ rad,
$\ell_{\max} = 30$, $1800$-point uniform-area grid on
$S^2$).
(a) Static correlator $C_\rho(\gamma, 0)$: empirical
KDE (blue) vs.\ the SO(3)
prediction~\eqref{eq:Crho_reduced} built from the
pooled $c_\ell$ (dashed). The characteristic double
peak at $\gamma \to 0$ (same-ring) and
$\gamma \to \pi$ (antipodal on the same great circle)
is reproduced.
(b) Time-decay slices $C_\rho(\gamma_k, \tau)$ at four
angles; points are empirical, solid curves are the SO(3)
prediction with $D_{\text{rot}} = 0.0035$ from the pooled
$q(\tau)$.
(c) Pooled orientation autocorrelation
$\langle q(\tau)\rangle_{\text{disorder}}$ (blue) vs.\
the Markovian $e^{-2D_{\text{rot}}\tau}$ (orange) using
the pooled-fit value $D_{\text{rot}} = 0.0035$; agreement
is near-perfect across the full $\tau = 0$--$40$ window,
a much cleaner check than the single-long-run estimator
$\widehat{q}(\tau)$~\eqref{eq:q_estimator}.}
\label{fig:density_correlator_100}
\end{figure}

\emph{Empirical
$\langle C_\rho(\gamma,\tau)\rangle$ from the particle
data.} For each snapshot we reconstruct the smooth
lab-frame density field $\rho(x, t)$ by Fisher-von Mises
kernel density estimation (bandwidth
$\sigma_{\text{KDE}} = 0.1$ rad, normalized so that
$\int_{S^2}\rho\,d\Omega = 1$) on a uniform-area grid of
$1800$ points, and expand it in real spherical
harmonics up to $\ell_{\max} = 30$,
$\rho(x, t) = \sum_{\ell m} a_{\ell m}(t)\,Y_{\ell m}(x)$.
Because the disorder-averaged ensemble is SO(3)-isotropic,
the correlator reduces to its Legendre form
\begin{equation}
\label{eq:Crho_ensemble}
\langle C_\rho(\gamma, \tau)\rangle
= \sum_{\ell = 0}^{\ell_{\max}}
\frac{2\ell + 1}{4\pi}\,A_\ell(\tau)\,
P_\ell(\cos\gamma),
\qquad
A_\ell(\tau) = \frac{1}{2\ell + 1}
\sum_m\overline{a_{\ell m}(t)\,a_{\ell m}(t + \tau)},
\end{equation}
where the overline denotes the joint time-origin and
disorder average over the $100$ runs. Equation
\eqref{eq:Crho_ensemble} is what we plot as the
empirical curves in
Fig.~\ref{fig:density_correlator_100}(a,b); no SO(3)
simulation is run at this stage.

\emph{Theoretical
$C_\rho^{\mathrm{SO(3)}}(\gamma,\tau)$.}
For the $\mathbb{RP}^2$-NLSM prediction we do \emph{not} run a separate $\mathbb{RP}^2$-NLSM simulation and compute its density correlator
empirically. Instead, we evaluate the closed-form
expression for $\langle C_\rho(\gamma,\tau)\rangle$ that
was derived analytically from the $\mathbb{RP}^2$ NLSM in
Section~\ref{subsec:corr_orientation} (the reduction
formula~\eqref{eq:Crho_reduced}),
\begin{equation}
\label{eq:Crho_O3_check}
C_\rho^{\mathrm{SO(3)}}(\gamma, \tau)
= \sum_{\ell = 0}^{\ell_{\max}}
\frac{c_\ell^{\,2}}{2\ell + 1}\,
e^{-\ell(\ell + 1)\,D_{\text{rot}}\,|\tau|}\,
P_\ell(\cos\gamma),
\end{equation}
directly on the same $(\gamma, \tau)$ grid as the
empirical estimator~\eqref{eq:Crho_ensemble}. The
theoretical content entering
Fig.~\ref{fig:density_correlator_100} is therefore the
functional form of~\eqref{eq:Crho_O3_check}: given
$\{c_\ell, D_{\text{rot}}\}$, the formula fully
determines the shape of $C_\rho^{\mathrm{SO(3)}}(\gamma,
\tau)$ in both $\gamma$ and $\tau$. An alternative route
(simulating the $\mathbb{RP}^2$-NLSM SDE~\eqref{eq:nhat_ito} and
dressing each sampled $\hat{\mathbf{n}}(t)$ with the
ring profile $\rho(x, t) = f_0(\hat{\mathbf{n}}(t)\cdot
\mathbf{x})$ before computing the density correlator via
the same KDE/SH pipeline) yields the same answer up to
Monte Carlo noise, since~\eqref{eq:Crho_O3_check} is
exactly the prediction of that procedure.

The two inputs $\{c_\ell,\,D_{\text{rot}}\}$ of
\eqref{eq:Crho_O3_check} are measured directly from the
particle data of the same $100$-realization ensemble (not
fit to Fig.~\ref{fig:density_correlator_100}).
Specifically:
(i) the Legendre coefficients
$c_\ell = \tfrac{2\ell + 1}{2}\int_{-1}^{1} f_0(u)\,
P_\ell(u)\,du$ come from the histogram of the ring-frame
coordinate $u = \hat{\mathbf{n}}(t)\cdot\mathbf{x}_n(t)$
pooled over all $100 \times 160 \times 400$
particle-snapshot samples, which gives a dense
estimate of the stationary (in the ring frame) profile
$f_0$;
(ii) the rotational diffusion coefficient
$D_{\text{rot}}$ is the log-slope fit of the pooled
orientation autocorrelation
$\langle q(\tau)\rangle_{\text{disorder}}$, and yields
the value $D_{\text{rot}} = 0.0035$ shown in
Fig.~\ref{fig:density_correlator_100}(c). The comparison
between \eqref{eq:Crho_ensemble} and
\eqref{eq:Crho_O3_check} in
Fig.~\ref{fig:density_correlator_100}(a,b) is therefore
a parameter-free test of the functional form predicted
by the $\mathbb{RP}^2$-NLSM effective theory: once $f_0$ and
$D_{\text{rot}}$ are fixed from two independent
orientation-sector diagnostics (the ring profile and the
orientation autocorrelation), the full spatial
$(\gamma, \tau)$ dependence of $\langle C_\rho\rangle$
is predicted by~\eqref{eq:Crho_O3_check} with no
additional fit parameters.

The pooled fit yields $D_{\text{rot}} = 0.0035$,
consistent with the ensemble mean of
Table~\ref{tab:Drot_disorder}
($\langle D_{\text{rot}}\rangle = 0.003 \pm 0.002$) and
with the single long run ($0.0031$), but now
disorder-averaged with an order-of-magnitude more
statistics than the earlier $10$-realization ensemble.
The on-ring decay slices of panel (b) agree with the $\mathbb{RP}^2$-NLSM prediction at the few-percent level over the full
accessible range of $\tau$; the small residual at the
$\gamma \to 0, \pi$ peaks of the static correlator
(panel a) is a finite-bandwidth artifact of the KDE
(FWHM $\approx 5.7^\circ$) combined with the finite
resolution of the pooled profile $f_0$ measured from
the ring-frame histogram. Crucially, the density sector
is reproduced \emph{without any free parameters beyond
those already fixed by the orientation-sector
diagnostics}: the same $f_0$ and $D_{\text{rot}}$ that
describe the scalar collective coordinate also determine
the full spatial density dynamics, confirming the SO(3)
effective theory at this stronger level.

\emph{Why the agreement is non-trivial.}
The closed-form prediction~\eqref{eq:Crho_O3_check} was
derived in Section~\ref{subsec:corr_orientation} under
the rigid-ring ansatz
$\rho(x, t) = f_0(\hat{\mathbf{n}}(t)\cdot\mathbf{x})$,
in which the static profile $f_0$ is passively advected
by the stochastic orientation $\hat{\mathbf{n}}(t)$. The
particle simulation does not satisfy this ansatz
exactly: $400$ particles produce finite-$N$ shot noise
at every snapshot, the ring has internal (breathing,
azimuthal) fluctuations of the gapped $\delta\rho$
modes, and disorder heterogeneity across realizations
gives a per-sample spread in $D_{\text{rot}}$ of roughly
a factor of two (Table~\ref{tab:Drot_disorder}). Any of
these effects could have contaminated $C_\rho$ at the
level of the signal, and each one failing separately
would have produced a distinct diagnostic signature.
The agreement in
Fig.~\ref{fig:density_correlator_100} therefore tests
four things simultaneously. First, the
Born-Oppenheimer decoupling of the gapped $\delta\rho$
modes from the slow orientation, which is the premise
of the collective-coordinate reduction of
Section~\ref{sec:low_energy}: if $\epsilon =
D_{\text{rot}}/m^2_{\text{gap}}$ were not small, the
$\ell$-th Legendre mode would acquire additional decay
channels and the single exponential
$e^{-\ell(\ell+1)D_{\text{rot}}\tau}$ would break.
Second, the full Fokker-Planck spectrum
$\lambda_\ell = \ell(\ell+1)D_{\text{rot}}$ of free
diffusion on $S^2$: the measurement of $D_{\text{rot}}$
at $\ell = 1$ (through $q(\tau)$) is extrapolated to all
$\ell \le \ell_{\max} = 30$ by the formula, and the
off-center $\gamma$ slices of
Fig.~\ref{fig:density_correlator_100}(b) sample
different combinations of these modes, providing an
independent check of the higher-$\ell$ rates. Third,
suppression of the finite-$N$ shot noise at the few
percent level in the pooled correlator, which would
otherwise appear as a $\tau$-independent offset at large
lags. Fourth, the commutativity of the quenched
disorder average with the $\mathbb{RP}^2$-NLSM reduction at $N = 400$:
a single pooled $D_{\text{rot}} = 0.0035$ describes the
full $(\gamma, \tau)$ dependence, so the sample-to-sample
heterogeneity in $D_{\text{rot}}$ does not distort the
functional form of $\langle C_\rho\rangle$. The structure
of the test is the same as a chiral-Lagrangian
consistency check in QCD, in which $f_\pi$ measured from
one observable (pion decay) is required to reproduce a
qualitatively different observable
(e.g.\ $\pi\pi$-scattering) without further fitting;
here $f_0$ and $D_{\text{rot}}$ measured from the
orientation sector reproduce the full density-sector
correlator~\eqref{eq:Crho_O3_check} through an exponent
chain $\ell(\ell+1)$ that is not present in the input.

\paragraph{Overall consistency assessment.}
After $D_{\text{rot}}$ is fixed from the particle
simulation (Tier~2 of the three-tier framework,
Section~\ref{subsec:three_tiers}), the Markovian $\mathbb{RP}^2$ NLSM
model~\eqref{eq:NLSM_action} reproduces the tested
observables of the orientation dynamics, within numerical
resolution:
(i) exponential autocorrelation
$C(\tau) = e^{-2D_{\text{rot}}\tau}$
(Fig.~\ref{fig:o3_model_comparison}, panel~d);
(ii) MSD $= 2(1 - e^{-2D_{\text{rot}}\tau})$
(panel~e);
(iii) Markovian power spectrum $\omega^{-2}$ (panel~f);
(iv) isotropic exploration of $S^2$ with no preferred
direction (panels~a--c, the three components $n_x, n_y,
n_z$ all showing similar diffusive volatility);
(v) zero deterministic drift, confirmed by the azimuthal
velocity distribution
$|\text{mean}/\text{std}| = 10^{-4}$
(Fig.~\ref{fig:ring_frame}, panel~b);
(vi) Markovian velocity (white noise,
$C_v(\Delta t)/C_v(0) \approx 0$;
Fig.~\ref{fig:memory_kernel}, panel~a);
(vii) constant memory kernel
$\hat{\kappa}(\omega) \approx \kappa$
(Fig.~\ref{fig:memory_kernel}, panel~c);
and (viii) the disorder-averaged density-density
correlator $\langle C_\rho(\gamma, \tau)\rangle$ from a
100-realization ensemble
(Fig.~\ref{fig:density_correlator_100}), which agrees
with the parameter-free SO(3)
prediction~\eqref{eq:Crho_reduced} at the few-percent
level over the full $(\gamma, \tau)$ domain.
The single parameter $D_{\text{rot}}$ (equivalently
$\kappa = 1/(2D_{\text{rot}})$), measured from one
observable (e.g., the autocorrelation decay rate),
predicts all others with no free parameters.

This realizes in concrete form the Tier-2 strategy of
Section~\ref{subsec:three_tiers}: the form of the
action is fixed by the SO(3)$\to$SO(2) symmetry
breaking, and the single low-energy constant
$D_{\text{rot}}$, measured from one observable, controls
all others. The QCD/$f_\pi$ analogy and the open Tier-3
computation are discussed there. The measured
$N$-scaling
$D_{\text{rot}} \sim N^{-1.7}$
(Table~\ref{tab:Drot_N}, Fig.~\ref{fig:n_scaling}) is
steeper than the mean-field $1/N$ and remains to be
explained at the microscopic level.

\subsection{Universality: what interaction potentials produce rings on $S^2$}
\label{subsec:coulomb}

The $\mathbb{RP}^2$-NLSM effective theory was derived from the F2 model
on $S^2$ with the linear geodesic-distance potential
$U_{ij} = \phi_{ij}\,d(x_i,x_j)$. To test whether the
ring formation and the resulting $\mathbb{RP}^2$-NLSM dynamics are
specific to this potential or reflect a more general
phenomenon, we simulated systems with logarithmic
potentials on $S^2$.

\paragraph{Soft Coulomb model.}
We consider the potential
$U_{ij} = \phi_{ij}\,\log(d_{\text{geo}}(x_i,x_j)
+ d_{\min})$, where $d_{\text{geo}}$ is the geodesic
distance, $\phi_{ij} = \pm 1$ (discrete random
couplings), and $d_{\min}$ is a soft regularization
parameter. The force is
$1/(d_{\text{geo}} + d_{\min})$, with ratio
$F_{\max}/F_{\min}
= (d_{\max} + d_{\min})/d_{\min}$ (where
$d_{\max} = \pi$ on $S^2$). For $d_{\min} \to 0$, the
force is strongly distance-dependent (short-range
dominated), and no ring formation is observed. For
larger $d_{\min}$, the force becomes more uniform,
approaching the constant-force regime of the F2 model.

\begin{figure}[H]
\centering
\includegraphics[width=\textwidth]{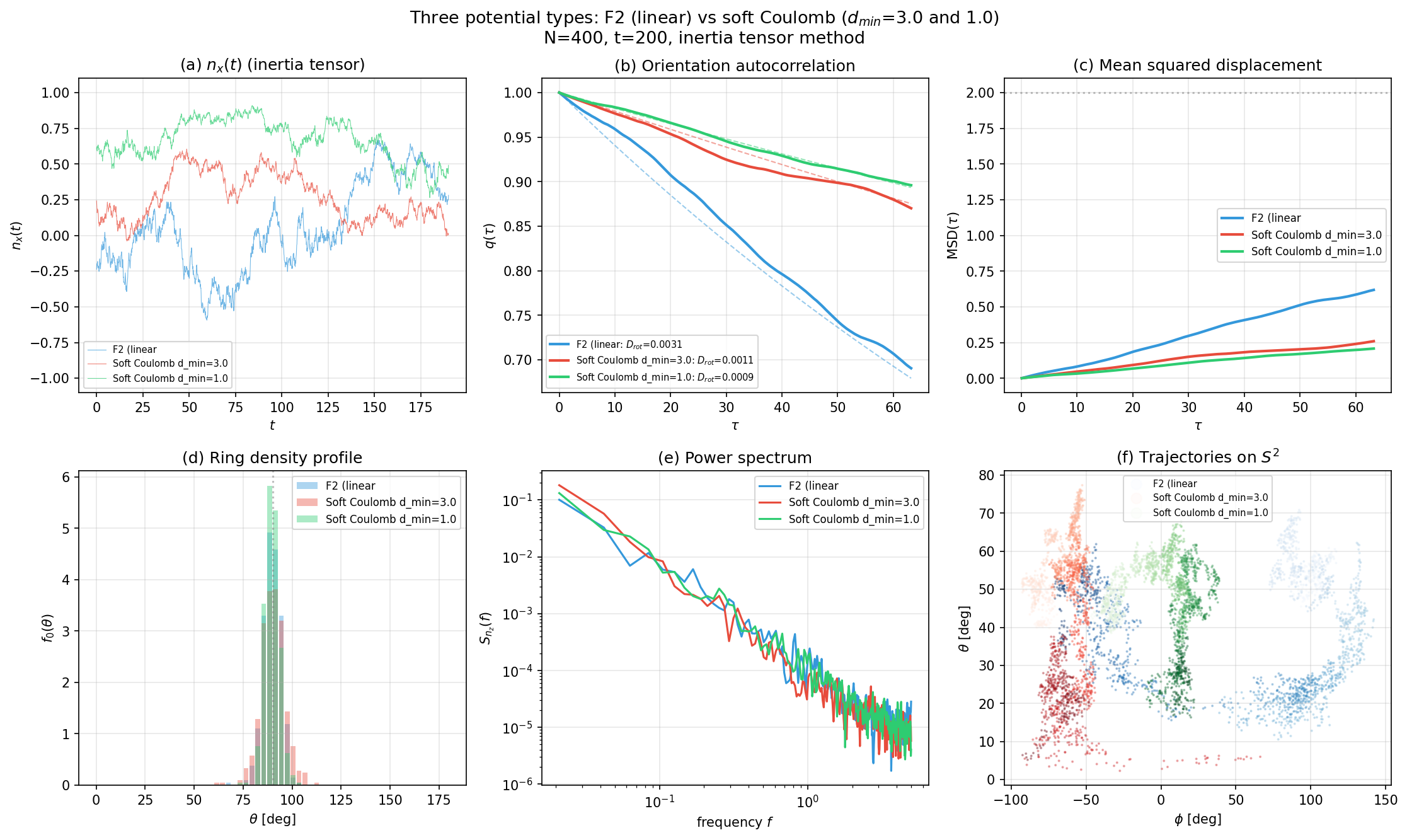}
\caption{Comparison of three potential types:
F2 model (blue, $U \propto d$, $T = 0.4$),
soft Coulomb with $d_{\min} = 3.0$ (red,
$T = 0.2$), and soft Coulomb with $d_{\min} = 1.0$
(green, $T = 0.2$). All at $N = 400$, $t = 200$,
inertia tensor method.
(a) $n_x(t)$: all show diffusive wandering.
(b) Autocorrelation $q(\tau)$: all decay exponentially.
(c) MSD.
(d) Ring density profile: the $d_{\min} = 1.0$ case
(green) produces the narrowest ring
(FWHM $= 9^\circ$).
(e) Power spectrum: all follow $\omega^{-2}$.
(f) Trajectories on $S^2$.}
\label{fig:coulomb_comparison}
\end{figure}

\paragraph{Three-model comparison.}
Figure~\ref{fig:coulomb_comparison} compares three
models at $N = 400$, $t = 200$, all analyzed with the
inertia tensor method:
\begin{table}[H]
\centering
\begin{tabular}{l|c|c|c|c|c}
Model & $T$ & Force ratio & Ring & FWHM
& $D_{\text{rot}}$ \\
\hline
F2 (linear) & 0.4 & 1:1 & YES & $11^\circ$
& 0.0031 \\
Soft Coulomb, $d_{\min}\!=\!3$ & 0.2 & 2:1
& YES & $13^\circ$ & 0.0011 \\
Soft Coulomb, $d_{\min}\!=\!1$ & 0.2 & 4.1:1
& YES & $9^\circ$ & 0.0009 \\
\end{tabular}
\caption{Comparison of three potential types with
long-run inertia tensor analysis ($N = 400$, $t = 200$).
The two soft Coulomb runs use $T = 0.2$ rather than $T =
0.4$, so absolute values of $D_{\text{rot}}$ are not
directly comparable across rows; the qualitative point
is that all three potentials form rings and exhibit
Markovian $\mathbb{RP}^2$ diffusion.}
\label{tab:three_models}
\end{table}

All three models form rings and exhibit the same SO(3)
diffusion dynamics: exponential autocorrelation, MSD
saturating at $2$, and Markovian $\omega^{-2}$ power
spectrum. The $d_{\min} = 1.0$ case produces the
tightest ring (FWHM $= 9^\circ$, anisotropy ratio $50$)
and the slowest diffusion ($\tau_c = 562$), despite
having the largest force ratio ($4.1$:$1$). Ring formation
therefore persists across a range of force profiles within
the soft-kernel family, not just the constant-force (F2)
limit; a kernel-family-dependent reading of the universality
criterion is given in the ``Universality of the SO(3)
reduction'' paragraph below.

\paragraph{Truncated geodesic-log potential.}
A fourth model uses the geodesic-distance logarithmic
potential with a hard floor:
$V = \max(U_{\min},\,\log(d_{\text{geo}}))$, where
$U_{\min}$ is a tunable parameter. The force is
$1/d_{\text{geo}}$ for $d_{\text{geo}} > e^{U_{\min}}$
and zero for $d_{\text{geo}} < e^{U_{\min}}$ (flat
bottom). This model has the exact $1/d$ Coulomb force at
all distances beyond the cutoff $e^{U_{\min}}$.

Ring formation does \emph{not} occur for
$U_{\min} \geq 0$ (cutoff $\geq 1$, force ratio
$\lesssim 3$:$1$), despite the force ratio being
comparable to the successful soft Coulomb cases. Ring
formation is observed for $U_{\min} = -2.15$ (cutoff
$\approx 0.12$, force ratio $\approx 27$:$1$), but the
resulting ring is much broader (FWHM $= 24^\circ$,
anisotropy ratio $3.7$) than the soft Coulomb or F2
rings ($10$--$12^\circ$). The difference is that the
hard-floor truncation creates a zero-force region
($d < e^{U_{\min}}$) where nearby particles can cluster
without any restoring force, partially disrupting the
collective ring ordering. The soft Coulomb model
$\log(d + d_{\min})$ avoids this by maintaining a
nonzero (though reduced) force at all distances.

\begin{table}[H]
\centering
\begin{tabular}{l|c|c|c|c}
Model & Force ratio & FWHM & Frac.\ eq.
& Ring \\
\hline
F2 (linear) & 1:1 & $11^\circ$ & 1.00 & YES \\
Soft Coulomb $d_{\min}\!=\!1$ & 4.1:1 & $10^\circ$
& 0.98 & YES \\
Soft Coulomb $d_{\min}\!=\!3$ & 2.0:1 & $12^\circ$
& 0.99 & YES \\
Trunc log $U_{\min}\!=\!{-}2.15$ & 27:1 & $24^\circ$
& 0.91 & YES \\
Trunc log $U_{\min}\!=\!0$ & 3.1:1 & --- & ---
& NO \\
Chord Coulomb (2D), $\log c$ & $\infty$ & --- & --- & NO \\
Chord Coulomb (3D), $1/c$ & $\infty$ & --- & --- & NO \\
\end{tabular}
\caption{Ring formation across all seven potential types
tested ($N = 400$, $T = 0.2$--$0.4$, discrete $\pm 1$
couplings except F2 which uses Gaussian). The column
``Frac.\ eq.'' is the fraction of particles within the
fitted ring band (Gaussian within $\pm 2\sigma$ of the
ring centre). Ring formation requires a force that
remains effective across the full sphere; both vanishing
antipodal force (chord Coulombs, whether 2D $\log c$ or
3D $1/c$) and hard-floor truncation at moderate cutoff
(truncated log, $U_{\min} \geq 0$) prevent it.}
\label{tab:all_potentials}
\end{table}

\paragraph{Chord-distance Coulomb and the role of
geometry.}
The exact 2D Coulomb potential on $S^2$ (the Green's
function of the Laplace-Beltrami operator) is
$V(\theta) = -(1/2\pi)\log(2\sin(\theta/2))$, where
the argument is the \emph{chord distance}
$c = 2\sin(\theta/2) = |\mathbf{x}_i - \mathbf{x}_j|$
(the 3D Euclidean distance between two points on the
unit sphere), as discussed by Caillol~\cite{caillol81}
and by Forrester, Jancovici, and
Madore~\cite{forrester92}. This is distinct from
$\log(d_{\text{geo}})$, where $d_{\text{geo}} = R\theta$
is the geodesic distance (equal to $\theta$ on the unit
sphere $R = 1$). The chord distance arises from the
intrinsic geometry of $S^2$ through the Laplace-Beltrami
operator, regardless of whether the electric field
propagates through the interior or along the surface.

The force (per unit coupling) for the chord-distance
potential is $F(s) = -dV/ds$, which on the unit sphere
gives $F = \cot(s/2)/2$ (where $s = d_{\text{geo}}$ is
the geodesic distance). This force \emph{vanishes at the
antipodal point} ($s = \pi$): particles on opposite
sides of the sphere do not interact. This makes the
chord-distance Coulomb potential effectively short-ranged
on $S^2$, and simulations confirm that it does
\emph{not} produce ring formation. The soft Coulomb
model $\log(d_{\text{geo}} + d_{\min})$ avoids this
problem because its force $1/(d_{\text{geo}} + d_{\min})$
remains nonzero at all distances, including the
antipodal point.

\paragraph{3D Coulomb on the sphere.}
The same antipodal-vanishing obstruction also rules out
the ordinary three-dimensional Coulomb interaction
$V(c) = 1/c$ for particles constrained to $S^2$, with
$c$ the embedding chord distance. The tangent-plane
component of the 3D Coulomb force between two surface
points is $|F_\perp(s)| = \cos(s/2)/(4\sin^2(s/2))$,
which has the same $\cos(s/2)$ factor as the 2D
chord-log potential and therefore vanishes at the
antipodal point $s = \pi$. Simulations using this
potential with a hard-core cutoff $c > 2r_p$ confirm
that no ring forms, consistent with the general
criterion: it is the vanishing of the force at the cut
locus, not the specific functional form of the
potential, that breaks the ring-forming mechanism. Both
the 2D chord Coulomb ($\log c$, the Green's function of
the Laplace-Beltrami operator) and the 3D chord Coulomb
($1/c$, the interaction mediated by the embedding
$\mathbb{R}^3$) therefore fall into the same
``effectively short-ranged on $S^2$'' class.

\paragraph{Universality of the $\mathbb{RP}^2$-NLSM reduction.}
The scan suggests an empirical criterion for ring
formation: the interaction should remain effective at
antipodal-scale separations on $S^2$. The features common
to the ring-forming runs are:
(i) particles confined to $S^2$;
(ii) quenched random pairwise couplings;
(iii) an interaction whose force does not vanish at the
cut locus of the manifold.
A naive global force-ratio threshold is insufficient as a
single number across kernel families. Within the soft-kernel
family (F2 with $V = d$, soft Coulomb with various
$d_{\min}$), the ring-forming runs sit at force ratios up to
$4.1$:$1$ in the present scan; a hard-floor truncated
logarithm can still form a (broader) ring at much larger
nominal force ratio ($\approx 27$:$1$), because the
hard-floor cutoff changes the near-field clustering dynamics
and so changes which configurations are stable. We therefore
treat the force-ratio threshold as kernel-family dependent:
within a smooth kernel family it provides a useful empirical
indicator (with order $\sim 5$:$1$ separating the present
soft-kernel ring-forming and non-forming cases), but it does
not transfer across families. Both chord Coulombs (the 2D
$\log c$ and the 3D $1/c$, $F_{\max}/F_{\min} \to \infty$)
and the unregularized geodesic Coulomb
($F_{\max}/F_{\min} \to \infty$) violate the cut-locus
criterion and show no ring formation. We state these
observations as numerical conjectures: an analytic
stability analysis of the uniform state and of the ring
saddle under general kernels, which would turn the empirical
indicators into derived selection rules, is left to future
work. Within the ring-forming subclass tested, the form of
the effective low-energy theory appears universal; only
$D_{\text{rot}}$ (the analog of $f_\pi$) is model-dependent.

%==============================================================================
\section{Discussion and Future Directions}
\label{sec:discussion}
%==============================================================================

We now turn to physical realizations of the F2 ring,
connections to other areas of physics, and open questions.

\subsection{Physical Realizations}
\label{subsec:physical}

Three ingredients are needed to realize the F2 ring on a
physical platform: (i) a closed 2D surface (or thin shell) on
which the constituent objects (particles, atoms, or
electrons) are confined; (ii) quenched random pairwise
couplings between them; and (iii) a pairwise force whose
magnitude remains nonzero across the full diameter of the
surface (Section~\ref{subsec:coulomb}). We sketch three
classes of candidates, grouped by whether the underlying
dynamics is classical stochastic or genuinely quantum.

\paragraph{Classical platform: charged colloids on a soft
membrane.}
A spherical polymer vesicle of radius $R \sim 1$--$10\;\mu$m,
decorated with two species of charged colloidal nanoparticles
($r \sim 10$--$100\;$nm, charges $q = \pm 1$), realizes the
microscopic F2 model directly at the level of classical
stochastic dynamics. For $r \sim 50\;$nm on
$R \sim 1\;\mu$m, the dimensionless contact distance
$2r/R \approx 0.1$ places the truncated-log potential of
Section~\ref{subsec:coulomb} at $U_{\min} \approx -2.3$,
inside the ring-forming regime. The ring orientation
$\hat{\mathbf{n}}(t)$ can be tracked by fluorescence
microscopy, and $D_{\text{rot}}$ extracted from the
orientation autocorrelation $q(\tau)$.

The main experimental constraint is the range of the
effective surface interaction: the bare 2D Coulomb on $S^2$
is effectively short-ranged in our sense (its force vanishes
at the antipodal point; Section~\ref{subsec:coulomb}), so
the pair potential must be engineered. Three mechanisms can
produce a force that stays effective across the sphere.
(i)~\emph{Membrane-mediated fluctuation interactions.}
Inclusions that deform or stiffen the membrane couple
through its thermal height fluctuations; the resulting
interaction decays only algebraically with geodesic distance
and, on a sphere whose bending rigidity is not much larger
than $k_B T$, is long-ranged in our sense.
(ii)~\emph{Thick-shell electrostatics.} For a spherical
conducting shell of thickness $h$, the surface-to-surface
potential interpolates between the chord form
$\log c$ (for $h \ll R$, the thin-shell Laplace-Beltrami
Green's function) and a geodesic form (for $h \gg R$, where
the field is expelled from the interior), with the force
ratio $F_{\max}/F_{\min}$ tunable by $h/R$ and the dielectric
contrast. (iii)~\emph{Screened bulk electrostatics combined
with surface binding.} Screening the 3D Coulomb field (by
embedding the vesicle in a high-permittivity or ionic
solvent) suppresses through-bulk interactions so the
effective coupling is controlled by surface terms whose
profile can be tuned via the joint screening length and
shell thickness. In each case the effective pair potential
is a physically distinct interaction, not a regularization
of the bare Laplace-Beltrami Coulomb; the criterion of
Section~\ref{subsec:coulomb} then selects which
implementations form rings.

\paragraph{Quantum platforms.}
A genuine quantum realization replaces Brownian particles by
atoms, ions, or electrons constrained to a 2D shell. Three
concrete directions are:
\begin{itemize}
    \item \emph{Cold atoms on a shell trap.} Optically
    trapped or "bubble-trapped" Bose gases can be confined
    to thin spherical shells. Spatial random potentials
    (laser speckle, optical disorder) and
    tunable contact interactions (Feshbach resonances)
    could produce a quantum analog of the F2 model; the
    orientation $\hat{\mathbf{n}}(t)$ of the resulting
    ring condensate would be a genuine quantum coordinate.
    \item \emph{Electronic shells.} The $\pi$-electron
    systems of fullerenes, carbon nanoshells, or thin
    films of 3D topological insulators shaped into closed
    surfaces carry electronic degrees of freedom confined
    to an approximately 2D curved geometry. Random
    dopants or chemisorbed impurities provide the quenched
    couplings; the screened electron-electron interaction
    projected onto the shell supplies the long-range pair
    force.
    \item \emph{Skyrmion-like textures on curved
    magnetic surfaces.} Magnetic thin films deposited on
    spherical or ellipsoidal substrates, with engineered
    random exchange couplings (frustrated magnetism),
    could carry a ring-like magnetization texture whose
    orientation is a quantum collective coordinate.
\end{itemize}
In all three quantum platforms, the $\mathbb{RP}^2$ NLSM of
Section~\ref{sec:low_energy} emerges as the low-energy
Hamiltonian of the orientation sector, not merely as the
Euclidean action of a classical stochastic process.
The rotational diffusion coefficient $D_{\text{rot}}$ is
replaced by a genuine kinetic coefficient
$\hbar^2/(2\mathcal{I})$ with $\mathcal{I}$ an effective
moment of inertia, and the combinations
$\ell(\ell+1)D_{\text{rot}}$ are true energy levels of the
orientation rather than Fokker-Planck relaxation rates.

\subsection{Connections to Other Physics Phenomena}
\label{subsec:connections}

\paragraph{Classical relaxation vs.\ quantum energy spectrum.}
In the classical stochastic realization studied here, the
orientation dynamics is a Fokker-Planck diffusion on
$\mathbb{RP}^2$ (lifted to $S^2$), and
$\lambda_\ell = \ell(\ell+1)\,D_{\text{rot}}$ is the
\emph{relaxation rate} of the $Y_{\ell m}$ component of the
orientation distribution, not an energy. A genuine quantum
implementation would replace the Fokker-Planck operator by
the rotor Hamiltonian $\mathbf{L}^2/(2\mathcal{I})$, with
energy levels
$E_\ell = \ell(\ell+1)\,\hbar^2/(2\mathcal{I})$ and
$\mathcal{I}$ a platform-dependent moment of inertia. The
two formulae have the same $\ell(\ell+1)$ dependence on
representation but encode different physical content; we use
the quantum analogy below only to motivate experimental
directions, never to claim that a relaxation rate is an
energy.

\paragraph{Quantum-realization outlook: nematic-type
ring-orientation qubit on $\mathbb{RP}^2$.}
Because the physical orientation manifold is
$\mathbb{RP}^2 = S^2/\mathbb{Z}_2$
(Section~\ref{subsec:low_energy_ansatz}), the single-valued
quantum Hilbert space of the orientation sector is the
even-$\ell$ subspace of $L^2(S^2)$, spanned by spherical
harmonics with $\ell = 0, 2, 4, \ldots$. The lowest two
levels are the $\ell = 0$ singlet and the $\ell = 2$
quintuplet, separated by the rotor gap
$\Delta_{02} = E_2 - E_0 = 6\,\hbar^2/(2\mathcal{I}) =
3\hbar^2/\mathcal{I}$ in a quantum implementation along the
lines of Section~\ref{subsec:physical}. The $\ell = 1$
triplet is odd under $\hat{\mathbf{n}}\to-\hat{\mathbf{n}}$
and is excluded from the physical sector on $\mathbb{RP}^2$
(it can be reached only by the $\theta = \pi$
quantization sector of
Section~\ref{subsec:low_energy_ansatz}, ``Topological
aspects of the $\mathbb{RP}^2$ target'', a spin-1/2-like
sector that we do not consider here).

The natural qubit on $\mathbb{RP}^2$ is therefore a
nematic-type two-level system built from the $\ell = 0$
singlet and one component of the $\ell = 2$ quintuplet
isolated by a quadrupolar (tensor) symmetry-breaking field.
A traceless symmetric coupling $H_{ij}\,n^i n^j$, which is
$\mathbb{Z}_2$-invariant in $\hat{\mathbf{n}}$ and therefore
admissible on $\mathbb{RP}^2$, splits the $\ell = 2$
quintuplet (axial vs.\ planar anisotropy) and selects a
logical pair with energy gap of order $\Delta_{02}$ shifted
by the quadrupolar splitting. A scalar Zeeman-like coupling
$\vec{h}\cdot\hat{\mathbf{n}}$ is forbidden on
$\mathbb{RP}^2$ (it is parity-odd in $\hat{\mathbf{n}}$ and
therefore couples only across the trivial / non-trivial
$\mathbb{Z}_2$-sectors); the appropriate experimental probes
are quadrupolar / strain / dielectric-anisotropy fields,
exactly the setup that drives nematic-liquid-crystal
director dynamics. The spectral gap $m^2_{\text{gap}}$ to
internal ring excitations
(Section~\ref{subsec:zero_modes}) protects the orientation
sector against leakage to bulk modes in the same way as in
the skyrmion-helicity qubit of Psaroudaki and
Panagopoulos~\cite{psaroudaki2021skyrmion}; we refer the
reader to that work for the analogous analysis in the
magnetic-skyrmion context, with the caveat that the
$\mathrm{U}(1)$ helicity coordinate of a magnetic skyrmion
is replaced here by the projective director on
$\mathbb{RP}^2$ and the natural coupling family changes from
Zeeman-like to quadrupolar accordingly. Calibrating the
classical (Fokker-Planck) tensor response of the orientation
on a colloidal platform before a quantum implementation is
the natural experimental sequence.

\paragraph{Adiabatic rotational reductions vs.\ true
symmetry breaking.}
The F2 reduction from a microscopic field theory to an
effective $\mathrm{SO}(3)$ classical-stochastic theory for a slow
orientation has direct physical precedent in
adiabatic-rotation problems: the Bohr-Mottelson picture of
nuclear rotational bands~\cite{bohr_mottelson}, where a
deformed nucleus breaks rotational symmetry in the body
frame and the slow lab-frame orientation produces the
spectrum $E_\ell = \ell(\ell+1)\hbar^2/(2\mathcal{I})$,
and the Born-Oppenheimer treatment of molecular
rotation~\cite{born_oppenheimer,landau_lifshitz_qm}, where
fast electronic and vibrational modes are integrated out to
yield the same $\ell(\ell+1)$ ladder. The F2 ring is the
classical-stochastic counterpart: the moment of inertia
plays the role of the stiffness $\kappa$, and the
rotational quanta $\hbar^2/(2\mathcal{I})$ are replaced by
the relaxation rates $\ell(\ell+1)D_{\text{rot}}$ of the
spherical harmonics. The kinematic action
$\propto \int|\dot{\hat{\mathbf{n}}}|^2\,dt$ also
coincides with that of the Haldane $\mathrm{O}(3)$
NLSM~\cite{haldane1983nlsm,haldane1983pla} and the Frank
elastic theory of nematic liquid
crystals~\cite{chaikin_lubensky}, but with a conceptual
distinction: in those systems rotational symmetry is
spontaneously broken in the thermodynamic limit and the slow
modes are genuine Nambu-Goldstone modes (continuous gapless
spectrum), whereas in the F2 model the orientation is broken
only adiabatically and is restored by the rotational diffusion
of $\hat{\mathbf{n}}(t)$. The slow F2 mode is not a Goldstone
in the strict sense but a discrete pair of soliton zero modes
(in the sense of Rajaraman~\cite{rajaraman1982}, Sec.~5.5)
promoted to a collective orientation coordinate, overdamped
(type-A) rather than propagating
(type-B, as in the symplectic Hamiltonian cases above).

\paragraph{Mean-field glasses and SYK-type dynamics.}
The disorder-averaged action of Section~\ref{sec:large_N},
with its non-local temporal kernel and time-reparametrization
quasi-invariance, shares its formal structure with the
dynamics of spherical $p$-spin models and SYK-type systems
\cite{cugliandolo1993, facoetti2019}. The analogy is not
merely formal: in the glassy regime the density two-point
function is expected to exhibit aging and a nontrivial
Parisi-type structure, and the mapping between classical
Langevin dynamics and quantum Hamiltonians established by
Facoetti et al.~\cite{facoetti2019} suggests that the F2
density sector may inherit the same SYK-like correlation
structure. The geometric setting, a physical 2D surface
embedded in 3D space rather than an abstract hypersphere in
spin space, raises the prospect of probing these phenomena in
a system where the order parameter is directly observable.

\paragraph{Non-perturbative field theory and adiabatic
dimension reduction.}
Ring formation is itself a non-perturbative effect: the
density profile $f_0(\hat{\mathbf{n}}\cdot\mathbf{x})$ is a
saddle point that cannot be reached by any finite-order
expansion around the uniform state. We refer to the
emergence of such configurations as \emph{adiabatic
dimension reduction}: at any instant the density
concentrates on a lower-dimensional submanifold of the
background (a great circle on $S^2$, a pair of small circles
on $T^2$) with a definite orientation, while that
orientation itself performs a slow stochastic motion, so
that the long-time-averaged state restores the rotational
invariance of the underlying manifold. The slowly moving
collective excitations seen in the FBP particle model and
reproduced by the F2 field theory, rings on $S^2$ and pairs
of rings on $T^2$ whose centers drift coherently while the
internal density supports finite-frequency breathing and
azimuthal fluctuations, are reminiscent of the breather
solutions of relativistic and integrable field theories, in
which a localized non-perturbative profile carries a slow
collective coordinate alongside oscillatory internal modes
(Rajaraman~\cite{rajaraman1982}, Ch.~2). In this respect the
F2 model provides a tractable laboratory for studying
adiabatic dimension reduction, instanton-like transitions
between orientations, and the coupling of the collective
coordinate to its gapped fluctuations through the
Faddeev-Popov sector of
Section~\ref{subsec:faddeev_popov}.

\subsection{Open Questions}

Several questions remain open beyond the universal SO(3)
description. Does the disorder-averaged field theory exhibit
replica symmetry breaking in the density sector, and if so,
what is the structure of the Parisi order parameter in this
geometric setting? What is the phase diagram in the
$(T, J, \lambda)$ parameter space, and at what control-parameter
values do phase transitions between uniform and ring states
occur? Beyond the mean-field treatment developed here, how do
fluctuations affect the formation of the dimension-reduced
structure, and what is the nucleation barrier for creating a
ring? Beyond these conceptual questions, the technical
program of computing $D_{\text{rot}}$ from the microscopic
theory and extending the construction to other geometries
and to a dynamical metric defines the next stage of this
work.

%==============================================================================
\section{Summary}
\label{sec:conclusion}
%==============================================================================

We set out to determine the statistical field theory that
governs the thermodynamic limit of frustrated Brownian
particles on a two-dimensional Riemannian manifold and to
extract its universal low-energy content. The main
achievements of the paper are the following.

\paragraph{(1) The F2 statistical field theory.}
Starting from the $N$-body overdamped Langevin dynamics of
Ref.~\cite{halperin2026frustrated}, the large-$N$ limit
yields a covariant statistical field theory for a smooth
density field $\rho(x,t)$, which we call the F2 model.
Cast as an MSRJD path integral and averaged over quenched
disorder, the theory produces a single-field effective
action with local and non-local (in space and time)
self-interactions characteristic of spin-glass dynamics.

\paragraph{(2) The effective Dean-Kawasaki equation of
the F2 model.} Saddle-point analysis of the F2 generating
functional produces the covariant \emph{effective}
Dean-Kawasaki equation of
Section~\ref{subsec:large_N_saddle}: a nonlinear functional
Langevin equation for the smooth density field
$\rho(x,t)$, in which the bare local Dean-Kawasaki noise
is dressed into a self-consistent two-time kernel
inherited from the disorder average. This is distinct
from the exact Dean-Kawasaki SPDE for the singular
empirical density, which enters only as an intermediate
step in the construction. The mean-field reduction of the
effective equation governs the evolution of
$\langle\rho(x,t)\rangle$ on the background manifold and
is the field-theoretic counterpart of the
Cugliandolo-Kurchan equations of mean-field glass theory,
adapted to a conserved density on a curved surface.

\paragraph{(3) The $\mathbb{RP}^2$ nonlinear sigma model in
(0+1) dimensions.} Combining the symmetry analysis of the F2
action with the ring-formation pattern observed in the
particle simulations of
Ref.~\cite{halperin2026frustrated}, we carried out a low-energy
reduction. The $\mathrm{SO}(3) \to \mathrm{SO}(2)$ symmetry
breaking singles out the ring orientation as the slow degree
of freedom, and integrating out the gapped density
fluctuations by the Gervais-Jevicki-Sakita collective-coordinate
construction with Faddeev-Popov gauge fixing yields the
nonlinear sigma model (NLSM) on the real projective plane
$S^2/\mathbb{Z}_2 = \mathbb{RP}^2$ (the $\mathbb{RP}^2$ NLSM
on the projective rotor space) in (0+1) dimensions; the
orientation is a director, since $\hat{\mathbf{n}}$ and
$-\hat{\mathbf{n}}$ describe the same density of the
even-profile ring. The model is governed by a single
low-energy constant $D_{\text{rot}}$, with no potential and
no Berry phase.
A simulation of the $\mathbb{RP}^2$ NLSM with
$D_{\text{rot}}$ fitted from a single observable reproduces
the tested orientation- and density-sector diagnostics within
numerical resolution after fixing $D_{\text{rot}}$ and, where
needed, $f_0$.

\paragraph{(4) Universality of ring formation and of the
$\mathbb{RP}^2$-NLSM dynamics.}
On $S^2$, the same ring geometry and the same
$\mathbb{RP}^2$-NLSM effective dynamics arise from a broader
class of pairwise potentials than the linear geodesic
form $V(d) = d$ of the original F2 model. Soft Coulomb
potentials $V(d) = \log(d + d_{\min})$ at several values
of $d_{\min}$ and the truncated geodesic-logarithm
$V(d) = \max(U_{\min}, \log d)$ in its ring-forming
regime all produce rings that undergo $\mathbb{RP}^2$-NLSM
orientation dynamics indistinguishable from the F2 model
in their scalar diagnostics
(Section~\ref{subsec:coulomb}). What these tests select
is a pairwise force that remains nonzero across the full
diameter of $S^2$. Both genuine Coulomb-type potentials
on the sphere, namely the intrinsic 2D chord-logarithm
$V(d) = \log(2\sin(d/2))$ (the Green's function of the
Laplace-Beltrami operator) and the restriction of the
ordinary 3D Coulomb
$V = 1/|\mathbf{x}_i - \mathbf{x}_j|$ to the surface, fail
this criterion: their forces vanish at the antipodal
point, where all geodesics from a source refocus, and
neither produces ring formation in our simulations. 

The
effective low-energy theory is insensitive to the
microscopic form of the pairwise potential within the
ring-forming subclass; only $D_{\text{rot}}$ is
model-dependent. We therefore conjecture, and provide
numerical evidence for, a universality class on $S^2$
whose defining feature is not any specific functional
form of the potential but a pairwise force that remains
effective across the cut locus of the sphere. Whether the
same mechanism selects ring-forming potentials on
two-dimensional manifolds of lower symmetry (torus,
cylinder, smoothly deformed spheres) is a natural
extension that is left for future work.

\paragraph{Future directions.}
The framework developed here admits several natural
extensions. The collective-coordinate construction should
be repeated on the other fixed manifolds treated in the
companion particle-simulation
study~\cite{halperin2026frustrated}, including the torus,
the cylinder, and smoothly deformed spheres. On these
manifolds the orientation target space and the selection
rules that enforce the zero-torque and zero-Berry-phase
properties on $S^2$ both change, so the exercise should
separate what is specific to the sphere from what is truly
geometry-independent. 

%Promoting the manifold metric to a
%dynamical variable coupled to the matter field (the F2M
%model, developed in a companion paper) turns the
%background surface into a degree of freedom of its own
%and is expected to couple the orientation dynamics to
%membrane-shape modes, producing a structure without a
%direct analog in standard quantum field theory.

Spatially extended versions of the F2 sigma model, in
which the orientation field varies over a 2D domain
(e.g., a soft membrane that supports multiple
ring-forming patches, or multi-ring configurations on the
torus), would activate the topological-defect content of
the $\mathbb{RP}^2$ NLSM laid out in
Section~\ref{subsec:low_energy_ansatz}: half-integer
disclinations of the director field from $\pi_1$ and
hedgehog charges from $\pi_2$, with the $\mathbb{Z}_2$
self-annihilation rule familiar from the nematic-defect
literature~\cite{lammert1995, degennesprost}. These
spatially-textured extensions are the most direct route
to making the $\mathbb{RP}^2$ topology of the orientation
manifold experimentally observable beyond the
selection-rule level resolved by the present (0+1)D
analysis. 

A second topological extension in a different
direction is to add quantum spin-1/2 internal degrees of
freedom to the FBP particles, with body-frame coupling to
the ring orientation through a spin-orbit term. A
$\hat{\mathbf{n}} \to -\hat{\mathbf{n}}$ flip is then a
$\pi$-rotation of the body frame and produces a
$(-1)^{N_s}$ Berry phase from the $N_s$ body-frame-coupled
spinors; for odd $N_s$ this realizes the
$\theta = \pi$ (spinor) sector of the $\mathbb{RP}^2$ NLSM,
inaccessible in the present spinless construction. The
mechanism is the rotor analog of the Haldane
$\theta$-term distinction between integer-spin and
half-integer-spin antiferromagnets, and connects to the
skyrmion-qubit construction of
Ref.~\cite{psaroudaki2021skyrmion} with the helicity
coordinate replaced by the projective ring orientation.

Computing
$D_{\text{rot}}$ from first principles through the
self-consistency loop of
Section~\ref{subsec:self_consistent_eqs} remains open and
is the natural next step toward an ab initio prediction of
the single low-energy constant that controls the effective
theory. The non-Markovian regime, in which the memory
kernel $\kappa(\tau)$ acquires nontrivial time structure,
is the regime where the connection between the F2 model
and glassy/SYK-type dynamics should become quantitatively
testable, with the ring orientation providing a
geometrically well-defined order parameter for aging
measurements. 

Finally, an experimental realization of the
ring-orientation dynamics, for example on a soft spherical
membrane decorated with charged colloidal particles
(Section~\ref{sec:discussion}), would make the
relaxation spectrum $\lambda_\ell = \ell(\ell+1)D_{\text{rot}}$
of the orientation distribution directly accessible. A
genuine quantum implementation would replace these
relaxation rates by the rotor energy levels
$E_\ell = \ell(\ell+1)\hbar^2/(2\mathcal{I})$ with
$\mathcal{I}$ a platform-dependent moment of inertia.
Because the orientation manifold is $\mathbb{RP}^2$, the
single-valued physical levels are even-$\ell$ only, and the
natural qubit candidate is a nematic-type pair built from
the $\ell = 0$ singlet and one component of the $\ell = 2$
quintuplet split by a quadrupolar (tensor) field, in analogy
with the skyrmion-helicity qubit of
Ref.~\cite{psaroudaki2021skyrmion} but with the
$\mathrm{U}(1)$ helicity replaced by the projective
director on $\mathbb{RP}^2$ and the natural coupling family
shifted from Zeeman-like to quadrupolar.

%==============================================================================
% APPENDICES
%==============================================================================
\appendix
\numberwithin{equation}{section}
\addtocontents{toc}{\protect\setcounter{tocdepth}{1}}

\section{Covariant Langevin dynamics and MSRJD path integrals on Riemannian manifolds}
\label{app:covariant_langevin}

This appendix develops the general theory of covariant Langevin
dynamics on a Riemannian manifold $\M$ (we follow standard
references on Riemannian geometry~\cite{nakahara2003geometry} for
the differential-geometric setup), largely following Zinn-Justin
\cite{zinnjustin2002} (Sections~4.8 and 17.4). We derive the
covariant Langevin equation, the associated Fokker-Planck equation
and path integral, and then construct the MSRJD field theory with
its supersymmetric structure. The framework is general and applies
to any covariant Langevin equation with state-dependent noise.
Appendix~\ref{app:dean_kawasaki} applies this machinery to the
Dean-Kawasaki equation for the empirical density $\rho_N(x,t)$.
Throughout, $x = (x^1, x^2)$ denotes a field point in intrinsic
coordinates, the same type of coordinates as the particle positions
$q_n^i$ ($i = 1,2$).

\subsection{Covariant Langevin Equation on the Manifold}
\label{subsec:covariant_langevin}

We derive the covariant Langevin equation for particle motion on
$\M$ by projecting the dynamics from the embedding space onto the
manifold, following Zinn-Justin \cite{zinnjustin2002}
(Sections~4.8, 17.4, and Appendix~A15.3.2). In this subsection we adopt Zinn-Justin's index conventions
(Sections~22.1 and~22.6):
$\alpha, \beta = 1, \ldots, N$ label the embedding-space coordinates
$X_\alpha$; $i, j, k = 1, \ldots, D$ are \emph{curved} (coordinate)
indices on the manifold, with metric $g_{ij}$; and
$a, b, c = 1, \ldots, D$ are \emph{flat} (frame) indices labelling
the orthonormal basis (vielbein) in the tangent plane, with
Euclidean metric $\delta_{ab}$ (ZJ, Eq.~22.69).
Both sets of manifold indices run over $1, \ldots, D$, but they
transform differently: curved indices transform under coordinate
changes $\varphi \mapsto \varphi'$ via the Jacobian $T^i_j$
(ZJ, Eq.~22.2), while flat indices transform under local $O(D)$
rotations of the frame (ZJ, Eq.~22.72). The vielbein $e^a_i$
converts between the two: $V^a = e^a_i\,V^i$
(ZJ, Eq.~22.83).
For our physical setting $N = 3$ and $D = 2$, but the
construction is general. The main text uses the same index
conventions as this appendix: $i,j,k$ for curved (coordinate)
indices and $a,b$ for flat (frame) indices.

\paragraph{Embedding and constraint.}
Consider a $D$-dimensional Riemannian manifold $\M$ embedded in
$\R^N$. The manifold is defined by $N - D$ constraint equations:
\begin{equation}
    E^s(X_\alpha) = 0, \qquad s = 1, \ldots, N - D
    \label{eq:zj_constraint}
\end{equation}
where $X_\alpha$ ($\alpha = 1, \ldots, N$) are the
embedding-space coordinates (corresponding to Zinn-Justin
Eq.~17.45). In our case $N = 3$ and $D = 2$, so there is a single
constraint ($s = 1$), but we keep the presentation general. We
solve the constraints locally and split the embedding coordinates
into $D$ independent components $\varphi^i$ ($i = 1, \ldots, D$)
and $N - D$ dependent components $\chi^s(\varphi)$
(Zinn-Justin Eq.~17.46):
\begin{equation}
    X_\alpha \equiv \{\chi^s(\varphi),\; \varphi^i\}
    \label{eq:zj_split}
\end{equation}

\paragraph{Coordinate gauge freedom on $S^2$.}
The splitting \eqref{eq:zj_split} requires choosing intrinsic
coordinates $\varphi^i$ on $\M$. For a fixed polar axis, this
is simply a choice of chart and raises no gauge issue. However,
when the dynamics produces structures with a preferred
orientation $\hat{\mathbf{n}}(t)$ (such as equatorial rings on
$S^2$), it is natural to introduce \emph{adapted} coordinates
that track this orientation by continuously rotating the
two-dimensional coordinate frame through the embedding
three-dimensional space. This rotation
$R(t) \in \mathrm{SO}(3)$ has three parameters
$(\alpha, \beta, \gamma)$, of which only two are determined by
$\hat{\mathbf{n}} = R\,\hat{\mathbf{e}}_3$. The third,
$\gamma$, is a gauge degree of freedom removed by the
Faddeev-Popov procedure. The rotation can be
parameterized using Cayley-Klein parameters
\cite{pennestri2016,cottingham2001}.

\paragraph{Induced metric, Christoffel symbols, and curvature.}
The embedding functions $\chi^s(\varphi)$ determine all
intrinsic geometry (Zinn-Justin, Appendix~A15.3.2). The
induced metric is
$g_{ij} = \delta_{ij}
+ \partial_i\chi^s\partial_j\chi^s$
(ZJ Eq.~A15.20), with determinant $g = \det g_{ij}$ and
covariant volume element
$d\mu_g = \sqrt{g}\prod_i d\varphi^i$. The torsion-free
Christoffel connection is
$\Gamma^i_{jk}
= \frac{1}{2}g^{il}(\partial_j g_{lk}
+ \partial_k g_{lj} - \partial_l g_{jk})$
(ZJ Eq.~22.41). The covariant derivative on a vector is
\begin{equation}
    \nabla_i V^j = \partial_i V^j
    + \Gamma^j_{ik}\,V^k
    \label{eq:covariant_derivative_app}
\end{equation}
(ZJ Eq.~22.26), on a covector
$\nabla_i V_j = \partial_i V_j
- \Gamma^k_{ji}V_k$~(ZJ Eq.~22.30), and on mixed tensors
by adding $+\Gamma$ per upper and $-\Gamma$ per lower
index. The contracted Christoffel symbol satisfies
$\Gamma^k_{ki}
= (\partial_i\sqrt{g})/\sqrt{g}$~(ZJ Eq.~22.66),
giving the covariant divergence
\begin{equation}
    \nabla_i V^i
    = \frac{1}{\sqrt{g}}\partial_i(\sqrt{g}\,V^i)
    \label{eq:covariant_divergence}
\end{equation}
(ZJ Eq.~22.67). Throughout this paper, $\nabla_i$
denotes the manifold covariant derivative.

\paragraph{Physical Langevin equation.}
The starting point is the overdamped Langevin equation in $\R^N$:
\begin{equation}
    \gamma\,\frac{dX_\alpha}{dt} = -\frac{\partial U^{(\phi)}}{\partial X_\alpha}
    + \sqrt{2\gamma k_B T}\,\nu'_\alpha(t)
    \label{eq:langevin_physical}
\end{equation}
where $\gamma$ is the friction coefficient (proportional to the
relaxation time), $T$ is the temperature, $k_B$ is the Boltzmann
constant, and $\nu'_\alpha$ is Gaussian white noise with
unit-normalized covariance
$\langle\nu'_\alpha(t)\,\nu'_\beta(t')\rangle
= \delta_{\alpha\beta}\,\delta(t-t')$.
The noise has $N$ components while the manifold has $D$ degrees of
freedom; the $N - D$ normal components are projected out below.

\paragraph{Dimensionless time and diffusion coefficient.}
We set $k_B = 1$ henceforth and introduce the dimensionless time
$\tau = t/\gamma$ together with the diffusion coefficient
\begin{equation}
    \Omega \;=\; 2\gamma T
    \label{eq:omega_def}
\end{equation}
which plays the role of $\hbar$ in the stochastic quantization
analogy. Defining the rescaled noise
$\nu_\alpha(\tau) = \sqrt{\Omega}\,\nu'_\alpha(\tau)$, the Langevin
equation in dimensionless time becomes:
\begin{equation}
    \frac{dX_\alpha}{d\tau} = -\frac{\partial U^{(\phi)}}{\partial X_\alpha}
    + \nu_\alpha(\tau)
    \label{eq:langevin_dimless}
\end{equation}
with covariance (Zinn-Justin Eq.~17.48):
\begin{equation}
    \langle\nu_\alpha(\tau)\,\nu_\beta(\tau')\rangle
    = \Omega\,\delta_{\alpha\beta}\,\delta(\tau-\tau')
    \label{eq:noise_embedding}
\end{equation}

\paragraph{Connection to Zinn-Justin notation.}
We revert to writing $t$ for the dimensionless time $\tau$
(in all final formulas, one substitutes $t \to t/\gamma$ to
recover physical time). The Langevin equation
\eqref{eq:langevin_dimless} can be written in the form of
Zinn-Justin Eqs.~16.104 and~17.1:
\begin{equation}
    \dot{X}_\alpha = -\frac{\Omega}{2}\,
    \frac{\delta\mathcal{A}}{\delta X_\alpha} + \nu_\alpha
    \label{eq:langevin_embedding}
\end{equation}
with the identification
$\frac{\Omega}{2}\frac{\delta\mathcal{A}}{\delta X_\alpha}
= \frac{\partial U^{(\phi)}}{\partial X_\alpha}$, i.e.\
$\mathcal{A} = 2U^{(\phi)}/\Omega = U^{(\phi)}/T$ is the
dimensionless action ($= \beta U^{(\phi)}$).

\paragraph{Projection onto the tangent plane.}
On the manifold, variations $\delta X_\alpha$ are constrained to
lie in the tangent plane:
$(\partial E^s / \partial X_\alpha)\,\delta X_\alpha = 0$.
To project the dynamics, we introduce an orthonormal basis
$e^\alpha_a$ ($a = 1, \ldots, D$) for the tangent plane,
satisfying (corresponding to Zinn-Justin Eqs.~17.50--17.51):
\begin{equation}
    \frac{\partial E^s}{\partial X_\alpha}\,e^\alpha_a = 0,
    \qquad
    e^\alpha_a\,e^\alpha_b = \delta_{ab}
    \label{eq:zj_ortho_basis}
\end{equation}
In terms of the intrinsic coordinates $\varphi^i$, the first
condition in \eqref{eq:zj_ortho_basis} becomes
(Zinn-Justin Eq.~17.52):
\begin{equation}
    e^s_a = \partial_i\chi^s\,e^i_a
    \label{eq:zj_es_relation}
\end{equation}
Substituting \eqref{eq:zj_es_relation} into the orthonormality
condition in \eqref{eq:zj_ortho_basis} and using
the induced metric $g_{ij}$ gives (Zinn-Justin Eq.~17.53):
\begin{equation}
    e^i_a\,g_{ij}\,e^j_b = \delta_{ab}
    \label{eq:zj_vielbein_ortho}
\end{equation}
so the matrix $e^i_a$ is the inverse vielbein (zweibein) of the
metric $g_{ij}$. It follows that the inverse metric is
(Zinn-Justin Eq.~17.58):
\begin{equation}
    g^{ij} = e^i_a\,e^j_a
    \label{eq:zj_inverse_metric}
\end{equation}

We also define the tangent vectors $t^\alpha_i$ in the embedding
space (Zinn-Justin Eq.~17.60):
\begin{equation}
    t^j_i = \delta_{ij}, \qquad t^s_i = \partial_i \chi^s
    \label{eq:zj_tangent_vectors}
\end{equation}
From the induced metric $g_{ij}$ and \eqref{eq:zj_tangent_vectors} one
verifies $t^\alpha_i\,t^\alpha_j = g_{ij}$
(Zinn-Justin Eq.~17.61). The identity
$\partial_i\,t^\alpha_j\,t^\alpha_k = g_{kl}\,\Gamma^l_{ij}$
(Zinn-Justin Eq.~17.62), which can be rewritten covariantly as
$\nabla_i\,t^\alpha_j\,t^\alpha_k = 0$
(Zinn-Justin Eq.~17.63), encodes the relation between the
Christoffel symbols the Christoffel symbols $\Gamma^i_{jk}$ and the embedding.

\paragraph{Projected Langevin equation.}
Projecting \eqref{eq:langevin_embedding} onto the tangent plane
using $e^\alpha_a$ from \eqref{eq:zj_ortho_basis} gives
(corresponding to Zinn-Justin Eq.~17.55):
\begin{equation}
    \dot{X}_\alpha = e^\alpha_a e^\beta_a\bigl(-\tfrac{\Omega}{2}\delta\mathcal{A}/\delta X_\beta + \nu_\beta\bigr)
    \label{eq:projected_langevin}
\end{equation}
To rewrite this in intrinsic coordinates,
note that $\dot{X}_\alpha$ has components
$\dot{\varphi}^i$ and
$\dot{\chi}^s = \partial_i\chi^s\,\dot{\varphi}^i$. Extracting
the $\varphi^i$ components and using the chain rule identity
(Zinn-Justin Eq.~17.57):
\begin{equation}
    e^j_a\frac{\partial U^{(\phi)}}{\partial\varphi^j} + \partial_j\chi^s e^j_a\frac{\partial U^{(\phi)}}{\partial\chi^s} = e^j_a\partial_j U^{(\phi)}(\varphi,\chi(\varphi))
    \label{eq:zj_chain_rule}
\end{equation}
together with the inverse metric \eqref{eq:zj_inverse_metric}, one
obtains the intrinsic Langevin equation (corresponding to
Zinn-Justin Eq.~17.59):
\begin{equation}
    \dot{\varphi}^i = -\tfrac{\Omega}{2}g^{ij}\partial_j \mathcal{A} + g^{ij}t^\alpha_j\nu_\alpha
    \label{eq:langevin_intrinsic_strat}
\end{equation}
The projected noise
$\eta^i \equiv g^{ij}\,t^\alpha_j\,\nu_\alpha$ has covariance
$\langle\eta^i(t)\,\eta^j(t')\rangle
= \Omega\,g^{ij}\,\delta(t-t')$,
which follows from \eqref{eq:noise_embedding} and
\eqref{eq:zj_inverse_metric}. The $N - D$ normal components of
$\nu_\alpha$ drop out of \eqref{eq:langevin_intrinsic_strat}
entirely.

\paragraph{Standard Brownian motion and state-dependent volatility.}
Writing \eqref{eq:langevin_intrinsic_strat} in increment form with
$d\eta^i(t) \equiv \eta^i(t)\,dt$, the Stratonovich SDE reads:
\begin{equation}
    d\varphi^i = -\tfrac{\Omega}{2}g^{ij}\partial_j \mathcal{A}\,dt + \sqrt{\Omega}\,d\eta^i(t) \quad\text{(Stratonovich)}
    \label{eq:langevin_standard_bm}
\end{equation}
where the noise increments $d\eta^i$ have the covariant correlation
$\langle d\eta^i(t)\,d\eta^j(t)\rangle = g^{ij}(\varphi)\,dt$.
The noise coefficient is $\sqrt{\Omega}$, where
$\Omega = 2\gamma T$ (Eq.~\eqref{eq:omega_def}).
The inverse metric $g^{ij}$ in the noise covariance reflects the
projection of isotropic embedding-space noise onto the tangent
plane. The position dependence of $g^{ij}$ is what makes the noise
multiplicative on the curved manifold. The midpoint discretization
inherited from the embedding-space projection corresponds to the
Stratonovich convention \cite{zinnjustin2002}.

To express $d\eta^i$ in terms of $D$ independent standard Wiener
processes $W_1(t), \ldots, W_D(t)$ with
$\langle dW_i(t)\,dW_j(t)\rangle = \delta_{ij}\,dt$, one needs a
``square root'' of the inverse metric, i.e.\ a state-dependent
volatility matrix $\sigma^{ij}(\varphi)$ satisfying:
\begin{equation}
    \sigma^{ik}(\varphi)\,\sigma^{jk}(\varphi) = g^{ij}(\varphi)
    \label{eq:cholesky_metric}
\end{equation}
The noise increments are then expressed as
$d\eta^i(t) = \sigma^{ij}(\varphi)\,dW_j(t)$. Equation
\eqref{eq:cholesky_metric} is the Cholesky factorization of the
positive-definite matrix $g^{ij}$. The explicit dependence of the
volatility $\sigma^{ij}(\varphi)$ on the state variable $\varphi$
is always assumed but sometimes suppressed in formulas for
compactness.

The vielbein $e^i_a$ from \eqref{eq:zj_vielbein_ortho}
satisfies $e^i_a\,e^j_a = g^{ij}$
\eqref{eq:zj_inverse_metric} and therefore provides a
geometrically natural realization of this Cholesky factor:
\begin{equation}
    \sigma^{ij}(\varphi) = e^i_j(\varphi)
    \label{eq:volatility_vielbein}
\end{equation}
The decomposition is not unique (any rotation
$\sigma^{ij} \to \sigma^{ik}O_{kj}$ with $O \in O(D)$ gives
the same $g^{ij}$), but the vielbein is tied to the embedding.

\paragraph{Stratonovich form.}
Rewriting the Langevin equation
\eqref{eq:langevin_standard_bm} in terms of the volatility
$\sigma^{ij}$:
\begin{equation}
    dq^i = -\tfrac{\Omega}{2}g^{ij}\partial_j \mathcal{A}\,dt + \sqrt{\Omega}\sigma^{ij}(q)dW_j(t) \quad\text{(Stratonovich)}
    \label{eq:langevin_stratonovich_app}
\end{equation}

\paragraph{Connection to Zinn-Justin Section~4.8.}
The Langevin equation on the manifold has state-dependent noise
of the form analyzed by Zinn-Justin in Section~4.8. In the
continuum limit, the general markovian Langevin equation reads
(Zinn-Justin Eq.~4.63):
\begin{equation}
    \dot{q}^i = -\tfrac{1}{2}\,f^i(q) + e^i_a(q)\,\nu_a(t)
    \label{eq:zj_general_langevin}
\end{equation}
with $\langle\nu_a(t)\,\nu_b(t')\rangle
= \Omega\,\delta_{ab}\,\delta(t-t')$ (Zinn-Justin Eq.~4.64).
Here $e^i_a$ is the vielbein \eqref{eq:zj_inverse_metric},
playing the role of the volatility matrix $\sigma^{ij}$
\eqref{eq:volatility_vielbein}. For our problem the force is
$f^i = \Omega\,g^{ij}\partial_j\mathcal{A}$.

In the main text, the Langevin drift is expressed as
$(1/\gamma)\,g^{ij}\partial_j U^{(\phi)}$ rather than
$(\Omega/2)\,g^{ij}\partial_j\mathcal{A}$, since $\gamma$ and $T$
are the natural physical parameters for the soft-matter applications
we consider.

\paragraph{Discretized form and It\^{o} conversion.}
The discretized form of \eqref{eq:zj_general_langevin}
evaluated at the initial point $q = q(t)$
reads (Zinn-Justin Eq.~4.52):
\begin{equation}
    q^i(t{+}\epsilon) - q^i(t) = -\tfrac{\epsilon}{2}f^i(q) + e^i_a(q)\nu_a + \tfrac{1}{2}d^i_{ab}(q)\nu_a\nu_b
    \label{eq:langevin_discretized}
\end{equation}
where $\nu_a$ ($a = 1,\ldots,D$) are independent Gaussian variables
with $\langle\nu_a\,\nu_b\rangle
= \epsilon\,\Omega\,\delta_{ab}$
(Zinn-Justin Eqs.~4.52--4.53). The third term, quadratic in noise,
is needed because $\nu_a$ is of order $\sqrt{\epsilon}$, so
$d^i_{ab}\,\nu_a\,\nu_b$ contributes at order $\epsilon$.

Expanding the vielbein at the midpoint
$\tfrac{1}{2}[q(t) + q(t{+}\epsilon)]$ gives
(Zinn-Justin Eq.~4.67):
\begin{equation}
    e^i_a\bigl\{\tfrac{1}{2}[q(t){+}q(t{+}\epsilon)]\bigr\} = e^i_a[q(t)] + \tfrac{1}{2}e^j_b[q(t)]\nu_b\partial_j e^i_a[q(t)] + O(\epsilon)
    \label{eq:midpoint_expansion}
\end{equation}
which identifies the drift tensor (Zinn-Justin Eq.~4.68):
\begin{equation}
    d^i_{ab}(q) = e^j_b(q)\,\frac{\partial}{\partial q^j}\,e^i_a(q)
    \label{eq:drift_correction}
\end{equation}
The choice $\varepsilon(0) = 0$ (initial point) corresponds to the
It\^{o} prescription, while $\varepsilon(0) = \tfrac{1}{2}$
(midpoint) gives the Stratonovich convention.

In the It\^{o} prescription, the term quadratic in noise is
replaced by its average
$\tfrac{1}{2}\,\epsilon\,\Omega\,d^i_{aa}$
(Zinn-Justin Eqs.~4.65--4.66). The resulting It\^{o} Langevin
equation is:
\begin{equation}
    \dd q^i = \bigl[-\tfrac{\Omega}{2}g^{ij}\partial_j\mathcal{A} + \tfrac{\Omega}{2}e^j_a\partial_j e^i_a\bigr]\dd t + e^i_a\dd\tilde{W}_a(t) \quad\text{(It\^{o})}
    \label{eq:langevin_ito_zj}
\end{equation}
where $\dd\tilde{W}_a$ are independent Wiener increments with
variance $\Omega\,\dd t$ and the second term in the bracket
is the It\^{o} drift correction
$\tfrac{\Omega}{2}\,d^i_{aa}$~\eqref{eq:drift_correction}.

\paragraph{Fokker-Planck equation and path integral.}
The Fokker-Planck equation for the probability density
$P(q,t)$ on $\M$ and the corresponding path integral
representation follow from the discretized Langevin
equation~\eqref{eq:langevin_discretized} by standard
methods (Zinn-Justin, Sections~4.7--4.8). The equilibrium
distribution is $P_{\text{eq}} \propto e^{-U/T}$, and the
partition function $Z = 1$ (the path integral is
normalized by the Gaussian noise measure). We do not
reproduce these derivations here, as our construction
proceeds directly through the MSRJD
formalism (Section~\ref{subsec:msrjd_general} below).

\subsection{MSRJD Action for Covariant Langevin Dynamics}
\label{subsec:msrjd_general}

We now derive the MSRJD path integral for a general covariant
Langevin equation of the form \eqref{eq:langevin_ito_zj} on a
Riemannian manifold, with arbitrary potential and state-dependent
vielbein $e^a_b(q)$. This construction applies to
the Dean-Kawasaki equation (Appendix~\ref{app:dean_kawasaki},
Section~\ref{subsec:dk_as_langevin})
via the substitutions $q^a \to \rho(x)$,
$\tilde{q}_a \to \tilde{\rho}(x)$, with the noise covariance
inherited from the DK noise structure. The derivation follows
Zinn-Justin \cite{zinnjustin2002} (Sections~4.8, 16.5--16.6,
and~17.4). The noise strength parameter $\Omega = 2\gamma T$ was
introduced in the preceding subsection
(Eq.~\eqref{eq:omega_def}).

\paragraph{Constraint formulation.}
Consider a Langevin equation of the general form
\begin{equation}
    \dot{q}^a = F^a(q) + e^a_b(q)\,\xi_b(t)
    \label{eq:general_langevin}
\end{equation}
where $F^a$ is the deterministic drift (incorporating both the
potential force and any noise-induced drift) and $\xi_b$ is
Gaussian white noise with $\langle\xi_a(t)\xi_b(t')\rangle =
\delta_{ab}\,\delta(t - t')$. For a given noise realization, the
probability of observing a trajectory $q^a(t)$ is enforced by a
constraint (Zinn-Justin, Eqs.~4.78 and~17.64):
\begin{equation}
    P[q|\xi] = \delta\bigl[\dot{q}^a - F^a - e^a_b\xi_b\bigr]\cdot|\det M|
    \label{eq:constraint_general}
\end{equation}
where $M^a{}_b(t,t') =
\delta(\dot{q}^a - F^a - e^a_c\,\xi_c)/\delta q^b(t')$ is the
operator obtained by varying the equation of motion with respect to
the trajectory (Zinn-Justin, Eq.~4.80). In the Stratonovich
(midpoint) discretization, $M$ is not lower-triangular in time
and its determinant is non-trivial.

\paragraph{Response field.}
The functional delta function is represented via a Fourier integral
over an auxiliary response field $\hat{q}_a(t)$ (Zinn-Justin,
Eqs.~4.79 and~16.16):
\begin{equation}
    \delta\bigl[\dot{q}^a - F^a - e^a_b\xi_b\bigr] = \int \mathcal{D}\hat{q}\,\exp\!\left(i\!\int\!\dd t\;\hat{q}_a(\dot{q}^a - F^a - e^a_b\xi_b)\right)
    \label{eq:response_field_intro}
\end{equation}
Following the MSRJD convention, we redefine
$\hat{q}_a \to i\tilde{q}_a$, converting the oscillatory integral
into a convergent one. Classically, the new response field
$\tilde{q}_a$ is purely imaginary, but it is a fully functional
object in the MSRJD path integral: it enables the computation of
correlation and response functions via functional differentiation.
Under certain circumstances (for example, saddle-point solutions
such as instantons), $\tilde{q}_a$ can acquire expectation values
with a non-vanishing real part.

\paragraph{Dynamic action on Riemannian manifolds.}
Zinn-Justin constructs the dynamic action for the Langevin equation
on Riemannian manifolds in Section~17.4. The starting point is the
covariant Langevin equation on the manifold (Eq.~17.59, equivalent
to our Eq.~\eqref{eq:langevin_ito_zj}), which he rewrites in the
convenient form (Eq.~17.64)
\begin{equation}
    g_{ij}\dot{\varphi}^j + \tfrac{1}{2}\Omega\partial_i\mathcal{A} - t^\alpha_i\nu_\alpha = 0
    \label{eq:zj_langevin_rewritten}
\end{equation}
where $t^\alpha_i$ are the tangent vectors in the embedding space
defined in our Eq.~\eqref{eq:zj_tangent_vectors}, satisfying
$t^\alpha_i\,t^\alpha_j = g_{ij}$
(Eq.~the induced metric $g_{ij}$), and $\nu_\alpha$ are independent
Gaussian noises. The tangent vectors $t^\alpha_i$ should not be
confused with the vielbeins $e^i_a$ introduced in
Eq.~\eqref{eq:zj_ortho_basis}: the index $\alpha$ runs over all
$D$ embedding dimensions, while the flat index $a$ runs over the
$d$ intrinsic dimensions of the manifold. In this paragraph,
$\varphi^i$ corresponds to our $q^a$.

The transition from Eq.~\eqref{eq:zj_langevin_rewritten} to the
dynamic action follows the general MSRJD construction presented in
Zinn-Justin Sections~4.8 and~16.6--16.8: one enforces the equation of
motion via a functional delta function, introduces a Lagrange
multiplier (response field) $\tilde{\varphi}^i$ and Grassmann
ghost fields $c^i$, $\bar{c}^j$ for the Jacobian determinant. Before
integration over the noise, the dynamic action splits as
(Eq.~17.65)
\begin{equation}
    \mathcal{S} = \mathcal{S}_0 + \mathcal{S}_1\,.
    \label{eq:zj1765}
\end{equation}
Here $\mathcal{S}_0$ collects all terms involving the noise
$\nu_\alpha$ (Eq.~17.66):
\begin{equation}
    \mathcal{S}_0 = \int\!\dd x\,\dd t\;\left(
    \frac{1}{2\Omega}\,\nu_\alpha\nu_\alpha
    - \frac{2}{\Omega}\,t^\alpha_i\,\nu_\alpha\,\tilde{\varphi}^i
    + \frac{2}{\Omega}\,c^i\,\partial_j t^\alpha_i\,
    \bar{c}^j\,\nu_\alpha\right),
    \label{eq:zj1766}
\end{equation}
and $\mathcal{S}_1$ is the noise-independent part (Eq.~17.67):
\begin{equation}
    \mathcal{S}_1 = \int\!\dd x\,\dd t\;\frac{2}{\Omega}\,
    \Bigl[\tilde{\varphi}^i\bigl(g_{ij}\dot{\varphi}^j
    + \tfrac{1}{2}\Omega\,\partial_i\mathcal{A}\bigr)
    + c^i\,g_{ij}\,\dot{\bar{c}}^j
    - c^i\,\partial_k g_{ij}\,\dot{\varphi}^j\,\bar{c}^k
    - \tfrac{1}{2}\Omega\,c^i\,\partial_i\partial_j\mathcal{A}\,
    \bar{c}^j\Bigr].
    \label{eq:zj1767}
\end{equation}

\paragraph{Noise integration.}
Since $\mathcal{S}_0$ is quadratic in $\nu_\alpha$, the Gaussian
integration over the noise can be performed exactly. After
completing the square and integrating, one obtains the
noise-averaged action (Eq.~17.68):
\begin{equation}
    \mathcal{S} = \mathcal{S}_1
    + \int\!\dd x\,\dd t\;\frac{2}{\Omega}\,\bigl(
    -\tilde{\varphi}^i\,g_{ij}\,\tilde{\varphi}^j
    + 2\tilde{\varphi}^i\,g_{il}\,\Gamma^l_{jk}\,c^j\,\bar{c}^k
    + c^i\,\bar{c}^j\,c^k\,\bar{c}^l\,
    \partial_i\partial_j\,g_{kl}\bigr).
    \label{eq:zj1768}
\end{equation}
The first term is the standard noise-averaging contribution
$\propto \tilde{\varphi}^i g_{ij}\tilde{\varphi}^j$. The second
term, involving the Christoffel symbol $\Gamma^l_{jk}$, couples
the response field to the ghost bilinear. The last term is a
quartic ghost interaction induced by the curvature of the manifold.

% The superfield formulation, SUSY/BRST structure,
% partition function Z=1, and the generating functional
% are presented in Appendix~\ref{app:dean_kawasaki}
% (Sections on MSRJD generating functional and BRST
% symmetry).

%% =============================================================
%% Appendix B: Dean-Kawasaki
%% =============================================================

\section{Derivation of the Dean-Kawasaki equation for fixed disorder}
\label{app:dean_kawasaki}

This appendix derives the Dean-Kawasaki equation for the empirical
density $\rho_N(x,t)$ on a Riemannian manifold, using the covariant
Langevin equation established in
Appendix~\ref{app:covariant_langevin}. The DK equation is then
identified as a field Langevin equation, so the MSRJD construction
of Appendix~\ref{app:covariant_langevin} applies to the density
field. Throughout this appendix we use the same index convention
as Appendix~\ref{app:covariant_langevin}: Latin indices $i, j, k$
denote curved (coordinate) indices on the manifold, $a, b$ denote
flat (frame) indices, and particle labels use $n$.

The equation bears the names of Kawasaki \cite{kawasaki1994} and
Dean \cite{dean1996}, who arrived at closely related results by
different routes. Kawasaki derived a Fokker-Planck equation for the
probability distribution functional $\mathcal{P}(\{\hat{\rho}\},t)$
of a locally coarse-grained density $\hat{\rho}$
\cite{kawasaki1994}:
\begin{equation}
    \frac{\partial}{\partial t}\mathcal{P}(\{\hat{\rho}\},t)
    = -\frac{\Omega}{2}\!\int\!\dd x\,
    \frac{\delta}{\delta\hat{\rho}(x)}\nabla\!\cdot\!\left\{
    \hat{\rho}(x)\nabla\!\left[
    \frac{\delta}{\delta\hat{\rho}(x)}
    + \frac{1}{k_B T}\frac{\delta F[\hat{\rho}]}
    {\delta\hat{\rho}(x)}\right]
    \mathcal{P}(\{\hat{\rho}\},t)\right\}
    \label{eq:kawasaki_fp}
\end{equation}
(Here $\nabla$ denotes the ordinary flat-space gradient; on the
manifold this is replaced by the covariant derivative $\nabla_i$
defined in \eqref{eq:covariant_derivative_app}.)
Dean instead derived a stochastic Langevin equation for the
microscopic (empirical) density
$\rho(\mathbf{x},t) = \sum_{\alpha=1}^N
\delta(\mathbf{r}^\alpha(t) - \mathbf{x})$,
defined path-wise for each noise realization, in flat
space with a common pairwise potential $V$
\cite{dean1996}. The Dean equation reads
\begin{equation}
\label{eq:dean_flat}
\partial_t\rho(\mathbf{x},t)
= D\nabla^2\rho(\mathbf{x},t)
+ \nabla\!\cdot\!\!\left[\boldsymbol{\xi}\sqrt{2D\rho}\right]
+ \mu\nabla\!\cdot\!\!\left[\rho(\mathbf{x},t)\!
\int\!\dd\mathbf{y}\,\rho(\mathbf{y},t)\,
\nabla V(\mathbf{x} - \mathbf{y})\right],
\end{equation}
where $D = k_B T/(m\gamma)$ is the bare diffusion
coefficient, $\mu = 1/(m\gamma)$ the mobility, and
$\boldsymbol{\xi}(\mathbf{x},t)$ is a Gaussian white
noise with
$\langle\xi_i(\mathbf{x},t)\,
\xi_j(\mathbf{x}',t')\rangle
= \delta_{ij}\,\delta(t-t')\,
\delta(\mathbf{x} - \mathbf{x}')$.
The three terms are: free diffusion, multiplicative
noise (scaling as $\sqrt{\rho}$), and the mean-field
drift from pairwise interactions.
The two formulations are related in the same way as
Fokker-Planck and Langevin descriptions of a stochastic
process: Kawasaki's equation~\eqref{eq:kawasaki_fp}
governs the probability of observing a given density
profile, while Dean's equation~\eqref{eq:dean_flat}
gives the stochastic evolution of a single realization
(see \cite{illien2025}, Section~II, and
\cite{archerRauscher2004} for a discussion of the
relationship between the two formulations).

We partially follow Dean's path-wise approach, using the compact presentation
in Illien's review \cite{illien2025} (Section~II.B) as a convenient
reference for the flat-space calculation; see
also~\cite{illien2024, lebon2025} for recent MSRJD-based
treatments of related disordered and non-Gaussian density
dynamics. Our derivation adapts each
step to intrinsic manifold coordinates
\cite{castrovillarreal2023}, with the main modifications arising
from the state-dependent vielbein $e^i_b(q)$ and the curved
geometry. Each step parallels Illien's flat-space construction; the
differences are noted as we proceed. We note that our derivation follows the original derivation of \cite{dean1996} only up to the step of constucting a single-particle SPDE. The final step of obtaining a single SPDE for the mean particle density in the limit $ N \rightarrow \infty $ in our setting is substantially more complex than in the original setting considered by Dean \cite{dean1996} where all two-particle pairwise potentials are identical and their values depend only on distances between particles. In our setting, we need to perform avaraging over disorder in order to arrive at a version of the DK equation for our model. 

Three remarks on the mathematical status of the DK equation are in
order. First, for any finite $N$ the empirical density $\rho_N$ is a
sum of singular delta functions, so the DK equation should be
understood in the sense of distributions rather than as a classical
PDE \cite{illien2025}. Second, the mathematical well-posedness of
the DK equation (existence and uniqueness of solutions) has been
investigated only recently \cite{konarovskyi2019}. Third, the
noise enters the DK equation under a covariant divergence,
$\nabla_i\xi^i$, and taking the gradient of a distributional noise
field is not a well-defined pointwise operation. This ambiguity is
resolved in the MSRJD path integral construction
(Section~\ref{subsec:dk_as_langevin}): the divergence
$\nabla_i\xi^i$ is never evaluated in isolation but is always
paired against the response field $\tilde{\rho}$, and integration
by parts transfers the gradient onto $\tilde{\rho}$, leaving only
a well-defined distributional pairing with $\xi^i$.
Our calculations are made at the level of rigor commonly accepted
in the theoretical physics literature. The final model is well
defined in the large-$N$ limit, where the noise term scales as
$1/\sqrt{N}$ and becomes subdominant, which is the regime targeted
by our theory (the companion paper \cite{halperin2026frustrated}
uses $N = 400$ particles).

\subsection{Setup and Definitions}

Define the single-particle empirical density for particle $n$:
\begin{equation}
    \rho_n(x,t) = \frac{\delta^{(2)}(x - x_n(t))}{\sqrt{g(x)}}
    \label{eq:single_particle_density}
\end{equation}
and the total empirical density:
\begin{equation}
    \rho_N(x, t) = \frac{1}{N} \sum_{n=1}^{N} \rho_n(x,t)
    \label{eq:empirical_density_dk}
\end{equation}
The factor $1/\sqrt{g(x)}$ ensures covariance under coordinate changes.

For any smooth test function $\varphi: \M \to \R$, we have the
fundamental identity:
\begin{equation}
    \varphi(x_n(t)) = \int_\M \varphi(x) \, \rho_n(x,t) \, \dd\mu_g(x)
    \label{eq:test_function_identity}
\end{equation}
where $\dd\mu_g(x) = \sqrt{g(x)}\,\dd^2 x$. For the total density, this gives
\begin{equation}
    \langle \rho_N, \varphi \rangle \equiv \int_\M \rho_N(x,t) \, \varphi(x) \, \dd\mu_g(x)
    = \frac{1}{N} \sum_{n=1}^{N} \varphi(x_n(t))
    \label{eq:test_function_pairing}
\end{equation}

\subsection{Langevin Equation: Stratonovich and It\^{o} Forms}

We begin by restating the covariant Langevin equation derived in
Appendix~\ref{app:covariant_langevin}. In the Stratonovich
convention, the equation of motion for particle $n$ reads
(Eq.~\eqref{eq:langevin_stratonovich_app}):
\begin{equation}
    \dd x_n^i = f_n^i\,\dd t + \sqrt{\Omega}\,e^i_a(x_n) \circ \dd\tilde{W}_n^a(t) \quad\text{(Stratonovich)}
    \label{eq:langevin_strat_dk}
\end{equation}
where $e^i_a(x)$ is the vielbein satisfying
$e^i_a\,e^j_a = g^{ij}$,
$\Omega$ is the diffusion parameter (Eq.~\eqref{eq:omega_def}), and
$\tilde{W}_n^a$ ($a = 1,\ldots,D$) are independent standard
Wiener processes. The deterministic force on particle $n$ is
\begin{equation}
    f_n^i(x) \equiv -\frac{1}{\gamma}\,g^{ij}(x)\,\partial_j U^{(\phi)}(x)
    = -\frac{1}{\gamma}\,g^{ij}(x) \sum_m \phi_{nm} \partial_j d_g(x,x_m)
    \label{eq:force_fi_def}
\end{equation}
with $U^{(\phi)}$ the fixed-disorder potential.
In Zinn-Justin's notation \cite{zinnjustin2002} (Eq.~4.63),
the Langevin drift is written
$-\frac{1}{2}f^i_{\text{ZJ}}$ with
$f^i_{\text{ZJ}} = \Omega\,g^{ij}\partial_j\mathcal{A}$, so
$f_n^i = -\frac{1}{2}f^i_{\text{ZJ}}$; we use $f_n^i$ and the
noise strength $\Omega = 2\gamma T$ throughout.

Converting to It\^{o} form as in
Section~\ref{subsec:covariant_langevin}
(Eq.~\eqref{eq:langevin_ito_zj}) adds the noise-induced drift
$(\Omega/2)\,e^j_a\,\partial_j\,e^i_a
= (\Omega/2)\,d^i_{aa}$
from the Stratonovich-to-It\^{o} correction
\cite{ito1962, zinnjustin2002, castrovillarreal2023}:
\begin{equation}
    \dd x_n^i = \bigl[f_n^i + \tfrac{\Omega}{2}e^j_a\partial_j e^i_a\bigr]\dd t + \sqrt{\Omega}\,e^i_a(x_n)\dd\tilde{W}_n^a(t) \quad\text{(It\^{o})}
    \label{eq:langevin_ito_dk}
\end{equation}
The It\^{o} noise has quadratic variation
$\langle\dd x_n^i,\dd x_n^j\rangle = \Omega\,g^{ij}\dd t$.
In flat space the vielbein reduces to the identity ($e^i_a
= \delta^i_a$), the drift correction vanishes, and the two forms
coincide; the force becomes
$f_n^i = -\mu\sum_\beta \nabla V(r^\alpha - r^\beta)$
(with $\mu = 1/\gamma$), which is Illien's starting point
\cite{illien2025} (Eq.~(3)).

In our analysis below, we choose to proceed with the 
Stratonovich SDE (\ref{eq:langevin_strat_dk}) rather than with the It\^o's form (\ref{eq:langevin_ito_dk}). This is because we use the resulting SPDE for the particle density to construct a MSRJD path integral, which is more conveniently done using the Stratonovich calculus.

\subsection{A single-particle density SPDE}

A test function $\varphi$ satisfying Eq.(\ref{eq:test_function_identity}) serves as a device for extracting an
equation for the distributional density $\rho_n$. Since $\rho_n$
is a Dirac delta, it is not a smooth function and its time
derivative cannot be computed directly. Instead, we work in the
weak (distributional) sense: any equation that holds when
integrated against every smooth test function $\varphi$ determines
$\rho_n$ uniquely as a distribution.

As the proceed with the Stratonovich rather than with It\^o's rule, the differential of the test function satisfies the regular chain rule:
\begin{equation}
\label{Stratonovich_chain_rule_B}
\frac{d \varphi(x_n)}{dt} = (\nabla_i \varphi) \frac{d x_n^i}{dt} =  (\nabla_i \varphi) \left( f_n^i + \sqrt{\Omega}\,e^i_a(x_n) \eta^{a}(x,t) \right)
\end{equation}
Here we wrote increments of the Brownian motion as $ d W^a(x,t) = \eta^a(x,t) dt $, and $  \Delta_i $ stands for the covariant derivative on the manifold that acts on a vector $ V^{j} $ as defined in Appedix A, i.e. $ \nabla_i V^j = \partial_i V^j + \Gamma_{ik}^{j}V^k $.

Now we use the definition of the single-particle density
(\ref{eq:single_particle_density}) to write
\begin{equation}
\label{varphi_identity}
\varphi(x_n) = \int d \mu_g(x) \rho_n(x,t) \varphi(x)
\end{equation}
which implies that Eq.(\ref{Stratonovich_chain_rule_B}) can be equivalently written as follows:
\begin{equation}
\label{Stratonovich_chain_rule_B_2} 
\frac{d \varphi(x_n)}{dt} = \int d \mu_g(x) \rho_n(x,t) \left[  \left( f_n^i + \sqrt{\Omega}\,e^i_a(x_n) \eta^a(x,t) \right) 
(\nabla_i \varphi) \right]
\end{equation}
On the other hand, as $ \varphi(x) $ does not explicitly depend on time, Eq.(\ref{varphi_identity}) implies that
\begin{equation}
\label{identity_implication}
\frac{d \varphi(x_n)}{dt} = \int d \mu_g(x) \varphi(x) 
\frac{\partial \rho_n}{\partial t}
\end{equation}  
Integrating in Eq,(\ref{Stratonovich_chain_rule_B_2}) by parts, comparing with (\ref{identity_implication}) and noting that the test functon $ \varphi(x) $ is arbitrary, we obtain the SPDE for a single-particle density $ \rho_n(x,t) $:
\begin{equation}
\label{single_particle_SPDE}
\frac{\partial \rho_n}{\partial t} = - \nabla_i \left[ 
\left(  f_n^i + \sqrt{\Omega}\,e^i_a \eta^a \right) \rho_n \right]
\end{equation}  
Unlike the case of interacting particles with identical two-body potential that only depend on the distance between particles, our case is more complex as particle interactions depend on particles themselves with quenched random coefficients $ \phi_{nm} $. Therefore, we cannot proceed in our case the same way as in \cite{dean1996} by summing over 
all particles in the single-particle SPDE (\ref{single_particle_SPDE}). While the proper approach based on the MSRJD path integral will be introduced shortly below, we will first pause to compare our SPDE (\ref{single_particle_SPDE}) with a single-particle that arises in \cite{dean1996}. 

\subsection{Equivalence to It\^{o} SPDE}

The single-particle SPDE (\ref{single_particle_SPDE}) has a similar gradient form to both a single-particle and joint particle densities in \cite{dean1996, illien2025}, which reflect the particle number conservation in our system.
Unlike the original derivation in \cite{dean1996} that was based on It\^o's calculus in a flat space, we work with the Stratonovich calculus on a Riemannian manifold.

To establish the equivalence between our Stratonovich SPDE and the It\^o-based results of Dean \cite{dean1996} (see also Illien \cite{illien2025} for a review), we derive the spurious drift arising from the space-time white noise field $\eta^i(x,t)$. In the flat-space limit ($g_{ij} = \delta_{ij}$), the stochastic part of our single-particle density equation is:
\begin{equation}
    \left( \frac{\partial \rho_n}{\partial t} \right)_{\text{noise}} = -\sqrt{\Omega} \, \partial_i \left( \rho_n(x,t) \circ \eta^i(x,t) \right),
\end{equation}
where $\langle \eta^i(x,t) \eta^j(y,t') \rangle = \delta^{ij} \delta^{(2)}(x-y) \delta(t-t')$. In field theory, the Stratonovich-to-It\^o drift correction $\Delta_{\text{corr}}(x)$ for a noise term of the form $\mathcal{F}_i[\rho] \circ \eta^i$ is given by the functional contraction:
\begin{equation}
    \Delta_{\text{corr}}(x) = \frac{1}{2} \int d^2y \sum_{i,j} \left\langle \frac{\delta \left( -\sqrt{\Omega} \partial_{x,i} [\rho_n(x) \eta^i(x)] \right)}{\delta \rho_n(y)} \left( -\sqrt{\Omega} \partial_{y,j} [\rho_n(y) \eta^j(y)] \right) \right\rangle,
\end{equation}
where the expectation $\langle \dots \rangle$ is taken over the noise realizations. The functional derivative of the term at $x$ with respect to the density at $y$ is:
\begin{equation}
    \frac{\delta}{\delta \rho_n(y)} \left[ -\sqrt{\Omega} \partial_{x,i} (\rho_n(x) \eta^i(x)) \right] = -\sqrt{\Omega} \, \partial_{x,i} \left[ \delta^{(2)}(x-y) \eta^i(x) \right].
\end{equation}
Substituting this into the integral and evaluating the noise correlation $\langle \eta^i(x) \eta^j(y) \rangle = \delta^{ij} \delta^{(2)}(x-y)$, we find:
\begin{equation}
    \Delta_{\text{corr}}(x) = \frac{\Omega}{2} \int d^2y \sum_{i} \left[ \partial_{x,i} \delta^{(2)}(x-y) \right] \left[ \partial_{y,i} (\rho_n(y) \delta^{(2)}(y-x)) \right].
\end{equation}
Using the property of the Dirac distribution $\partial_{y,i} \delta^{(2)}(y-x) = -\partial_{x,i} \delta^{(2)}(y-x)$ and integrating by parts with respect to $y$, the integral collapses to:
\begin{equation}
    \Delta_{\text{corr}}(x) = \frac{\Omega}{2} \partial_{x,i} \partial_{x,i} \int d^2y \delta^{(2)}(x-y) \rho_n(y) = \frac{\Omega}{2} \nabla^2 \rho_n(x).
\end{equation}
By identifying the diffusion coefficient $D = \Omega/2$, the It\^o-equivalent drift is $D \nabla^2 \rho_n - \partial_i (f_n^i \rho_n)$. This matches exactly the result obtained by Dean using It\^o's lemma on particle trajectories. In our Stratonovich formulation, the explicit Laplacian is absent from the drift because the midpoint discretization of the conservative noise $\partial_i (\rho \circ \eta^i)$ naturally accounts for the thermal diffusion of the density field.

\subsection{MSRJD generating functional}
\label{subsec:dk_as_langevin}

For any functional $ A[\rho_n] $ of the dynamic variable $ \rho_n$, its expectation with respect to realizations of the thermal noise is given by the Wiener path integral
\begin{equation}
\langle A[ \rho_n] \rangle \equiv D \left[ \eta_n \right]  e^{- \frac{1}{2} \int dt d^2x \eta^{a}(x,t) \eta^{a}(x,t) } 
A[ \rho_n]_{\eta}	
\end{equation}
where $ D \left[ \eta_n \right] $ is the Wiener path integral measure
\begin{equation}
D \left[ \eta_n \right] = \prod_{a, t,x} \frac{d \eta_n^a(x,t)}{\sqrt{2 \pi}} e^{- \frac{1}{2} \int dt d^2x \eta_n^{a}(x,t) \eta_n^{a}(x,t) }
\end{equation}
Our objective is to change the Wiener path integral to a path integral with respect to fields $ \rho_n$ that satisfies the Langevin equation (\ref{single_particle_SPDE}) which we will write as a constraint 
\begin{equation}
\label{single_particle_Langevin}
\mathcal{Q}_{x,t}^{(\phi)}(\rho_n, e_{a}^i, \eta^a) \equiv \frac{\partial \rho_n}{\partial t} - \mathcal{L}^{\phi}[\rho_n] 
+ \sqrt{\Omega}\nabla_i\left[\rho_n e_{a}^i  \eta^a\right] = 0
\end{equation} 
Here $ \mathcal{L}^{\phi}[\rho_n] $ is the differential operator for a fixed disorder $ \{ \phi_{nm} \} $ appearing in 
Eq.(\ref{single_particle_SPDE}). 
\begin{equation}
\label{L_operator}
\mathcal{L}^{\phi}[\rho_n(x,t)]  \equiv \frac{1}{\gamma} \sum_{m} \phi_{nm}
 \int d y  \rho_{m}(y,t)  \nabla_i \left[ g^{ij}(x) \rho_{n}(x,t) \nabla_j d_g(x,y) \right]   
\end{equation}
where the covariant derivative in $ \nabla_j d_g(x,y) $ acts only on the first argument. 

Next we introduce the partition function
\begin{equation}
\label{Z_for_start}
	Z \equiv \int \prod_{n} D \left[ \eta \right] D \left[ \rho_n \right]
	\delta \left( \frac{\partial \rho_n}{\partial t} - \mathcal{L}^{\phi}[\rho_n]
+ \sqrt{\Omega}\nabla_i\left[\rho_n e_{a}^i  \eta^a\right] \right) \mathcal{J} \left[ \rho_n \right] 
\end{equation}
Here the Dirac delta-function ensures that the functional integration over all fields $ \rho_n $ in (\ref{Z_for_start}) only includes fields satisfying the Langevin equation (\ref{single_particle_Langevin}), and $ \mathcal{J} \left[ \rho_n \right] $is a Jacobian of transformation from $ \rho_n $ to the Langevin equation (\ref{single_particle_Langevin}):
%\begin{eqnarray}
%\label{Jacobian_MSRJD}
%\mathcal{J} \left[ \rho_n \right] \hspace{-0.2cm} &\equiv & %\hspace{-0.2cm} \text{det} \left[ \frac{\delta \mathcal{Q}_{x,t}^{(\phi)}%(\rho_n, e_{a}^i, \eta^a)}{\delta \rho_n(x',t')} \right] =  
%\text{det} \left[
%\left( \frac{\partial}{\partial t} - \frac{\delta \mathcal{L}%[\rho_n(x,t)]}{\delta \rho_n(x',t')} \delta^{(2)}(x-x')  
%\right) \delta(t-t') \right. \nonumber \\ 
%\hspace{-0.2cm} &+& \hspace{-0.2cm} \left. \sqrt{\Omega}%\nabla_i\left[ \left( \delta^{(2)}(x- x') 
%e_{a}^i(x,t) \eta^a(x,t)  \right] \right)
%\delta(t-t') \right]
%\end{eqnarray}
\begin{equation}
\label{Jacobian_MSRJD}
\mathcal{J} \left[ \rho_n \right] \equiv \text{det} \left[ \frac{\delta \mathcal{Q}_{x,t}^{(\phi)}(\rho_n, e_{a}^i, \eta^a)}{\delta \rho_n(x',t')} \right] 
\end{equation}
Note that due to the multiplicative noise in the Langevin equation (\ref{single_particle_Langevin}), the Jacobian (\ref{Jacobian_MSRJD}) depends on the noise $ \eta_n$. This implies, in particular, that unlike a Langevin equation with additive noise, the procedure of changing variables in the partition function (\ref{Z_for_start}) from $ \eta_n $ to $ \rho_n $ cannot be done separately from handling the Jacobian $\mathcal{J} \left[ \rho_n \right] $. The Jacobian 
(\ref{Jacobian_MSRJD}) will be computed in terms of the 
a fermion path integral below.
  
The delta-function can be represented by a Fourier integral\footnote{We use the convention that all constant factors such as $ 1/(2 \pi)$ etc. are 
absorbed in corresponding functional measures.}
\begin{eqnarray}
\label{delta_function_via_response_field}
&& \delta \left( \frac{\partial \rho_n}{\partial t} - \mathcal{L}[\rho_n] 
+ \sqrt{\Omega}\nabla_i\left[\rho_n e_{a}^i  \eta^a\right]
\right) = \int D \tilde{\rho}_n
e^{ i \tilde{\rho}_n \left(\frac{\partial \rho_n}{\partial t} - \mathcal{L}[\rho_n] 
+ \sqrt{\Omega}\nabla_i\left[\rho_n e_{a}^i  \eta^a\right]
\right)}  \\
&&= 
\int D \hat{\rho}_n
e^{ - \hat{\rho}_n \left(\frac{\partial \rho_n}{\partial t} - \mathcal{L}[\rho_n] + \sqrt{\Omega}\nabla_i\left[\rho_n e_{a}^i  \eta^a\right]
\right)} = 
\int D \hat{\rho}_n e^{ - \hat{\rho}_n \left(\frac{\partial \rho_n}{\partial t} - \mathcal{L}[\rho_n] \right) + \sqrt{\Omega}\rho_n e_{a}^i \eta^a \nabla_i \hat{\rho}_n
} \nonumber 
\end{eqnarray}
where we change the integration variable $ \tilde{\rho}_n = i \hat{\rho}_n$
in the second equation, and integrated by parts in the third equation.

Note that by construction, we have $ Z = 1$, irrespective of a realization of a quenched disorder in the interaction potential. This observation 
drastically simplifies the procedure of averaging over disorder in the dynamic approach based on the MSRJD formalism in comparison to a static 
analysis that typically requires replica methods or their equivalents to perform such averaging \cite{dedominicis1978, sompolinsky1982, castellani2005, hertz2016}. 

\subsection{Fermion path integral representation of the Jacobian}

%The transition from the path integral over the thermal noise $\{\eta_n^a\}%$ to the path integral over the single-particle density fields $\{\rho_n\}%$ requires the inclusion of a functional Jacobian $\mathcal{J}_n = %\det[\mathcal{M}_n]$ for each particle $n$. The matrix kernel of this %operator in coordinate space is defined as the variational derivative of %the $n$-th Langevin constraint \eqref{single_particle_Langevin} with %respect to its own density field $\rho_n$:
%\begin{equation}
%    \mathcal{M}_{n}(x,t; y,t') \equiv \frac{\delta \mathcal{Q}_{n}(x,t)}%{\delta \rho_n(y,t')} 
%\end{equation}
To proceed, we need to evaluate the Jacobian (\ref{Jacobian_MSRJD}).
By varying the constraint $\mathcal{Q}_n = \partial_t \rho_n - \mathcal{L}^\phi[\rho_n] + \sqrt{\Omega} \nabla_i (\rho_n e^i_a \eta^a_n)$, we obtain the explicit kernel:
\begin{equation}
    \mathcal{M}_{n} = \left[ \delta_g(x,y) \partial_t - \frac{\delta \mathcal{L}^\phi[\rho_n(x,t)]}{\delta \rho_n(y,t)} + \sqrt{\Omega} \nabla_i^{(x)} \left( \delta_g(x,y) e_a^i(x) \eta_n^a(x,t) \right) \right] \delta(t-t')
\end{equation}
where $\delta_g(x,y) = \delta^{(2)}(x-y)/\sqrt{g(x)}$ is the covariant delta function. Given that the operator $\mathcal{L}^\phi[\rho_n]$ is linear in $\rho_n$ for a fixed background of other particles, its functional derivative acting on a a field $\psi_n$ amounts to the same 
expression with the argument $\psi_n$: 
%We define the linearized kernel acting on a field $\psi_n$ as:
\begin{equation}
    (\hat{\mathcal{K}}_n \psi_n)(x) \equiv \int d\mu_g(y) \frac{\delta \mathcal{L}^\phi[\rho_n(x)]}{\delta \rho_n(y)} \psi_n(y) = \mathcal{L}^\phi[\psi_n(x)] 
\end{equation}
To represent $\det[\mathcal{M}_n]$ via a fermion path integral, we introduce anticommuting ghost fields $\psi_n(x,t)$ and $\bar{\psi}_n(x,t)$, and use the Berezin representation of a functional determinant $\det[\mathcal{M}_n]$:
\[
\det[\mathcal{M}_n] = \int D [\psi_n] D [\bar{\psi}_n] e^{-\int dt d \mu_g(x) \bar{\psi}_n \mathcal{M}_n \psi_n} \equiv 
\int D [\psi_n] D [\bar{\psi}_n] e^{-S_{gh}^{(n)}( \psi_n,\bar{\psi}_n)}
\]
% The ghost action for particle $n$ is given by $S_{gh}^{(n)} = \int %\bar{\psi}_n \mathcal{M}_n \psi_n$. 
 Integrating the noise term and the drift term by parts to move the covariant divergence $\nabla_i$ onto the adjoint ghost $\bar{\psi}_n$, we obtain:
\begin{equation}
\label{S_ghost}
    S_{gh}^{(n)} = \int dt \int d\mu_g(x) \left[ \bar{\psi}_n \partial_t \psi_n - \bar{\psi}_n \mathcal{L}^\phi[\psi_n] - \sqrt{\Omega} (\nabla_i \bar{\psi}_n) e_a^i \eta_n^a \psi_n \right]
\end{equation}
Combining the last term in this expression with the last term in the exponent in Eq.(\ref{delta_function_via_response_field}), we integrate our the noise term to obtain the following noise-induced term in the total MSRJD action for particle $ n $:
%Defining the single-particle ghost current $g_{n,i}^{gh} \equiv (\nabla_i %\bar{\psi}_n) \psi_n$, we combine $S_{gh}^{(n)}$ with the response field %terms and integrate out the Gaussian noise $\eta_n^a$. The resulting %fluctuating part of the MSRJD action for particle $n$ is:
\begin{equation}
\label{S_noise}
S_{noise}^{(n)} = - \frac{\Omega}{2} \int dt \int d\mu_g(x) \, g^{ij}(x) G_{i}^{(n)}(x,t)  G_{j}^{(n)}(x,t)   
\end{equation}
where we defined the composite boson-fermion current 
\begin{equation}
\label{G_current}
G_{i}^{(n)}(x,t) 
%= \left. G_{i}^{(n)}(\rho_n, \hat{\rho}_n, \psi_n, \bar{\psi}_n) \right|_{x,t} 
\equiv  \left. \rho_n \nabla_i \hat{\rho}_n + (\nabla_i \bar{\psi}_n) \psi_n \right|_{x,t}
\end{equation}
%The ghost fields thus effectively track the linearized dynamics of the %density fluctuations, ensuring the normalization $Z=1$ for the %Stratonovich generating functional.
Combining the remaining terms from Eqs.(\ref{delta_function_via_response_field}) and (\ref{S_ghost}), we obtain the MSRJD action for particle $ n $ for a fixed disorder in our model:
\begin{equation}
\label{S_MSRJD_fixed_disorder}
S_n^{(\phi)} \equiv \int dt \int d\mu_g(x) \left[ 
\hat{\rho}_n \left(\frac{\partial \rho_n}{\partial t} - \mathcal{L}[\rho_n] \right) 
+ \bar{\psi}_n \left( \partial_t \psi_n - \mathcal{L}^\phi[\psi_n] 
\right)  - \frac{\Omega}{2} g^{ij} G_{i}^{(n)} G_{j}^{(n)} \right]  
\end{equation}

\section{Disorder Averaging of the MSRJD Action}
\label{app:disorder_averaging_details}

This appendix provides the full derivation of the disorder-averaged
action (\ref{S_MSRJD_fixed_disorder}). 

\paragraph{Gaussian integration over disorder.}
The disorder enters the MSRJD action through the interaction term
\eqref{eq:S_int_phi}, which is linear in the couplings $\phi_{nm}$.
Averaging $\exp(-S_{\text{int}}^{(\phi)})$ over the Gaussian
distribution of $\phi_{nm}$ with zero mean and covariance
$\E[\phi_{nm}\phi_{lp}] = (J^2/N)(\delta_{nl}\delta_{mp}
+ \delta_{np}\delta_{ml})$ produces
\begin{equation}
    \E_\phi\!\left[ e^{-S_{\text{int}}^{(\phi)}} \right]
    = \exp\!\left( \frac{1}{2}
    \E_\phi\!\left[\left(S_{\text{int}}^{(\phi)}\right)^2\right]
    \right)
    \label{eq:gaussian_avg_app}
\end{equation}
since all cumulants beyond the second vanish for Gaussian
disorder, and the first cumulant vanishes by
$\E[\phi_{nm}] = 0$.

We collect the terms in the action \eqref{S_MSRJD_fixed_disorder} depending on the interaction drift $\mathcal{L}^{\phi}$. Integrating the covariant divergence by parts, we identify the interaction action for a fixed disorder realization as:
\begin{equation}
    S_{\text{int}}^{(\phi)} = \frac{1}{\gamma} \sum_{n,m} \phi_{nm} \int dt d\mu_g(x)  d\mu_g(y) \, G_i^{(n)}(x,t) g^{ij}(x) \nabla_j^{(x)} [d_g(x,y)] \rho_m(y,t) = 
\frac{1}{\gamma} \sum_{n,m} \phi_{nm} \int dz  
X_n(z)  Y_m(z)   
\end{equation}
%where $G_i^{(n)}(x,t)$ is the composite current defined in Eq.(\ref{G_current}),  
%and we defined
%\begin{eqnarray}
%&& \overline{d_m(x,t)} \equiv \int dy d_g(x,y) \rho_m(y,t) \nonumber \\
%&& C^{i}_m (x,t) \equiv g^{ij}(x) \nabla_j^{(x)} \overline{d_m(x,t)},
%\end{eqnarray}
%is the average geodesic distance from the $ n $ particle at point $ x $ to the $ m$-th particle.

Here in the last equation we defined the short notation $z := (x,y,t)$ with
$ dz = dt d \mu_g(x) d \mu_g(y) $ and  $X_n(z) \equiv G_i^{(n)}(x,t) g^{ij}(x) \nabla_j^{(x)} [d_g(x,y)]$ and $Y_n(z) \equiv \rho_n(y,t)$. The interaction is $S_{\text{int}}^{(\phi)} = \frac{1}{\gamma} \sum_{n,m} \phi_{nm} I_{nm}$ with $I_{nm} = \int dz X_n(z) Y_m(z)$. Evaluating the average in \eqref{eq:gaussian_avg_app}, we find:
\begin{equation}
\label{squared_terms}
    \frac{1}{2} \E_\phi \left[ \left( S_{\text{int}}^{(\phi)} \right)^2 \right] = \frac{\alpha^2}{N} \sum_{n,m} \left( I_{nm}^2 + I_{nm} I_{mn} \right) 
    = \alpha^2  N \frac{1}{N^2}\sum_{n,m} \left(A_n^{(1)} B_m^{(1)} + A_n^{(2)} B_m^{(2)} \right)
\end{equation}
where we have defined the constant $\alpha^2 = \frac{J^2}{2 \gamma^2}$ and two sets of bi-local operators $A_n^{(1,2)}$ and $B_n^{(1,2)}$:
\begin{align}
    A_n^{(1)} = X_n(z) X_n(z'), \quad B_n^{(1)} = Y_n(z) Y_n(z'), \\
    A_n^{(2)} = X_n(z) Y_n(z'), \quad B_n^{(2)} = Y_n(z) X_n(z').
\end{align}
Applying the identity 
\[
\frac{1}{N^2}\sum_{n,m} A_n B_m = \frac{1}{4} \left( \frac{1}{N} \sum_n (A_n + B_n) \right)^2 - \frac{1}{4} \left( \frac{1}{N} \sum_n (A_n - B_n) \right)^2
\]
to both terms, the exponent takes the form of four squared single sums.

%\paragraph{Hubbard-Stratonovich Linearization.}

To linearize these terms, we introduce four bi-local Hubbard-Stratonovich (HS) fields $\hat{Q}_a(z,z')$ ($a=1\dots 4$). Two HS transformation needed in our case take the following forms (here $f(x,t) $ is an arbitrary function):
\begin{eqnarray}
\label{HS_transforms}
&& e^{ \alpha^2 N/4 \int dt d \mu_g(x) f^2(x,t)} = \int D \left[ \hat{Q} \right]
e^{ \int dt d \mu_g(x) \left( - \frac{1}{4} \hat{Q}(x,t)^2 + 
\frac{\alpha}{2}  \sqrt{N} \hat{Q}(x,t) f(x,t) \right) } \nonumber \\
&& e^{ - \alpha^2 N/4 \int dt d \mu_g(x) f^2(x,t)} = \int D \left[ \hat{Q} \right]
e^{ \int dt d \mu_g(x) \left( - \frac{1}{4} \hat{Q}(x,t)^2 + i 
\frac{\alpha}{2}  \sqrt{N} \hat{Q}(x,t) f(x,t) \right) }   
\end{eqnarray}
Using this relation to linearize all four squared terms in Eq.(\ref{squared_terms}), the generating functional for the disorder part becomes:
\begin{eqnarray}
\label{gen_functional_02}
e^{-\bar{S}_{dis}} \hspace{-0.2cm} &=& \hspace{-0.2cm}  \int \prod_{a=1}^4 \mathcal{D}\hat{Q}_a \exp \left[ -\frac{1}{4} \int dz dz' \sum_a \hat{Q}_a^2 + \int dz dz' \left[ 
 \hat{Q}_1 \frac{\alpha}{2 \sqrt{N}} 
\sum_n (A_n^{(1)} + B_n^{(1)})  
\right. \right. \\
 \hspace{-0.2cm} &+&\hspace{-0.2cm} \left. \left.  
i \hat{Q}_2 \frac{\alpha}{2 \sqrt{N}} \sum_n (A_n^{(1)} - B_n^{(1)}) 
+ \hat{Q}_3 \frac{\alpha}{2 \sqrt{N}} \sum_n (A_n^{(2)} + B_n^{(2)}) 
+ i\hat{Q}_4 \frac{\alpha}{2 \sqrt{N}} \sum_n (A_n^{(2)} - B_n^{(2)}) 
\right] \right] \nonumber 
\end{eqnarray}
%In the large-$N$ limit, we evaluate the functional integral via the %saddle-point method. 
%\begin{equation}
%    Q_a = \bar{Q}_a + O(1/N), \quad \langle \dots \rangle = \langle \dots %\rangle_{MF} + O(1/\sqrt{N})
%\end{equation}
%where the $O(1/N)$ terms represent the Gaussian fluctuations of the HS %fields, and $O(1/\sqrt{N})$ represents the finite-size corrections to the %field expectations.
In the large-N limit, all fields $ \hat{Q}_i $ can be replaced by their vacuum expectations. This produces the following set of relations
\begin{eqnarray}
\label{vacuum_averages}
&&  \hat{Q}_1 = \alpha \sqrt{N} \frac{1}{N} \sum_n \langle A_n^{(1)} + B_n^{(1)} \rangle, \; \; \; 
\hat{Q}_2 = i \alpha \sqrt{N} \frac{1}{N} \sum_n \langle A_n^{(1)} - B_n^{(1)} \rangle,  \nonumber \\
&&  \hat{Q}_3 = \alpha \sqrt{N} \frac{1}{N} \sum_n \langle A_n^{(2)} + B_n^{(2)} \rangle, \; \; \; 
 \hat{Q}_4 = i \alpha \sqrt{N} \frac{1}{N} \sum_n \langle A_n^{(2)} - B_n^{(2)} \rangle
\end{eqnarray}
%The saddle values for the integration variables are $\hat{Q}_1 = \alpha N Q_1$, $\hat{Q}_2 = i \alpha N Q_2$, $\hat{Q}_3 = \alpha N Q_3$, and $\hat{Q}_4 = i \alpha N Q_4$, where the mesoscopic order parameters $Q_a$ are defined as:
%\begin{equation}
%    Q_1 = \frac{1}{N} \sum_n \langle A_n^{(1)} + B_n^{(1)} \rangle, \quad Q_2 = \frac{1}{N} \sum_n \langle A_n^{(1)} - B_n^{(1)} \rangle, \quad Q_3 = \frac{1}{N} \sum_n \langle A_n^{(2)} + B_n^{(2)} \rangle, \quad Q_4 = \frac{1}{N} \sum_n \langle A_n^{(2)} - B_n^{(2)} \rangle.
%\end{equation}
Substituting these solutions back, re-arranging and skipping constant terms, we obtain:
%\begin{equation}
%    e^{-\bar{S}_{dis}} \sim \exp \left[ -\frac{J^2}{4 N \gamma^2} \int dz dz' \sum_n \left\{ A_n^{(1)}(Q_2 - Q_1) - B_n^{(1)}(Q_2 + Q_1) + A_n^{(2)}(Q_4 - Q_3) - B_n^{(2)}(Q_3 + Q_4) \right\} \right].
%\end{equation}
%Using the relations $Q_2 - Q_1 = -2 \langle B^{(1)} \rangle$, $Q_2 + Q_1 = 2 \langle A^{(1)} \rangle$, $Q_4 - Q_3 = -2 \langle B^{(2)} \rangle$, and $Q_3 + Q_4 = 2 \langle A^{(2)} \rangle$, where $\langle \dots \rangle$ denotes the average over all particles, we arrive at the decoupled effective action:
\begin{equation}
    e^{-\bar{S}_{dis}} = \exp \left[ \alpha^2 \int dz dz' \sum_n \left( \langle B^{(1)} \rangle A_n^{(1)} + \langle A^{(1)} \rangle B_n^{(1)} + \langle B^{(2)} \rangle A_n^{(2)} + \langle A^{(2)} \rangle B_n^{(2)} \right) \right].
\end{equation}
%This completes the dynamic mean-field result: in the large $N$-limit, the resulting effective actions for all particles become identical and describe a single representative process where the quenched disorder manifests as a self-consistent memory kernel and potential mediated by the manifold's geometry.

To provide a clear representation of the factorized dynamics, we explicitly express the bi-local single-particle operators at coordinates $(z, z')$ in terms of the current $G_i^{(n)}$, the density $\rho_n$, and the force kernel $V^i(x,y) \equiv g^{ij}(x) \nabla_j^{(x)} d_g(x,y)$:
\begin{align}
    A_n^{(1)}(z,z') &= \left[ G_i^{(n)}(x,t) V^i(x,y) \right] \left[ G_j^{(n)}(x',t') V^j(x',y') \right], \\
    B_n^{(1)}(z,z') &= \rho_n(y,t) \rho_n(y',t'), \\
    A_n^{(2)}(z,z') &= \left[ G_i^{(n)}(x,t) V^i(x,y) \right] \rho_n(y',t'), \\
    B_n^{(2)}(z,z') &= \rho_n(y,t) \left[ G_j^{(n)}(x',t') V^j(x',y')  
    \right] = A_n^{(2)}(z',z).
\end{align}
The coefficients $\langle A^{(1,2)} \rangle$ and $\langle B^{(1,2)} \rangle$ appearing in the decoupled action are the mesoscopic order parameters determined by the self-consistent vacuum expectation values:
\begin{align}
\label{VEVs_for_AB}
    \langle A^{(1)}(z,z') \rangle &= \frac{1}{N} \sum_{k=1}^N \left\langle A_k^{(1)}(z,z') \right\rangle, \quad \langle B^{(1)}(z,z') \rangle = \frac{1}{N} \sum_{k=1}^N \left\langle B_k^{(1)}(z,z') \right\rangle, \\
    \langle A^{(2)}(z,z') \rangle &= \frac{1}{N} \sum_{k=1}^N \left\langle A_k^{(2)}(z,z') \right\rangle, \quad \langle B^{(2)}(z,z') \rangle = \frac{1}{N} \sum_{k=1}^N \left\langle B_k^{(2)}(z,z') \right\rangle \nonumber 
\end{align}
In the limit $N \to \infty$, these averages are identical for all $n$ and depend only on the spacetime-manifold coordinates $(z,z')$. Substituting these into the generating functional, the full MSRJD path integral factorizes into a product of $N$ identical terms:
\begin{equation}
\label{Z_after_averaging}
    Z[J,\bar{J}] = \int \prod_{n=1}^N \mathcal{D}[\rho_n, \hat{\rho}_n, \psi_n, \bar{\psi}_n] e^{ - \sum_n S_n^{\text{eff}}[\rho_n, \hat{\rho}_n, \psi_n, \bar{\psi}_n] + \langle J_n \rho_n \rangle 
    + \langle \bar{J}^i  G_i^{(n)} \rangle}
\end{equation}
where the effective single-particle action is given by:
\begin{eqnarray}
\label{eq:S_n_eff}
S_n^{\text{eff}} \hspace{-0.2cm} 
&=& \hspace{-0.2cm} 
= \int dt d \mu_g(x) \left[ \hat{\rho}_n \partial_t \rho_n + \bar{\psi}_n \partial_t \psi_n - \frac{\Omega}{2} g^{ij} G_i^{(n)} G_j^{(n)} \right] 
\nonumber \\
\hspace{-0.2cm} &-& \hspace{-0.2cm} 
\alpha^2  \iint dz dz' \left( \langle B^{(1)} \rangle A_n^{(1)} + \langle A^{(1)} \rangle B_n^{(1)} 
+ \langle B^{(2)} \rangle A_n^{(2)} + \langle A^{(2)} \rangle B_n^{(2)} 
\right)
\end{eqnarray}
Source fields $ J(x,t), \bar{J}^i(x,t) $ are added in the generating functional (\ref{Z_after_averaging}) in order to compute correlators 
(\ref{VEVs_for_AB}).

Eqs.(\ref{VEVs_for_AB}) and (\ref{eq:S_n_eff}) show that
after averaging over quenched disorder, in the large-$N$ limit, each particle satisfies the same effective dynamics. The quenched disorder has been transformed into a self-interaction and non-Markovian dynamics with memory.

\paragraph{From single-particle densities to the mean density.}
The factorized generating functional (\ref{Z_after_averaging}) implies that
the single-particle densities $\rho_1, \ldots, \rho_N$ become
independent and identically distributed random fields in the
large-$N$ limit. 
%This is the phenomenon of \emph{propagation of
%chaos} \cite{mezard1987, castellani2005}: the correlations
%between distinct particles, originally mediated by the quenched
%couplings $\phi_{nm}$, are replaced by a self-consistent
%mean-field coupling that renders the particles statistically
%independent.
As a consequence, the empirical (mean) particle density
$\rho_N(x,t) = (1/N)\sum_{n=1}^N \rho_n(x,t)$
(Eq.~\ref{eq:empirical_density_dk}) converges to the expectation
$\langle \rho(x,t) \rangle$ of a single representative particle
as $N \to \infty$, by the law of large numbers. At the
saddle-point level, this expectation is the stationary solution
$\rho_*(x)$. The self-consistent correlators $C_\rho$, $K_{ij}$,
$R_i$ defined in Eqs.(\ref{eq:two_point_functions}) are the
two-point functions of the representative particle, and they
coincide with the two-point functions of $\rho_N$ up to $O(1/N)$
corrections. The self-consistent PDE (\ref{PDE_rho_n}) with the
effective drift $\hat{G}^i[G,\rho]$ therefore plays the role of the
disorder-averaged Dean-Kawasaki equation for the mean particle
density.

It is worth emphasizing the status of this construction at finite
$N$. The original Dean-Kawasaki equation
\cite{dean1996, illien2025} is an exact identity at every finite
$N$: it is a tautological rewriting of $N$ coupled Langevin
equations as a single stochastic PDE for the empirical density
$\rho_N = (1/N)\sum_n \delta(x - x_n(t))$. No approximation is
involved, but the empirical density is a sum of Dirac measures,
making expressions like $\sqrt{\rho}$ purely formal. The MSRJD
path-integral reformulation introduces no additional approximation
\cite{velenich2008, kim2014}; it recasts the multiplicative noise
as a vertex in the action. The large-$N$ limit enters only when
evaluating the path integral by saddle-point methods. At finite
$N$, the noise action $S_{\text{noise}} = O(1)$ provides the
leading correction to the $O(N)$ deterministic and disorder
terms. A systematic $1/N$ expansion can be organized around the
saddle point, with Gaussian (one-loop) fluctuations at order
$O(1/N)$, and higher-loop corrections at successively higher
orders in $1/N$. 
%Each order captures finer aspects of the
%discrete-particle structure. In particular, the density statistics
%are intrinsically non-Gaussian at any finite $N$
%\cite{velenich2008}: even for non-interacting Brownian particles,
%the equal-time correlations are Poissonian, a non-perturbative
%feature of the singular empirical measure.

In its current form (\ref{Z_after_averaging}), 
(\ref{eq:S_n_eff}), the generating functional $ Z[J,\bar{J}] $ may be inconvenient to work with, as it involves the composite current field $ G_i^{(n)} $ and correlation function involving it. We now want to simplify it by promoting the composite field $ G_i^{(n)} $ to an {\it independent} field and integrating out the response field $ \hat{\rho}_n $.

To this end, we fist re-write the effective action  (\ref{eq:S_n_eff})
in the following equivalent form:
\begin{eqnarray}
\label{eq:S_n_eff_2}
S_n^{\text{eff}} 
\hspace{-0.2cm} &=& \hspace{-0.2cm} \int dt dx \left[ \hat{\rho}_n \partial_t \rho_n + \bar{\psi}_n \partial_t \psi_n - \frac{\Omega}{2} g^{ij} G_i^{(n)} G_j^{(n)}  
\right]   \\
\hspace{-0.2cm} &-& \hspace{-0.2cm} \alpha^2 
\iint dx dx' dt dt' 
\left( 
G_{i}^{(n)}(x,t) \hat{\mathcal{C}}_{n}^{ij}(x,x',t,t') G_{j}^{(n)}(x',t')
+ \rho_{n}(x,t) \hat{\mathcal{K}}_{n}(x,x',t,t') \rho_n(x',t')  
\right. \nonumber \\
\hspace{-0.2cm} &+& \hspace{-0.2cm} \left. 
2 G_i^{(n)}(x,t) \hat{\mathcal{R}}^i (x,t,x',t') \rho_n(x',t')
\right) 
\nonumber 
\end{eqnarray}
where we defined the following set of two-point functions (here
 $V^i(x,y) \equiv g^{ij}(x) \nabla_j^{(x)} d_g(x,y) $:
\begin{eqnarray}
\label{eq:two_point_functions}
&& C_{\rho}(x,t,x',t') \equiv \frac{1}{N} \sum_{n} \langle \rho_n(x,t) \rho_n(x',t')
\rangle \nonumber \\
&& K_{ij}(x,t,x',t') \equiv \frac{1}{N} \sum_{n} \langle G_i^{(n)}(x,t) G_j^{(n)}(x',t')
\rangle \nonumber \\
&& R_i(x,t,x',t') \equiv \frac{1}{N} \sum_{n} \langle \rho_{n}(x,t) G_i^{(n)}(x',t') \rangle \nonumber \\
&& \hat{\mathcal{C}}_n^{ij}(x,t,x',t') \equiv
\int dy dy' V^i (x,y)  C_{\rho}(y,t,y',t') V^j (x',y') \nonumber \\
&& \hat{\mathcal{R}}^{i}(x,t,x',t') \equiv  \int dy dy'  V^i(x,y') R_j(y',t,y,t') V^j(y,x')   \nonumber \\
&& \hat{\mathcal{K}}_{n}(x,t,x',t') \equiv \int dy dy' V^i(y,x)
 K_{ij}(y,t,y',t') V^j(y',x')
\end{eqnarray} 
Next we introduce the following identity
\begin{equation}
\label{identity_for_G}
    1 = \int \mathcal{D}[G_i^{(n)}] \mathcal{D}[\hat{G}_n^i] e^{- \int dt d\mu_g(x) \, \hat{G}_{n}^i(x,t) \left[ G_i^{(n)}(x,t) - \left( \rho_n \nabla_i \hat{\rho}_n + (\nabla_i \bar{\psi}_n) \psi_n \right) \right] }
\end{equation}
where $\hat{G}_n^i$ is an auxiliary response current field. Inserting this identity into Eq,(\ref{eq:S_n_eff_2}), we  
write the generating functional in the following form:
\begin{eqnarray}
\label{eq:gen_functional_0}
Z[J, \bar{J} ] = \int \prod_{n=1}^N \mathcal{D}[\rho_n, \hat{\rho}_n, \psi_n, \bar{\psi}_n,  G_i, \hat{G}^i] ] 
e^{- \sum_n S(\rho_n, \hat{\rho}_n, \psi_n, \bar{\psi}_n, G_i^{(n)}, \hat{G}_n^i) + \langle \rho_n J_n  \rangle 
+ \langle G_i^{(n)} \bar{J_n}^i  \rangle }
\end{eqnarray} 
where 
\begin{eqnarray}
\label{eq:action_0} 
S(\rho, \hat{\rho}, \psi, \bar{\psi}, G_i, \hat{G}^i) 
\hspace{-0.2cm} &=& \hspace{-0.2cm} \int dt dx \left[ \hat{\rho}
\left( \partial_t  
+ \nabla_i\hat{G}^i \right) \rho  
+ \bar{\psi} \left( \partial_t  + \nabla_i \hat{G}^i \right) \psi  
+ \hat{G}^i G_{i} \right]  \nonumber \\
\hspace{-0.2cm} &-& \hspace{-0.2cm}     
 \iint dx dx' dt dt' 
\left( 
\frac{1}{2} G_{i}(x,t) \hat{\mathcal{A}}^{ij}(x,t,x',t') G_{j}(x',t') \right.  
\\
\hspace{-0.2cm} &+& \left. \hspace{-0.2cm} \alpha^2 \rho(x,t) \hat{\mathcal{K}}(x,t,x',t') \rho(x',t')  
+ 2 \alpha^2 G_i(x,t) \hat{\mathcal{R}}^i (x,t,x',t') \rho(x',t')
\right) \nonumber 
\end{eqnarray}
where we defined the kernel $ \hat{\mathcal{A}}^{ij}(x,t',x',t') $ as follows:
\begin{equation}
\label{tilde_mathcal_C}
\hat{\mathcal{A}}^{ij}(x,t,x',t') \equiv  \Omega g^{ij}(x) \delta(x-x') \delta(t-t') +  2 \alpha^2  \hat{\mathcal{C}}^{ij}(x,t,x',t')
\end{equation}

Varying the action with respect to the field $ G_i $ produces the following relation:
\begin{equation}
\label{hat_G_from_G}
\hat{G}^{i}(x,t) 
= \int dx' dt' \left[ \hat{\mathcal{A}}^{ij}(x,t,x',t') G_j(x',t')
+ 2 \alpha^2 \hat{\mathcal{R}}^i (x,t,x',t') \rho_n(x',t') \right]
\end{equation}
%This equation can be inverted:
%\begin{equation}
%G_i(x,t) = \int dx' dt' \mathcal{W}_{ij}(x,t,x',t') 
%\left[ \hat{G}^j(x',t') - 2 \alpha^2 \int dx'' dt'' \hat{\mathcal{R}}^i (x',t',x'',t'') \rho_n(x'',t'') \right]
%\end{equation}
%where $ \mathcal{W}_{ij} $ is the inverse kernel defined by the following equation:
%\begin{equation}
%\label{inverse_kernel_W}
%\int dz'' \hat{\mathcal{A}}^{ik}(z,z'') W_{kj}(z'',z') = \delta_j^i \delta(z-z')
%\end{equation}
Plugging this back to Eqs.(\ref{eq:gen_functional_0}), (\ref{eq:action_0}), we obtain:
\begin{eqnarray}
\label{eq:gen_functional_1}
Z[J, \bar{J} ] = \int \prod_{n=1}^N \mathcal{D}[\rho_n, \hat{\rho}_n, \psi_n, \bar{\psi}_n,  G_i^{(n)}] ] 
e^{- \sum_n S(\rho_n, \hat{\rho}_n, \psi_n, \bar{\psi}_n, G_i^{(n)}) + \langle \rho_n J_n  \rangle 
+ \langle G_i^{(n)} \bar{J_n}^i  \rangle }
\end{eqnarray} 
where 
\begin{eqnarray}
\label{eq:action_1} 
S(\rho, \hat{\rho}, \psi, \bar{\psi}, G_i) 
\hspace{-0.2cm} &=& \hspace{-0.2cm} \int dt dx \left[ \hat{\rho}
\left( \partial_t  
+ \nabla_i\hat{G}^i[G, \rho] \right) \rho  
+ \bar{\psi} \left( \partial_t  + \nabla_i \hat{G}^i[G, \rho] \right) \psi  
 \right]  \nonumber \\
\hspace{-0.2cm} &+& \hspace{-0.2cm}     
 \iint dx dx' dt dt' 
\left( 
\frac{1}{2} G_{i}(x,t) \hat{\mathcal{A}}^{ij}(x,t,x',t') G_{j}(x',t') \right.  
\\
\hspace{-0.2cm} &-& \left. \hspace{-0.2cm} \alpha^2 \rho(x,t) \hat{\mathcal{K}}(x,t,x',t') \rho(x',t')  
%+ 2 \alpha^2 G_i^{(n)}(x,t) \hat{\mathcal{R}}^i (x,t,x',t') \rho_n(x',t')
\right) \nonumber 
\end{eqnarray}
where $ \hat{G}^i[G, \rho] $ is defined by Eq.(\ref{hat_G_from_G}).

Now we can formally integrate out the response field and the ghost fields to obtain
\begin{eqnarray}
\label{integrate_psi_n}
&& \int D [\hat{\rho}_n ] D [ \psi_n] D[ \bar{\psi}_n] 
e^{ - \int dx dt \left[ 
\hat{\rho}_n
\left( \partial_t  
+ \nabla_i\hat{G}_n^i[G, \rho] \right) \rho_n  
+ \bar{\psi}_n \left( \partial_t  + \nabla_i \hat{G}_n^i[G, \rho] \right) \psi_n \right] } = \nonumber \\
\hspace{-0.2cm} & & \hspace{-0.2cm}   
\delta \left( \partial_t \rho_n + \nabla_i( \hat{G}_n^{i}[G, \rho] \rho_n) \right) \text{det} \left| \partial_t + 
\nabla_i \hat{G}_n^i[G, \rho] \right| 
%\nonumber \\ 
%\hspace{-0.2cm} & \equiv & \hspace{-0.2cm}  
\equiv \delta \left( \rho_n - \rho_n^{G} \right)  
\end{eqnarray}
The last expression is a symbolic expression enforcing the condition that integration with respect to $ \rho $ in (\ref{eq:gen_functional_0}) is performed over fields $ \rho^{G} $ that satisfy the PDE
\begin{equation}
\label{PDE_rho_n}
\frac{\partial \rho}{\partial t}    
+ \nabla_i \left[ \hat{G}^i[G, \rho] \rho \right] = 0
\end{equation}
%where $ \hat{G}_n^i $ is defined in terms of the field $ G_i^{n} $
%by Eq.(\ref{hat_G_from_G}).
 
Eq.(\ref{PDE_rho_n}) is analogous to the single-particle SPDE (\ref{single_particle_SPDE}) which we repeat here for convenience:
\begin{equation}
\label{single_particle_SPDE_2}
\frac{\partial \rho_n}{\partial t} = - \nabla_i \left[ 
\left(  f_n^i + \sqrt{\Omega}\,e^i_a \eta^a \right) \rho_n \right]
\end{equation}  
Comparing Eqs.(\ref{PDE_rho_n}) and (\ref{single_particle_SPDE_2}), 
we find that while the effective mean particle density PDE 
retains the gradient form of the single-particle SPDE (and thus automatically conserves the particle number, as needed),
now the field $ \hat{G}^i $ plays the role of the stochastic 
force $  f_n^i + \sqrt{\Omega}\,e^i_a \eta^a $ that arises in the single-particle SPDE (\ref{single_particle_SPDE_2}).

Using (\ref{integrate_psi_n}), we finally write  
the generating function (\ref{eq:gen_functional_1}) as follows:
\begin{eqnarray}
\label{eq:gen_functional_2}
Z[J, \bar{J} ] = \prod_{n} \int D [\rho_n] D[ G_i^{(n)}]
\delta \left( \rho_n - \rho_n^{G} \right) 
e^{-\sum_n S_n(\rho_n, G_i^{(n)}) + \langle \rho_n J_n  \rangle 
+ \langle G_i^{(n)} \bar{J}_n^i  \rangle }
\end{eqnarray} 
%\begin{eqnarray}
%\label{eq:gen_functional_1}
%Z[J, \bar{J} ] = \int D [\rho] D[\hat{G}^i]
%\delta \left( \rho - \rho_n^{\hat{G}} \right) 
%e^{-N S(\rho, \hat{G}^i, \bar{J}) + \langle \rho J  \rangle 
%}
%\end{eqnarray} 
where 
%\begin{eqnarray}
%\label{eq:action_1} 
%S(\rho, G_i)
%\hspace{-0.2cm} &= & \hspace{-0.2cm}  \iint dx dx' dt dt' 
%\left[ 
%\frac{1}{2} \left( \bar{J}^i_{x,t} - \hat{G}^{i}_{x,t} 
%- 2 \alpha^2 \hat{\mathcal{R}}^i \rho_{x,t})  \right) 
%\mathcal{W}_{ij} \left( \bar{J}^j_{x',t'} - \hat{G}^{j}_{x',t'} 
%- 2 \alpha^2 \hat{\mathcal{R}}^j \rho_{x',t'}  \right) \right. \nonumber \\
%\hspace{-0.2cm} &- & \left. \hspace{-0.2cm} 
%\alpha^2 \rho_{x,t} \hat{\mathcal{K}} \rho_{x',t'}  \right] 
\begin{equation}
\label{eq:action_2} 
S(\rho,  G_i) 
= \iint dx dx' dt dt' 
\left( 
\frac{1}{2} G_{i}(x,t) \hat{\mathcal{A}}^{ij}(x,t,x',t') G_{j}(x',t')   
- \alpha^2 \rho(x,t) \hat{\mathcal{K}}(x,t,x',t') \rho(x',t')  
\right)  
\end{equation}
%where we used the short notations $ \hat{\mathcal{R}}^i \rho_{x,t} \equiv \int dx'' dt'' \hat{\mathcal{R}}^j (x,t,x'',t'') \rho_n(x'',t'') $ and  $ \rho_{x,t}
%\hat{\mathcal{K}} \rho_{x',t'} \equiv \rho(x,t) 
%\hat{\mathcal{K}}(x,x',t,t') \rho(x',t') $, and similarly for other terms, and $ \mathcal{W}_{ij} $ is the inverse kernel defined in Eq.(\ref{inverse_kernel_W}).  

This representation shows that the current fields $ G_{i}$ acts as a noise variable whose correlation is determined by the kernel $ \hat{\mathcal{A}}_{n}^{ij} $.
This kernel incorporates both the bare thermal fluctuations and the self-consistent current-current correlations induced by the quenched disorder on the manifold. Furthermore, the second term in \eqref{eq:action_1} is a non-local quadratic self-interaction term that drives correlation functions of the density field $ \rho_n $.

\paragraph{Consistency conditions.}
Using the generating functional (\ref{eq:gen_functional_2}) we can compute the correlator of current $ G_i $:
\begin{equation}
\langle G_i(x,t) G_j(x',t') \rangle = \frac{\delta^2  Z[J, \bar{J} ] }{\delta \bar{J}^i(x,t') 
\delta \bar{J}^j(x',t')} =  \mathcal{W}_{ij}(x,t',x',t')
\end{equation}
where $ \mathcal{W}_{ij} $ is the inverse kernel of kernel $ \hat{\mathcal{A}}^{ij} $ defined by the following equation:
\begin{equation}
\label{inverse_kernel_W}
\int dz'' \hat{\mathcal{A}}^{ik}(z,z'') W_{kj}(z'',z') = \delta_j^i \delta(z-z')
\end{equation}
where $ z = (x,t) $, and similar for $ z', z'' $. Comparing this with second of Eqs.(\ref{eq:two_point_functions}), we obtain the constraint $ 
\mathcal{W}_{ij} = K_{ij} $. Combining this with 
Eq.(\ref{inverse_kernel_W}), we obtain the first constraint
\begin{equation}
\label{first_constraint}
\int dz'' \hat{\mathcal{A}}^{ik}(z,z'') K_{kj}(z'',z') = \delta_j^i \delta(z-z')
\end{equation}
Substituting here Eqs.(\ref{eq:two_point_functions}) and (\ref{tilde_mathcal_C}), we write this more explicitly as follows:
\begin{equation}
\label{first_constraint_1}
2 \alpha^2 \int dz'' dy dy' V^{i}(x,y) V^{k}(x'',y') C_{\rho}(y,t,y',t'') K_{kj}(z'',z') = \delta_{j}^{i} \delta(z - z') - 
\Omega g^{ik}(x) K_{kj}(z,z')
\end{equation}
Computing the two-point function of the density field, we find
\begin{equation}
\langle \rho(x,t) \rho(x',t') \rangle = \frac{\delta^2  Z[J, \bar{J} ] }{\delta J(x,t') 
\delta J(x',t')} =  \frac{1}{2 \alpha^2} \left[ \hat{\mathcal{K}} \right]^{-1} (x,t',x',t')
\end{equation}
Comparing this expression with the first of 
Eqs.(\ref{eq:two_point_functions}), we obtain $ \left[ C_{\rho} \right]^{-1} = 2 \alpha^2 \hat{\mathcal{K}} $. This produces the second constraint:
\begin{equation}
\label{second_constraint}
2 \alpha^2 \int dz'' C_{\rho}(z,z'') \hat{\mathcal{K}}(z'',z') = \delta(z-z')
\end{equation} 
In terms of the original correlators Eqs.(\ref{eq:two_point_functions}), this constraint can be written as follows:
\begin{equation}
\label{second_constraint_1}
2 \alpha^2 \int dz'' dy dy' V^{i}(x'',y') V^{j}(y,x') C_{\rho}(x,t,y',t'') K_{ij}(x'',t'',y,t') = \delta(z - z') 
\end{equation}
Other constraints on correlators (\ref{eq:two_point_functions}) are dynamic. Differentiating the first equation in  (\ref{eq:two_point_functions}) with respect to time, using (\ref{PDE_rho_n}) and simplifying, we obtain the following ODE:
\begin{eqnarray}
\label{ODE_for_C_rho}
\partial_t C_{\rho}(z,z') 
\hspace{-0.2cm} &= & \hspace{-0.2cm}    - \Omega g^{ij}(x) \nabla_i R_j(x',t',x,t) \\
\hspace{-0.2cm} &+& \hspace{-0.2cm}    2 \alpha^2 \int dz'' dy dy' \left( \nabla_i V^i(x,y) \right) \left\{ 
V^j(x',y') C_{\rho}(y,t,y',t') R_j(x',t',x'',t'') \right.
\nonumber \\
\hspace{-0.2cm} &+& \hspace{-0.2cm} \left.
V^j(x'',y') C_{\rho}(x',t,y',t'') R_j(y,t',x'',t'') \right\}
\nonumber 
\end{eqnarray} 
Similarly we derive another ODE for the correlator $ R_i $:
\begin{eqnarray}
\label{ODE_for_C_R_i}
\partial_t R_{i}(z,z') 
\hspace{-0.2cm} &= & \hspace{-0.2cm}    - \Omega g^{jl}(x) \nabla_l K_{ij}(x,t,x',t') \\
\hspace{-0.2cm} &+& \hspace{-0.2cm}    2 \alpha^2 \int dz'' dy dy' \left( \nabla_l V^l(x,y') \right) \left\{ 
V^j(x',y) C_{\rho}(y',t,y,t') K_{ji}(x',t',x'',t'') \right.
\nonumber \\
\hspace{-0.2cm} &+& \hspace{-0.2cm} \left.
V^j(y,x'') R_j(y',t,y,t'') R_{i}(x',t',x'',t'') \right\}
\nonumber 
\end{eqnarray}

\paragraph{Comparison with standard Schwinger-Dyson equations.}
The self-consistent
equations~\eqref{ODE_for_C_rho}--\eqref{ODE_for_C_R_i}
involve three families of two-point functions ($C_\rho$,
$K_{ij}$, $R_i$), richer than the standard $C$--$R$
structure of mean-field spin glasses
\cite{castellani2005, cugliandolo1993}, where only two
scalar functions appear. The additional correlators arise
from promoting the current $G_i$ to an independent field
via~\eqref{identity_for_G}; integrating $G_i$ back out
recovers the standard two-function structure. The mixed
correlator $R_i$ carries a vector index built from the
displacement between two points ($R_i \propto (x'_i -
x_i)\,h(|x-x'|,t,t')$ in flat space). In fully
connected models without spatial structure, $R_i$
vanishes (no displacement vector available) and the
system reduces to scalar $C$--$R$ equations. On the
manifold, $R_i$ captures the directional asymmetry of
density-current fluctuations between separated points.

%This shows that $ \tilde{G}^{i}(x,t) $ is a random field with $ \langle \tilde{G}^{i}(x,t) \rangle  = 0 $ and
%\begin{eqnarray}
%\label{corr_tilde_G}
%&& \hspace{-0.2cm}  \langle \tilde{G}^{i}(x,t)  \tilde{G}^{i}(x',t') \rangle 
%%\hspace{-0.2cm} &=& \hspace{-0.2cm} 
%=   \int dx'' dt'' dx''' dt''' \tilde{\mathcal{C}}^{ij}(x,t,x'',t'') 
%\tilde{\mathcal{C}}^{ij}(x',t',x''',t''') \langle G_j(x'',t'')
%G_j(x''',t''') \rangle \nonumber \\
%\hspace{-0.2cm} &=& \hspace{-0.2cm} 
%\int dx'' dt'' dx''' dt''' \tilde{\mathcal{C}}^{ij}(x,t,x'',t'') 
%\tilde{\mathcal{C}}^{ij}(x',t',x''',t''') \mathcal{K}_{ij}(x'',t'',x''',t''')  \nonumber \\
%\hspace{-0.2cm} &=& \hspace{-0.2cm} 
%\Omega^2 g^{il}(x) g^{jk}(x')  K_{lk}(x,t,x',t') +
%4 \alpha^2 \Omega \int dx'' dt'' K_{lm}(x,t, x'',t'') 
%\hat{\mathcal{C}}^{jm}(x',t',x'',t'') \nonumber \\
%%\hspace{-0.2cm} &+& \hspace{-0.2cm} 
%&&+
%4 \alpha^4 \int dx'' dt'' dx''' dt''' \hat{\mathcal{C}}^{il}(x,t,x'',t'') \hat{\mathcal{C}}^{im}(x',t',x''',t''') K_{jm}(x'',t'',x''',t''')  
%\end{eqnarray}

\subsection{Effective mean-field equation for the density}
\label{subsec:effective_rho_derivation}

This subsection derives the effective equation for the density
$\rho(x,t)$ from the constrained path integral. We use the
Lagrange multiplier method to enforce the PDE constraint,
derive the Euler-Lagrange equations, and obtain the
disorder-averaged Dean-Kawasaki equation.

%\paragraph{Lagrange multiplier formulation.}
The generating functional~\eqref{eq:gen_functional_2} involves
the action $S(\rho, G_i)$~\eqref{eq:action_2} subject to the
PDE constraint~\eqref{PDE_rho_n}. To enforce the constraint for
all $(x,t)$, introduce a Lagrange multiplier field $\lambda(x,t)$
and define the augmented action:
\begin{equation}
\label{eq:augmented_action}
S_\lambda[\rho, G_i, \lambda]
= S(\rho, G_i)
+ \int\!dx\,dt\;\lambda(x,t)\left[
\partial_t\rho + \nabla_i(\hat{G}^i[G,\rho]\,\rho)\right]
\end{equation}
where $\hat{G}^i[G,\rho]$ is the effective
drift~\eqref{hat_G_from_G}:
\begin{equation}
\label{eq:Ghat_repeated}
\hat{G}^i(x,t) = \int\!dx'\,dt'\left[
\hat{\mathcal{A}}^{ij}(x,t,x',t')\,G_j(x',t')
+ 2\alpha^2\hat{\mathcal{R}}^i(x,t,x',t')\,\rho(x',t')
\right]
\end{equation}
The generating functional becomes:
\begin{equation}
\label{eq:Z_lambda}
Z = \int\mathcal{D}\rho\,\mathcal{D}G_i\,\mathcal{D}\lambda\;
e^{-S_\lambda[\rho, G_i, \lambda]}
\end{equation}

%\paragraph{Euler-Lagrange equations.}
The saddle-point Euler-Lagrange equations are obtained by varying
$S_\lambda$ with respect to each field.
In particular,  $\delta S_\lambda/\delta\lambda(x,t) = 0$ gives the
PDE constraint:
\begin{equation}
\label{eq:EL_lambda}
\partial_t\rho + \nabla_i[\hat{G}^i[G,\rho]\,\rho] = 0
\end{equation}

%(ii) $\delta S_\lambda/\delta G_k(x'',t'') = 0$ gives:
%\begin{equation}
%\label{eq:EL_G}
%\int\!dx'\,dt'\;\hat{\mathcal{A}}^{kj}(x'',t'',x',t')\,
%G_j(x',t')
%= -\int\!dx\,dt\;\lambda(x,t)\,
%\nabla_i^{(x)}\!\left[
%\hat{\mathcal{A}}^{ik}(x,t,x'',t'')\,\rho(x,t)\right]
%\end{equation}
%The LHS is the variation of $\frac{1}{2}G\hat{\mathcal{A}}G$;
%the RHS comes from the $\lambda$-term, using
%$\delta\hat{G}^i(x,t)/\delta G_k(x'',t'')
%= \hat{\mathcal{A}}^{ik}(x,t,x'',t'')$.
%
%(iii) $\delta S_\lambda/\delta\rho(x'',t'') = 0$ gives the
%equation for $\lambda$:
%\begin{equation}
%\label{eq:EL_rho}
%-2\alpha^2\!\int\!dx'\,dt'\;
%\hat{\mathcal{K}}(x'',t'',x',t')\,\rho(x',t')
%+ \partial_t\lambda(x'',t'')
%+ \hat{G}^i\nabla_i\lambda(x'',t'')
%+ 2\alpha^2\!\int\!dx\,dt\;\lambda(x,t)\,
%\nabla_i^{(x)}\!\left[\hat{\mathcal{R}}^i(x,t,x'',t'')
%\,\rho(x,t)\right] = 0
%\end{equation}

\paragraph{The stochastic PDE for the density.}
The constraint equation~\eqref{eq:EL_lambda} is the PDE for
$\rho$ given a realization of $G$. Substituting the
drift~\eqref{eq:Ghat_repeated} and separating the
$G$-dependent and $G$-independent parts:
\begin{equation}
\label{eq:SPDE_appendix}
\partial_t\rho
+ \nabla_i\!\left[\left(
\int\!dx'\,dt'\;\hat{\mathcal{A}}^{ij}(x,t,x',t')\,
G_j(x',t')\right)\rho\right]
+ \nabla_i\!\left[F^i[\rho]\,\rho\right] = 0
\end{equation}
where the self-consistent disorder drift is
\begin{equation}
\label{eq:F_appendix}
F^i[\rho](x,t)
= 2\alpha^2\int\!dx'\,dt'\;
\hat{\mathcal{R}}^i(x,t,x',t')\,\rho(x',t')
\end{equation}
(note the integration over both $x'$ and $t'$: the drift is
non-local in both space and time through the response kernel
$\hat{\mathcal{R}}^i$).

\paragraph{Decomposition of the noise.}
Using the decomposition~\eqref{tilde_mathcal_C} of
$\hat{\mathcal{A}}^{ij}$ into local and non-local parts:
\begin{equation}
\label{eq:SPDE_decomposed_app}
\partial_t\rho
+ \nabla_i\!\left[\Omega\,g^{ij}G_j(x,t)\,\rho\right]
+ \nabla_i\!\left[\left(
2\alpha^2\!\int\!dz'\;\hat{\mathcal{C}}^{ij}(z,z')\,
G_j(z')\right)\rho\right]
+ \nabla_i\!\left[F^i[\rho]\,\rho\right] = 0
\end{equation}
The first noise term ($\Omega\,g^{ij}G_j\,\rho$) is local
in spacetime and reproduces the Dean-Kawasaki noise structure.
The second involves the non-local kernel
$\hat{\mathcal{C}}^{ij}$ and couples the density to the noise
at other spacetime points, generating temporal memory.

\paragraph{Noise statistics.}
The Gaussian weight $\exp(-\frac{1}{2}\int G\hat{\mathcal{A}}G)$
in the generating functional determines the correlator of $G$:
\begin{equation}
\label{eq:G_correlator_app}
\langle G_i(z)\,G_j(z')\rangle_G
= \mathcal{W}_{ij}(z,z')
\end{equation}
where $\mathcal{W}_{ij}$ is the inverse of
$\hat{\mathcal{A}}^{ij}$~\eqref{inverse_kernel_W}:
$\int dz''\;\hat{\mathcal{A}}^{ik}(z,z'')\,
\mathcal{W}_{kj}(z'',z') = \delta_j^i\,\delta(z-z')$.
%For the local part of $\hat{\mathcal{A}}^{ij}$
%($\Omega\,g^{ij}\delta(z-z')$), the corresponding piece of
%$\mathcal{W}$ is:
%\begin{equation}
%\label{eq:W_local_app}
%\mathcal{W}_{ij}^{(\text{local})}(z,z')
%= \frac{1}{\Omega}\,g_{ij}(x)\,\delta(z-z')
%\end{equation}
%So the local noise $G_j(x,t)$ is white in both space and time,
%with variance $1/\Omega$.

\paragraph{The effective DK equation (Stratonovich form).}
Equation~\eqref{eq:SPDE_decomposed_app} is a stochastic PDE in
Stratonovich form (inherited from the Stratonovich convention of
the underlying Langevin
dynamics~\eqref{single_particle_SPDE}). It is the
disorder-averaged Dean-Kawasaki equation for the F2 model:
\begin{equation}
\label{eq:DK_stratonovich}
\partial_t\rho
= - \nabla_i\!\left[\left(\Omega\,g^{ij}G_j
+ 2\alpha^2\!\int\!dz'\;
\hat{\mathcal{C}}^{ij}(z,z')G_j(z')
+ F^i[\rho]\right)\rho\right]
\end{equation}
with $G_j$ a Gaussian noise with correlator
$\mathcal{W}_{ij}$~\eqref{eq:G_correlator_app} and $F^i[\rho]$
the self-consistent drift~\eqref{eq:F_appendix}. This equation
has the divergence form $\partial_t\rho
+ \nabla_i J^i = 0$ and thus conserves the total density
$\int\rho\,d\mu_g$.

\paragraph{It\^o form (for numerical simulation).}
For numerical integration via the Euler-Maruyama scheme, it is
convenient to convert the Stratonovich
equation~\eqref{eq:DK_stratonovich} to It\^o form. The local
noise term $\nabla_i[\Omega g^{ij}G_j\,\rho]$ is multiplicative
(the noise $G_j$ multiplies $\rho$). The standard
Stratonovich-to-It\^o correction on a Riemannian manifold
\cite{zinnjustin2002, castrovillarreal2023} produces an
additional diffusion term $(\Omega/2)\Delta_g\rho$. The
It\^o form is:
\begin{equation}
\label{eq:effective_rho_appendix}
\partial_t\rho
= \frac{\Omega}{2}\,\Delta_g\rho
- \nabla_i\!\left[F^i[\rho]\,\rho\right]
- \nabla_i\!\left[\Omega\,g^{ij}G_j\,\rho\right]
- \nabla_i\!\left[\left(
2\alpha^2\!\int\!dz'\;\hat{\mathcal{C}}^{ij}G_j(z')
\right)\rho\right]
\end{equation}
%The first term is the It\^o diffusion correction (from the
%local noise). The second is the deterministic disorder drift.
%The third and fourth are the noise terms (local and non-local).
%The mean-field equation (obtained by taking
%$\langle\cdots\rangle$ and using $\langle G_j\rangle = 0$,
%valid at leading order in $1/N$) is:
%\begin{equation}
%\label{eq:FP_appendix}
%\partial_t\langle\rho\rangle
%= \frac{\Omega}{2}\,\Delta_g\langle\rho\rangle
%- \nabla_i\!\left[F^i[\langle\rho\rangle]\,
%\langle\rho\rangle\right]
%\end{equation}

\paragraph{Consistency checks.}
For $\alpha = 0$ (no disorder), $F^i = 0$ and
$\hat{\mathcal{A}}^{ij} = \Omega\,g^{ij}\delta$.
The Stratonovich equation~\eqref{eq:DK_stratonovich} reduces to
$\partial_t\rho + \nabla_i[\Omega\,g^{ij}G_j\,\rho] = 0$,
which is the DK equation for free Brownian motion on the
manifold. The It\^o form gives
$\partial_t\rho = (\Omega/2)\Delta_g\rho
- \nabla_i[\Omega g^{ij}G_j\rho]$, and the mean-field equation
is $\partial_t\langle\rho\rangle
= (\Omega/2)\Delta_g\langle\rho\rangle$ (the heat equation
on $(\M, g)$), as expected.

For finite $\alpha$, the Stratonovich
equation~\eqref{eq:DK_stratonovich} reproduces the structure of
the original single-particle DK
equation~\eqref{single_particle_SPDE}: the deterministic force
$f_n^i$ is replaced by the self-consistent drift $F^i[\rho]$,
and the noise $\sqrt{\Omega}\,e^i_a\eta^a$ is represented by
the Gaussian field $\int\hat{\mathcal{A}}^{ij}G_j$.

%==============================================================================
% Appendix D (Cayley-Klein parameterization) removed.
% The rotating-frame construction is referenced in the
% main text with citations to \cite{pennestri2016,cottingham2001}.
%==============================================================================

%==============================================================================
\section{Two-time correlator of Legendre polynomials on $S^2$}
\label{app:legendre_two_time}
%==============================================================================

This appendix derives Eq.~\eqref{eq:Legendre_two_time},
namely
\begin{equation}
\label{eq:Legendre_two_time_app}
\left\langle P_\ell(\hat{\mathbf{n}}(t)\cdot\mathbf{x})
\,P_{\ell'}(\hat{\mathbf{n}}(t')\cdot\mathbf{x}')
\right\rangle
= \frac{\delta_{\ell\ell'}}{2\ell+1}\,
[q(\tau)]^{\ell(\ell+1)/2}\,
P_\ell(\mathbf{x}\cdot\mathbf{x}'),
\end{equation}
with $\hat{\mathbf{n}}(t)$ performing isotropic diffusion
on $S^2$ at rate $D_{\text{rot}}$,
$\tau = t - t'$, and $q(\tau) = e^{-2 D_{\text{rot}}|\tau|}$
the $\ell = 1$ orientation correlator. The calculation is
carried out on the lifted $S^2$ (the double cover of the
physical orientation manifold $\mathbb{RP}^2 = S^2/\mathbb{Z}_2$,
Section~\ref{subsec:low_energy_ansatz}), where the
spherical-harmonic addition theorem and the full $Y_{\ell m}$
basis are most directly applicable. Projection back to the
physical $\mathbb{RP}^2$ at the level of observables retains
only the even-$\ell$ contributions
(Eq.~\eqref{eq:rp2_selection_rule}); the closed form for the
density correlator $C_\rho$ in
Eq.~\eqref{eq:Crho_reduced} embodies that projection. We use standard
orthonormal spherical harmonics $Y_{\ell m}$ on $S^2$ and
proceed in four steps.

\paragraph{1. Addition theorem.}
For any two unit vectors $\hat{\mathbf{a}},
\hat{\mathbf{b}} \in S^2$ the spherical-harmonic
addition theorem reads~\cite{arfken2013}
\begin{equation}
\label{eq:addition_thm}
P_\ell(\hat{\mathbf{a}}\cdot\hat{\mathbf{b}})
= \frac{4\pi}{2\ell+1}\sum_{m = -\ell}^{\ell}
Y_{\ell m}^*(\hat{\mathbf{a}})\,Y_{\ell m}(\hat{\mathbf{b}}).
\end{equation}

\paragraph{2. Expansion of each Legendre factor.}
Applying \eqref{eq:addition_thm} to both $P$-factors on
the left of \eqref{eq:Legendre_two_time_app} separates
the $\hat{\mathbf{n}}$- and $\mathbf{x}$-dependences:
\begin{equation}
\label{eq:Plx_expand}
P_\ell(\hat{\mathbf{n}}(t)\cdot\mathbf{x})
= \frac{4\pi}{2\ell+1}\sum_{m}
Y_{\ell m}(\hat{\mathbf{n}}(t))\,Y_{\ell m}^*(\mathbf{x}),
\end{equation}
and similarly for the primed factor. The average on
$\hat{\mathbf{n}}$ of the product reduces to the
two-time correlator of spherical harmonics,
$\langle Y_{\ell m}(\hat{\mathbf{n}}(t))\,
Y_{\ell' m'}^*(\hat{\mathbf{n}}(t'))\rangle$, with the
$\mathbf{x}$- and $\mathbf{x}'$-dependent factors
factored outside.

\paragraph{3. Stationary two-time $Y_{\ell m}$
correlator.}
Isotropic diffusion on $S^2$ has stationary measure
$d\mu_{\text{stat}}(\hat{\mathbf{n}})
= d\hat{\mathbf{n}}/(4\pi)$ and transition kernel
\begin{equation}
\label{eq:diff_semigroup}
p(\hat{\mathbf{n}}', \tau\,|\,\hat{\mathbf{n}})
= \sum_{L, M} Y_{LM}(\hat{\mathbf{n}}')\,
Y_{LM}^*(\hat{\mathbf{n}})\,
e^{-L(L+1)D_{\text{rot}}|\tau|}.
\end{equation}
Using \eqref{eq:diff_semigroup} and the orthonormality
of the $Y_{\ell m}$'s, the stationary two-time
correlator collapses to a single mode:
\begin{equation}
\label{eq:Y_two_time}
\left\langle Y_{\ell m}(\hat{\mathbf{n}}(t))\,
Y_{\ell' m'}^*(\hat{\mathbf{n}}(t'))\right\rangle
= \frac{1}{4\pi}\,
\delta_{\ell\ell'}\,\delta_{mm'}\,
e^{-\ell(\ell+1)D_{\text{rot}}|\tau|}.
\end{equation}

\paragraph{4. Contraction with the addition theorem.}
Substituting \eqref{eq:Y_two_time} back into the
$\hat{\mathbf{n}}$-averaged product of \eqref{eq:Plx_expand}
collapses the double sum $\sum_{m, m'}\to\sum_m$ and
leaves a single residual sum over $\mathbf{x},\mathbf{x}'$
harmonics, which \eqref{eq:addition_thm} re-sums back to a
Legendre polynomial:
\begin{equation}
\label{eq:YY_to_P}
\sum_m Y_{\ell m}^*(\mathbf{x})\,Y_{\ell m}(\mathbf{x}')
= \frac{2\ell+1}{4\pi}\,
P_\ell(\mathbf{x}\cdot\mathbf{x}').
\end{equation}
Combining the pieces gives
\begin{equation}
\label{eq:PP_exp}
\left\langle P_\ell(\hat{\mathbf{n}}(t)\cdot\mathbf{x})
\,P_{\ell'}(\hat{\mathbf{n}}(t')\cdot\mathbf{x}')
\right\rangle
= \frac{\delta_{\ell\ell'}}{2\ell+1}\,
e^{-\ell(\ell+1)D_{\text{rot}}|\tau|}\,
P_\ell(\mathbf{x}\cdot\mathbf{x}').
\end{equation}
Finally, rewriting the exponent as
$e^{-\ell(\ell+1)D_{\text{rot}}|\tau|}
= [e^{-2 D_{\text{rot}}|\tau|}]^{\ell(\ell+1)/2}
= [q(\tau)]^{\ell(\ell+1)/2}$ reproduces
\eqref{eq:Legendre_two_time_app} and expresses each
mode's decay as a power of the $\ell = 1$ orientation
correlator $q(\tau)$. $\blacksquare$

%==============================================================================
\section{Faddeev-Popov determinant and one-loop corrections}
\label{app:FP_expansion}
%==============================================================================

This appendix derives the explicit $\delta\rho$-expansion
of the Faddeev-Popov Jacobian determinant referenced in
Section~\ref{subsec:faddeev_popov}. The key results, all
SO(3)-invariant and stated in the main text, are: (i) at
the saddle $\delta\rho = 0$, the FP Jacobian is the
$\hat{\mathbf{n}}$-independent constant $G_0^2$ (Gram
matrix determinant of the zero modes); (ii) the
$\delta\rho$-dependent piece, expanded via $\ln\det
= \mathrm{Tr}\ln$, produces a linear source $J$ and a
quadratic kernel renormalization $K$; (iii) the kernel
renormalization $\tfrac{1}{2}\mathrm{Tr}(\mathbf{M}^{-1}K)$
is at the same one-loop order as the leading
$\tfrac{1}{2}\log\det'\mathbf{M}$, while the source-shift
$\tfrac{1}{2}\!\int\!J\,\mathbf{M}^{-1}J$ is two-loop;
(iv) all are functions of the SO(3)-invariants of
$(\hat{\mathbf{n}}, \dot{\hat{\mathbf{n}}})$ and so
renormalize the moment-of-inertia coefficient $\kappa$
of the effective $\mathrm{SO}(3)$ action without generating an
$\hat{\mathbf{n}}$-dependent potential.

We use the notation introduced in
Section~\ref{subsec:low_energy_ansatz}:
$\mathbf{x}\in\mathbb{R}^3$ is the lab-frame
embedding-space position vector with $|\mathbf{x}|^2 = 1$;
$u \equiv \hat{\mathbf{n}}\cdot\mathbf{x}$ is the polar
height; in $(u, \phi)$ coordinates adapted to
$\hat{\mathbf{n}}$, with $u = \cos\theta$ and $\phi$ the
azimuthal angle measured in the basis
$\{\hat{\mathbf{e}}_1, \hat{\mathbf{e}}_2\}$ of the
tangent plane $T_{\hat{\mathbf{n}}}S^2$ (orthonormal,
$\hat{\mathbf{e}}_a\cdot\hat{\mathbf{n}} = 0$,
$\hat{\mathbf{e}}_a\cdot\hat{\mathbf{e}}_b = \delta_{ab}$),
the round measure factorizes as $d\mu_g(\mathbf{x})
= du\,d\phi = \sin\theta\,d\theta\,d\phi$, and
$\mathbf{x} = u\,\hat{\mathbf{n}} + \sqrt{1-u^2}\,
(\cos\phi\,\hat{\mathbf{e}}_1 + \sin\phi\,
\hat{\mathbf{e}}_2)$.

\paragraph{FP Jacobian: full form.}
Differentiating the gauge condition
$F_a[\rho;\hat{\mathbf{n}}] = \int\psi_a[\hat{\mathbf{n}}]
(\rho - \rho_0[\hat{\mathbf{n}}])\,d\mu_g$
\eqref{eq:FP_gauge} at fixed $\rho$, using
$\partial\rho_0/\partial n^b = \psi_b$ (zero-mode
defining property),
\begin{equation}
\label{eq:FPapp_jacobian}
\frac{\partial F_a}{\partial n^b}\bigg|_\rho
= \int\!\frac{\partial\psi_a}{\partial n^b}\,
\delta\rho\,d\mu_g \;-\;\int\!\psi_a\,\psi_b\,d\mu_g.
\end{equation}
The second term is the (constant) Gram matrix; the first
is linear in $\delta\rho$ and produces the
$\hat{\mathbf{n}}$-fluctuation corrections analyzed
below.

\paragraph{Gram matrix at the saddle.}
With $\psi_a = f_0'(u)\,(\hat{\mathbf{e}}_a\cdot\mathbf{x})$
and the $(u,\phi)$ parametrization above,
\begin{equation}
\hat{\mathbf{e}}_1\cdot\mathbf{x} = \sqrt{1-u^2}\,
\cos\phi,
\qquad
\hat{\mathbf{e}}_2\cdot\mathbf{x} = \sqrt{1-u^2}\,
\sin\phi.
\end{equation}
The integrand $(\hat{\mathbf{e}}_a\cdot\mathbf{x})
(\hat{\mathbf{e}}_b\cdot\mathbf{x})$ is $(1-u^2)$ times
$\cos^2\phi$, $\sin^2\phi$, or $\cos\phi\sin\phi$
according to $(a,b)$; the $\phi$-integration over
$[0, 2\pi)$ gives $\pi$, $\pi$, or $0$, so
$\int_0^{2\pi}\!d\phi\,(\hat{\mathbf{e}}_a\cdot\mathbf{x})
(\hat{\mathbf{e}}_b\cdot\mathbf{x}) = \pi\,(1-u^2)\,
\delta_{ab}$ (the $\mathrm{SO}(2)_{\hat{\mathbf{n}}}$
orbit-average projects onto the
$\mathrm{SO}(2)_{\hat{\mathbf{n}}}$-singlet component).
Hence
\begin{equation}
\label{eq:FPapp_Gram}
G_{ab} = \int\psi_a\psi_b\,d\mu_g = G_0\,\delta_{ab},
\qquad
G_0 = \pi\!\int_{-1}^{1}\!f_0'(u)^2\,(1-u^2)\,du,
\end{equation}
manifestly $\hat{\mathbf{n}}$-independent. The constant
$\det G_{ab} = G_0^2$ absorbs into the overall
normalization of $Z$.

\paragraph{Tr-log expansion.}
Using $\ln\det M = \mathrm{Tr}\ln M$ and~\eqref{eq:FPapp_Gram},
the FP Jacobian determinant in~\eqref{eq:FPapp_jacobian}
expands as
\begin{equation}
\label{eq:FPapp_logdet}
\ln\det\!\left(\frac{\partial F_a}{\partial n^b}\right)
= \ln G_0^2 + \mathrm{Tr}\ln(I - G_0^{-1}L)
= \ln G_0^2 - G_0^{-1}\mathrm{Tr}\,L
- \tfrac{1}{2}G_0^{-2}\mathrm{Tr}(L^2) + O(L^3),
\end{equation}
with $L_{ab}[\delta\rho]
= \int(\partial\psi_a/\partial n^b)\,\delta\rho\,d\mu_g$.

\paragraph{Computation of $\partial\psi_a/\partial n^b$.}
From $\psi_a = f_0'(u)\,(\hat{\mathbf{e}}_a\cdot\mathbf{x})$
with $u = \hat{\mathbf{n}}\cdot\mathbf{x}$, using
$\partial\hat{\mathbf{n}}/\partial n^b = \hat{\mathbf{e}}_b$
and
$\partial\hat{\mathbf{e}}_a/\partial n^b
= -\delta_{ab}\hat{\mathbf{n}}$ (in normal coordinates
with parallel-transported frame at $\hat{\mathbf{n}}$;
the $\hat{\mathbf{n}}$-component of
$\partial\hat{\mathbf{e}}_a/\partial n^b$ is fixed by
differentiating $\hat{\mathbf{e}}_a\cdot\hat{\mathbf{n}}
= 0$, the tangential component is the spin-connection
of the frame and vanishes at $\hat{\mathbf{n}}$ in this
choice). The orientation manifold (the lifted $S^2$
that double-covers $\mathbb{RP}^2$) admits no global smooth
frame, so $\hat{\mathbf{e}}_a$ is defined only locally; final observables (the source $J$, the kernel
$K$, the Gram determinant $G_0$) are scalar contractions
in $a, b$, and a different local frame at the same point
is related by a frame rotation that leaves these
contractions invariant. We therefore work in this local
normal-coordinate frame at the chosen $\hat{\mathbf{n}}$
without loss of generality, with frame independence of
the end results understood.
\begin{equation}
\label{eq:FPapp_dpsi_dn}
\frac{\partial\psi_a}{\partial n^b}
= f_0''(u)\,(\hat{\mathbf{e}}_a\cdot\mathbf{x})
(\hat{\mathbf{e}}_b\cdot\mathbf{x})
- u\,f_0'(u)\,\delta_{ab}.
\end{equation}

\paragraph{Linear source: spherical Laplacian.}
Tracing~\eqref{eq:FPapp_dpsi_dn} over $a = b$ and using
$\sum_a(\hat{\mathbf{e}}_a\cdot\mathbf{x})^2 = 1 - u^2$,
\begin{equation}
\label{eq:FPapp_trace}
\sum_a\frac{\partial\psi_a}{\partial n^a}
= (1-u^2)\,f_0''(u) - 2u\,f_0'(u)
= \frac{d}{du}\!\bigl[(1-u^2)\,f_0'(u)\bigr]
= \nabla^2_{S^2}\,f_0(u),
\end{equation}
the Legendre operator (the radial part of the spherical
Laplacian on axisymmetric functions). The linear piece
of~\eqref{eq:FPapp_logdet} is therefore a source for
$\delta\rho$,
\begin{equation}
\label{eq:FPapp_source}
-G_0^{-1}\mathrm{Tr}\,L
= -\!\int\! J(\mathbf{x})\,\delta\rho(\mathbf{x},t)\,
d\mu_g(\mathbf{x})\,dt,
\qquad
J(\mathbf{x}) = G_0^{-1}\,\nabla^2_{S^2}f_0
(\hat{\mathbf{n}}\cdot\mathbf{x}),
\end{equation}
manifestly an $\mathrm{SO}(2)_{\hat{\mathbf{n}}}$-singlet
(a function of $u$ alone).

\paragraph{Quadratic kernel.}
The quadratic piece of~\eqref{eq:FPapp_logdet} is
$-\tfrac{1}{2}G_0^{-2}\mathrm{Tr}(L^2) = -\tfrac{1}{2}\!
\iint\!\delta\rho\,K\,\delta\rho$ with
$K(\mathbf{x};\mathbf{y}) = G_0^{-2}
\sum_{ab}(\partial\psi_a/\partial n^b)(\mathbf{x})
(\partial\psi_b/\partial n^a)(\mathbf{y})$.
Substituting~\eqref{eq:FPapp_dpsi_dn} and using the
tangent-plane completeness identity $\sum_a
(\hat{\mathbf{e}}_a\cdot\mathbf{x})(\hat{\mathbf{e}}_a\cdot
\mathbf{y}) = \mathbf{x}\cdot\mathbf{y} - u_x u_y$
(which follows from $I = \hat{\mathbf{n}}
\hat{\mathbf{n}}^T + \sum_a\hat{\mathbf{e}}_a
\hat{\mathbf{e}}_a^T$ sandwiched between $\mathbf{x}$
and $\mathbf{y}$),
\begin{equation}
\begin{aligned}
K(\mathbf{x};\mathbf{y}) = G_0^{-2}\,\Bigl[
&\, f_0''(u_x)f_0''(u_y)\,(\mathbf{x}\cdot\mathbf{y}
 - u_x u_y)^2
- f_0''(u_x)f_0'(u_y)\,u_y(1-u_x^2) \\
&\, - f_0'(u_x)f_0''(u_y)\,u_x(1-u_y^2)
+ 2u_x u_y\,f_0'(u_x)f_0'(u_y)
\Bigr],
\end{aligned}
\end{equation}
symmetric in $\mathbf{x}\leftrightarrow\mathbf{y}$ and
depending only on the SO(3)-invariants $u_x, u_y,
\mathbf{x}\cdot\mathbf{y}$.

\paragraph{Gaussian completion and effective FP
contribution.}
Combining the linear source and the quadratic kernel
with the leading Gaussian quadratic form, $\tilde{\mathbf{M}}
\equiv \mathbf{M} + K$, the constrained Gaussian integral
gives
\begin{equation}
\label{eq:FPapp_Gauss}
\int\!\mathcal{D}'\delta\rho\,
\exp\!\Bigl(-\tfrac{1}{2}\delta\rho\,\tilde{\mathbf{M}}\,
\delta\rho - \!\int\! J\,\delta\rho\Bigr)
= (\det'\tilde{\mathbf{M}})^{-1/2}\,
\exp\!\Bigl(\tfrac{1}{2}\!\int\! J\,
\tilde{\mathbf{M}}^{-1}\,J\Bigr).
\end{equation}

\paragraph{Loop order of FP corrections.}
Expanding $\tfrac{1}{2}\log\det'\tilde{\mathbf{M}}
= \tfrac{1}{2}\log\det'\mathbf{M}
+ \tfrac{1}{2}\mathrm{Tr}(\mathbf{M}^{-1}K) + O(K^2)$,
the kernel renormalization
$\tfrac{1}{2}\mathrm{Tr}(\mathbf{M}^{-1}K)$ is a one-loop
tadpole ($V=1, I=1, L=1$) at the \emph{same loop order}
as the leading $\tfrac{1}{2}\log\det'\mathbf{M}$. The
source-shift $\tfrac{1}{2}\!\int J\,\tilde{\mathbf{M}}^{-1}J$
is tree-level in $\delta\rho$ topology, but its $J$
vertices come from $\ln\det$ rather than from $S/\hbar$,
so in the standard $\hbar$-counting it carries an extra
power of $\hbar$ relative to one-loop and sits at
two-loop order. Both carry the explicit $G_0^{-2}$
prefactor.

By the SO(3)-equivariance argument of
Section~\ref{subsec:faddeev_popov}, both contributions
are SO(3)-invariant functions of $(\hat{\mathbf{n}},
\dot{\hat{\mathbf{n}}})$. They renormalize SO(3)-invariant
kinetic-term coefficients of the effective $\mathrm{SO}(3)$ NLSM
(in particular, the moment-of-inertia $\kappa$); they cannot
generate an $\hat{\mathbf{n}}$-dependent potential. Since
$\kappa$ is treated phenomenologically in our paper (fit
to simulation, Section~\ref{subsec:comparison}), the
$G_0^{-2}$-suppressed FP renormalizations are
automatically absorbed into the physical value of
$\kappa$ rather than computed diagrammatically.

\end{document}